%% file: ms.tex
\documentclass[12pt]{ociamthesis}

\usepackage[toc, page]{appendix}
\usepackage[dvipsnames]{xcolor}
\usepackage[utf8x]{inputenc}
\usepackage{iftex}
\usepackage{graphicx}
\usepackage{collect}
\usepackage{etex}
\usepackage{textgreek}
\usepackage{tipa}
\usepackage{float}
\usepackage{environ}
\usepackage{courier}
\usepackage{bbm}
\usepackage{stackengine}
\usepackage{etoolbox}
\usepackage{setspace}
\usepackage[most]{tcolorbox}
\usepackage{amsmath}
\usepackage{amsthm}
\usepackage{amssymb}
\usepackage{amsfonts}
\usepackage{nccmath}
\usepackage{pst-node}
\usepackage{pst-plot}
\usepackage{stackengine}
\usepackage{mathtools}
\usepackage{stmaryrd}
\usepackage{bussproofs}
\usepackage{caption}
\usepackage{chapterfolder}
\usepackage{makecell}
\usepackage{soul}
\makeatletter
\let\th@plain\relax
\usepackage{tikz}
\usetikzlibrary{arrows}
\usetikzlibrary{positioning}
\usetikzlibrary{calc,intersections}
\usepackage{tikz-cd}
\usepackage{cmll}

\ifXeTeX
\usepackage{fontspec}
\fi

\usepackage{hyperref}
\usepackage{syntax}
\usepackage[shortlabels]{enumitem}

\newcommand{\mathcolorbox}[2]{\colorbox{#1}{$\displaystyle #2$}}

\input{\treeroot latex/notation}
\input{\treeroot coq}

\newtheorem{theorem}{Theorem}[section]
\newtheorem{corollary}{Corollary}[section]
\newtheorem{lemma}{Lemma}[section]

\theoremstyle{plain}
\newtheorem{proposition}{Proposition}[section]
\newtheorem{definition}{Definition}[section]
\newtheorem{remark}{Remark}[section]
\newtheorem{example}{Example}[section]
\newtheorem*{notation}{Notation}

\DeclareMathOperator*{\cansum}{\boxplus}
\DeclareMathOperator*{\boxsum}{\boxplus}

\newif\ifreview\reviewfalse
\newcommand{\rff}[1]{\ifreview \red{[REQUEST FOR FEEDBACK] {#1}}\fi}

\newcommand{\DF}[1]{\mathsf{D}\left[ {#1} \right]}

\newcommand{\CDCAx}[1]{{\bf [CDC.{#1}]}}
\newcommand{\CdCAx}[1]{{\bf [C$\d$C.{#1}]}}
\newcommand{\DlCAx}[0]{{\bf [D$\lambda$C.1]}}

\newlength{\alignEqShift}
\renewcommand{\eq}[1][=]{&#1}
\newcommand{\continued}[1][30pt]{&\hspace{#1}}

\NewEnviron{salign*}{
  \par\vspace{-\parskip}
  \small
  \begin{align*}
      \BODY
  \end{align*}
}

\newcommand{\seq}[1]{\left[ {#1} \right]}

\newcommand\Item[1][]{%
  \ifx\relax#1\relax  \item \else \item[#1] \fi
  \abovedisplayskip=0pt\abovedisplayshortskip=0pt~\vspace*{-\baselineskip}
}

\ifXeTeX
\renewcommand{\T}[0]{\ftor{T}}
\else
\newcommand{\T}[0]{\ftor{T}}
\fi

\newcommand{\sem}[1]{\left \llbracket {#1} \right \rrbracket}

\definecolor{indigo}{HTML}{332288} 
\definecolor{cyan}{HTML}{88ccee}
\definecolor{teal}{HTML}{44aa99}   
\definecolor{green}{HTML}{117733}  
\definecolor{olive}{HTML}{999933} 
\definecolor{sand}{HTML}{ddcc77}   
\definecolor{rose}{HTML}{cc6677}   
\definecolor{wine}{HTML}{882255}   
\definecolor{purple}{HTML}{aa4499} 

\apptocmd{\thebibliography}{\csname phantomsection\endcsname\addcontentsline{toc}{chapter}{\bibname}}{}{}

\makeatletter
\patchcmd{\chapter}{\if@openright\cleardoublepage\else\clearpage\fi}{}{}{}
\makeatother

\title{Change actions
  \\\vspace{18pt}
  {\Large
  From incremental computation to discrete derivatives
  }}
\author{Mario Alvarez-Picallo}
\college{Wolfson College}

\degree{Doctor of Philosophy}
\degreedate{Michaelmas Term 2019/2020}

\begin{document}

\baselineskip=18pt plus1pt

\setcounter{secnumdepth}{3}
\setcounter{tocdepth}{3}

\pagenumbering{gobble}
\maketitle

\include{acknowledgements}
\include{abstract}
\newpage

\pagenumbering{arabic}

\begin{romanpages}          
\tableofcontents            
\listoffigures              
\end{romanpages}            

\def\chapterroot{./}

\newcommand{\includechapter}[2]{\clearpage\chapter{#1}\input{.//#2}}

\clearpage\chapter{Introduction}\input{.//introduction}

\clearpage\chapter{Preliminaries}\input{.//preliminaries}

\clearpage\chapter{Change actions}\input{.//change-actions}

\clearpage\chapter{Incremental computation with change actions}\input{.//incremental-computation}

\clearpage\chapter{Change action models}\input{.//change-action-models}

\clearpage\chapter{From differentials to differences}\input{.//difference-categories}

\clearpage\chapter{A calculus of finite differences}\input{.//simply-typed-calculus}

\clearpage\chapter{Conclusions}\input{.//conclusions}

\newpage

\bibliographystyle{splncs04}
\bibliography{biblio}

\appendix

\clearpage\chapter{Mechanised Proofs}\input{.//mechanized-proofs}

\end{document}

%% file: latex/notation.tex
\renewcommand{\epsilon}[0]{\varepsilon}
\newcommand{\bdot}[0]{\boldsymbol{\cdot}}

\newcommand{\LHS}[0]{\text{LHS}}
\newcommand{\RHS}[0]{\text{RHS}}
\newcommand{\PROOFSTEP}[0]{\displaybreak[0]\\}

\newcommand{\NN}[0]{\mathbb{N}}
\newcommand{\RR}[0]{\mathbb{R}}
\newcommand{\ZZ}[0]{\mathbb{Z}}
\newcommand{\BB}[0]{\mathbb{B}}
\newcommand{\KK}[0]{\mathbb{K}}

\renewcommand{\vec}[1]{\mathbf{#1}} 
\newcommand{\D}[0]{\text{D}}

\newcommand{\brak}[1]{\left[ #1 \right]}

\newcommand{\pa}[1]{\left( #1 \right)}

\newcommand{\ra}[0]{\rightarrow}

\newcommand{\nlra}[0]{\nleftrightarrow}

\newcommand{\da}[0]{\downarrow}

\newcommand{\Ra}[0]{\Rightarrow}

\newcommand{\defeq}[0]{\coloneqq}

\newif\ifdraft
\draftfalse

\renewcommand{\red}[1]{{\color{red} {#1}}}

\newcommand{\todo}[1]{
  \red{[TODO: #1]}
}

\newcommand{\cat}[1]{\mathbf{#1}}
\newcommand{\cart}[0]{\cat{Cat}_\times}
\newcommand{\ftor}[1]{\mathrm{#1}}

\newcommand{\homset}[3]{{#1}\left[{#2}, {#3}\right]}

\newcommand{\arr}[1]{\mathbf{#1}}
\newcommand{\A}[1]{\arr{#1}}
\newcommand{\Id}[1]{\arr{id}_{#1}}
\newcommand{\Ev}[0]{\arr{ev}}

\newcommand{\tensor}[0]{\otimes}

\newcommand{\terminal}[0]{\mathbf{1}}
\newcommand{\pair}[2]{\left\langle {#1}, {#2} \right\rangle}
\newcommand{\four}[4]{\pair{\pair{#1}{#2}}{\pair{#3}{#4}}}

\newcommand{\initial}[0]{\mathbf{0}}

\newcommand{\inj}[0]{\A{i}}

\makeatletter
\newsavebox{\@brx}
\newcommand{\llangle}[1][]{\savebox{\@brx}{\(\m@th{#1\langle}\)}%
  \mathopen{\copy\@brx\kern-0.5\wd\@brx\usebox{\@brx}}}
\newcommand{\rrangle}[1][]{\savebox{\@brx}{\(\m@th{#1\rangle}\)}%
  \mathclose{\copy\@brx\kern-0.5\wd\@brx\usebox{\@brx}}}
\makeatother

\newcommand{\changes}[0]{\Delta}
\newcommand{\change}[0]{\delta}
\renewcommand{\d}[0]{\partial}
\newcommand{\codiff}[0]{\partial^\dagger}
\newcommand{\der}[2]{\frac{\d\,{#1}}{\d{#2}}}

\newcommand{\cact}[1]{\overline{#1}}
\renewcommand{\ul}[1]{\underline{#1}}
\newcommand{\us}[1]{#1}

\newcommand{\alter}[0]{\ |\ }
\newcommand{\reducesTo}[0]{\leadsto}
\newcommand{\parTo}[0]{
  \mathrel{\mathrlap{\raisebox{2pt}{$\reducesTo$}}
    \raisebox{-2pt}{$\reducesTo$}}}
\newcommand{\subst}[3]{{#1}\left[{#3}/{#2}\right]}
\newcommand{\diffS}[3]{\frac{\partial {#1}}{\partial {#2}}\pa{{#3}}}
\DeclarePairedDelimiter{\ceil}{\lceil}{\rceil}

\newcommand{\ts}[0]{\vdash}
\newenvironment{bprooftree}
  {\leavevmode\hbox\bgroup}
  {\DisplayProof\egroup}
  
\newcommand{\bpt}[1]{
  \begin{bprooftree}
    #1
  \end{bprooftree}
}

\newcommand{\set}[1]{\left\lbrace #1 \right\rbrace}
\newcommand{\for}[0]{\mid}

%% file: coq.tex
\ifXeTeX
\setmonofont{DejaVu Sans Mono}

\else

\fi

\lstdefinelanguage{Coq}{ 
%
mathescape=true,
%
texcl=false, 
%
morekeywords=[1]{Section, Module, End, Require, Import, Export,
  Variable, Variables, Parameter, Parameters, Axiom, Hypothesis,
  Hypotheses, Notation, Local, Tactic, Reserved, Scope, Open, Close,
  Bind, Delimit, Definition, Let, Ltac, Fixpoint, CoFixpoint,
  Morphism, Relation, Implicit, Arguments, Unset, Contextual,
  Strict, Prenex, Implicits, Inductive, CoInductive, Record,
  Structure, Canonical, Coercion, Context, Class, Global, Instance,
  Program, Infix, Theorem, Lemma, Corollary, Proposition, Fact,
  Remark, Example, Proof, Goal, Save, Qed, Defined, Hint, Resolve,
  Optimize,
  Rewrite, View, Search, Show, Print, Printing, All, Eval, Check,
  Projections, inside, outside, Def},
%
morekeywords=[2]{forall, exists, exists2, fun, fix, cofix, struct, match,
  lazymatch, context, with, end, as, in, return, let, if, is, then, else, for, of,
  nosimpl, when},
%
morekeywords=[3]{Type, Prop, Set, true, false, option, nat, list},
%
morekeywords=[4]{pose, set, move, case, elim, apply, clear, hnf,
  intro, intros, generalize, rename, pattern, after, destruct,
  induction, using, refine, inversion, injection, rewrite, congr,
  unlock, compute, ring, field, fourier, replace, fold, unfold,
  change, cutrewrite, simpl, have, suff, wlog, suffices, without,
  loss, nat_norm, assert, cut, trivial, revert, bool_congr, nat_congr,
  symmetry, transitivity, auto, eauto, eapply, split, left, right, autorewrite,
  by, done, exact, reflexivity, tauto, romega, omega,
  assumption, solve, contradiction, discriminate},
morekeywords=[5]{},
morekeywords=[6]{Var, App, Zero, Add, E, Dif, D},
%
morecomment=[s]{(*}{*)},
%
showstringspaces=false,
%
morestring=[b]",
morestring=[d],
%
tabsize=3,
%
extendedchars=false,
%
sensitive=true,
%
breaklines=false,
%
basicstyle=\footnotesize\ttfamily,
%
captionpos=b,
%
keepspaces=true,
%
identifierstyle={\ttfamily\color{black}},
keywordstyle=[1]{\ttfamily\color{Plum}},
keywordstyle=[2]{\ttfamily\color{OliveGreen}},
keywordstyle=[3]{\ttfamily\color{ProcessBlue}},
keywordstyle=[4]{\ttfamily\color{RoyalBlue}},
keywordstyle=[5]{\ttfamily\color{Maroon}},
keywordstyle=[6]{\ttfamily\color{Bittersweet}},
stringstyle=\footnotesize\ttfamily,
commentstyle={\ttfamily\color{OliveGreen}},
%
literate=
    {\\forall}{{\color{dkgreen}{$\forall\;$}}}1
    {\\exists}{{$\exists\;$}}1
    {<-}{{$\leftarrow\;$}}1
    {=>}{{$\Rightarrow$}}1
    {==}{{\code{==}\;}}1
    {==>}{{\code{==>}\;}}1
    {->}{{$\rightarrow$}}1
    {<->}{{$\leftrightarrow\;$}}1
    {<==}{{$\leq\;$}}1
    {\#}{{$^\star$}}1 
    {\\o}{{$\circ\;$}}1 
    {\@}{{{\color{Bittersweet}$\cdot$}}}1 
    {\/\\}{{$\wedge\;$}}1
    {\\\/}{{$\vee\;$}}1
    {++}{{\code{++}}}1
    {TILDE}{{\ }}1
    {\@\@}{{$@$}}1
    {\\mapsto}{{$\mapsto\;$}}1
    {\\hline}{{\rule{\linewidth}{0.5pt}}}1
    {LAM}{{{\color{Bittersweet}\textLambda}}}1
    {EPSILON}{{{\color{Bittersweet}\textepsilon}}}1
    {CDOT}{{{\color{Bittersweet}$\cdot$}}}1
    {+}{{{\color{Bittersweet}+}}}1
    {0}{{{\color{Bittersweet}0}}}1
    {TILDE}{{$\sim$}}1
    {\{}{{{\color{Bittersweet}\{}}}1
    {\}}{{{\color{Bittersweet}\}}}}1
    {|}{{{\color{Bittersweet}|}}}1
}[keywords,comments,strings]

\lstnewenvironment{coq}{\lstset{language=Coq}}{}

\def\coqe{\lstinline[language=Coq, basicstyle=\footnotesize\ttfamily]}

%% file: acknowledgements.tex
\section*{Acknowledgements}

This thesis represents the fruition of many long years of work and, as such, I
feel a profound personal attachment to it. But it is also true that, like most
scientific endeavours, it would have been a barren effort were it not for the
support and collaboration of many people along the way. 

First and foremost, I would like to thank my doctoral supervisors Samson
Abramsky and, especially, Luke Ong, without whose support and guidance the very
idea for this thesis would have never sparked, let alone flourished. I would
also like to acknowledge the valuable suggestions and remarks of my examiners
throughout my DPhil: Hongseok Yang and Sam Staton for overseeing my transfer
of status, Sam Staton and Andrzej Murawski for overseeing my confirmation, and
finally Neel Krishnaswami and once again Sam Staton, who kindly agreed to review 
this thesis in its entirety.

I am also grateful to the UK Engineering and Physical Sciences Research Council,
as well as the Department of Computer Science, for providing the necessary
funding for this research, as well as Semmle Ltd. and IOHK, two companies in
which I had the privilege of working as an intern. My experiences through these
industrial escapades greatly expanded my perspectives on research. 

During my work as a DPhil student, I have been lucky to collaborate with some
excellent researchers. I must give particular credit to Michael Peyton Jones,
whom I had the pleasure to meet while at Semmle, and with whom I would
later work again during my internship at IOHK. I would also like to thank
Jean-Simon Lemay for some very productive discussions culminating in a joint
publication that has informed my work ever since.

Although it is impossible to name them all, I am indebted to many researchers
around me. My good friend Alex Kavvos cannot go without mention as, despite my
best efforts, his patience with the most outlandish of my ideas has been
inexhaustible so far. I owe a good deal to my conversations with Marie Kerjean
and her contagious enthusiasm. I should also like to mention in passing all the
merry fellows who toiled with me in Office 347, and offer my encouragement
to those who remained there after me.

Finally, on a more personal note, I wish to sincerely thank Iris Tom\'e, my
closest friend through these years and, of course, to my mother, Elena Picallo,
who not once failed to support me.

%% file: abstract.tex
\section*{Abstract}

It seems to be a piece of folkloric knowledge among the incremental computation
community that incremental programs behave, in some way, like derivatives.
Indeed, they track the effect of a function on finite differences in the
input, much like derivatives in calculus track the effect of a function on
infinitesimal differences. This idea has recently come to the forefront
when Kelly, Pearlmutter and Siskind proposed reinterpreting Cai's incremental
lambda-calculus as a basis to understand automatic differentiation.

On the other hand, the differential lambda-calculus, an extension of the
lambda-calculus equipped with a differential operator that can differentiate
arbitrary higher-order terms, has been shown to constitute a model for
differentiation in the traditional sense -- that is, there is a model of the
differential lambda-calculus where function spaces correspond to spaces of
smooth functions, and the term-level derivative operator corresponds to the
usual notion of derivative of a multivariate function.

The goal of this thesis is threefold: first, to provide a general semantic
setting for reasoning about incremental computation. Second, to establish and
clarify the connection between derivatives in the incremental sense and
derivatives in the analytic sense, that is to say, to provide a common
definition of derivative of which the previous two are particular instances.
Third, to give a theoretically sound calculus for this general setting.

To this end we define and explore the notions of change actions and
differential maps between change actions and show how these notions
relate to incremental computation through the concrete example of the semi-naive
evaluation of Datalog queries. We also introduce the notion of a change
action model as a setting for higher-order differentiation, and exhibit some
interesting examples. Finally, we show how Cartesian difference
categories, a family of particularly well-behaved change action models,
generalise Cartesian differential categories and give rise to a calculus in the
spirit of Ehrhard and Regnier's differential lambda-calculus.

%% file: introduction.tex

Consider a piece of input data on which we wish to perform some expensive
computation. Accomplishing this is simple enough -- it suffices to insert a call
to the relevant function in our code \footnote{Or submit a job to a computing
centre, or spin up an AWS instance, depending on your choice of scale.}, pass it
our input data and wait until the calculation ends. But what if our input data
isn't quite as static as we would like it to be? For example, if we are
calculating the clustering coefficient of the graph of friendships on a social
network we might expect this graph to change over time -- should we, then,
restart the whole computation from scratch every time two teenagers bicker? Or
should we search for a cleverer algorithm \cite{liao2017incremental}?

Even if the input to our program does not change under our noses of its own
volition, many algorithms work by iteratively updating the same piece of data
with new information (usually until a fixed point is reached). The famous
PageRank algorithm is such an example: even if one assumes that the graph of
connections between pages remains static, a naive implementation would require
updating the weights of every edge in each iteration. A more cunning algorithm
\cite{mihaylov2012rex} would only recompute the weight of edges that depend on
other edges whose weight changed in the last iteration, thus avoiding redundant
effort.

Both of these examples are but particular cases of the broader topic of
\emph{incremental computation}. At its most general, incremental computation is
a family of techniques for updating the output of a program as its input changes
(either over time or from an iteration to the next). By its nature, approaches
to incrementalising programs tend to be ad-hoc and domain-specific, but general
patterns arise: one such common theme is that the incrementalised version of a
function is usually referred to as a ``derivative'' obtained by a process of
``differentiation''
\cite{paige1981formal,bancilhon1986naive,bancilhon1986amateur,cai2014changes}.

It seems to be a piece of folkloric knowledge among the incremental computation
community that incremental programs behave, in some way, like derivatives.
Indeed, they track the effect of a function on \emph{finite differences} in the
input, much like derivatives in calculus track the effect of a function on
\emph{infinitesimal differences}. This idea has recently come to the forefront
when Kelly, Pearlmutter and Siskind proposed reinterpreting Cai's incremental
$\lambda$-calculus as a basis to understand automatic differentiation
\cite{kelly2016evolving}.

On the other hand, the differential $\lambda$-calculus, an extension of the
$\lambda$-calculus equipped with a differential operator that can differentiate
arbitrary higher-order terms, has been shown to constitute a model for
differentiation in the traditional sense -- that is, there is a model of the
differential $\lambda$-calculus where function spaces correspond to spaces of
smooth functions, and the term-level derivative operator corresponds to the
usual notion of derivative of a multivariate function
\cite{blute2010convenient,kerjean2016mackey}.

In full generality, the notion of derivative induced by the differential
$\lambda$-calculus is captured by Cartesian differential categories, an abstract
categorical model for differentiation in calculus and differential geometry
\cite{blute2006differential,blute2009cartesian,cruttwell2017cartesian}.

The goal of this thesis is threefold: first, to provide a more general version
of Cai and Giarrusso's change structures, and investigate its properties.
Second, to establish and clarify the connection between derivatives in the
incremental sense and derivatives in the analytic sense, that is to say, to
provide a common definition of derivative of which the previous two are
particular instances. Third, to give a theoretically sound calculus for this
general setting.

To this effect, this thesis is structured as follows: in
Chapter~\ref{chp:preliminaries} we give a more formal account of incremental
computation and change structures, the differential $\lambda$-calculus and its
history and models, together with an in-depth description of Cartesian
differential categories. In Chapter~\ref{chp:change-actions} we set the basic
definitions of a \emph{change action} and a \emph{differential map} between
change actions -- these being the main conceptual contributions of this work. In
Chapter~\ref{chp:incremental-computation} we develop the theory of change
actions for the practical application of incrementalising the evaluation of
Datalog queries. In Chapter~\ref{chp:change-action-models} we introduce the
notion of a \emph{change action model} as a setting for higher-order
differentiation, and exhibit some interesting examples -- notably, following our
intuition of changes being differences, we show how the calculus of finite
differences appears naturally in the context of change actions.
In Chapter~\ref{chp:difference-categories} we define Cartesian difference
categories, a generalisation of Cartesian differential categories and a
particularly well-behaved class of change action models, for which we develop a
higher-order calculus in Chapter~\ref{chp:simply-typed-calculus}.

%% file: preliminaries.tex

\newcommand{\dlc}[0]{differential $\lambda$-calculus }
\label{chp:preliminaries}

This thesis deals with the convergence of two separate notions of derivative:
the differentiation of higher-order functions that appears in the differential
$\lambda$-calculus and its models, and the incrementalisation of programs by a
transformation akin to syntactic differentiation. To contextualise the research
that is to follow, we offer here a bird's eye view of the corresponding
notions. In Section \ref{sec:geometric-models} we give the historical
progression from linear to differential logic, leading up to the development of
the differential $\lambda$-calculus, of which we give a brief overview in
Section \ref{sec:differential-lambda-calculus}, followed by a discussion of
Cartesian differential categories in Section
\ref{sec:cartesian-differential-categories}. 

On the other hand, Section \ref{sec:incremental-computation-with-derivatives}
presents some trends in incremental computation, emphasising Cai and
Giarrusso's principled approach to incrementalisation as formal differentiation,
from which much of our work is derived. Finally, Section
\ref{sec:automatic-differentiation} closes with a brief introduction to
automatic differentiation, and a discussion on how both incremental and
differential calculi are connected through it.

\section{Geometric Models of Linear Logic}
\label{sec:geometric-models}

While work on the differential $\lambda$-calculus would not start until later,
the seeds of its development could already be found in earlier works on the
semantics of proofs in linear logic \cite{Girard1987, Girard1995}.
These started as an attempt to formalize the connection between linearity
in logic and linearity in the sense of linear algebra, and
would culminate in Ehrhard's work on K\"othe space semantics
\cite{EhrhardKothe} and finiteness space semantics \cite{Ehrhard2005Finiteness},
which provided the semantic motivation for the development of the differential
$\lambda$-calculus.

An early glimpse at the connection between linear logic and vector spaces
was already present in the foundational work by Girard \cite{Girard1987,
  Girard1995} on the semantics of linear logic, where a denotational semantics
is given for proof terms for linear logic which interprets propositions as
\textit{coherence spaces}, or reflexive graphs, and proofs as cliques in those
graphs. The category of coherence spaces with stable linear maps, i.e. maps
that send cliques to cliques and preserve sums (disjoint unions) of cliques can
then be endowed with enough structure to interpret every operator in linear
logic, including exponentials. As Ehrhard \cite{EhrhardKothe} observes, these
spaces bear a strong similarity to vector spaces, with the web of a coherence
space playing the role of a basis and stable linear maps corresponding to linear
(in the algebraic sense) transformations.

This connection was made clear by Blute, Panangaden and Seely \cite{BluteFock}
when they showed how to model exponential linear logic on a category of vector
spaces by using the so called \textit{Fock space construction} \cite[pp. 114-119]
{GerochFock} from quantum physics to model exponentials. Fock spaces are
construced as infinite sums of iterated symmetric (or antisymmetric) tensor
products $\tensor_s$:
\begin{align*}
  \mathcal{F}_s(A) \defeq 1 \oplus A \oplus (A \tensor_s A) \oplus \ldots
  \oplus (\tensor_s^n A) \oplus \ldots
\end{align*}
where $\tensor_s$ is the coequalizer of the maps $\A{id}, \tau : A \tensor A$
$\ra A \tensor A$. Elements of these vector spaces can be thought of as
polynomials where variables are guaranteed to commute. The authors then show that
the category of Banach spaces \cite{Megginson1998} together with contractive maps
acts as a model of linear logic, with exponentials $!A$ given precisely by the
(antisymmetric) Fock spaces $\mathcal{F}_A(A^\bot)^\bot$. Furthermore, they remark,
in this particular category the Fock space construction on some Banach space $B$
corsponds to a space of holomorphic (non-linear) functions described by power
series on $B$ \cite{BluteFock}, perhaps the first instance of non-linear function
spaces arising as a result of modeling exponentials in linear logic.

The idea of using Banach spaces and, more generally, vector spaces,
to model linear logic resulted quite successful: indeed, further work by 
Girard \cite{Girard1999} showed that Banach spaces can be extended into
\textit{coherent Banach spaces}, consisting of a pair of (complex) Banach spaces
$E, E^\bot$ equipped with a bilinear operator $\pair{\cdot}{\cdot}$ into
$\mathbb{C}$ where the following equations hold:
\begin{align*}
  |x| \equiv \text{sup}\{|{\pair{x}{y}}|; y \in E^\bot, ||y|| \leq 1\}\\
  |y| \equiv \text{sup}\{|{\pair{x}{y}}|; x \in E^\bot, ||x|| \leq 1\}
\end{align*}
As noted by Girard, this amounts to requiring that each of the $E, E^\bot$ be
isomorphic to a subspace of the dual of the other.

These spaces, together with multilinear maps, can be endowed with enough
structure to interpret the usual $\tensor, \parr, +, \&, \cdot^\bot, $
$\multimap$ operators accounting for the corresponding operators in linear logic
and coherence spaces. As an interesting note, intuitionistic implication
$A \Ra B \equiv{} !A \multimap B$ happens to coincide with the set of non-linear
functions from $A$ to $B$ that are holomorphic on some neighbourhood of 0\footnote{
  We remind the reader that a complex-valued function is holomorphic on some open set $U$
  whenever it admits a complex derivative at every point of $U$ which in particular 
  implies that it is (locally) infinitely differentiable and coincides (locally) with its
  own Taylor expansion.
},
indicating that exponentials, were they to exist, would allow for non-linearity in
the analytic, as well as the logical, sense.

Other geometric models with more structure have been developed, notably K\"othe
space semantics, due to Thomas Ehrhard. This semantics is based on
\textit{K\"othe spaces} \cite{Kothe, Dieudonne1951Kothe, EhrhardKothe}, pairs 
$X = (|X|, E_X)$
of a countable set $|X|$ of indices and a set $E_X$ of $|X|$-indexed sequences which is
\emph{self-dual}. That is to say, $E_X$ satisfies $(E_X^\bot)^\bot = E_X$, where by 
$(E_X)^\bot$ we mean the set $F_X$ of $|X|$-indexed sequences $y$ such that for all 
$x \in E_X$ the following sum:
\begin{align*}
  \sum_{i \in |X|} |{x_i y_i}|
\end{align*}
converges absolutely. The $E_X$ that satisfy this condition can be endowed
with the structure of a vector space and again in this setting the linear
implication $X \multimap Y$ is given by a space of linear functions from
$E_X$ to $E_Y$. Intuitionistic implication, however, corresponds to
non-linear, entire (i.e. holomorphic everywhere) functions (as opposed to
functions that are holomorphic on some open set). This has the important consequence of
compositionality, since the composition of entire functions is guaranteed to be
an entire function.

The niceties of the K\"othe space semantics already allow a differential
structure to arise. In \cite[Section~4.1]{EhrhardKothe}, Ehrhard remarks that
given any entire function $f$ between K\"othe spaces $X, Y$, a corresponding
arrow $\mathsf{D}f\ :\ !X\tensor X \ra Y$ can be found which is a multilinear map
from the K\"othe spaces $E_X, E_{!X}$ into $E_Y$ that corresponds to the standard
notion of derivative, thus driving the intuitions behind the development of the
differential $\lambda$-calculus.

Shortly after the advent of the differential $\lambda$-calculus, this model would
be adapted into the more ``discrete'' notion of \textit{finiteness spaces}
\cite{Ehrhard2005Finiteness}. A finiteness space can be thought of as a
generalisation of a coherence space and is given by a pair $X \equiv (I, F)$
where $I = |{X}|$ is some set and $F(X) = F \subseteq \mathcal{P}(I)$ a family of
subsets of $|{X}|$ called the ``finitary'' sets (although note that they need not
be finite) with the property that $F^{\bot\bot} = F$ (where $F^\bot$ denotes the
family of subsets of $|{X}|$ that intersect each element of $F(X)$ in at most a
finite number of points).

Morphisms between $X$ and $Y$ in the corresponding category of finiteness spaces
are given by relations $f \subseteq X \times Y$ relating finitary sets in $X$ to
finitary sets in $Y$, and finitary sets in $Y^\bot$ into finitary sets in
$X^\bot$. Again, under the adequate definitions, this category is a model for
multiplicative additive linear logic with exponentials and, similarly to the
construction of exponentials in coherence semantics, the exponential $!(X, F)$
is given by $(\mathcal{M}_{fin}(X), F(!X)$, where
\begin{align*}
  F(!X) &\defeq \{ u^!~|~u \in F(X) \}^{\bot\bot}\\
  u^!   &\defeq \{ \mu \in |{!X}|~|~|{\mu}| \subseteq \mu \}
\end{align*}

Finiteness spaces also admit a geometric interpretation (namely, given a
finiteness space $X$ and a ring $R$, a module can be constructed $R\langle X\rangle$
of $|X|$-indexed sequences of elements of $R$ whose support is a
finitary set). Note the similarity to K\"othe spaces, which are given by a set
$X$ and some space of $X$-indexed sequences of complex numbers.

\section{The Differential \texorpdfstring{$\lambda$}{Lambda}-Calculus}
\label{sec:differential-lambda-calculus}

Motivated by the discovery of this series of strongly ``analytically-flavoured''
models, Thomas Ehrhard and Laurent R\'egnier develop the differential $\lambda$-calculus
\cite{ehrhard2003differential, ehrhard2003differential} to give a computational,
syntactic account of these continuous models that also reflects the syntactic notion
of linearity as seen in e.g. \cite{DanosHerbelin1996GameSemanticsAbstractMachines},
wherein a \textit{linear substitution} is a substitution that affects only the
leftmost occurrence of a given variable. It will be in this sense that the
differential operator will produce linear functions.

The linear structure is thus baked into the syntax of the differential
$\lambda$-calculus: terms are closed under linear combinations with coefficients
on some given commutative and unital semiring $\mathbb{K}$. This can be regarded as a
generalisation of the $\lambda$-calculus with multiplicities 
\cite{boudol1993multiplicities}.
A differential application operator $\D s\cdot t$ is
also introduced, representing the derivative of the function $s$ at $t$ (or the
application of function $s$ to exactly one copy of $t$). Terms are divided into classes
$\Lambda^s, \Lambda^d$ of \emph{simple} and \emph{differential} terms (or simply terms),
defined inductively as follows:
\[
\begin{array}{rccl}
  \Lambda^s: &s, t& \defeq & x \alter \lambda x . s \alter (s\ T) \alter \D (s)\cdot t\\
  \Lambda^d: &S, T& \defeq & \sum_{i = 1}^n a_i s_i \quad\text{where $a_i \in \mathbb{K}$}
\end{array}
\]

While the original formulation \cite{ehrhard2003differential} did not define separate
syntactic categories $\Lambda^s, \Lambda^d$ and instead relied on a notion of congruence
to ensure linearity of the relevant syntactic constructs, later presentations
\cite{manzonetto2012categorical, mak2016} enforce linearity directly on the syntax
by defining some constructions as syntactic sugar, e.g.
\begin{align*}
  \pa{\sum_i a_i s_i}T &\defeq \sum_i \pa{a_i s_i T}\\
  \lambda x . \pa{\sum_i a_i s_i} &\defeq \sum_i a_i \pa{\lambda x . s_i}\\
  \D \pa{\sum_i a_i s_i} \cdot \pa{\sum_j b_j t_j} &\defeq \sum_{k,j} a_ib_j \D s_i \cdot t_j
\end{align*}
Importantly, we note that application is linear in the function being applied, but 
\textit{not} in the argument, as it might be used more than once, whereas differential
application is linear in both.

Reduction for the differential $\lambda$-calculus is based around a new notion
of substitution: the \textit{partial derivative of s with respect to x along u},
denoted by the familiar notation from calculus $\diffS{s}{x}{u}$, given by the
rules in Figure~\ref{fig:differential-substitution-standard}.
\begin{figure}[ht]
  \begin{center}
  \[\def\arraystretch{1.7}
    \begin{array}{cccl}
      \diffS{x}{x}{u} &\defeq& u\\
      \diffS{y}{x}{u} &\defeq& 0 &\text{ if $x \neq y$}\\
      \diffS{(\lambda y . t)}{x}{u} &\defeq& \lambda y . \pa{\diffS{t}{x}{u}}
                      &\text{ if $y \not\in \text{FV}(t)$}\\
      \diffS{(t\ e)}{x}{u} &\defeq& \brak{\D(t) \cdot \pa{\diffS{e}{x}{u}}\ e} +
                                    \brak{\diffS{t}{x}{u}\ \pa{e}}\\
      \diffS{(\D(t) \cdot e)}{x}{u} &\defeq& 
        \D(t) \cdot \pa{\diffS{e}{x}{u}} 
        + \D\pa{\diffS{t}{x}{u}} \cdot (e)\\
      \diffS{(t + e)}{x}{u} &\defeq& \pa{\diffS{t}{x}{u}} + \pa{\diffS{e}{x}{u}}\\
      \diffS{0}{x}{u} &\defeq& 0
    \end{array}
  \]
  \caption{Differential substitution in the differential $\lambda$-calculus}
  \label{fig:differential-substitution-standard}
  \end{center}
\end{figure}
Operationally, the differential substitution $\diffS{s}{x}{u}$ can be understood
as substituting one occurrence of $x$ in $s$, chosen non-deterministically, by
$u$. 

The one-step reduction relation for the differential $\lambda$-calculus is then
induced by the rules in Figure~\ref{fig:reduction-standard} below.
\begin{figure}[H]
  \[\def\arraystretch{1.3}
  \begin{array}{rcl}
    (\lambda x . t)\ s &\reducesTo_\beta& \subst{t}{x}{s}\\
    \D(\lambda x . t) \cdot s &\reducesTo_\partial& \lambda x . \pa{\diffS{t}{x}{s}}\\
  \end{array}
  \]
  \caption{One-step reduction rules for the differential $\lambda$-calculus}
  \label{fig:reduction-standard}
\end{figure}

In the same work, the authors also study a typed version of the calculus
\cite[Section 4]{ehrhard2003differential}, by using a straightforward extension
of simple types for the $\lambda$-calculus. In order to recover some
meta-theoretical properties, however, further constraints are required on the
underlying semiring $R$: to prove subject reduction \cite[Lemma
22]{ehrhard2003differential}, $R$ must be \textit{positive}, i.e. it must lack
any negative elements, otherwise for any type $A$ and (not necessarily
well-typed) term $t$, the sequent $\vdash 0 : A$ is derivable and $0 \ra_\beta
(t - t)$ but $\vdash (t - t) : A$ may not be provable.

Strong normalization cannot be proven in all generality, either, and Ehrhard and
R\'egnier \cite[Theorem 35]{ehrhard2003differential} prove it only for the case
of the natural numbers. Although it is noted that the results can be extended
to positive rings without zero divisors as long as every element admits only a
finite number of decompositions, Vaux \cite{vaux2007diffmu} would later show
stronger conditions are necessary.

One notable result about the differential $\lambda$-calculus is that the usual Taylor
expansion from calculus does hold:
\begin{align*}
  s\ u = \sum_{n = 0}^\infty{\frac{1}{n!}(\D^n (s) \cdot u^n)}0
\end{align*}
whenever the term $s\ u$ reduces to some distinguished variable $\star$
\cite[Theorem 39]{ehrhard2003differential}. This theorem was later generalised
in \cite{ehrhard2008differential}, where it is used to encode the full
$\lambda$-calculus into a simplified subset of the differential
$\lambda$-calculus featuring differential application but not ordinary
application. While this simplified calculus is not Turing-complete (as every
function is linear), adding infinite sums to the language allows one to encode
the full $\lambda$-calculus in a way which, while considerably more involved,
is reminiscing of how arbitrary (analytic) functions can be represented by a
power series. We believe these results show that infinite sums of terms may play
a pivotal role in interpreting fixpoints.

Further work has been carried out on the differential $\lambda$-calculus. On one
hand, the theory of proof nets and, more generally, interaction nets
\cite{lafont1995proof, girard1996proof}, graphical formalisms for studying
proofs and proof reduction in linear logic, have been extended to the
differential case by enriching the structure of usual interaction nets with
addition \cite{ehrhard2006interaction} and, more recently, with exponential
boxes \cite{tranquilli2011nets}.

Moreover, some work has gone into extending the differential $\lambda$-calculus
with additional primitives, most notably Lionel Vaux's differential
$\lambda\mu$-calculus \cite{vaux2007diffmu}. This calculus subsumes both the
differential $\lambda$-calculus and Parigot's $\mu$-calculus
\cite{parigot1992mu}, an extension of the polymorphic $\lambda$-calculus that
corresponds to natural deduction for classical (as opposed to intuitionistic)
logic. This extension is mostly straightforward, with the exception of the
interaction between differential application and $\mu$-abstraction $\text{D}(\mu
\alpha . \nu) \cdot t$ which, intuitively, feeds the $t$ linearly to every
subterm of $\nu$ which is labelled by $\alpha$. 

Vaux's work also revises and extends the results in
\cite{ehrhard2003differential}: some extra notational convenience is provided
and some of Ehrhard and R\'egnier's claims are revisited and expanded upon.
Notably, Vaux shows that stronger conditions are necessary for strong
normalization \cite[Theorem 35]{ehrhard2003differential} to hold than was
believed originally.

\subsection{Algebraic Calculi}
\label{ref:algebraic-calculi}

Extensions of the $\lambda$-calculus endowed with some form of vector space
structure on terms have been studied independently of the \dlc, particularly,
particularly for applications in quantum computation. Despite these calculi
being different from the \dlc, however, a significant amount of useful
literature has been written about them from which many insights can be directly
applied to the differential $\lambda$-calculus, particularly concerning more
operational aspects such as termination and confluence. 

The first development in this direction was due to Pablo Arrighi and Gilles
Dowek \cite{arrighi2005computational}, who studied the theory of
$\mathcal{K}$-vector spaces as rewrite systems and tackled the problem of
providing a sound, terminating and confluent rewrite theory, given some
reasonable rewrite system for the underlying field $\mathcal{K}$. The main
result of this work is \cite[Proposition 2.14]{arrighi2005computational}, which
tells us that a certain algebraic structure is a vector space if and only if it
is a model of the rewrite system constructed by the authors. Furthermore, this
reduction system is terminating (\cite[Proposition
2.4]{arrighi2005computational}) and confluent in semi-open terms
(\cite[Proposition 2.10]{arrighi2005computational}), that is, terms that do not
contain any free variables of scalar sort. This is particularly interesting in a
differential context, since the \dlc forbids syntactically the presence of
variables in a scalar context.

Building on this algorithm, Arrighi and Dowek \cite{arrighi2005lambda}
developed an extension of the $\lambda$-calculus equipped with a vectorial
structure with the intention of providing a model for quantum computation. While
this language does not feature any differential constructs, it is far more
powerful than the differential $\lambda$-calculus when it comes to expressing
vectorial operations, including primitives for computing tensor products and dot
products, as well as a form of pattern maching. The vectorial aspects of this
calculus are, however, limited to boolean values and one cannot take linear
combinations of arbitrary terms.
    
Simpler calculi have later been constructed by Arrighi and Dowek
\cite{arrighi2008linear} and Vaux
\cite{vaux2006lambda,vaux2007linear,vaux2009algebraic} that are significantly
closer to the \dlc - indeed, they can be seen as a restriction of the former to
terms without differential application. The study of these systems was partly
motivated by the reduction issues that arise in the differential
$\lambda$-calculus if the underlying scalar structure is not ``nice'' enough. In
Ehrhard's original work \cite{ehrhard2003differential} it is already noted that
strong normalization in the differential $\lambda$-calculus would fail to hold
if coefficients are taken to be the (positive) rational numbers, since the
following reduction sequence would be possible (whenever $t \rightarrow t'$):
\begin{align*}
  t = \frac{1}{2}t + \frac{1}{2}t \rightarrow \frac{1}{2}t + \frac{1}{2}t' = \frac{1}{4}t + \frac{1}{4}t + \frac{1}{2}t'
  \rightarrow \frac{1}{4}t + \frac{1}{4}t' + \frac{1}{2}t' = \frac{1}{4}t + \frac{3}{4}t' \ldots
\end{align*}

Indeed Vaux
\cite{vaux2007diffmu,vaux2006lambda,vaux2007linear,vaux2009algebraic} shows that
strict restrictions on the semiring of scalars have to be imposed for the
resulting calculus to be strongly normalizing. Particularly, the semiring $R$ of
scalars must be \textit{finitely splitting} \cite[Theorem 3]{vaux2007linear},
i.e. every element $a \in R$ can be expressed as a sum $b_1 + \ldots + b_n$ in
only a finite number of ways, although for weak normalization to hold it is only
required that $R$ be positive, whereas confluence is shown without any
assumptions on the semiring $R$ \cite[Theorem 1]{vaux2007linear}.
    
Another relevant issue that has been studied in the context of the algebraic
$\lambda$-calculus is that of fixpoints. Indeed, it is well known
\cite{ehrhard2003differential,vaux2006lambda} that if the semiring $R$ contains
negative elements the reduction system collapses entirely in the untyped case.
To see why, consider some fixpoint combinator $\A{Y}$ such that $\A{Y}(\A{f}) =
f(\A{Y}(\A{F}))$ and, given some term $\A{s}$, define the following auxiliary
combinator:
\begin{align*}
  \A{Y_s} \defeq \A{Y}(\lambda \A{x} . \A{s} + \A{x})
\end{align*}
Then $\A{Y_s} \ra^* s + \A{Y_s}$ and therefore $\A{s} \equiv \A{s} + \A{Y_s} -
\A{Y_s} + \A{Y_t} - \A{Y_t} \ra^+ \A{s} - \A{s} + \A{t} = \A{t}$. As this
problem only depends on the existence of a fixpoint combinator, it will also
arise in any typed extension of the \dlc equipped with such. This behavior
corresponds to the old adage that $\infty - \infty$ does not have a determinate
value.

While most work avoids this problem by restricting $R$ to be positive, i.e. $a +
b = 0$ implies $a = 0$ and $b = 0$ for any $a, b \in R$, Arrighi and Dowek
\cite{arrighi2008linear} tackle the issue in a completely different way, by
considering a system where algebraic equations such as $at + bt = (a + b)t$ are
seen not as an equivalence on terms but rather as rewrite rules in the same
style as \cite{arrighi2005computational}. Then a series of side conditions are
added to the resulting rewrite system so as to eliminate the pathological cases:
for example, the term $t$ is required to be in normal form for the rule $at + bt
\rightarrow (a + b)t$ to apply, which renders the previous pathologic reduction
sequence impossible. It is an open question, however, whether a denotational
semantics can be obtained for such a calculus and what its models are.

\section{Cartesian Differential Categories}
\label{sec:cartesian-differential-categories}

Since the discovery of the correspondence between the simply-typed
$\lambda$-calculus and Cartesian closed categories due to Lambek
\cite{lambek1972deductive}, an important area of research within type theory has
been the search for categorical models for any new calculi that appear, and the
differential $\lambda$-calculus would be no different.

Indeed, not long after Ehrhard and R\'egnier's unveiling of their new calculus,
Blute, Cockett and Seely put forth the idea of differential categories
\cite{blute2006differential}, i.e. monoidal categories enriched over some
commutative monoid (modelling addition of terms) equipped with a coalgebra
modality that plays a similar role to the familiar bang modality from linear
logic, plus a differential operator $D$ mapping arrows $f :\ !A \longrightarrow
B$ into arrows $D [f] : A\ \tensor \ !A \longrightarrow B$. In particular, this
means that the hom-sets of a differential category $\cat{C}$ are equipped with a
commutative operation $+ : \cat{C}(A, B) \ra \cat{C}(A, B) \ra \cat{C}(A, B)$
which admits a neutral element $0 : \cat{C}(A, B)$.

This differential operator $\mathsf{D}[\cdot]$ must, of course, satisfy some coherence
conditions: it must be linear and natural, but most importantly it must ``behave
like a derivative'', in the sense that the differential satisfies a categorical
analogue of the product and chain rules, sends linear maps to themselves and
constant maps to the $0$ arrow (see \cite[Equations
D.1-D.4]{blute2006differential} for details). These conditions are also stated
in a graphical setting as equations between proof nets, where the differential
is represented as a kind of ``box''.

In \cite{blute2006differential}, Cockett and Seely also identify various
differential categories. As an example, $\A{Vec}_K^{op}$, the dual of the
category of $K$-vector spaces, is a differential category, with the coalgebra
modality $!V$ being given by the ring of polynomials $K[X]$ (where $X$ denotes a
basis of $V$). What is more, they vastly generalise this construction: given any
commutative semiring $R$, it is shown that every \textbf{polynomial theory}
(i.e. a Lawvere theory $\cat{T}$ whose constants $\cat{T}(0, 1)$ are exactly the
elements of $R$ and that includes functions $+, \cdot : \cat{T}(2, 1)$
satisfying the equations for a commutative $R$-algebra, see
\cite[Section~3.1]{blute2006differential} for details) induces a coalgebra
modality $S_\cat{T}(V)$ in $\A{Mod}_R^{op}$ given by the space of functions $h$
mapping linear forms $u : V \multimap R$ into scalars in $R$, such that $h(u)$
corresponds to the action of some arrow $\alpha : \cat{T}(n, 1)$ on some finite
set of vectors $v_1, \ldots, v_n \in V$. This can be thought of as a more
general version of the Fock construction previously used by Blute, Panangaden
and Seely to model exponentials \cite{BluteFock} (which is obtained when
$\cat{T}$ is taken to be the theory of polynomials).

Furthermore, if the theory $\cat{T}$ admits the existence of some terms behaving
like partial derivatives (in a sense that is made precise in \cite[Equations
pd.1-pd.5]{blute2006differential}), plus some extra technical conditions (that
roughly amount to requiring that independent vectors in each $R$-module $V$ can
be separated by a linear functional), then $\A{Mod}_K^{op}$ with the previously
defined coalgebra $S_\cat{T}$ is indeed a differential category. This is quite a
powerful result, as it allows one to construct differential categories ``\'a la
carte''.

The notion of differential category was, however, still tied to a monoidal
category of linear maps, from which differential operators arose from the
interaction between the linear structure and a coalgebra modality, with smooth
maps being coKleisli arrows. To give an account of smooth maps directly, Blute,
Cockett and Seely introduced the notion of \textbf{Cartesian differential
categories} \cite{blute2009cartesian}, or CDCs, which replace the monoidal tensor
with Cartesian product.

In a CDC, hom-sets are taken to be formed of smooth, rather than linear,
functions. They are also enriched with a monoidal structure, but Cartesian
differential categories are \textit{not} additive (which would entail that every
arrow is ``linear''), but only left-additive.

\begin{definition}
  \label{def:left-additive-category} 
  A \textbf{left-additive category} \cite[Definition 1.1.1.]{blute2009cartesian}
  is a category $\cat{C}$ together with a choice of a commutative monoid
  structure $(\cat{C}(A, B), +, 0)$ on each hom-set $\cat{C}(A,B)$, subject to
  the constraint that pre-composition (but \emph{not} post-composition) respects
  the additive structure. That is to say, for all $f, g, h$:
  \begin{align*}
    (f + g) \circ h &= (f \circ h) + (g \circ h)\\
    0 \circ h &= 0
  \end{align*}

  A certain map $f : A \to B$ in a left additive category is \textbf{additive}
  whenever post-composition by $f$ preserves the additive structure. That is to
  say, for all $g$:
  \begin{align*}
    f \circ (g + h) &= (f \circ g) + (f \circ h)\\
    f \circ 0 &= 0
  \end{align*}
\end{definition} 

A Cartesian differential category is also equipped with a differential operator,
which mimics the behaviour of the differential of a smooth function in
multi-variate calculus.

\begin{definition}
  \label{def:differential-category}
  A \textbf{Cartesian differential category} \cite[Definition
  2.1.1.]{blute2009cartesian} is a Cartesian left-additive category $\cat{C}$
  equipped with a \textbf{differential combinator} $\mathsf{D}$ of the
  form\footnote{
    We might expect the type of $\mathsf{D}[f]$ to have the form 
    $A \times \mathsf{T}_0 A \to B$. The theory of differential categories assumes that
    every space is ``Euclidean'', in the sense that it is equal to its tangent space at
    any point.
    Note also that we flip the convention found in
    \cite{blute2009cartesian}, so that the linear argument is in the second
    argument, as usual in calculus, rather than in the first argument.
  }: 
  \[ \frac{f : A \to B}{\mathsf{D}[f]: A \times A \to B} \]
  verifying the following coherence conditions: 
  \begin{enumerate}[\CDCAx{\arabic*},ref={\CDCAx{\arabic*}},align=left]
    \item $\mathsf{D}[f+g] = \mathsf{D}[f] + \mathsf{D}[g]$ and
    $\mathsf{D}[0]=0$ \label{CDC1}

    \item $\mathsf{D}[f] \circ \langle x, u+v \rangle= \mathsf{D}[f] \circ
    \langle x, u \rangle + \mathsf{D}[f] \circ \langle x, v \rangle$ and
    $\mathsf{D}[f] \circ \langle x, 0 \rangle=0$ \label{CDC2}

    \item $\mathsf{D}[\Id{A}]=\pi_2$ and $\mathsf{D}[\pi_1] = \pi_1 \circ \pi_2$
    and $\mathsf{D}[\pi_2] = \pi_2 \circ \pi_2$ \label{CDC3}

    \item $\mathsf{D}[\langle f, g \rangle] = \langle  \mathsf{D}[f],
    \mathsf{D}[g] \rangle$ and $\mathsf{D}[!_A]=!_{A \times A}$ \label{CDC4}

    \item $\mathsf{D}[g \circ f] = \mathsf{D}[g] \circ \langle f \circ \pi_1,
    \mathsf{D}[f] \rangle$ \label{CDC5}

    \item $\mathsf{D}\left[\mathsf{D}[f] \right] \circ \left \langle \langle x,
    u \rangle, \langle 0, v \rangle \right \rangle=  \mathsf{D}[f] \circ \langle
    x, v \rangle$ \label{CDC6}

    \item $\mathsf{D}\left[\mathsf{D}[f] \right] \circ \left \langle \langle x,
    u \rangle, \langle v, 0 \rangle \right \rangle =
    \mathsf{D}\left[\mathsf{D}[f] \right] \circ \left \langle \langle x, v
    \rangle, \langle u, 0 \rangle \right \rangle$ \label{CDC7}

  \end{enumerate}
\end{definition}

We highlight that by \cite[Proposition 4.2]{cockett2014differential},
axioms \ref{CDC6} and \ref{CDC7} have an equivalent alternative
expression. 
\begin{lemma}
  In the presence of the other axioms, \ref{CDC6} and \ref{CDC7} are
  equivalent to:
 \begin{enumerate}[\CDCAx{\arabic*(a)}, align=left]
 \setcounter{enumi}{5}
  \item $\mathsf{D}\left[\mathsf{D}[f] \right] 
  \circ \four{x}{0}{0}{w} = \mathsf{D}[f] \pair{x}{w}$
  \label{CDC6a}

  \item $\mathsf{D}\left[\mathsf{D}[f] \right] 
    \circ \four{x}{u}{v}{w}
    \rangle
    = \mathsf{D}\left[\mathsf{D}[f] \right] 
    \circ \four{x}{u}{v}{w}
    $
  \label{CDC7a}
\end{enumerate}
\end{lemma}

The above axioms are somewhat abstruse and so we give an informal explanation.
\ref{CDC1} states the sum rule for the operator $\mathsf{D}$. \ref{CDC2}
states that every derivative is additive in its second argument. \ref{CDC3}
states that the identity function and projections are ``linear'', that is, their
derivative is equal to the function itself. \ref{CDC4} states that the
derivative of a vector-valued map is the vector of the derivatives of each
component, and \ref{CDC5} is simply the chain rule. Perhaps the least intuitive
of the lot, \ref{CDC6} can be understood as saying that every derivative is
``linear'' in its second argument, not in the sense that it commutes with the
sum (as \ref{CDC2} states), but in the sense that taking the partial 
derivative with respect to its second argument is the function itself. Finally,
\ref{CDC7} is an abstract formulation of Schwarz's theorem, which states that
the matrix of second derivatives of any smooth function is symmetric.

We refer the reader to \cite[Section 4]{blute2009cartesian} for further explanations 
and an exposition of a term calculus which may help better understand the
axioms of a Cartesian differential category.

Intuitively it might seem like a CDC is much like a differential category where
we forget about the difference between linear and non-linear functions. Indeed
this intuition can be made precise: it turns out that the coKleisli category of
a differential storage category \cite{blute2006differential} is a Cartesian
differential category \cite[Proposition 2.3.1]{blute2009cartesian}, thus
providing us with a simple way of constructing CDCs.

The canonical example of a
Cartesian differential category is the category of real smooth functions, which
we will discuss in more detail in Section \ref{smoothex}. 
Other interesting examples can be found throughout the literature such as the
subcategory of differential objects of a tangent category
\cite{cockett2014differential}.

The notion of a Cartesian differential category has more recently been generalised by
Cruttwell \cite{cruttwell2017cartesian}, based on the observation that, while a
differential operator as usually defined maps arrows $f : A \ra B$ to arrows
$\mathsf{D}[f] : A \times A \ra B$, the two intstances of $A$ in the codomain of 
$\mathsf{D}[f]$
play radically different roles, as one usually talks about the derivative of a
function at a point and along a certain vector. It so happens that in many
common cases, as in functions from $\cat{R}^m$ to $\cat{R}^n$, both points and
vectors are elements of the same vector space (as the usual Euclidean spaces
$\cat{R}^m$ are their own tangent spaces), but this need not be the case in
general.

For this reason, Cruttwell proposes to relax the definition of Cartesian
differential category: rather than requiring it to be a left-additive category,
the monoidal structure is restricted only to certain ``spaces of vectors''. A
\textbf{generalised Cartesian differential category}, or GCDC, is then a
Cartesian category $\cat{C}$ equipped with a commutative monoid $L(X)$ for each
$X$ (intuitively corresponding to the tangent space of $X$) which is involutive
(i.e. $L(L(X)) = L(X)$) and Cartesian ($L(A \times B) = L(A) \times L(B)$). In
addition, a generalised Cartesian differential category is equipped with a
differential operator $\mathsf{D}$ satisfying obvious analogues to
\hyperref[def:differential-category]{\bf [CDC.1-7]}, with the difference that,
if a map $f$ has type $f : A \ra B$ then its derivative has type $\mathsf{D}[f]
: L(A) \times A \ra B$.

\begin{definition}
  \label{def:gcdc}
  A \textbf{generalised Cartesian differential category} \cite[Definition
  2.1.]{cruttwell2017cartesian} is a Cartesian category $\cat{C}$ such that:
  \begin{enumerate}[\roman*.]
    \item For each object $A$ of $\cat{C}$, a choice of a commutative monoid $L(A) =
    (L_0(A), +_A, 0_A)$ internal to $\cat{C}$, satisfying
      \[ L(L_0(A)) = L(A) \quad\quad\quad L(A \times B) = L(A) \times L(B) \]
    \item For each map $f : A \to B$ in $\cat{C}$, a map $\mathsf{D}[f] : A
    \times L_0(A) \to L_0(B)$ satisfying 
    axioms \hyperref[def:differential-category]{\CDCAx{1-7}}.
  \end{enumerate}
\end{definition}

An immediate consequence of the previous definition is that any CDC is trivially
a GCDC, with $L(X) = X$. Not only that, but important categories that are not
CDCs can be seen as generalised Cartesian differential categories (an example
being the category formed by finitely-dimensional differentiable manifolds and
smooth functions, which is not Cartesian differential since an arbitrary
differentiable manifold is not its own tangent space). A term calculus for
generalised Cartesian differential categories is not yet known, and we believe
it to be an important open problem and a first step in addressing some technical
problems in certain calculi \cite{manzyuk2012simply}.

While Blute, Cockett and Seely already provided a term calculus which is sound
and complete with respect to Cartesian differential categories
\cite[Section~4.1]{blute2009cartesian}, the full power of Ehrhard's original
differential $\lambda$-calculus requires more structure: in particular,
exponentials need to be added to the Cartesian differential structure if one is
to model differential $\lambda$-terms. To this end, Bucciarelli, Ehrhard and
Manzonetto developed
\textbf{differential $\lambda$-categories}
\cite{bucciarelli2010categorical,manzonetto2012categorical}, which are
essentially Cartesian closed differential categories where the exponential is
well-behaved with respect to the differential structure. 

\begin{definition}
  \label{def:differential-lambda-category}
  A Cartesian left-additive category $\cat{C}$ is \textbf{Cartesian closed
  left-additive} whenever it is Cartesian closed and verifies the following
  conditions:
  \begin{enumerate}[i.]
    \item $\Lambda(f + g) = \Lambda(f) + \Lambda(g)$
    \item $\Lambda(0) = 0$
  \end{enumerate}

  A \textbf{differential $\lambda$-category} 
  \cite[Definition 4.4.]{bucciarelli2010categorical}
  is a Cartesian closed left-additive category
  that verifies the following additional axiom:
  \begin{enumerate}[\DlCAx{},ref={\DlCAx{}},align=left]
    \item 
    $\mathsf{D}[\Lambda(f)] = \Lambda(\mathsf{D}[f] \circ 
    \pair{\pi_1 \times \Id{}}{\pi_2 \times 0})$
    \label{DlC}
  \end{enumerate}
\end{definition}

The above amounts to saying that taking the partial derivative of an arrow with
respect to its first argument is equivalent to currying the arrow, then taking
the derivative with respect to its only argument. It can then be shown that
differential $\lambda$-categories are sound with respect to the differential
$\lambda$-calculus \cite{bucciarelli2010categorical}. Lately, a completeness result has been proven by Mak
\cite{mak2016}.

More recent work by Manzyuk \cite{manzyuk2012tangent} and Cockett and Cruttwell
\cite{cockett2014differential} has also unearthed a deep connection between Cartesian
differential categories and the so-called categories with tangent structure,
initially developed by Rosick\'y \cite{rosicky1984tangent} in order to give an
abstract characterization of the tangent bundle construction familiar from
differential geometry. Essentially, a \textbf{category with tangent structure}
is a category $\cat{C}$ equipped with some endofunctor $\T : \cat{C} \ra \cat{C}$
equipped with a projection $p_A : \T A \to A$ and an additive structure, which,
roughly speaking, amounts to natural transformations:
\begin{align*}
  +_A &: \T A \times_A \T A \Ra \T A\\
  0_A &: A \Ra \T A
\end{align*}
where $\T A \times_A \T A$ denotes the pullback of the projection $p_A$ along itself
(we use the pullback along $p_A$ and not the product because vectors of the
tangent space at some point $x$ can be added together, but they cannot be added
to vectors of the tangent space of some other point $y$). Some extra
conditions must hold, but we omit them here, instead referring the reader to
\cite[Definition 2.3]{cockett2014differential}. A \textbf{Cartesian tangent category}
is precisely a Cartesian category with tangent structure which preserves
products, i.e. $T(A \times B) \cong T(A) \times T(B)$ and the corresponding
natural isomorphism $\alpha$ preserves the additive structure of $T$.

Most importantly, one can also show that every Cartesian differential category
induces a Cartesian tangent category via the \textbf{tangent bundle} functor $TA
\defeq A \times A$ acting on arrows by $Tf \defeq \pair{\D f}{f \circ \pi_1}$
\cite[Proposition 4.7]{cockett2014differential}. Furthermore, given a Cartesian
tangent category $\cat{C}$, the full subcategory of differential objects
(objects $A$ whose tangent structure is given by the product $A\times A$, see
\cite[Definition 4.8]{cockett2014differential}) is in fact a Cartesian differential
category. These transformations define a pair of adjoint functors between the
category of Cartesian differential categories and the category of Cartesian
tangent categories, thus making precise the connection between differential
categories and categories with tangent structure.

This connection is further explored in Cockett and Cruttwell's recent work
\cite{cockett2016differential}, via the more general notion of
\textbf{differential bundle}. It is then shown that a Cartesian tangent category
is itself Cartesian differential if and only if one can choose a differential
bundle for each object in a coherent way \cite[Lemma 3.13, Theorem
3.14]{cockett2016differential}. An area where we believe further work is
neccesary is generalising all these helpful results on Cartesian differential
categories to GCDCs or to differential $\lambda$-categories.

\subsection{Convenient Models}

The theory of differential categories provides a solid categorical foundation
for the semantics of differential linear logic, with the \dlc: terms are
interpreted as arrows in some differential $\lambda$ category $\cat{C}$, with
differential application $\D t \cdot u$ corresponding to the differential
operator in $\cat{C}$. The question may arise, however, as to whether there are
differential $\lambda$-categories whose objects are some family of vector spaces
and whose arrows are smooth functions in some geometric sense, especially since
the examples presented when the concept was first introduced
\cite{bucciarelli2010categorical} (corresponding to the relational and
finiteness space semantics) are strongly combinatorial in nature.

This question was answered affirmatively by Blute, Ehrhard and Tasson
\cite{blute2010convenient}, when they showed that so-called \textbf{convenient
vector spaces} \cite{frolicher1988linear, kriegl1997convenient}, together with
linear maps form a differential category which $\A{Con}$ which is also symmetric
monoidal closed. Furthermore, an exponential modality $!$ can be given in
$\A{Con}$ where $!E$ is given as the Mackey closure (a generalisation of the
notion of Cauchy completion) of the linear span of the Dirac delta distributions
$\delta(v)$ for $v \in E$ (where $\delta(v)$ sends every smooth curve $f : E \ra
\mathbb{R}$ to $f(v)$). The coKleisli category for the comonad $!E$ then happens
to be precisely $\mathcal{C}^\infty$, the category of convenient vector spaces
and smooth functions, and thus, by \cite[Proposition
2.3.1]{blute2009cartesian}, it follows that $\mathcal{C}^\infty$ is, in fact,
a Cartesian differential category.

This approach has been later applied by Kerjean and Tasson
\cite{kerjean2016mackey} to \textbf{Mackey-complete vector spaces}
\cite{kriegl1997convenient}, which slightly extend convenient vector spaces
by not requiring a bornological structure as convenient vector spaces do. These
vector spaces together with linear maps form a symmetric monoidal closed
category $\A{Lin}$ which is also Cartesian (but not Cartesian closed). The
authors then show that Mackey-complete vector spaces with power series form a
$\lambda$-category \cite{bucciarelli2010categorical} $\A{Series}$ which arises
from an exponential modality on $\A{Lin}$ constructed just as in
\cite{blute2010convenient}. The category $\A{Series}$ constitutes, thus, a
reasonably ``geometric'' concrete model of the differential $\lambda$-calculus.

\section{Incremental Computation With Derivatives}
\label{sec:incremental-computation-with-derivatives}

Computational applications are rarely one-shot. As much as functional
programmers praise immutable data structres, change is the only constant in the
real world, and so it is often the case that one needs to repeat some 
calculation on an input that changes over time, be it because the corresponding
data has actually mutated or simply because the output of our algorithm prompts
us to re-evaluate the computation.

We have already mentioned in the introduction the usefulness of applying incremental
computation to graph algorithms such as PageRank.
A similar case, and the one we will focus on, is the evaluation of Datalog queries.
In this field, it has long been
known that some degree of incrementalisation results in asymptotic performance
gains. This procedure is so widespread that it is usually known as semi-naive
evaluation (the implication being that such a trick is so basic and essential
that it is only one step above ``naive'' evaluation). In
Chapter~\ref{chp:incremental-computation} we will give a more detailed
explanation of semi-naive evaluation, but we refer the interested reader to one
of the standard textbooks \cite[Chapter 13.1]{abiteboul1995foundations}.

While incremental computing is pervasive \cite{ramalingam1993categorized}, there
is very little material on an underlying, unifying theory, with most of the
application-agnostic work focusing on general-purpose implementations like
\cite{carlsson2002monads} or \cite{acar2009self}, instead of its mathematical
underpinnings.

It is partly to address this lack that recent work by Cai and Giarrusso
\cite{cai2014changes,giarrusso2019incremental} introduced the notion of
\emph{change structures} and the \emph{incremental $\lambda$-calculus}. While
novel, these concepts have already found applications to the incrementalisation
of Datalog \cite{mario2019fixing} and Datafun \cite{arntz2016datafun} programs.

\begin{definition}
  A \textbf{change structure} \cite[Definition 2.1]{cai2014changes}
  is a tuple $\widehat{V} = (V, \Delta, \oplus, \ominus)$ such that:
  \begin{enumerate}[\roman*.]
    \item $V$ is a set.
    \item $\Delta : V \to \mathsf{Set}$ is a $V$-indexed family of sets. We
    write $\Delta_v$ for the set associated to the element $v \in V$.
    \item $\oplus, \ominus$ are dependently-typed functions:
    \begin{align*}
      \oplus &: \forall (v : V),\ \Delta_v \to V\\
      \ominus &: V \to \forall (v : V),\ \Delta_v
    \end{align*}
    \item For all $u, v$, we have $u \oplus (v \ominus u) = u$.
  \end{enumerate}
\end{definition}
Intuitively, the set $\Delta_v$ is to be understood as a space of possible
``changes'' or ``updates'' that can be applied to some value $v$, with $v \oplus \delta_v$
denoting the result of updating the value $v$ with the change $\delta_v$. A crucial
assumption of Cai and Giarrusso's model is that, given any two values $u, v$, there will
necessarily be a change $\delta_u \in \Delta_u$ satisfying $u \oplus \delta_u = v$. Or,
more concisely, every value can be updated into any other value. The map $\ominus$ is
responsible for selecting such a change (although in general there may be more than
one change with the desired property).

The challenge
of incremental computation can then be framed in the language of change structures as the
problem of propagating changes from the input to the output of a function.

Cai and Giarrusso go on to give concrete change structures on some sets.
Notably, they give an explicit construction for a change structure on the
function space $A \Rightarrow B$, given change structures on $A$ and $B$
respectively. More importantly, they give an algorithm for incrementalising
functions between change structures by ``syntactic differentiation''. Although
they do not frame it in such terms, this algorithm amounts to an iterated
application of the chain rule from standard calculus (see the 
the definition of $\mathit{Derive}$ in \cite[Figure 4]{cai2014changes})

\newcommand{\derp}[1]{\frac{\partial}{\partial #1}}
\newcommand{\derpf}[2]{\frac{\partial #1}{\partial #2}}
\section{Automatic Differentiation}
\label{sec:automatic-differentiation}

An interesting area where the differential $\lambda$-calculus has tentatively
been applied is automatic differentiation. Automatic differentiation, or AD for
short, is a family of methods that allow for the efficient computation of the
derivative of arbitrary programs operating on numeric values. Given some program
$f : \mathbb{R} \rightarrow \mathbb{R}$ and some point $x_0 \in \mathbb{R}$, an
AD algorithm will compute the value of the derivative $\frac{\partial}{\partial
x}f$ evaluated at point $x_0$. 

While an in-depth exposition of these techniques is outside the scope of this
work (we refer to \cite{griewank2008evaluating} for a more complete treatment of
the topic), the fundamental intuition driving them is that such a program $f$
must necessarily be a composition of sums, products and some primitive functions
($\A{sin}$, $\A{cos}$, $\A{log}$, \ldots) whose derivatives are already known.
The derivative of $f$ can then be computed by a repeated application of the
chain rule, which states that:
\begin{align*}
  \derpf{(g \circ f)(x)}{x} \Bigr|_{x = x_0} = \derpf{g(y)}{y} \Bigr|_{y = f(x_0)} \times 
  \derpf{f(x)}{x} \Bigr|_{x_0}
\end{align*}
The chain rule is applied iteratively until the derivative of $f$ is expressed
in terms of the derivatives of elementary functions, which are already known and
can be easily evaluated. This process is quite efficient and can be performed in
linear time with respect to the complexity of the input program.

Automatic differentiation is significantly more precise than numeric
differentiation, as it allows for completely accurate computations (up to the
precision of the underlying numeric types). It is also more efficient than
approaches based on symbolic differentiation as the derivatives computed by
symbolic differentiation are, in general, exponentially larger than the input.

One may observe that the expansion of the expression $\derp{x} f$ into partial
derivatives can be computed in more than one order. For example:
\begin{align*}
  \derpf{y}{x} = \derpf{y}{w_1} \derpf{w_1}{x} = \derpf{y}{w_1}\left(\derpf{w_1}{w_2}\derpf{w_2}{w_3}\right)\\
  \derpf{y}{x} = \derpf{y}{w_1} \derpf{w_1}{x} = \left(\derpf{y}{w_1}\derpf{w_1}{w_2}\right)\derpf{w_2}{w_3}
\end{align*}
Likewise, automatic differentiation can be carried in a number of semantically
equivalent (but operationally different) ways. The first evaluation order leads
to what is known as forward-mode automatic differentiation, which works by
computing the derivatives of successively larger subexpressions $w_i$ in terms
of the input variable $x$. The second evaluation order, known as reverse-mode
automatic differentiation, proceeds backwards by finding the derivative of the
output $y$ in terms of derivatives of successively smaller subexpressions $w_j$.
The difference between these methods lies in their complexity: if $f :
\mathbb{R}^n \rightarrow \mathbb{R}^m$, forward-mode AD will take time
proportional to $n$ whereas reverse-mode AD will take time proportional to $m$.

A common formulation of forward-mode AD is in terms of \textbf{dual numbers}.
The dual numbers are essentially an extension of the real numbers with an
element $\epsilon$ with the property that $\epsilon \times \epsilon = 0$. The
derivative of the polynomial function $f(x) = x^2 + 3x + 1$ at point $2$ can
then be computed by evaluating $f(x)$ at $2 + \epsilon$:
\begin{align*}
  f(1 + \epsilon) &= (2 + \epsilon)(2 + \epsilon) + 3(2 + \epsilon) + 1\\
                  &= 4 + 4\epsilon + \epsilon^2 + 3\epsilon + 7\\
                  &= 11 + 7\epsilon 
\end{align*}
Note that $11 = f(2)$ and $7 = f'(2)$. In general, if $f(x)$ is a polynomial,
its derivative at $x_0$ is given by $\mathcal{E}(f(x_0 + \epsilon))$, where
$\mathcal{E}(x + y\epsilon) \defeq y$ (the term $\epsilon$ is usually called a
\textit{perturbation} and understood as some kind of infinitesimal). This
approach can be extended to more functions by defining the action of each extra
primitive on dual numbers.

While it seems quite natural to talk about AD using a functional language (after
all, the derivative is a textbook example of a higher-order function), this
approach runs into some subtle issues as soon as we consider the derivative
$\mathcal{D}(f)(x) \defeq \mathcal{E}(x + \epsilon)$ as a first-class operator,
which have been discussed by Siskind and  Pearlmutter \cite{siskind2008nesting}.
The main problem deals with the confusion of different perturbations
corresponding to different (but nested) calls of a differential operator: if
$\mathcal{D}(f)(x) \defeq \mathcal{E}(f(x + \epsilon))$, then
$\mathcal{D}(\lambda x . x \times (\mathcal{D}(\lambda y . x) 2)) 1 = 1$,
whereas we would expect this value to be $0$.

These issues are solved by Siskind and Pearlmutter in their system
$\textsc{Stalin}\nabla$ \cite{siskind2008stalin}, which efficiently compiles a
dialect of $\textsc{Scheme}$ code with a derivative operator $\mathcal{D}$ into
performant code. The machinery involved is, however, rather complicated and
requires a significant amount of introspection in the shape of the
$\texttt{map-closure}$ operator introduced by the authors in
\cite{siskind2007closures}. Furthermore, as noted by Manzyuk
\cite{manzyuk2012simply}, they present no proof that $\textsc{Stalin}\nabla$
correctly handles derivatives of higher-order functions.

It is in an attempt to provide a solid basis for the theory of higher-order
automatic differentiation that Oleksandr Manzyuk defines the perturbative
$\lambda$-calculus \cite{manzyuk2012simply}, a language closely resembling the
\dlc but based on categories with tangent structure, rather than differential
$\lambda$-categories. In the perturbative $\lambda$-calculus, the derivative of
a function is computed by a \textit{pushforward} operator that, given some
function $f : A \rightarrow B$, produces a function $Tf : A \times A \rightarrow
B \times B$ which is precisely the functorial action of the tangent bundle
functor on some differential $\lambda$-category.

While Manzyuk's calculus is sound with respect to differential
$\lambda$-categories, and thus theoretically sound, it suffers from the flaw of
being non-confluent \cite{manzyuk2012blog}. We believe Manzyuk's general idea to
be solid, but the flaws in his calculus must be addressed before any further
work can take place. If this can be done, however, two interesting directions
for further research await: extending the calculus with general recursion to
give a more practical perspective, and providing a similar foundation for
reverse-mode automatic differentiaiton. 

This last development would be particularly exciting: reverse-mode AD subsumes
backpropagation \cite{pearlmutter2008reverse}, which is the most common method
of training a neural network \cite{rumelhart1988learning}. A higher-order
calculus featuring AD could then constitute a basis for the efficient, modular
development of neural networks, as it has been postulated by researchers in the
field \cite{blogprog,blogdiff}.

On the other hand, Cai and Giarrusso's theory of change structures has recently
been suggested by Kelly, Pearlmutter and Siskind \cite{kelly2016evolving} as a
candidate for both a semantic setting in which to reason about higher-order
automatic differentiation and an executable calculus that can be of practical
use in end-to-end differentiable programming. Kelly et al. contend that it
should be possible to understand the sets of changes $\Delta_v$ as power series
on some distinguished variable $\varepsilon$, which is then truncated into an
infinitesimal perturbation by letting $\epsilon^2 = 0$. 

From a geometric perspective, when interpreted in this way a change structure
becomes a representation for a geometric space, with $\Delta_v$ being an
analogue for the tangent space at a point $v$, and the syntactic
$\mathit{Derive}$ transformation from \cite[Figure 4]{cai2014changes} is the
chain rule in meaning as well as in form. The operator $\ominus$ is the only
obstacle to this reading of change structures -- given two points $u, v$ in a
manifold\footnote{
A manifold in a setting that admits infinitesimal elements, such as synthetic
differential geometry\cite{Kock06,Kock09}.
}, it may well be the case that there is no infinitesimal change that
can transform $u$ into $v$, and so there is no candidate for $v \ominus u$.

%% file: change-actions.tex

\label{chp:change-actions}
In this chapter we introduce the central notion of this thesis: that of a change
action. This definition was originally proposed as an attempt to formalise
incremental computation of higher order programs based on Yufei Cai and Paolo
Giarrusso's notion of a \emph{change structure}, while addressing some of the
issues with their original formulation.

This chapter is divided in three parts. In Section~\ref{sec:change-actions-set} we
introduce the basic definitions of change action and derivative of a function
for the case when the underlying category is $\cat{Set}$. We propose two
different categories of change actions, one where morphisms are maps for which a
derivative exist and another where maps are pairs of a function together with a
derivative for it and we show ways to construct change actions that give rise to
a series of adjunctions between the category of sets and the category of change
actions.

In Section~\ref{sec:change-actions-arbitrary} we generalise the previous
definition of a category of change actions to a functor that maps an arbitrary
Cartesian category to a corresponding category of change actions internal to it.
We show that this category has products and coproducts and we discuss some
important subcategories.

Finally, Section~\ref{sec:change-actions-as-categories} shows that change
actions can be understood entirely as a particularly well-behaved sort of
internal categories. Differential maps correspond precisely to functors, thus
justifying and motivating our definition.

\section{Change Actions in \texorpdfstring{$\cat{Set}$}{Set}}
\label{sec:change-actions-set}

In the simplest possible terms, a change action is merely a monoid acting on a
set. This is a rather unimpressive definition - indeed, as we shall see, the
main interest of change actions is not the change actions themselves but the
associated idea of differentiable maps.

\begin{definition}
  A \textbf{change action} is a tuple $\cact{A} =
	(\us{A}, \changes{A}, \oplus_A, +_A, 0_A)$ where $\us{A}$ and $\changes{A}$
	are sets, $(\changes{A}, +_A, 0_A)$ is a monoid, and $\oplus_A : \us{A} \times
	\changes{A} \ra \us{A}$ is an action of the monoid on $\us{A}$.
	
	We omit the subscript from $\oplus_A, +_A$ and $0_A$ whenever it can be
	deduced from the context. We will often refer to the set $\changes{A}$ as the
	\textbf{change set} of $\cact{A}$, and to the set $A$ as its \textbf{base set}
	or \textbf{underlying set}.
\end{definition}

\begin{remark}
  Change actions are closely related to the notion of \emph{change structures}
  introduced in \cite{cai2014changes} but differ from the latter in neither
  being dependently typed nor assuming the existence of an $\ominus$ operator.
  The importance of this will become apparent later in
  Chapter~\ref{chp:change-action-models} when we introduce important families of
  change actions where no such operators exist (for example, this is true of
  change actions derived from differential categories and 
  Kleene algebras, see Sections~
  \ref{sec:change-action-models:examples-differential} and~
  \ref{sec:change-action-models:examples-kleene}).

  On the other hand, change actions require the change set $\changes{A}$ to have
  a monoid structure compatible with the map $\oplus$, hence neither notion is
  strictly a generalisation of the other. Whenever one has a change structure,
  however, one can easily obtain a change action by taking $\changes{A}$ to be
  the free monoid generated by the set of changes of the original change
  structure.
\end{remark}

Even if it is straightforward, some intuition for the above definition is in
order. One should understand the set $\changes A$ as a set of \emph{changes}
that can be applied to objects of type $A$. For example, the base space could
be the set of all text files, with elements of $\changes A$ being \emph{diffs}.
Or $A$ could be the space of all database configurations, with elements of
$\changes A$ representing possible updates to the database. In
Chapter~\ref{chp:incremental-computation} we will work with a change action
where elements of $A$ are sets of facts and changes $\delta \in \changes A$
are sets of new facts to be added and old facts to be discarded.

\begin{figure}
  \begin{center}
    \begin{tikzcd}
      x_i \arrow[rr, "f"] \arrow[ddd, >=stealth, bend right, "\delta x_i"{name=Left}, dashed]
      &&y_i \arrow[ddd, bend left, >=stealth, swap, "\delta y_i"{name=Right}, dashed]
      \arrow[from=Left,to=Right,"\d f", Rightarrow]
      \\
      \\
      \\
      x_{i+1} \arrow[rr, "f"]&& y_{i+1}
    \end{tikzcd}
  \end{center}
  \caption{Incremental computation in the abstract}
  \label{fig:incremental-computation}
\end{figure}
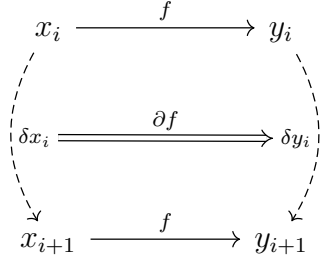

The connection with incremental computation is evident: the problem of
incrementalising some function $f : A \to B$, assuming change actions $\cact{A}$
and $\cact{B}$, we can formalise the problem of incrementalising $f$ by assuming
an input $x_i$ that results from the application of successive changes $\delta
x_i \in \Delta A$ to an initial value $x_0$. An incremental version of $f$ would
then somehow transform these into changes to the output $\delta y_i \in \Delta
B$ in such a way that $f(x_{i + 1})$ can be obtained by updating $f(x_{i})$ with
$\delta y_i$, as shown in Figure~\ref{fig:incremental-computation}.

A change action can also be understood in a more geometric way, by thinking of a
change $\delta a$ as a directed path (or, more properly, a family of paths, one
for every base point) in $A$, with the point $a \oplus_A \delta a$ corresponding
to the end-point of the path of $\delta a$ starting at point $a$. Representing
elements of $a$ as points and changes as arrows, we can picture the condition
that $\oplus_A$ be an action by the ``commutative diagram'' below\footnote{
Diagrams of this sort are of immense help when reasoning about
change actions -- as we shall see, their use is justified by the double nature
of change actions as categories.
}:

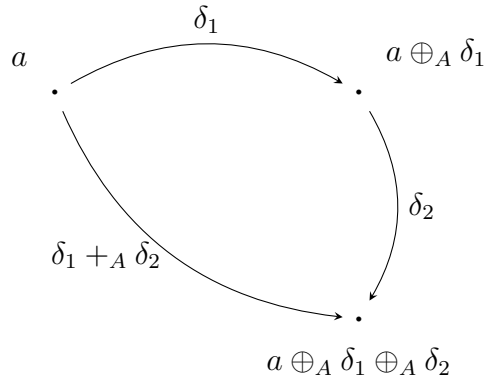
\begin{figure}[H]
  \begin{center}
    \begin{tikzpicture}
      \node[label=135:{$a$}] (a) at (0, 0) {$\bdot$};
      \node[label=45:{$a \oplus_A \delta_1$}] (a1) at (4, 0) {$\bdot$};
      \node[label=-90:{$a \oplus_A \delta_1 \oplus_A \delta_2$}] (a2) at (4, -3)
      {$\bdot$};
      \draw[->, bend left, >=stealth] (a) to node[above] {$\delta_1$} (a1);
      \draw[->, bend left, >=stealth] (a1) to node[right] {$\delta_2$} (a2);
      \draw[->, bend right, >=stealth] (a) to node[left] {$\delta_1 +_A \delta_2$} (a2);
    \end{tikzpicture}
  \end{center}
  \label{fig:oplus-action}
  \caption{The action property of $\oplus_A$}
\end{figure}

Between any two change actions, there is an obvious notion of
``structure-preserving'' map which is just that of a homomorphism of the
corresponding multi-sorted algebras. That is to say, a homomorphism from
$\cact{A}$ to $\cact{B}$ would be a function $f : A \to B$ and a monoid
homomorphism $\Delta f : \Delta A \to \Delta B$ with the property that $f(x
\oplus \delta x) = f(x) \oplus \Delta f(\delta x)$. 

It is a natural question whether such a $\Delta f$ gives us the right notion for
an ``incremental'' version of $f$ -- but it turns out to be much too stringent,
as it requires that the ``incremental'' version of $f$, $\Delta f$, does not
depend on the current ``state'' $x_i$ but merely on the incoming update $\delta
x_i$. In practice this does not seem to be the case (see
Chapter~\ref{chp:incremental-computation} for a relevant counterexample).

Instead of working with homomorphisms of change actions, we will work with a
much weaker notion, that of a \emph{derivative}, and the associated notion of a
\emph{differentiable} function.

\begin{definition}
  Let $\cact{A}$ and $\cact{B}$ be change
  actions. A function $f : \us{A} \ra \us{B}$ is \textbf{differentiable} if
  there is a function $\d f : \us{A} \times \changes{A} \ra \changes{B}$
  satisfying
  \(
    f(a \oplus_A \change{a}) = f(a) \oplus_B \d f (a, \change{a}),
  \)
  for all $a \in \us{A}, \change{a} \in \changes{A}$. We call $\d f$ a
  \textbf{derivative} for $f$, and write $f : \cact{A} \ra \cact{B}$ whenever $f$
  is differentiable.
\end{definition}

The nomenclature may seem overly ambitious, but it is nothing new: the Datalog
literature already discusses derivatives of programs for semi-naive evaluation
\cite{bancilhon1986amateur,bancilhon1986naive}, a kind of incremental
computation which we will explore in depth in
Chapter~\ref{chp:incremental-computation}, and this was also the name chosen by
Cai and Giarrusso in their work on change structures \cite{cai2014changes}. We
will later provide more instances which show that the name ``derivative'' is
warranted.

The most immediate (and most prominent) parallel with derivatives as they are
defined in calculus or differential geometry, and a vital result about change
actions, is that derivatives compose via the familiar chain rule.

\begin{theorem}[Chain rule] Given differentiable functions $f : \cact{A} \ra
  \cact{B}$ and $g : \cact{B} \ra \cact{C}$ with derivatives $\d f$ and $\d g$
  respectively, the function $\d (g \circ f) : \us{A} \times \changes{A} \to
  \changes{C}$ defined by
  \(
  \d(g \circ f)(a, \change a) \defeq \d g(f(a), \d f(a, \change a))
  \)
  is a derivative for $g \circ f : \us A \to \us C$.
\end{theorem}
\begin{proof}
  Unpacking the definition, we have
  \begin{salign*}
    (g \circ f)(a) \oplus_C \d (g \circ f)(a, \change a)
    &= g(f(a)) \oplus_C \d g(f(a), \d f(a, \change a))\\
    &= g(f(a) \oplus_B \d f(a, \change a))\\
    &= g(f(a \oplus_A \change a)) \qedhere
  \end{salign*}
\end{proof}

\begin{example}
  \label{ex:monoidal-change-action}
  Whenever $(A, +, 0)$ is a monoid, the action of this monoid on itself defines
  a change action $(A, A, +, +, 0)$ which we call the \textbf{monoidal} action.
  
  Of particular importance is the case when the underlying monoid is the set of
  natural numbers with addition. We denote this change action by
  $\cact{\NN} \defeq (\NN, \NN, +, +, 0)$. A function $f : \NN \ra \NN$ is
  differentiable according to $\cact{\NN}$ if and only if it is monotone.
  
  When the monoid $(A, +, 0)$ is in fact a group, every function $f : B \ra A$
  is differentiable, with exactly one derivative given by $\d f(x, u) \defeq
  -f(x) + f(x \oplus_B u)$. 
\end{example}

\begin{example}
  \label{ex:discrete-change-action}
  For any set $A$, there is a \textbf{discrete} change action $\cact{A}_{\star}
  \defeq (A, \{\star\}, \pi_1, \pi_1, \star)$ whose change set $\Delta A$ is the
  one-element set $\{\star\}$ and whose action ``does nothing''.
 
  Given any other change action $\cact{B}$, every function $f : A \ra B$ is
  differentiable as a function from $\cact{A}_\star$ into $\cact{B}$ with
  derivative $\d f(a, \star) \defeq 0_B$, which trivially satisfies the
  derivative condition.
\end{example}

\begin{example}
  \label{ex:functional-change-action}
  Let $A \Ra B$ denote the set of functions from $A$ from $B$, and $\arr{ev}_{A,
  B} : A \times (A \Ra B) \ra B$ be the usual evaluation map. Then
  $\cact{A}_{\Ra} \defeq (A, A \Ra A, \arr{ev}_{A,A}, \circ, \Id{A})$ is a
  change action.

  More generally, whenever a family of functions $U \subseteq (A \Ra A)$
  contains the identity and is closed under composition, $(A, U, \arr{ev}_{A, A}
  \restriction_{A \times U}, \circ \restriction_{U \times U}, \Id{U})$ is a
  change action. 
\end{example}

The derivative condition can also be represented graphically: a differentiable
function is precisely one that, fixing a point of origin $a$, preserves
``paths'' (i.e. changes) that start at $a$.

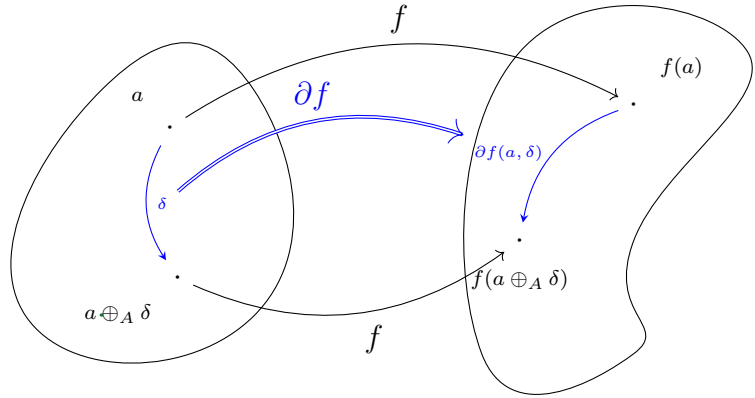
\begin{figure}[h]
  \begin{center}
  \begin{tikzpicture}
    \node[color=green] (Origin) at (0, 0) {$\cdot$};

    \draw[smooth cycle,tension=1.] plot coordinates{(-1, 0)(2, 0)(2, 3)(0, 3)};
    \draw[smooth cycle,tension=.9] plot coordinates{(5, -.5)(7,
    -.5)(7,1)(8.5,3.5)(5.5,3.5)};

    \node[label=135:{\scriptsize$a$}] (A) at (0.9, 2.5) {$\cdot$};
    \node[label=45:{\scriptsize$f(a)$}] (B) at (7.0, 2.8) {$\cdot$};

    \node[label=-135:{\scriptsize$a \oplus_A \delta$}] (A') at (1, 0.5) {$\cdot$};
    \node[label=-90:{\scriptsize$f(a \oplus_A \delta)$}] (B') at (5.5, 1.0) {$\cdot$};

    \draw (A) edge[->, bend left, "$f$"] (B);
    \draw (A') edge[->, bend right, "$f$", swap] (B');
    \draw[blue] (A) edge[->,>=stealth, bend right] node[right] (da) {\tiny$\delta$} (A');
    \draw[blue] (B) edge[->,>=stealth, bend right] node[left] (db) {\tiny$\d
    f(a, \delta)$}(B');
    \draw[blue] (da) edge[->, double, bend left] node[above] {$\d f$} (db);
  \end{tikzpicture}
  \end{center}
  \caption{A derivative acting on changes}
  \label{fig:coderivative}
\end{figure}

\subsection{Regular derivatives}

The preceding definitions neither assume nor guarantee a derivative to be
additive (i.e. they may not satisfy $\d f(x, \changes{a} + \changes{b}) = \d
f(x, \changes{a}) + \d f(x, \changes{b})$), as they are in standard differential
calculus. A strictly weaker condition that we will now require is
\emph{regularity}: if a derivative is additive in its second argument then it is
regular, but not vice versa. As we shall show later, the converse is true under
certain conditions.

\begin{definition}\label{def:regular}
  Given a differentiable map $f : \cact{A} \ra \cact{B}$, a derivative $\d f$
  for $f$ is \textbf{regular} if, for all $a \in \us{A}$ and $\change{a},
  \change{b} \in \changes{A}$, we have:
  \begin{align}
     \d f(a, 0) &= 0\\
     \d f(a, \change{a} + \change{b}) &= \d f(a, \change{a}) + \d f(a \oplus \change{a}, \change{b})
  \end{align}
\end{definition}

While we will often just require the regularity conditions to hold, one can show
that they follow from moderately weak premises. Indeed, it should be remarked
that it is always possible to ``pretend'' that a derivative is regular. That is
to say:
\begin{salign*}
  f(a) \oplus \left[\d f(a, \delta_1) a + \d f(a \oplus \delta_1, \delta_2) \right]
  &= f(a) \oplus \d f(a, \delta_1) \oplus \d f(a \oplus \delta_1, \delta_2)\\
  &= f(a \oplus \delta_1) \oplus \d f(a, \oplus \delta_1, \delta_2)\\
  &= f(a \oplus \delta_1 \oplus \delta_2)\\
  &= f(a \oplus (\delta_1 + \delta_2))\\
  &= f(a) \oplus \d f(a, \delta_1 + \delta_2)
\end{salign*}

That is to say, $\d f(a, \delta_1) + \d f(a \oplus \delta_1, \delta_2)$ has the
same effect on $f(a)$ as $\d f(a, \delta_1 + \delta_2)$. In any context where
derivatives are unique, then, it is provable that derivatives are necessarily
regular. It turns out there is a simple criterion to determine when this is the
case.

\begin{proposition}
  If $f : \cact{A} \ra \cact{B}$ is differentiable and has a unique derivative
  $\d f$ then this derivative is regular.
\end{proposition}
\begin{proof}
  When $A$ is empty, the property follows trivially. Otherwise, if $a \in \us{A}$ 
  and $\change{a}, \change{b} \in \changes{A}$, we have:
  \begin{salign*}
    f(a \oplus (\change{a} + \change{b}))
    &= f(a \oplus \change{a} \oplus \change{b})
    \\
    &= f(a \oplus \change{a}) \oplus \d f(a \oplus \change{a}, \change{b})
    \\
    &= \big(f(a) \oplus \d f(a, \change{a})\big) \oplus \d f(a \oplus \change{a}, \change{b})
    \\
    &= f(a) \oplus \big(\d f(a, \change{a}) + \d f(a \oplus \change{a}, \change{b})\big)
  \end{salign*}
  Thus we can define the following derivative for $f$
  $$
  \d f_a(x, \change{x}) \defeq
  \begin{cases}
    \d f(a, \change{a}) + \d f(a \oplus \change{a}, \change{b})
    &\text{ when $x = a, \change{x} = \change{a} + \change{b}$}\\
    \d f(x, \change{x})
    &\text{ otherwise}
  \end{cases}
  $$
  Since the derivative is unique, it must be the case that $\d f = \d f_a$ and
  therefore $\d f(a, \change{a} + \change{b}) = \d f(a, \change{a}) + \d f(a
  \oplus \change{a}, \change{b})$. By a similar argument, $\d f(a, 0) = 0$ and
  thus $\d f$ is regular.
\end{proof}

\begin{remark}\label{rem:reg-der-not-nec}
  One may wonder whether every differentiable function admits a regular
  derivative; the answer is no. Consider the change actions:
  \begin{align*}
    \cact{A_1} &= (\ZZ_2, \ZZ_2, +, +, 0)\qquad
    \cact{A_2} = (\ZZ_2, \NN, \left[ + \right], +, 0)
  \end{align*}
  where $\left[ m \right]\left[ + \right] n = \left[ m + n \right]$. The
  identity function $\Id{} : \cact{A_1} \ra \cact{A_2}$ admits infinitely many
  derivatives, none of which are regular. The characterisation of differentiable
  functions admitting a regular derivative is an open problem.
\end{remark}

\begin{proposition}
  Given $f : \cact{A} \ra \cact{B}$ and $g : \cact{B} \ra \cact{C}$ with regular
  derivatives $\d f$ and $\d g$ respectively, the derivative $\d (g \circ f)
  \defeq \d g \circ \pair{f \circ \pi_1}{\d f}$ is regular.
\end{proposition}
\begin{proof}
  Suppose $\d f$, $\d g$ are regular derivatives. It is straightforward to check
  that $\d (g \circ f)$ preserves the zero change:
  \begin{gather*}
    (\d g \circ \pair{f \circ \pi_1}{\d f})(a, 0)
    = \d g(f(a), \d f(a, 0))
    = \d g(f(a), 0)
    = 0
  \end{gather*}
  
  Now consider arbitrary changes $\delta_1, \delta_2 \in \Delta A$:
  \[
    \begin{array}{llr}
      & \d (g \circ f)(a, \delta_1 + \delta_2)\\
      = & (\d g \circ \pair{f \circ \pi_1}{\d f})(a, \delta_1 + \delta_2)\\ 
      = & \d g(f(a), \d f(a, \delta_1 + \delta_2))\\
      =& \d g(f(a), \d f(a, \delta_1) + \d f(a \oplus\delta_1, \delta_2))\\
      =& \d g(f(a), \d f(a, \delta_1))
         + \d g(f(a) \oplus \d f(a, \delta_1), \d f(a\oplus\delta_1, \delta_2))\\
      =& (\d g \circ \pair{f \circ \pi_1}{\d f})(a, \delta_1)
         + (\d f \circ \pair{f\circ \pi_1}{\d f})(a \oplus \delta_1, \delta_2)\\
      =& \d (g \circ f)(a, \delta_1) + \d (g \circ f)(a \oplus \delta_1, \delta_2) 
         &\qedhere
    \end{array}
  \]
\end{proof}

\begin{remark}
  If one thinks of changes (i.e.~elements of $\changes A$) as morphisms between
  elements of $\us A$, then the condition of regularity remarkably resembles
  functoriality. This intuition is explored in
  Section~\ref{sec:change-actions-as-categories}, where we show that categories
  of change actions organise themselves into 2-categories.
\end{remark}

\subsection{Two categories of change actions}
\label{sec:two-categories}

The study of change actions can be undertaken in two ways. First, one can
consider functions that are differentiable, without choosing a particular
derivative. Alternatively, the derivative itself can be considered part of the
morphism. The first option leads to the category $\cat{CAct^-}$, whose objects
are change actions and morphisms are all the differentiable maps\footnote{
  We have found the study of $\cat{CAct^-}$ to be laborious and not particularly
  enlightening, so it appears here only as a historical note. The reader in a rush is
  encouraged to skip to Definition~\ref{def:category-cact}.
}.

The category $\cat{CAct^-}$ was the category we originally proposed in
\cite{mario2019fixing}. It is well-behaved, possessing limits, colimits, and
exponentials, which is a trivial corollary of Theorem~\ref{thm:cact-preord} below.

  \newcommand{\reachOrder}[0]{\mathrel{\sqsubseteq_\oplus}}
\begin{theorem}
  \label{thm:cact-preord}
  The category $\cat{CAct^-}$ of change actions and differentiable morphisms is
  equivalent to $\cat{PreOrd}$, the category of preorders and monotone maps.
\end{theorem}

To prove this, we introduce the notion of the reachability order generated by a
change action $\cact{A}$.
\begin{definition}
  For $a, b \in \us A$, we define the \textbf{reachability preorder} $a
  \reachOrder b$ iff there is a $\change{a} \in \changes{A}$ such that $a \oplus
  \change{a} = b$. Then $\reachOrder$ defines a preorder on $\us A$.
\end{definition}

The intuitive significance of the reachability preorder induced by $\cact{A}$ is
that it contains all the information about differentiability of functions from
or into $A$. This is made precise in the following result:

\begin{lemma}
  A function $f : \us A \ra \us B$ is differentiable as a function from
  $\cact{A}$ to $\cact{B}$ iff it is monotone with respect to the reachability
  preorders on $\cact A$ and $\cact B$.
\end{lemma}
\begin{proof}
  Let $f$ be a differentiable function, with $\d f$ an arbitrary derivative, and
  suppose $a \reachOrder a'$. Hence there is some $\change a$ such that $a
  \oplus \change a = a'$. Then, by the derivative condition, we have $f(a') =
  f(a \oplus \change a) = f(a) \oplus \d f (a, \change a)$, hence $f(a)
  \reachOrder f(a')$.

  Conversely, suppose $f$ is monotone, and pick arbitrary $a \in A, \change a
  \in \changes A$. Since $a \reachOrder a \oplus \change a$ and $f$ is
  monotone, we have $f(a) \reachOrder f(a \oplus \change a)$ and therefore
  there exists a $\change b \in \changes B$ such that $f(a) \oplus \change b =
  f(a \oplus \change a)$. We define $\d f(a, \changes a)$ to be precisely such a
  change $\change b$ (note that the process of arbitrarily picking a $\change b$
  for every pair $a, \change a$ makes use of the Axiom of Choice).
\end{proof}

The correspondence between a change action and its reachability preorder gives
rise to a full and faithful functor $\ftor{Reach} : \cat{CAct^-} \ra
\cat{PreOrd}$ that acts as the identity on morphisms.

Conversely, any preorder $\leq$ on some set $\us A$ induces an equivalent
change action: indeed, let $(A \Ra_\leq A)$ denote the set of all
$\leq$-monotone functions from $A$ into itself. Since monotone functions
are closed under composition, we define the change action $\cact{A_\leq}$
as in Example~\ref{ex:functional-change-action} by setting
\[ 
  \cact{A_\leq} 
  \defeq (\us A, A \Ra_\leq A, \arr{ev}_{A,A}, \circ, \Id{A})
\] 

It is immediately apparent that the reachability order in $\cact{A_\leq}$ is
precisely the preorder $\leq$ (since $b$ is reachable from $a$ if and only if
there is some monotone function satisfying $f(a) = b$). Thus this correspondence
defines another full and faithful functor $\ftor{Act} : \cat{PreOrd} \ra
\cat{CAct^-}$ that is the identity on morphisms.

It remains to check that there exists a pair of natural isomorphisms $\mathcal{U} :
\ftor{Act} \circ{} \ftor{Reach} \ra \Id{\cat{CAct^-}}$ and $\mathcal{V} :
\ftor{Reach} \circ{} \ftor{Act} \ra \Id{\cat{PreOrd}}$. But these are trivial:
it suffices to set $\mathcal{U}_A = \Id{A}$ and $\mathcal{V}_{(\us A, \leq)} =
\Id{(\us A, \leq)}$. Hence $\ftor{Act}, \ftor{Reach}$ establish an equivalence
of categories between $\cat{PreOrd}$ and $\cat{CAct^-}$.

The explicit structure of the limits and colimits in $\cat{CAct^-}$ is, however,
not so satisfactory. One can, for example, obtain the limit of a certain diagram
in $\cat{CAct^-}$ by taking its limit \emph{qua} diagram in $\cat{PreOrd}$
 and turning it into a change action, but the corresponding change set
is not, in general, easily expressible in terms of those for $\cact{A}$
and $\cact{B}$. Even then, some limits constructed in this fashion do admit a 
``sensible'' presentation. For example, the previous construction suggests that the
product $\cact{A} \times \cact{B}$ should be given by $(A \times B, A\times B \Ra_\leq 
A\times B, \arr{ev}, \circ, \Id{A \times B})$, but the space of monotone maps 
$A \times B \Ra_\leq A \times B$ can be thought of as (a quotient of) the set of
arbitrary maps $A \times B \Ra \Delta A \times \Delta B$, by encoding a monotone map
by the change that it induces at each point.

Despite the above, derivatives of structure morphisms in $\cat{CAct^-}$ can still be hard
to obtain, as exhibiting e.g. $\pair{f}{g}$ as a morphism in $\cat{CAct^-}$ merely proves
it is differentiable but gives no clue as to how a derivative might be constructed. 
One way to understand this problem is that the change set $\Delta A$ is a set of
``witnesses'' for the reachability relation in that each $\delta$ witnesses the
property that $a \reachOrder a \oplus \delta$. Differentiable maps, however, are
monotone but do not provide a witness for this property; a choice of witness for
the monotonicity of a map would form precisely a derivative for it.

A more constructive approach is to take a morphism to be a function together
with a choice of a (regular) particular derivative for it. This has the
advantage that a choice of a morphism automatically determines a choice of
derivative.

\begin{definition}
  Given change actions $\cact{A}$ and $\cact{B}$, a \textbf{differential map}
  $\cact{f} : \cact{A} \ra \cact{B}$ is a pair $(\us{f}, \d f)$ where $\us{f} :
  \us{A} \ra \us{B}$ is an arbitrary function (which we refer to as the
  \textbf{underlying map} of $\cact f$) and $\d f : \us{A} \times \changes A \ra
  \changes B$ is a \emph{regular} derivative for $\us{f}$. 
\end{definition}

\begin{definition}
  \label{def:category-cact}
  The category $\cat{CAct}$ has change actions as objects and differential maps
  as morphisms. The identity morphisms are defined by $\cact{\Id{A}} \defeq
  (\Id{A}, \pi_1)$. Given morphisms $\cact{f} : \cact{A} \to \cact{B}$ and
  $\cact g : \cact B \to \cact C$, their composite $\cact{g} \circ \cact{f}$ is
  the differential map whose underlying map is $g \circ f$ along with the
  derivative $\d g \circ \pair{f \circ \pi_1}{\d f}$. That is:
  \[
    \cact g \circ \cact f \defeq (g \circ f, \d g \circ \pair{f \circ \pi_1}{\d f})
  \]
\end{definition}

It is trivial to show that pre- and post-composition with the maps
$\cact{\Id{A}}$ is indeed the identity, but the fact that composition in
$\cat{CAct}$ is associative might not be so clear. This is a consequence of the
fact that the chain rule itself is associative, which will be proven in more
generality in the following section (see
Proposition~\ref{prop:composition-associative}).

Finite products and coproducts exist in $\cat{CAct}$: see
Theorems~\ref{thm:cact-products} and~\ref{thm:cact-coproducts} for a more
general statement. However, unlike in $\cat{CAct^-}$, it is not clear whether
general limits and colimits exist.

\subsection{Adjunctions With \texorpdfstring{$\cat{Set}$}{Set}}

There is an obvious forgetful functor $\ftor{U} : \cat{CAct} \ra \cat{Set}$
that maps every change action $\cact{A}$ to its underlying set $\us{A}$ and
every differential map $\cact{f} : \cact{A} \ra \cact{B}$ to the function on the
underlying sets $\us f : \us A \ra \us B$.

Recall from Section~\ref{sec:two-categories} that a change action $\cact{A}$
induces a preorder $\reachOrder$ on $A$. Then we can define a quotient functor
$\ftor{Q} : \cat{CAct} \ra \cat{Set}$ that maps the change action $\cact{A}$ to
the set $\us{A} /{\sim}$, where $\sim$ is the transitive and symmetric closure
of $\reachOrder$. The action on morphisms $\cact{f} : \cact{A} \ra \cact{B}$ is
defined by $\ftor{Q}(\cact{f})(\left[ a \right]) = \left[\us{f}(a) \right]$,
where $[a]$ is the $\sim$-equivalence class of $a$. 

Finally, there is a functor $\ftor{D} : \cat{Set} \ra \cat{CAct}$ that maps
every set $A$ to the discrete change action $\cact{A}_\star \equiv (A, \{\star\},
\pi_1, \pi_1, \star)$ (where $\{\star\}$ denotes the set with a single element
$\star$). This functor $\ftor{D}$ sends every function $f$ to the
differential map $(f, !)$. Notice that both compositions $\ftor{Q} \circ
\ftor{D}$ and $\ftor{U} \circ \ftor{D}$ are in fact equal to the
identity endofunctor $\ftor{Id}_{\cat{Set}}$.

In what follows we will make use of the fact that
$\ftor{U} \circ \ftor{D}
= \ftor{Q} \circ \ftor{D}
= \ftor{Id}_{\cat{Set}}$.

\begin{lemma}
  \label{lem:forgetful-functor-is-right-adjoint}
  The forgetful functor $\ftor{U}$ is right-adjoint to the functor
  $\ftor{D}$
\end{lemma}
\begin{proof}
  Define the unit and counit as:
  \begin{align*}
    \epsilon &: \ftor{D} \circ \ftor{U} \Ra \ftor{Id}_{\cat{CAct}}\\
    \epsilon_{\cact{A}} &\defeq (\Id{A}, 0)\\
    \eta &: \ftor{Id}_{\cat{Set}} \Ra \ftor{U} \circ \ftor{D}
    \cong \ftor{Id}_{\cat{Set}}\\
    \eta_A &\defeq \cact{\Id{A}}
  \end{align*}
  We can check that the above definitions satisfy the zig-zag equations by
  simple calculation:
  {\small
  \begin{gather*}
    \epsilon_\ftor{D} \circ \ftor{D}(\eta)
    = (\Id{}, 0) \circ (\eta, !)
    = (\Id{} \circ \eta, 0)
    = (\Id{} \circ \Id{}, 0)
    = \Id{}\\
    \ftor{U}(\epsilon) \circ \eta_\ftor{U}
    = \ftor{U}(\Id{}, 0) \circ \Id{}
    = \Id{} \circ \Id{}
    = \Id{} \qedhere
  \end{gather*}
  }
\end{proof}

\begin{lemma}
  The functor $\ftor{D}$ is right adjoint to the quotient functor
  $\ftor{Q}$.
\end{lemma}
\begin{proof}
  Define the unit and counit as:
  \begin{align*}
    \epsilon &: \ftor{Id}_{\cat{Set}} \cong \ftor{Q} \circ \ftor{D}
    \ra \ftor{Id}_{\cat{Set}}\\
    \epsilon_A &\defeq \A{Id}_A\\
    \eta &: \ftor{Id}_{\cat{CAct}} \ra \ftor{D} \circ \ftor{Q}\\
    \eta_{\cact{A}} &\defeq (\left[ \A{Id}_{\cact{A}} \right], !)
  \end{align*}
  where $\left[ \A{Id}_{\cact{A}} \right]$ is the map that sends an element $a$ in
  $A$ to the equivalence class $\left[ a \right]$ of $a$ modulo $\sim_\leq$.
  Note that $\eta$ is well-defined since whenever $a \oplus \change a = b$ it
  is the case that $\left[ a \right] = \left[ b \right]$. The zig-zag equations
  follow by similar calculations as Lemma~\ref{lem:forgetful-functor-is-right-adjoint}.
\end{proof}

For the next result, we assume the Axiom of Choice -- or rather, we assume its
equivalent formulation that there is a map $\ftor{G}$ that sends each non-empty set $A$
to some group $\ftor{G}_A$ whose underlying set is $A$.

This map can be extended to a functor $\ftor{G} : \cat{Set} \ra \cat{CAct}$
which sends every (non-empty) set to the monoidal change action $(\ftor{G}_A,
\ftor{G}_A, +, +, 0)$, the empty set to the trivial change action 
$(\emptyset, \lbrace \star \rbrace, !, \pi_1, \star)$, and every 
function $f : A \ra B$ to the differential
map $(f, \d^1 f)$, with $\d^1f (x, \change x) = -f(x) + f(x + \change x)$. 
It is easy to see that $\ftor{U} \circ \ftor{G} = \ftor{Id}_{\cat{Set}}$,
and that $\ftor{G}$ is full and faithful.

\begin{lemma}
  \label{lem:set-to-cact}
  The functor $\ftor{G}$ is a right adjoint to the forgetful
  functor $\ftor{U}$.
\end{lemma}
\begin{proof}
  Using the hom-set isomorphism definition of adjunction, the result is an
  immediate consequence from the fact that every function into a change action
  of the form $\ftor{G}(A)$ has one and only one derivative. Thus there is a one-to-one
  correspondence between differential maps $ \cact{f} : \cact{A} \to
  \ftor{G}(B)$ and (arbitrary) functions $f : \ftor{U}(\cact{A}) \to B$.
\end{proof}

A version of this functor $\ftor{G}$ which acts directly on groups (rather than sets)
will become the key to interpreting the calculus of finite differences in the setting
of change actions, see Section~\ref{sec:discrete-calculus}.

\section{Change actions in arbitrary categories}

\label{sec:change-actions-arbitrary}
The definition of change actions makes no use of any properties of $\cat{Set}$
beyond the existence of products. Indeed, change actions can be characterised as
a multi-sorted algebra, which is definable in any category with
products (although we emphasise again that differential maps are \emph{not} the
same as homomorphisms of change actions \emph{qua} algebras).

\begin{definition}
  Given an ambient category $\cat{C}$ equipped
  with all finite products, a $\cat{C}$-\textbf{change action} is a tuple
  $\cact{A} = (\us{A}, \changes{A}, \oplus_A, +_A, 0_A)$ where 
  \begin{itemize}
  \item $A$ is an object in $\cat{C}$.
  \item $(\changes{A}, +_A, 0_A)$ is a monoid internal to $\cat{C}$.
  \item $\oplus_A : A \times \changes{A} \ra A$ is an action of $\changes{A}$ on
    $A$, i.e. it is a $\cat{C}$-morphism such that the following diagrams
    commute:
    \begin{center}
    \begin{tikzcd}
      \us A \arrow[rr, "\pair{\Id{}}{0_A}"] \arrow[rrd,swap, "\Id{}"]
      && \us A \times \changes{A} \arrow[d, "\oplus_A"]\\
      && \us A
    \end{tikzcd}\qquad
    \begin{tikzcd}
      \us A \times \changes{A} \times \changes{A}
      \arrow[d, swap, "\pair{\Id{}}{+_A}"] 
      \arrow[r, "\pair{\oplus_A}{\Id{}}"]
      & \us A \times \changes{A}
      \arrow[d, "\oplus_A"]\\
      \us A \times \changes{A}
      \arrow[r, swap, "\oplus_A"] & \us A
    \end{tikzcd}
    \end{center}
  \end{itemize}
\end{definition}

\begin{notation}
  For the sake of clarity, when composing with one of the ``operator'' maps
  $\oplus_A, +_A$ we will write $f +_A g$ for $+_A \circ \pair{f}{g}$ and $f \oplus_A
  g$ for $\oplus_A \circ \pair{f}{g}$. Both of these operators bind less tightly
  than function composition, that is, $f \circ g +_A h$ should be read as $(f
  \circ g) +_A h$. When there is no risk of confusion, we will abuse the notation
  and write $0_A : U \ra \changes{A}$ instead of $0_A \circ{}!$.
\end{notation}

\begin{definition}
  \label{def:derivative}
  Given two $\cat{C}$-change actions $\cact{A}, \cact{B}$ and a $\cat{C}$-map $f
  : A \ra B$, a \textbf{derivative} for $f$ is a $\cat{C}$-map $\d f : A \times
  \changes A \ra \changes B$ such that the following diagrams commute:
  \begin{center}
    \begin{tikzcd}
      \us A \times \changes{A}
      \arrow[rr, "\pair{\us f \circ \pi_1}{\d f}"]
      \arrow[d, swap, "\oplus_A"] 
      && \us B \times \changes{B} \arrow[d, "\oplus_B"]\\
      \us A \arrow[rr, swap, "\us{f}"] && \us B
    \end{tikzcd}
    \qquad
    \begin{tikzcd}
      \us A \arrow[d, swap, "\pair{\Id{}}{0_A \circ{} !}"] \arrow[r, "!"] & 1 \arrow[d, "0_B"]\\
      \us A \times \changes{A} \arrow[r, swap, "\d f"] & \changes{B}
    \end{tikzcd}
  \end{center}
  \begin{center}
    \begin{tikzcd}
      \us A \times (\changes{A} \times \changes{A})
      \arrow[rr, "\pair{\pair{\pi_1}{\pi_1 \circ \pi_2}}{\A{a}}"]
      \arrow[dd, "\d f \circ (\Id{} \times +_A)", swap]
      && (\us A \times \changes{A}) \times ((\us A \times \changes{A}) \times \changes{A})
      \arrow[dd, "\d f \times (\d f \circ (\oplus_A \times \Id{}))"]
      \\\\
      \changes{B}
      &&
      \changes{B} \times \changes{B}
      \arrow[ll, "+_B"]
    \end{tikzcd}
  \end{center}
  The first diagram states the derivative condition, while the other two
  diagrams amount to a diagrammatic version of the regularity of $\d f$. We will
  often refer to these properties independently (that is, we will say that a
  certain morphism ``satisfies the derivative condition'' or ``satisfies the
  regularity conditions'').
\end{definition}

\begin{notation}
  In many definitions and proofs we make statements about equalities that may
  involve some significant ``shuffling around'' of products. As these details
  are inconsequential and can significantly obscure the underlying ideas, we
  will prefer to leave these implicit and work in the internal language of
  Cartesian categories instead.

  For example, and using the more convenient infix notation, the regularity
  condition for derivatives can be rephrased as stating that the equation
  \[
    \d f \circ \pair{a}{\delta_1 + \delta_2} = (\d f \circ \pair{a}{\delta_1})
    +_B (\d f \circ \pair{a \oplus_A \delta_1}{\delta_2})
  \]
  holds for any choice of $a : U \ra A, \delta_1, \delta_2 : U \ra \Delta A$,
  instead of a much more unwieldy presentation using projections.
\end{notation}

\begin{lemma}[Chain rule] Given differential maps $(f, \d f) : \cact{A} \ra
  \cact{B}, (g, \d g) : \cact{B} \ra \cact{C}$, the pair $(g \circ f, \d g \circ
  \pair{f \circ \pi_1}{\d g})$ is a differential map from $\cact{A}$ into
  $\cact{C}$.
\end{lemma}

The chain rule can be elegantly summarised by pasting together two instances of
the diagram in Definition~\ref{def:derivative}:
\begin{center}
  \begin{tikzcd}
    \us{A} \times \changes{A}
    \arrow[d, swap, "\oplus_A"]
    \arrow[rr, "\pair{\us{f} \circ \pi_1}{\d f}"] 
    \arrow[rrrr, bend left,
    "\pair{(\us{g} \circ \us{f}) \circ \pi_1}{\d g \circ \pair{\us{f} \circ \pi_1}{\d f}}"]
    &&
    \us B \times \changes{B} \arrow[d, "\oplus_B"]
    \arrow[rr, "\pair{\us{g} \circ \pi_1}{\d \us{g}}"] &&
    \us C \times \changes{C}
    \arrow[d, "\oplus_C"] \\
    \us A \arrow[rr, "\us{f}"]
    \arrow[rrrr, bend right, swap, "\us{g} \circ \us{f}"]
    && \us B
    \arrow[rr, "\us{g}"] && \us C
  \end{tikzcd}
\end{center}

\begin{definition}
  A \textbf{differential map} from $\cact{A}$ to $\cact{B}$ is a pair $\cact{f}
  \equiv (f, \d f)$ where:
  \begin{itemize}
    \item $f : A \ra B$ is a $\cat{C}$-map, which we call the \textbf{underlying}
    map of $\cact{f}$.
    \item $\d f : A \times \Delta A \ra \Delta B$ is a derivative for $f$.
  \end{itemize}
  Given differential maps $\cact{f}: \cact{A} \ra \cact{B}, \cact{g} : \cact{B}
  \ra \cact{C}$, we define their composition $\cact{g} \circ \cact{f}$ to be the
  differential map whose underlying map is $g \circ f$ and whose derivative is
  the derivative obtained by applying the chain rule to $\partial g, \partial
  f$. That is:
  \begin{gather*}
    \cact{g} \circ \cact{f} \defeq (g \circ f, \d g \circ \pair{f \circ \pi_1}{\d f})
  \end{gather*}
\end{definition}

\begin{proposition}
  \label{prop:composition-associative}
  Composition of differential maps is associative. That is, given $\cact{f},
  \cact{g}, \cact{h}$, we have $(\cact{h} \circ \cact{g}) \circ \cact{f} =
  \cact{h} \circ (\cact{g} \circ \cact{f})$. Furthermore, the identity
  differential maps $\cact{\Id{}}_{\cact{A}} \defeq (\Id{A}, \pi_2)$ are
  identities for the composition operation.
\end{proposition}
\begin{proof}
  The second property is trivial. We prove associativity; for this, it suffices
  to show that the derivative of both composites is the same:
  \begin{salign*}
    \d ((\cact{h} \circ \cact{g}) \circ \cact{f})
    &= \d (\cact{h} \circ \cact{g}) \circ \pair{f \circ \pi_1}{\d f}\\
    &= \d \cact{h} \circ \pair{g \circ \pi_1}{\d g} \circ \pair{f \circ \pi_1}{\d f}\\
    &= \d \cact{h} \circ \pair{g \circ f \circ \pi_1}{\d g \circ \pair{f \circ \pi_1}{\d f}}\\
    &= \d \cact{h} \circ \pair{g \circ f \circ \pi_1}{\d g \circ \pair{f \circ \pi_1}{\d f}}\\
    &= \d \cact{h} \circ \pair{(g \circ f) \circ \pi_1}{\d (\cact{g} \circ \cact{f})}\\
    &= \d (\cact{h} \circ (\cact{g} \circ \cact{f})) \qedhere
  \end{salign*}
\end{proof}

\subsection{The Category \texorpdfstring{$\ftor{CAct}(\cat{C})$}{CAct(C)}}

Given any object $\cat{C}$ in $\cart$, the category of (small) Cartesian
categories and strong monoidal functors, it is straightforward enough to
define a category $\ftor{CAct}(\cat{C})$ of change actions and differential maps
internal to $\cat{C}$, in much the same way that we defined the category
$\cat{CAct}$ in the previous section.

This construction also allows one to lift any (strong monidal) functor
$\ftor{F} : \cat{C} \ra \cat{D}$ into a functor $\ftor{CAct}(\ftor{F}) :
\ftor{CAct}(\cat{C}) \ra \ftor{CAct}(\cat{D})$. Thus we can define a functor
$\ftor{CAct} : \cart \ra \cat{Cat}$ (as we shall later see, the category
$\ftor{CAct}(\cat{C})$ always has products and $\ftor{CAct}(-)$ is in fact an
endofunctor on $\cart$).

\begin{definition}
  \label{def:cact}
  The objects of $\ftor{CAct}(\cat{C})$ are all the change actions $\cact{A}$
  internal to $\cat{C}$.

  Given objects $\cact{A}, \cact{B}$ in $\ftor{CAct}(\cat{C})$, the morphisms of
  $\ftor{CAct}(\cact{A}, \cact{B})$ are differential maps $\cact{f} : \cact{A}
  \ra \cact{B}$.
  Given a functor $\ftor{F} : \cat{C} \ra \cat{D}$ and isomorphisms
  $\varepsilon : \terminal_\cat{D} \to \ftor{F}(\terminal_\cat{C}), \mu_{A,B} :
  F(A) \times F(B) \to F(A \times B)$, the functor
  $\ftor{CAct}(\ftor{F}) : \ftor{CAct}(\cat{C}) \ra \ftor{CAct}(\cat{D})$ is defined
  by:
  \begin{align*}
    \ftor{CAct}(\ftor{F})(\cact{A} )
    &\defeq (\ftor{F}(\us{A}), \ftor{F}(\changes{A}), \ftor{F}(\oplus_A) \circ \mu^{-1}_{A,\Delta A}, \ftor{F}(+_A) \circ \mu^{-1}_{\Delta A, \Delta A}, \ftor{F}(0_A) \circ \epsilon^{-1})\\
    \ftor{CAct}(\ftor{F})(\cact{f}) 
    &\defeq (\ftor{F}(\us {f}), \ftor{F}(\d f) \circ \mu^{-1}_{A, \Delta A})
  \end{align*}
\end{definition}

As is the case for $\cat{Set}$, the category $\ftor{CAct}(\cat{C})$ has
enough information to ``encode'' the original objects and maps of $\cat{C}$,
which embed into $\ftor{CAct}$ via the trivial change action.

\begin{theorem}
  The category $\cat{C}$ embeds fully and faithfully into the category
  $\ftor{CAct}(\cat{C})$ via the (strong monoidal) functor $(-)_\star$ which
  sends an object $A$ of $\cat{C}$ to the trivial change action $\cact{A}_\star =
  (A, \terminal, \pi_1, !, !)$ and every morphism $f : A \ra B$ of $\cat{C}$ to
  the differential map $(f, !)$.
  
  Furthermore, the family of (strong monoidal) functors $(-)_\star : \cat{C} \ra
  \ftor{CAct}(\cat{C})$ defines a natural transformation from the identity
  endofunctor $\ftor{Id}_{\cart}$ into the functor $\ftor{CAct}$.
\end{theorem}

Additionally, there is an obvious forgetful functor $\epsilon_{\cat{C}} :
\ftor{CAct}(\cat{C}) \ra \cat{C}$, which defines the components of a natural
transformation $\epsilon$ from the functor $\ftor{CAct}$ to the identity
endofunctor $\Id{}$.

\begin{theorem}
  The trivial change action embedding $(-)_\star : \cat{C} \ra
  \ftor{CAct}(\cat{C})$ is left-adjoint to the forgetful functor $\epsilon :
  \ftor{CAct}(\cat{C}) \ra \cat{C}$.
\end{theorem}

Given $\cat{C}$, we write $\xi_{\cat{C}}$ for the functor
$\ftor{CAct}(\epsilon_{\cat{C}}) : \ftor{CAct}(\ftor{CAct}(\cat{C})) \ra
\ftor{CAct}(\cat{C})$\footnote{One might expect $\ftor{CAct}$ to be a comonad
with $\epsilon$ as a counit. But if this were the case, we would have
$\xi_{\cat{C}} = \epsilon_{\ftor{CAct}(\cat{C})}$, which is, in general, not
true.}. Explicitly, this functor maps an object $(\cact{A}, \cact{B},
\cact{\oplus}, \cact{+}, \cact{0})$ in $\ftor{CAct}(\ftor{CAct}(\cat{C}))$ to
the object $(\us{A}, \us{B}, \us\oplus, \us+, \us 0)$. Intuitively,
$\epsilon_{\ftor{CAct}(\cat{C})}$ prefers the ``original'' structure on objects,
whereas $\xi_{\cat{C}}$ prefers the ``higher'' structure. The equaliser of these
two functors is precisely the category of change actions whose higher structure
is the original structure.

This forgetful functor $\epsilon$ is evidently not faithful, as a $\cat{C}$-map
may admit more than one derivative and so be the image of more than one
differential map. There is, however, a way to embed the category of change
actions faithfully back into $\cat{C}$.

\begin{definition}
  The \textbf{tangent bundle} functor $\ftor{T} : \ftor{CAct}(\cat{C}) \ra
  \cat{C}$ maps a change action $\cat{A}$ to the $\cat{C}$-object $A \times
  \Delta A$, and a differential map $\cact{f} : \cact{A} \ra \cact{B}$ to the
  $\cat{C}$-map $\pair{f \circ \pi_1}{\d f} : A \times \Delta A \ra B \times
  \Delta B$
\end{definition}

\begin{remark}
  So far we have not shown that products exist in $\ftor{CAct}(\cat{C})$, hence
  all of the functors we have defined above should be read as arrows in
  $\cat{Cat}$, not in $\cart$. As we shall show, $\ftor{CAct}(\cat{C})$ does
  have finite products, and it will be the case that all the previous functors are
  indeed product-preserving.
\end{remark}

An interesting property of the category of change actions is that, while
regularity is a much weaker notion than additivity, the category of monoids
internal to $\cat{C}$ can nonetheless be embedded fully and faithfully into the
category $\ftor{CAct}(\cat{C})$ of change actions - that is, monoid
homomorphisms correspond precisely to differentiable maps between carefully
chosen change actions.

\begin{theorem}
  The category $\ftor{Mon}(\cat{C})$ whose objects are monoids internal to
  $\cat{C}$ and whose morphisms are monoid homomorphisms embeds fully and
  faithfully into the category $\ftor{CAct}(\cat{C})$ via the functor
  $\mathcal{M} : \ftor{Mon}(\cat{C}) \ra \ftor{CAct}(\cat{C})$ defined as
  follows:
  \begin{align*}
    \mathcal{M}(A, +_A, 0_A) &\defeq (\terminal, A, !, +_A, 0_A)\\
    \mathcal{M}(f : A \ra B) &\defeq (!, f \circ \pi_2)
  \end{align*}
\end{theorem}
\begin{proof}
  First, we show the functor $\mathcal{M}$ is well-defined. Clearly it maps
  monoids to change actions, and $\mathcal{M}(f)$ trivially satisfies the
  derivative condition: it remains to show that its derivative is regular. This
  is an immediate consequence of $f$ being a monoid homomorphism:
  \begin{salign*}
    f \circ \pi_2 \circ \pair{!}{\delta_1 +_A \delta_2} 
    &= f \circ (\delta_1 +_A \delta_2)\\
    &= (f \circ \delta_1) +_B (f \circ \delta_2)\\
    &= (f \circ \pi_2 \circ \pair{!}{\delta_1}) +_B (f \circ \pi_2 \circ \pair{!}{\delta_2})
  \end{salign*}
  $\mathcal{M}(\Id{A})=(\Id{\terminal}, \Id{A} \circ \pi_2) = (\Id{\terminal},
  \pi_2)= \cact{\Id{}}_{\mathcal{M}A}$, therefore $\mathcal{M}$ preserves the
  identity. It also preserves composition since:
  \begin{salign*}
    \mathcal{M}(g) \circ \mathcal{M}(f)
    &= (\Id{\terminal}, g \circ \pi_2) \circ (\Id{\terminal}, f \circ \pi_2)\\
    &= (\Id{\terminal}, g \circ \pi_2 \circ \pair{\Id{\terminal} \circ \pi_1}{f \circ \pi_2})\\
    &= (\Id{\terminal}, g \circ f \circ \pi_2)\\
    &= \mathcal{M}(g \circ f)
  \end{salign*}
  
  It is evident that $\mathcal{M}$ is faithful. To see that it is full, consider
  a differential map $\cact{f} : \mathcal{M}(A, +_A, 0_A) \ra \mathcal{M}(B,
  +_B, 0_B)$. Clearly the underlying map is just the terminal map $f ={}!$.
  Similarly since $\terminal \times A \cong A$ the derivative $\d f$ has type
  $\d f : \terminal \times A \ra B$. It remains to show that $\d f \circ
  (\cong)$ is a monoid homomorphism and therefore $(\Id{\terminal}, \d f) =
  \cact{f} = \mathcal{M}(\partial f \circ (\cong))$, but this follows
  immediately from regularity of $\d f$.
\end{proof}

\subsection{Products and Coproducts in \texorpdfstring{$\ftor{CAct}(\cat{C})$}{CAct(C)}}

We have defined $\ftor{CAct}$ as a functor from $\cart$ into $\cat{Cat}$. This
can in fact be strengthened, as $\ftor{CAct}(\cat{C})$ inherits finite products
from the underlying category $\cat{C}$.

\begin{theorem}
  \label{thm:cact-products}
  Let $\cact{A} = (\us A, \changes A, \oplus_A, +_A, 0_A)$ and $\cact{B} = (\us
  B, \changes B, \oplus_B, +_B, 0_B)$ be change actions on $\cat{C}$. Then the
  following change action is the product of $\cact{A}$ and $\cact{B}$ in
  $\ftor{CAct}(\cat{C})$
  \begin{gather*}
    \cact{A} \times \cact{B} \defeq 
    (\us A \times \us B, \changes A \times \changes B, \oplus_{A \times B}, +_{A \times B}, \pair{0_A}{0_B})
  \end{gather*}
  where
  \begin{align*}
    (a, b) \oplus_{A \times B} (\delta a, \delta b)
    &\defeq (a \oplus_A \delta_1, b \oplus_B \delta_2)\\
    (\delta a_1, \delta b_1) +_{A \times B} (\delta a_2, \delta b_2)
    &\defeq (\delta a_1 +_A \delta a_2, \delta b_1 +_B \delta b_2)
  \end{align*}
  The projection maps are
  \begin{align*}
    \cact{\pi_1} &\defeq (\pi_1, \pi_1 \circ \pi_2)\\
    \cact{\pi_2} &\defeq (\pi_2, \pi_2 \circ \pi_2)
  \end{align*}
\end{theorem}
\begin{proof}
  Given any pair of differential maps $\cact{f_1} : \cact C \ra \cact A, \cact{f_2} : \cact C \ra 
  \cact B$, define
  \begin{gather*}
    \pair{\cact{f_1}}{\cact{f_2}} \defeq (\pair{\us{f_1}}{\us{f_2}},
    \pair
    {\d f_1 \circ \pair{\pi_1 \circ \pi_1}{\pi_1 \circ \pi_2}}
    {\d f_2 \circ \pair{\pi_2 \circ \pi_1}{\pi_2 \circ \pi_2}})
  \end{gather*}
  Then $\cact{\pi_i} \circ \pair{f_1}{f_2} = f_i$. Furthermore, given any map
  $\cact{h} : \cact C \ra \cact A \times \cact B$ whose projections coincide
  with $\pair{\cact{f_1}}{\cact{f_2}}$, by applying the universal property of
  the product in $\cat{C}$, we obtain $h = \pair{\cact{f_1}}{\cact{f_2}}$.
\end{proof}

\begin{theorem}
  \label{cact-terminal}
  The change action $\cact{\terminal} = (\terminal, \terminal, \pi_1, \pi_1,
  \Id{\terminal})$ is the terminal object in $\ftor{CAct}(\cat{C})$, where
  $\terminal$ is the terminal object of $\cat{C}$. Furthermore, if $\cact A$ is
  a change action every point $\us{f} : \terminal \ra \us{A}$ in $\cat{C}$ is
  differentiable, with (unique) derivative $0_A$.
\end{theorem}
\begin{proof}
  Given a change action $\cact A$, there is exactly one differential map
  $\cact{!} \defeq ({!}, {!}) : \cact A \ra \cact{\terminal}$. Now given a
  differential map $(\us f, \d f) : \cact{\terminal} \ra \cact A$, applying
  regularity we obtain:
  \begin{gather*}
    \d f = \d f \circ \pair{\Id{\terminal}}{\Id{\terminal}}
         = \d f \circ \pair{\Id{\terminal}}{0_\terminal}
         = 0_A
         \qedhere
  \end{gather*}
\end{proof}

It is a well-known result from differential calculus that a multivariate
function is (continuously) differentiable if and only if all its partial
derivatives exist and are continuous. It is perhaps surprising that a similar
result holds for change actions, given an adequate definition of a partial
derivative.

\begin{definition}\label{def:partial}
  Given a $\cat{C}$-map $f : A \times B \ra C$ and fixed change actions
  $\cact{A}, \cact{B}, \cact{C}$, a \textbf{partial derivative with respect to
  $A$} for $f$ is a map $\d_1 f : (A \times B) \times \changes A \ra \Delta C$
  such that the following three conditions hold:
  
  \begin{itemize}
    \item $f (a, b) \oplus_C \partial_1 f((a, b), \delta a)
        = f (a \oplus_A \delta a, b)$
    \item $\d_1 f ((a, b), 0_A) = 0_C$
    \item $\d_1 f ((a, b), \delta a_1 +_A \delta a_2) = \d_1 f((a, b), \delta
    a_1) +_C \d_2 f((a \oplus_A \delta a_1, b), \delta a_2)$
  \end{itemize}
  
  In the same manner, we write $\d_2 f$ for the partial derivative of $f$ with
  respect to $B$, with the obvious definition.
  
  In a simple, set-theoretic setting, a partial derivative for $f$ with respect
  to $A$ is precisely a function mapping each $b \in A$ to a derivative for the
  partially applied function $f(-, b) : A \ra C$.
\end{definition}

\begin{lemma}
  Let $f : A \times B \ra C$ be a $\cat{C}$-map and consider fixed change
  actions $\cact{A}, \cact{B}, \cact{C}$, with $+_A$ commutative. then $f$
  admits a derivative if and only if $f$ admits partial derivatives with respect
  to its first and second arguments.
\end{lemma}
\begin{proof}
  Suppose $\d f$ is a derivative for $f$. Then we define:
  \[ \d_1 f \defeq \d f \circ \pair{\pi_1}{\pair{\pi_2}{0_B}} 
    : (A \times B) \times \Delta A \ra \Delta C \] Clearly $\d_1 f$ is a first
  partial derivative for $f$. We explicitly check the first condition:
  \begin{salign*}
    f (a, b) \oplus_C (\d_1 f((a, b), \delta a))
    = f(a, b) \oplus_C \d f((a, b), (\delta a, 0_B))
    = f(a \oplus_A \delta a, b)
  \end{salign*}
  
  Conversely, suppose $\d_1 f, \d_2 f$ are partial derivatives for $f$. Then the
  following $\d f$ is a derivative for $f$:
  \[
    \d f ((a, b), (\delta a, \delta b)) \defeq 
    \d_1 f((a, b), \delta a) +_C \d_2 f((a \oplus_A \delta a, b), \delta b)
  \]
  We can easily check that the above verifies the derivative condition:
  \begin{salign*}
    f (a, b) \oplus_C \d f ((a, b), (\delta a, \delta b))
    &= f (a, b) \oplus_C \brak{ \d_1 f((a, b), \delta a) 
      +_C \d_2 f((a \oplus_A \delta a, b), \delta b) }\\
    &= f(a \oplus \delta a, b) \oplus_C \d_2 f((a \oplus \delta a, b), \delta b)\\
    &= f(a \oplus \delta a, b \oplus \delta b)
  \end{salign*}
  Regularity follows from straightforward calculation and commutativity of $+_C$.
\end{proof}

Furthermore, as a direct consequence of $\d f$ being regular, this process is
invertible. That is, whenever $\d f$ is a derivative for $f$, the derivative
obtained by ``adding together'' the partial derivatives $\d_1 f, \d_2 f$ induced
by $\d f$ is precisely itself (and vice versa).
\begin{corollary}\label{cor:partial-derivatives}
    Let $\cact f : \cact A \times \cact B \ra \cact C$ be a differential map,
    and let $\d_1 f, \d_2 f$ denote the partial derivatives for $f$ computed as
    in the above result. Then $\d f ((a, b), (\delta a, \delta b)) = 
    (\d_1 f ((a, b), \delta a) ) +_C (\d_2 f ((a \oplus_A \delta a, b), \delta b))$
\end{corollary} 

When applying derivatives to incremental computation, we will sometimes want to 
incrementalise a multivariate function only with respect to one of its arguments (for
example, because the other remains fixed, as in Section \ref{sec:fixpoints}). The
previous results guarantee that this sort of differentiation with respect to one 
variable only is nothing more and nothing less than our usual notion of 
differentiation. 

Whenever $\cat{C}$ has distributive coproducts, then these are also inherited by
$\ftor{CAct}(\cat{C})$, although in this case the structure is not as obvious:
one might expect that the change space of $\cact{A} \coprod \cact{B}$ would be
the coproduct $\Delta A \coprod \Delta B$. This is, however, not the case: given
the definition of a change action, a change $\delta : \Delta (\cact{A} \coprod
\cact{B})$ needs to be applicable to elements of the form $\inj_1 a$ as well
as elements of the form $\inj_2 b$. The natural choice for such a $\delta$ is in
fact a pair $(\delta a, \delta b) : \Delta A \times \Delta B$, which acts as
$\delta a$ on objects of the form $\inj_1 a$ and as $\delta b$ on objects of the
form $\inj_2 b$.

The category $\ftor{CAct}(\cat{C})$ also inherits coproducts and the initial
element from the base category: whenever $\cat{C}$ has (distributive)
coproducts, then so does $\ftor{CAct}(\cat{C})$. 

\begin{theorem}
  \label{thm:cact-coproducts}
  If $\cat{C}$ is distributive, with distributive law $\delta_{A, B, C} : (A
  \coprod B) \times C \ra (A \times C) \coprod (B \times C)$, the following change
  action is the coproduct of $A$ and $B$ in $\ftor{CAct}(\cat{C})$
  \begin{gather*}
    A \coprod B \defeq (\us A \coprod \us B, \changes A \times \changes B, 
    \oplus_{A \coprod B}, +_{A \coprod B}, \pair{0_A}{0_B})
  \end{gather*}
  where 
  \begin{align*}
    \oplus_{A \coprod B} 
    &\defeq \left[ 
      \oplus_A \circ (\Id{A} \times \pi_1), \oplus_B \circ (\Id{B} \times \pi_2) 
      \right] \circ \delta_{A, B, C}\\
    +_{A \coprod B} &\defeq \langle +_A \circ (\pi_1 \times \pi_1), +_B \circ (\pi_2 \times \pi_2)\rangle
  \end{align*}
  The injections are
  \( \overline{\inj_1} = (\inj_1, \pair{\pi_2}{0_B})
  \) 
  and
  \(\overline{\inj_2} = (\inj_2, \pair{0_A}{\pi_2})
  \).  
\end{theorem}
\begin{proof}
  Given any pair of differential maps $f_1 : A \ra C, f_2 : B \ra C$ define:
  \begin{align*}
    \left[ \cact {f_1}, \cact{f_2} \right] &\defeq (\left[ \us{f_1}, \us{f_2} \right], \d h)\\
    \d h &\defeq \left[ \d f_1 \circ (\A \times \pi_1), \d f_2 \circ (\Id{B} \times \pi_2) \right] 
           \circ \delta_{A, B, C}
  \end{align*}
  where $\delta_{A, B, C} : (A \coprod B) \times C \ra (A \times C) \coprod (B \times C)$ is the 
  distributive law of $\cat{C}$.

  We check that, indeed, the relevant diagram commutes since:
  \[
  \begin{array}{ll}
    &\left[ f_1, f_2 \right] \circ \overline{\inj_1}\\
    =& (\left[ \us{f_1}, \us{f_2} \right], \d h) \circ (\inj_1, \pair{\pi_2}{0_B})\\
    =& (\left[ \us{f_1}, \us{f_2} \right] \circ \inj_1,
      \d h \circ \pair{\inj_1 \circ \pi_1}{\pair{\pi_2}{0_B}})\\
    =& (\us{f_1},
      \left[ \d f_1 \circ (\A \times \pi_1), \d f_2 \circ (\Id{B} \times \pi_2) \right]
      \circ \delta_{A, B, C} \circ \pair{\inj_1 \circ \pi_1}{\pair{\pi_2}{0_B}})\\
    =& (\us{f_1},
      \d f_1 \circ (\A \times \pi_1) \circ \pair{\pi_1}{\pair{\pi_2}{0_B}})\\
    =& (\us{f_1},
      \d f_1 \circ \pair{\pi_1}{\pi_2}) \\
    =& (\us{f_1}, \d f_1)\\
    =& f_1
  \end{array}
  \]
  The universal property of the coproduct in $\cat{C}$ entails that if $\cact{h}
  = (\us{h}, \d h)$ is such that $\cact{h} \circ \overline{\inj_i} = f_i$, then
  $h = \left[ {f_1}, {f_2} \right]$. Furthermore, since 
  \[
  (\us A \coprod \us B) \times \changes A \times \changes B \ \cong \ (\us A \times \changes A \times 
  \changes B) \coprod (\us B \times \changes A \times \changes B),
  \]
  the universal property of the coproduct also shows that necessarily $\d h = \d
  f$.
\end{proof}

\begin{proposition}
  Whenever $\cat{C}$ has an initial object $\initial$, then the change action
  $\cact{\initial} \defeq (\initial, \terminal, \Id{\initial}, !, !)$ is an
  initial object in $\ftor{CAct}(\cat{C})$.
\end{proposition}

It follows from the results of this section that the category
$\ftor{CAct}(\cat{C})$ has all finite products and coproducts (provided that
$\cat{C}$ has distributive finite coproducts). The existence of more general
limits and colimits is an important open question that we have left for future
work.

\section{Change actions as categories}
\label{sec:change-actions-as-categories}
\newcommand{\incat}[1]{\mathbb{#1}}

Thus far we have developed the theory of change actions without providing much
motivation for our definitions. For example, Definition~\ref{def:derivative}
may seem like an ad-hoc artifice that does not correspond to any natural notion
of ``homomorphism of change actions''. Even if this were the case, we would
argue its the theoretical interest and practical applications suffice to warrant
its study.

No such excuses are necessary, however: as it turns out, our definition of
differential map arises elegantly from the fact that change actions are nothing
but a very special kind of categories, with differential maps corresponding
exactly to functors!

For the sake of convenience, and to fix the notation, we present here the
notions of internal category and internal functor which will appear in this
section. For a more detailed introduction to internal category theory, we refer
the reader to \cite[\S 7]{jacobs1999categorical}, whose definitions we are using
mostly verbatim.

\begin{definition}
  \label{definition:internal-category}
  Given an ambient category $\cat{C}$ equipped with all finite products, a
  \textbf{category internal to} $\cat{C}$ is a tuple $\incat{C} = (C_0, C_1, s,
  t, i, c)$ where:
  \begin{itemize}
    \item $C_0$ and $C_1$ are objects of $\cat{C}$ (the `object of objects' and
    the `object of morphisms' respectively).
    \item `Source' and `target' morphisms $s, t : C_1 \to C_0$ such that the
    following pullbacks exist:
    \begin{center}
      \begin{tikzcd}
        C_2 \defeq C_1 \times_{C_0} C_1
          \arrow[r, "\pi_2"]
          \arrow[d, "\pi_1"{swap}]
        & C_1
          \arrow[d, "s"] \\
        C_1 
          \arrow[r, "t"{swap}] 
        & C_0
      \end{tikzcd}
      \quad
      \begin{tikzcd}
        C_3 \defeq C_1 \times_{C_0} C_1 \times_{C_0} C_1
          \arrow[r]
          \arrow[d]
        & C_1
          \arrow[d, "s"] \\
        C_2
          \arrow[r, "t \circ \pi_2"{swap}]
        & C_0
      \end{tikzcd}
    \end{center}
    \item An `identity-assigning' morphism $i : C_0 \to C_1$ such that the
    following diagram commutes:
    \begin{center}
      \begin{tikzcd}
        & C_0
          \arrow[d, "i"]
          \arrow[dl, equal]
          \arrow[dr, equal]
        & 
        \\
        C_0
        & C_1
          \arrow[l, "s"]
          \arrow[r, "t"{swap}]
        & C_0
      \end{tikzcd}
    \end{center}
    \item A `composition' morphism $c : C_1 \times_{C_0} C_1 \to C_1$ making the
    following diagrams commute:
    \begin{center}
      \begin{tikzcd}
        C_1
          \arrow[d, "s"{swap}]
        & C_2
          \arrow[l, "\pi_1"{swap}]
          \arrow[r, "\pi_2"]
          \arrow[d, "c"]
        & C_1
          \arrow[d, "t"]
        \\
        C_0
        & C_1
          \arrow[l, "s"]
          \arrow[r, "t"{swap}]
        & C_0
      \end{tikzcd}\quad
    \begin{tikzcd}
      C_3
        \arrow[r, "\Id{} \times c"]
        \arrow[d, "c \times \Id{}"{swap}]
      & C_2
        \arrow[d, "c"]
      \\
      C_2
        \arrow[r, "c"{swap}]
      & C_1
    \end{tikzcd}\quad
    \begin{tikzcd}[column sep=large]
      C_1
        \arrow[r, "\pair{i \circ s}{\Id{}}"]
        \arrow[dr]
      & C_2
        \arrow[d, "c"]
      & C_1
        \arrow[l, "\pair{\Id{}}{i \circ t}"{swap}]
        \arrow[dl]
      \\
      & C_1
      & 
    \end{tikzcd}
    \end{center}
  \end{itemize}
\end{definition}

The aforementioned diagrams encode a point-free version of the usual axioms of
categories. As with change actions, they could also be given in the internal
language of $\cat{C}$. However, this is not as much help as it is in the case of
change actions, since the reliance on pullbacks complicates the expressions
significantly.

\begin{definition}
  \label{definition:internal-functor}
  Given internal categories $\incat{C}$ and $\incat{D}$ in $\cat{C}$, an
  \textbf{internal functor} from $\incat{C}$ to $\incat{D}$ is a pair of
  $\cat{C}$-morphisms, $f_0 : C_0 \ra D_0$ and $f_1 : C_1 \ra D_1$, such that
  all of the following diagrams commute:
  \[
    \begin{tikzcd}
      C_1
        \arrow[r, "f_1"]
        \arrow[d, "s"{swap}]
      & D_1
        \arrow[d, "s'"] \\
      C_0
        \arrow[r, "f_0"{swap}]
      & D_0
    \end{tikzcd}
    \quad
    \begin{tikzcd}
      C_1
        \arrow[r, "f_1"]
        \arrow[d, "t"{swap}]
      & D_1
        \arrow[d, "t'"] \\
      C_0
        \arrow[r, "f_0"{swap}]
      & D_0
    \end{tikzcd}
    \quad
    \begin{tikzcd}
      C_0
        \arrow[r, "f_0"]
        \arrow[d, "i"{swap}]
      & D_0
        \arrow[d, "i'"] \\
      C_1
        \arrow[r, "f_1"{swap}]
      & D_1
    \end{tikzcd}
    \quad
    \begin{tikzcd}
      C_1 \times_{C_0} C_1
        \arrow[r, "{f_1 \times_{C_0} f_1}"]
        \arrow[d, "m"{swap}]
      & D_1 \times_{D_0} D_1
        \arrow[d, "m'"] \\
      C_1
        \arrow[r, "f_1"{swap}]
      & D_1
    \end{tikzcd}
  \]
\end{definition}

These diagrams simply state the usual definition of a functor internally in
$\cat{C}$. It is easy to show that internal functors compose in a strictly
associative manner, and that internal categories and internal functors form a
category, which we denote by $\ftor{Cat}(\cat{C})$

\begin{remark}
  In order to define the category $\ftor{Cat}(\cat{C})$ above, we have not required that 
  all (or indeed any) pullbacks exist in the category $\cat{C}$ (in fact the category 
  $\ftor{Cat}(\cat{C})$ can be defined even when $\cat{C}$ lacks products). For example,
  if $\cat{C}$ only has trivial pullbacks (i.e. pullbacks of the identity
  along itself), the resulting $\ftor{Cat}(\cat{C})$ will
  likewise be made up of trivial internal categories, that is, those for which 
  $C_0 = C_1$ and $s, t, e = \Id{C_0}$.
\end{remark}

The question is, then: what is the relation between $\ftor{Cat}(\cat{C})$ and
$\ftor{CAct}(\cat{C})$? As we have remarked before, the changes in $\Delta A$
can be thought of as ``directed paths'', with the peculiarity that they are
\emph{free-floating}, that is to say, they don't start at a particular point,
but can be freely transported around. Since changes can
be applied to any point, a change $\delta$ represents both a path from $a$ to $a
\oplus \delta$ and a path from $b$ to $b \oplus \delta$ (perhaps a more accurate
description of changes would be to say that a change associates a path to every
point).

Now given a $\cat{C}$-change action $\cact{A} = (A, \Delta A, \oplus_A, +_A,
0_A)$ we can obtain a $\cat{C}$-category by choosing $A$ as the object of
objects, with the object of arrows being given by the product $A \times \Delta
A$. This might seem odd but, as changes are ``free-floating'' arrows, one can
understand this product as the space of changes with a fixed base. The pair $(a,
\delta) : A \times \Delta A$ is then to be interpreted as an arrow from the
``object'' $a$ into the ``object''$a \oplus_A \delta$.

\begin{theorem}
  Whenever $\cat{C}$ is equipped with a choice of finite products, there is a
  full and faithful functor \[\kappa_\cat{C} : \ftor{CAct}(\cat{C}) \ra
  \ftor{Cat}(\cat{C}) \] from the category of $\cat{C}$-change actions and
  differential maps into the category of $\cat{C}$-categories and
  $\cat{C}$-functors.

  Furthermore, whenever product by a constant is injective (that is to say, whenever the
  functor $X \mapsto A \times X$ is injective for all objects in $\cat{C}$) the
  functor $\kappa_\cat{C}$ is in fact an embedding, i.e. it is injective on objects.
\end{theorem}
\begin{proof}
  Let $\cact{A} = (A, \changes{A}, \oplus_A, +_A, 0)$ be a change action.
  Define \begin{align*}
    C_0 &\defeq A \qquad
    & C_1 &\defeq A \times \changes{A}  \\
    s &\defeq \pi_1 : A \times \changes{A} \rightarrow A \qquad
    & t &\defeq \oplus :A \times \changes{A} \ra A \\
    i &\defeq \pair{\Id{A}}{0_A} : A \ra A \times \changes{A}
  \end{align*} To define composition, notice that \[
    \begin{tikzcd}[column sep=large]
      A \times \changes{A} \times \changes{A}
        \arrow[r]
        \arrow[d, "{\oplus_A \times \Id{\Delta{A}}}"{swap}]
      & A \times \changes{A}
        \arrow[d, "t = \oplus_A"]
      \\
      A \times \changes{A}
        \arrow[r, "s = \pi_1"{swap}]
      &
      A
    \end{tikzcd}
  \] is a pullback, which is to say that $A \times \changes{A}
  \times \changes{A}$ is the object of `composable tuples.' We
  thus define composition by \[
    c \defeq 
      A \times \changes{A} \times \changes{A}
        \xrightarrow{\Id{A} \times +_A}
      A \times \changes{A} \] 
  
  It is easy to check that the diagrams in Definition
  \ref{definition:internal-category} commute (notably, the property that
  $\oplus_A$ is an action guarantees that composition is associative), and hence
  $\kappa_C(\cact{A}) \defeq (C_0, C_1, s, t, i, c)$ is a category internal to
  $\cat{C}$.

  Given a differential map $\cact{f} = (f, \d f) : \cact{A} \ra \cact{B}$, we
  claim that the pair $\kappa_{\cat{C}}(f, \d f) \defeq (f, \ftor{T}\cact{f})$
  is an internal functor.  Evidently each component has the right type. The
  source and target maps commute with $\ftor{T}f$, since $\pi_1$ and $\oplus$
  are natural transformations $\ftor{T} \Rightarrow U$.  Finally, the regularity
  conditions ensure that $\kappa_{\cat{C}}(f, \d f)$ preserves identities
  (zeros) and composition (addition).

  It is immediate that $\kappa_{\cat{C}}$ is injective on morphisms. To prove
  that it is surjective is not much harder: consider an internal functor $(f_0,
  f_1) : \kappa_{\cat{C}}\cact{A} \ra \kappa_{\cat{C}}\cact{B}$. Then $\pi_2
  \circ f_1 : A \times \Delta A \ra \Delta B$ is a derivative for $f_0$, hence
  $(f_0, \pi_2 \circ f_1)$ is a differential map, and $T(f_0, \pi_2 \circ f_1) =
  (f_0, f_1)$.
\end{proof}

\begin{remark}
  When working in $\cat{Set}$, the category $\kappa_{}\cact{A}$
  boils down to a particular instance of the Grothendieck construction: indeed, when
  considering the monoid $\Delta A$ as a single-object category, the data in the change
  action $\cact{A}$ can be encoded as a functor $\hat{\oplus_A} : \Delta A \to \cat{Set}$,
  with $A = \hat{\oplus_A}(\star)$ and $a \oplus_A \delta = \hat{\oplus_A}(\delta)(a)$.

  One can then construct the category of elements $\ftor{El}(\oplus_A)$ whose objects
  are pairs $(\star, a)$, with $a$ an element of $\hat{\oplus_A}(\star) \equiv A$
  and where the arrows from $(\star, a)$ to $(\star, b)$ are arrows $\delta$ in the one-object
  category $\Delta A$ satisfying $\hat{\oplus_A}(\delta)(a) = b$. That is to say, the
  arrows from $(\star, a)$ to $(\star, b)$ are precisely changes $\delta$ satisfying
  $a \oplus \delta = b$.
\end{remark}

%% file: incremental-computation.tex
\label{chp:incremental-computation}

\newcommand{\supd}[1]{\d_\uparrow {#1}}
\newcommand{\subd}[1]{\d_\downarrow {#1}}

Equipped with the basic machinery of change actions, we set out to show how it
can be applied to incremental computation, as we originally intended. For this,
our driving example will be the semi-naive (that is to say, incremental)
evaluation of Datalog queries, an area where incremental evaluation has immense
benefits and has been broadly applied. We will use the theory of change actions
and derivatives to derive, in a principled way, an incremental evaluation
strategy for Datalog. While this incremental evaluation procedure is not novel,
the contribution of this chapter lies in providing a solid mathematical
foundation for it, which can easily be generalised or extended with minimal
effort.

This chapter proceeds as follows: in Section~\ref{sec:incremental-1} we give a
brief overview of the evaluation mechanism for recursive Datalog queries, which
will motivate the need for incremental computation.
Section~\ref{sec:change-actions-datalog} introduces a family of convenient
change actions on Boolean algebras which Section~\ref{sec:nonRecursiveDatalog}
uses to build explicit derivatives for the incremental evaluation of the formula
semantics of Datalog. Section~\ref{sec:changesOnFunctions} explores the
possibility and difficulties of building change actions on function spaces.
Finally, Section~\ref{sec:dcpos} uses these ``functional'' change actions to
incrementalise the computation of fixed points and even the least fixed point
operator itself.

Many of the results in this chapter are the result of a collaboration with
Semmle Ltd. and have appeare in print in \cite{mario2019fixing}. Some passages
and figures have been reproduced verbatim with permission.

\section{Incremental Evaluation of Datalog Programs}
\label{sec:incremental-1}

Consider the following classic Datalog program \footnote{The syntax and informal
semantics of Datalog programs are outside the scope of this thesis -- we refer
the reader to \cite[part D]{abiteboul1995foundations} for an introduction to
Datalog.}, computing the transitive closure of an edge relation $e$:
\begin{align*}
  tc(x, y) &\leftarrow e(x, y)\\
  tc(x, y) &\leftarrow e(x, z) \wedge tc(z, y)
\end{align*}
This pattern of recursion appears very often in Datalog programming to compute
e.g. the list of nodes reachable from the root of a tree from the edge relation,
or the set of ancestors of a person when $e(x, y)$ denotes the relation ``$x$ is
the parent of $y$''.

The above Datalog program can be understood as a procedure mapping denotations
of $\sem{e}$ and $\sem{tc}$ for $e$ and $tc$ (given in extensional form as
finite sets of tuples) to a new denotation for $tc$. The semantics of the
relation $tc$ is then the least fixed point of this procedure which, in
accordance with Kleene's fixed point theorem, can be obtained by repeated
iteration starting from the empty relation. Formally, $\sem{tc}$ is the least
fixed point of the iterative procedure in Figure~\ref{fig:iterative-tc}.

\begin{figure}[h]
  \begin{align*}
    \sem{tc}_{0} &\leftarrow \emptyset\\
    \sem{tc}_{i+1} &\leftarrow \sem{e} \cup \big( \sem{e} \circ \sem{tc}_i \big)
  \end{align*}
  \captionof{figure}{Iterative computation of $tc$}
  \label{fig:iterative-tc}
\end{figure}

For example, suppose that $e = \{ (1, 2), (2, 3), (3, 4) \}$. Performing the
above iteration naively yields the following trace:

\begin{center}
    \begin{tabular} {c | p{14em} | p{14em}}
        \Xhline{3\arrayrulewidth}
        Iteration & Newly deduced facts & Accumulated data in $\sem{tc}$ \\
        \hline
        0 & $\emptyset$ & $\emptyset$\\
        1 & $\{ (1, 2), (2, 3), (3, 4) \}$ & $\{ (1, 2), (2, 3), (3, 4) \}$\\
        2 & $\{ (1, 2), (2, 3), (3, 4),$ $(1, 3), (2, 4) \}$ & $\{ (1, 2), (2, 3), (3, 4),$ $(1, 3), (2, 4) \}$\\
        3 & $\{ (1, 2), (2, 3), (3, 4),$ $(1, 3), (2, 4), (1, 4),(1, 4) \}$ & $\{ (1, 2), (2, 3), (3, 4),$ $(1, 3), (2, 4), (1, 4) \}$\\
        4 & (as above) & (as above) \\
        \Xhline{3\arrayrulewidth}
    \end{tabular}
    \captionof{figure}{Execution of a recursive Datalog program}
\end{center}

The process ends after the fourth iteration, when no new information has been
added to $tc$ and therefore a fixed point has been reached.

However, this approach is quite wasteful: for example, the fact $tc(1, 2)$ was
deduced at every single iteration, despite it being known since iteration 1.
More generally, if the relation $e$ has $n$ edges, the last iteration in the
corresponding fixed point calculation will deduce $O(n^2)$ redundant facts.

The standard improvement to this evaluation strategy is known as ``semi-naive'' 
evaluation (see \cite[section 13.1]{abiteboul1995foundations}), wherein
the above program is
transformed into an \emph{incrementalised} version, consisting of two steps:
\begin{itemize}
  \item A \emph{delta} step that computes only the \emph{new} facts at each
    iteration. 
  \item An \emph{accumulator} step that adds together the increments computed by
    the delta rula to obtain the final result.
\end{itemize}

In the transitive closure case the delta step is simple: the only new edges at
iteration $n+1$ are the ones that can be deduced from the transitive edges that
were newly deduced at iteration $n$. So, writing $\Delta \sem{tc}_i$ for the
facts newly deduced at iteration $i$, we can rewrite the procedure in
Figure~\ref{fig:iterative-tc}:
\begin{samepage}
\begin{align*}
  \Delta \sem{tc}_{0} &\leftarrow \sem{e}\\
  \Delta \sem{tc}_{i+1}(x, y) &\leftarrow \sem{e} \circ \Delta \sem{tc}_i\\
  \sem{tc}_{0} &\leftarrow \Delta \sem{tc}_0\\
  \sem{tc}_{i+1} &\leftarrow \sem{tc}_{i} \cup \Delta \sem{tc}_{i+1}
\end{align*}\captionof{figure}{An incrementalised transitive closure program}
\end{samepage}

Note in particular that computing $\Delta \sem{tc}_{i + 1}$ no longer requires
calculating the union of all the previously derived facts.

\begin{center}
  \begin{tabular} {c | p{10em} | p{12em}}
    \Xhline{3\arrayrulewidth}
    Iteration & $\Delta \sem{tc}_i$ & $\sem{tc}_i$ \\
    \hline
    0 & $\{ (1, 2), (2, 3), (3, 4) \}$ & $\{ (1, 2), (2, 3), (3, 4) \}$\\
    1 & $\{ (1, 3), (2, 4) \}$ & $\{ (1, 2), (2, 3), (3, 4),$ $(1, 3), (2, 4) \}$\\
    2 & $\{ (1, 4) \}$ & $\{ (1, 2), (2, 3), (3, 4),$ $(1, 3), (2, 4), (1, 4) \}$\\
    3 & $\{ \}$ & (as above) \\
    \Xhline{3\arrayrulewidth}
  \end{tabular}
  \captionof{figure}{The incrementalised transitive closure in action}
\end{center}
\medskip

The performance of this procedure is much better -- every iteration only adds
$O(n)$ facts, as we have pruned out the redundant ones. The delta transformation
is a specific case of incremental computation: at each stage we only compute
the change of the rule $tc$ given the previous changes to its inputs.

But this delta translation works only for traditional Datalog. It is common to
liberalise the formula syntax with additional features, such as disjunction,
existential quantification, negation, and aggregation\footnote{See, for
  example, LogiQL \cite{logicbloxWebsite,halpin2014logiql},
  Datomic \cite{datomicWebsite},
  Souffle \cite{souffleWebsite,scholz2016fast}, and
  DES \cite{saenz2011deductive}, which support all of the aforementioned
  features and more. We do not here explore supporting extensions to the syntax
  of rule heads, although as long as they allow for a denotational
  semantics in a similar style our techniques should be applicable.}.
This allows one to write programs like the following, which computes whether all the
nodes in a sub-tree (given by the relation $child$) have some property $p$:
\begin{align*}
  treeP(x) &\leftarrow p(x) \wedge \neg \exists y . (child(x,y) \wedge \neg treeP(y))
\end{align*}

The body of this predicate amounts to recursion through a universal
quantifier (encoded as $\neg \exists \neg$). One would like to be able to use the
same semi-naive strategy for this rule too, but the standard definition of
semi-naive transformation is not well defined for the extended program syntax,
and it is unclear how to extend it (and its correctness proof) to handle such
cases, since they hinge on the semantics of a recursive formula to be a
monotone procedure.

It is possible, however, to write a delta program for $treeP$ by hand; indeed,
here is a definition for the delta predicate (the accumulator is as
before):\footnote{This rule should be read as: we can newly deduce that $x$ is
in $treeP$ if $x$ satisfies the predicate, and we have newly deduced that one of
its children is in $treeP$, and we currently believe that all of its children
are in $treeP$.}
\begin{align*}
  \Delta_{i+1}treeP(x) \leftarrow & p(x)\\
  &\wedge \exists y. (child(x, y) \wedge \Delta_i treeP(y))\\
  &\wedge \neg \exists y. (child(x,y) \wedge \neg treeP_i(y))
\end{align*}

The above is a correct delta program: using it to iteratively compute
$\sem{treeP}$ yields the same answer as the non-incremental version. It is not
precise, in that it derives some facts repeatedly -- there is
still some degree of redundancy. In practice, there is often a trade-off between
overly-precise incremental rules (which derive fewer redundant facts but might
be harder to compute) and excessively redundant ones.

Handling extended Datalog is of more than theoretical interest -- part of the
research in this work was undertaken in collaboration with Semmle, a program
analysis company which makes heavy use of a commercial Datalog implementation to
implement large-scale static program analysis
\cite{semmleWebsite,avgustinov2016ql,sereni2008adding,schafer2010type}. Semmle's
implementation includes parity-stratified negation\footnote{Parity-stratified
negation means that recursive calls must appear under an even number of
negations. This ensures that the overall rules remain monotone, so the least
fixed point still exists.}, recursive aggregates \cite{demoor2013aggregates},
and other non-standard features which are not amenable to the traditional
semi-naive approach.

A similar scenario happens when considering \emph{maintenance} of Datalog
programs, that is to say, updating the semantics of a Datalog program when one
of the involved relations changes; for example, updating the value of $tc$ given
a change to $e$. Again, this is a natural fit for incremental computation, and
there are known solutions for traditional Datalog \cite{gupta1993maintaining},
but these break down when the language is extended.

There is a piece of folkloric knowledge in the Datalog community that hints at a
solution: the semi-naive translation of a rule corresponds to the
\emph{derivative} of that rule\cite[section
3.2.2]{bancilhon1986amateur,bancilhon1986naive}. This suggests that the
incremental version of a Datalog program should correspond to its derivative,
given a suitable change action on its semantics.

\section{Change Actions for Datalog}
\label{sec:change-actions-datalog}

For this case study, we aim to flesh out the folkloric idea that incremental
computation behaves like a derivative. We approach this goal in stages: first,
we establish some basic properties of change actions which will be of use later
(paying particular attention to change actions in partial orders and Boolean
algebras). Then, we apply use these constructions to build a change action on a
domain of relations where Datalog can be interpreted and show how to
differentiate (possibly non-monotone) Datalog formulae. Finally, we show how
these derivatives can be used to incrementalise the computation of fixed points.

\subsection{Transitive Change Actions and Difference Operators}

A particularly convenient class of change actions are those which are
\emph{transitive}, that is to say, where every element can be transformed into
any other element via some change (in the language of monoid actions, a change
action is transitive whenever it has a single \emph{orbit}).

\begin{definition}
  A change action $\cact{A}$ is \textbf{transitive} if for any $a, b \in A$,
  there exists a change $\delta \in \changes{A}$ such that $a \oplus \delta =
  b$.
\end{definition}

\begin{example}
    \label{ex:group-action-transitive}
    Whenever $(A, +, 0)$ is a group, then the monoidal change action $(A, A, +,
    +, 0)$ is transitive: given points $a_1, a_2$ in $A$, the change $-a_1 + a_2$ sends
    $a_1$ to $a_2$.
\end{example}

\begin{example}
    \label{ex:functional-action-transitive}
    For any set $A$, the change action $\cact{A}_{\Ra} \defeq (A, A \Ra A,
    \arr{ev}_{A, A}, \circ, \Id{A})$ is transitive: given points $a_1, a_2$,
    the constant function sending every point to $a_2$ in particular sends $a_1$
    to $a_2$ (as do many other changes).
\end{example}

An equivalent, and more useful, characterisation of transitive change actions is
that they are actions that allow for finding the ``difference'' $a \ominus b$
between two points.\footnote{
    Cai and Giarrusso's original definition of change structures was in fact given
    in terms of such operators.
}

\begin{definition}
  A \emph{difference operator} for $\cact{A}$ is a function $\ominus: A \times A
  \rightarrow \changes{A}$ such that $a \oplus (b \ominus a) = b$ for all $a, b
  \in A$.
\end{definition}

\begin{proposition}
  \label{prop:minusDerivatives}
  Given a minus operator $\ominus$, and a function $f$, let
  \begin{displaymath}
    \d{f}_\ominus(a, \delta) \defeq f(a \oplus \delta) \ominus f(a)
  \end{displaymath}
  Then $\d{f}_\ominus$ is a derivative for $f$.
\end{proposition}\rff{does this need a proof?}

\begin{proposition}
  Let $\cact{A}$ be a change action. Then the following are equivalent:
  \begin{enumerate}[\roman*.]
    \item $\cact{A}$ is transitive.
    \item There is a minus operator on $\cact{A}$.
    \item For any change action $\cact{B}$ all functions $f: {B} \rightarrow
    {A}$ are differentiable.
  \end{enumerate}
\end{proposition}
\begin{proof}\leavevmode
    \begin{itemize}
    \item ii. $\Rightarrow$ iii. is a straightforward consequence of
    Proposition~\ref{prop:minusDerivatives}.
    \item iii. $\Ra$ i.:
        Given $a, b$ in $A$, we define the path $f_{ab} : \NN \to A$ by:
        \begin{gather*}
            f_{ab}(n) \defeq \begin{cases}
                a&\text{ if $n = 0$}\\
                b&\text{ otherwise}
            \end{cases}
        \end{gather*}
        By hypothesis, $f_{ab}$ admits a derivative $\d [f_{ab}]$, and therefore
        $b = a \oplus \d[f_{ab}](0, 1)$, thus $\cact{A}$ is transitive.
    \item i. $\Ra$ ii.:
        Let $\cact{A}$ be transitive. We apply the Axiom of Choice to pick a
        $\delta_{ab}$ for every pair $a, b$ of elements of $A$ such that $a
        \oplus \delta_{ab} = b$. Then clearly $b \ominus a \defeq \delta_{ab}$
        is a difference operator. \qedhere
    \end{itemize}
\end{proof}

In practice this last property is of the utmost importance: since we are
concerned with the differentiability of functions, working with transitive
change actions will guarantee that derivatives always exist.

\subsection{Comparing Change Actions}

Much like topological spaces and partial orders, we can compare change actions
on the same base set according to coarseness. This is useful since
differentiability of functions between change actions is characterised entirely
by the coarseness of the actions.

\begin{definition}
  Let $\cact{A}_1$ and $ \cact{A}_2$ be change actions with the same underlying
  set $A$. We say that $\cact{A}_1$ is \textbf{coarser} than $\cact{A}_2$ (or
  that $\cact{A}_2$ is \textbf{finer} than $\cact{A}_1$) whenever for every $a
  \in A$ and change $\change_1 \in \changes{A}_1$, there is a change
  $\change_2 \in \changes{A}_2$ such that $a \oplus_{A_1} \change_1 = a
  \oplus_{A_2} \change_2$.
  
  We will write $\cact{A}_1 \leq \cact{A}_2$ whenever $\cact{A}_1$ is coarser
  than $\cact{A}_2$. If $\cact{A}_1$ is both finer and coarser than
  $\cact{A}_2$, we will say that $\cact{A}_1$ and $\cact{A}_2$ are equivalent.
\end{definition}

The relation $\leq$  defines a preorder (but not a partial order) on the set of
all change actions over a fixed set A. Least and greatest elements exist (up to
equivalence), and correspond respectively to the discrete change action
$\cact{A}_\top$ and any transitive change action, such as the ones in
Example~\ref{ex:group-action-transitive} or
Example~\ref{ex:functional-action-transitive}.

\begin{proposition}
  Let $\cact{A}_2 \leq \cact{A}_1$, $\cact{B}_1 \leq \cact{B}_2$ be change
  actions, and suppose the function $f : A \rightarrow B$ is differentiable as a
  function from $\cact{A}_1$ into $\cact{B}_1$. Then $f$ is differentiable as a
  function from $\cact{A}_2$ into $\cact{B}_2$.
\end{proposition}\rff{Is proof necessary?}

A consequence of this fact is that whenever two change actions are equivalent
they can be used interchangeably without affecting which functions are
differentiable. One last parallel with topology is the following result, which
establishes a simple criterion for when a change action is coarser than another:

\begin{proposition}
  Let $\cact{A}_1, \cact{A}_2$ be change actions on $A$. Then $\cact{A}_1$ is
  coarser than $\cact{A}_2$ if and only if the identity function $\Id{A} : A
  \rightarrow A$ is differentiable from $\cact{A}_1$ to $\cact{A}_2$.
\end{proposition}\rff{is proof necessary?}

The semantic domain of Datalog is a complete Boolean algebra, and so our next
step is to construct a suitable change action for Boolean algebras. Along the
way, we consider change actions over posets, which give us the ability to
entirely characterise the set of derivatives of a given function in terms of the
poset structure. This turns out to be desirable in practice, as it gives one a
broad range of derivatives from which an ``optimal'' one might be selected.

\subsection{Change Actions on Posets}

Ordered sets give us a constrained class of functions: monotone functions. We
can define \emph{ordered} change actions, which are those that are well-behaved
with respect to the order on the underlying set. -- that is to say, an ordered
change action is precisely a change action internal to the category of posets
and monotone functions.

\begin{definition}
  A change action $\cact{A}$ is \textbf{ordered} if
  \begin{itemize}
    \item $A$ and $\changes{A}$ are posets.
    \item $\oplus$ is monotone as a map from $A \times \changes{A} \rightarrow A$
    \item $+$ is monotone as a map from $\changes{A} \times \changes{A} \rightarrow \changes{A}$
  \end{itemize}
\end{definition}

In practice, instead of assuming a partial order on the change set $\Delta A$
and restricting ourselves to ordered change actions, we will often start with a
change action $\cact{A}$ over some poset $A$ and lift the partial order on $A$
to one on $\Delta A$ which is compatible with the change action structure.

\newcommand{\changeOrder}{\leq_\Delta}
\newcommand{\fineOrder}{\leq_D}
\newcommand{\minusOrder}{\leq_\cminus}
\begin{definition}
    Given a change action $\cact{A}$ and a partial order $\leq$ on $A$, we
    define a partial order $\changeOrder$ on $\Delta A$ by
    \[ \delta_1 \changeOrder \delta_2 \quad\Leftrightarrow\quad 
    \forall a \in A.\ a \oplus \delta_1
    \leq a \oplus \delta_2 \]
\end{definition}

\begin{proposition}
  \label{prop:changeOrderOrdered}
  Let $\cact{A}$ be a change action on a set $A$ equipped with a partial order
  $\leq$ such that $\oplus$ is monotone in its first argument. Then $\cact{A}$
  is an ordered change action when $\changes{A}$ is equipped with the partial
  order $\changeOrder$.
\end{proposition}
\begin{proof}
    Monotonicity of $\oplus$ is immediate. It remains to show that $+$ is monotone.
    Suppose $\delta_1 \leq \delta'_1$ and $\delta_2 \leq \delta'_2$. Then, for
    any $a \in A$ we have:
    \begin{align*}
        a \oplus (\delta_1 + \delta_2) &= (a \oplus \delta_1) \oplus \delta_2\\
        &\leq (a \oplus \delta'_1) \oplus \delta_2\\
        &\leq (a \oplus \delta'_1) \oplus \delta'_2\\
        &= a \oplus (\delta'_1 + \delta'_2) \qedhere
    \end{align*}
\end{proof}

In what follows, we will extend the partial order $\changeOrder$ on some change
set $\changes{B}$ pointwise to sets of functions (not necessarily
differentiable) from some $A$ into $\changes{B}$. This pointwise order implies
that the space of valid derivatives of a function is convex, that is to say, all 
functions between any two derivatives for $f$ are themselves derivatives for $f$.

\begin{theorem}[Sandwich lemma]
  \label{theorem:sandwich}
  Let $\cact{A}, \cact{B}$ be change actions, with $\cact{B}$ ordered,
  and let $f: {A} \rightarrow {B}$ admit derivatives
  $\subd{f}, \supd{f}$. Then any $g : A \times \Delta A \to \Delta B$ satisfying
  \[
    \subd{f} \changeOrder g \changeOrder \supd{f}
  \]
  is a derivative for $f$.
\end{theorem}
\begin{proof}
    Take arbitrary $a \in A, \delta \in \Delta A$. By hypothesis we have:
    \[
        \subd{f}(a, \delta) \changeOrder g(a, \delta) \changeOrder \supd{f}(a, \delta)
    \]
    Then the first inequality gives:
    \[
        f(a \oplus \delta) = f(a) \oplus \subd{f}(a, \delta) \leq f(a) \oplus g(a, \delta)
    \]
    Applying the second inequality we obtain:
    \[
        f(a) \oplus g(a, \delta) \leq f(a) \oplus \supd{f}(a, \delta) = f(a \oplus \delta)
    \]
    Hence $f(a) \oplus g(a, \delta) = f(a \oplus \delta)$ so $g$ is a derivative
    for $f$.
\end{proof}

Whenever unique minimal and maximal derivatives exist, the above theorem
suffices to completely characterise all the derivatives for a function as a closed
interval.

\begin{corollary}
\label{theorem:derivativeCharacterization}
  Let $\cact{A}$ and $\cact{B}$ be change actions, with $\cact{B}$ ordered, and let
  $f: {A} \rightarrow {B}$ be a function. If there exist $\d_\bot f$ and
  $\d_\top{f}$ which are unique minimal and maximal derivatives of $f$,
  respectively, then the derivatives of $f$ are precisely
  the functions $\d{f}$ such that
  \begin{displaymath}
    \d_\bot{f} \changeOrder \d{f} \changeOrder \d_\top{f}
  \end{displaymath}
\end{corollary}

The importance of this result will become clearer when we describe our choice of
a change action for Datalog programs. In that setting, unique minimal and
maximal derivatives do indeed exist and have particularly simple descriptions,
which we can then use to check whether a given strategy for incremental
evaluation is correct.

\subsection{Boolean Algebras}
\label{sec:booleanAlgebras}

The semantics of a Datalog predicate is usually taken to be a (finite) set of
tuples. Under this interpretation, conjunction and disjunction correspond to
intersection and union, and every plain Datalog program is a monotone map.

If we were to restrict ourselves to this setting, it would be reasonable to
focus on change actions of the form $\cact{D}_\cup \defeq (\mathcal{P}(D),
\mathcal{P}(D), \cup, \cup, \emptyset)$, with $D$ being some (finite) universe
containing all possible tuples. The semantics of a Datalog program with $n$
relations would then be the least fixed point of some monotone map $\phi :
\mathcal{P}(D)^n \to \mathcal{P}^n$.

This set-up would allow us to give an account of ``classic'' semi-naive
evaluation, that is, incremental evaluation of Datalog without non-monotone
extensions. This is because the above change action $\cact{D}_\sqcup$ has the
convenient property that any monotone endomorphism $f : \mathcal{P}(D) \to
\mathcal{P}(D)$ is differentiable -- in particular, so is every Datalog program.

Such an approach cannot, however, integrate such features as parity-stratified
negation or universal quantification: any Datalog program that uses such
extensions must remain monotone (otherwise its execution might fail to converge
to a fixed point), but it may not be possible to give its semantics as a
composition of monotone functions (that is, it may contain non-monotone
subterms).

Instead of working with the change action $\cact{D}_\sqcup$, we will show how to
endow any Boolean algebra with a much richer change action that allows for the
differentiation of arbitrary functions.

\newcommand{\twist}[0]{\bowtie}
\begin{proposition}
    \label{prop:booleanCAct}
  Let $L$ be a Boolean algebra. Define
  \[
    \cact{L}_{\twist} \defeq (L, L \twist L, \oplus_{\twist}, +, (\bot, \bot))
  \]
  where
  \begin{align*}
    L \twist L &\defeq \{ (a, b) \in L \times L \mid a \wedge b = \bot \}\\
    a \oplus_{\twist} (p, q) &\defeq (a \vee p) \wedge \neg q\\
    (p, q) \bowtie (r, s) &\defeq ((p \wedge \neg s) \vee r, (q \wedge \neg r) \vee s)
  \end{align*}
  with identity element $(\bot, \bot)$.
  Then $\cact{L}_{\twist}$ is a complete change action on $L$.
\end{proposition}
\begin{proof}
    First we need to check that the monoid addition $\bowtie$ is well-defined --
    that is, whenever $(p, q), (r, s) \in L \twist L$ then $(p, q) \bowtie (r,
    s) \in L \twist L$. Writing the above in disjunctive normal form and
    cancelling out trivial terms we obtain:
    \begin{align*}
        (p, q) \bowtie (r, s) &= (p \wedge \neg s \wedge q \wedge \neg r) \vee (r \wedge s)
    \end{align*}
    Since $(p, q), (r, s) \in L \twist L$, we have that both disjunctive terms
    in the right-hand side are equal to $\bot$.
    It remains to prove that $\oplus_{\bowtie}$ is indeed a monoid action. We
    show that it is compatible with $\bowtie$:
  \begin{align*}
    &a \oplus_{\bowtie} \left[(p, q) \bowtie (r, s)\right]\\
    &= a \oplus_{\bowtie} ((p \wedge \neg s) \vee r, (q \wedge \neg r) \vee s)\\
    &= \left(
      a \vee
      \left(
        \left(
          p \wedge \neg s
        \right)
        \vee r
      \right)
    \right)
    \wedge \neg
    \left(
      \left(
        q \wedge \neg r
      \right)
      \vee s
    \right)\\
    &= \left(
      \left(
        \left(
          a \vee p
        \right)
        \wedge
        \left(
          a \vee \neg s
        \right)
      \right)
      \vee r
    \right)
    \wedge
    \left(
      \neg q \vee r
    \right)
    \wedge
    \neg s
    \\
    &= \left(
      \left(
        \left(
          a \vee p
        \right)
        \wedge
        \left(
          a \vee \neg s
        \right)
        \wedge
        \neg q
      \right)
      \vee r
    \right)
    \wedge
    \neg s
    \\
    &= \left(
      \left(
        \left(
          a \vee p
        \right)
        \wedge
        \neg q
      \right)
      \vee r
    \right)
    \wedge
    \neg s
    \\
    &= a \oplus_{\bowtie} (p, q) \oplus_{\bowtie} (r, s) \qedhere
  \end{align*}
\end{proof}

We can think of $\cact{L}_{\twist}$ as tracking changes as pairs of
``additions'' and ``deletions'' to a piece of data, where the monoid action
simply applies one after the other. The condition on $\bowtie$ then stipulates
that changes never ``undo'' any of their own work.

\begin{example}
    In the powerset Boolean algebra $\mathcal{P}(\NN)$, a change to a set of numbers $\{ 1, 2 \}$
    might consist of \emph{adding} $\{3, 4 \}$ and \emph{removing} $\{ 1 \}$.
    In $\cact{\NN}_{\bowtie}$ this change corresponds to the element $(\{3, 4\},
    \{1\})$. Applying it to the set $\{1, 2, 3\}$ would yield:
    \[
        \{1, 2, 3\} \oplus_{\bowtie} (\{3, 4\}, \{1\}) = \{2, 3, 4\}
    \]
\end{example}

Boolean algebras are naturally endowed with the structure of a partial order,
where $a \leq b$ whenever $a \vee b = b$. Furthermore, we observe that, whenever
$a \leq b$, then for any change $(p, q)$ we have $a \oplus_{\bowtie} (p, q) \leq
b \oplus_{\bowtie} (p, q)$. Therefore, by
Proposition~\ref{prop:changeOrderOrdered}, we know that $\cact{L}_{\twist}$ is
in fact an ordered change action.

\begin{proposition}
    The change action $\cact{L}_{\twist}$ is ordered, when $L$ is endowed with
    its natural partial order as a Boolean algebra and $L \bowtie L$ is endowed
    with the order
    \[
        (p, q) \leq_{\twist} (r, s) \Leftrightarrow (p \leq r) \wedge (q \geq s)
    \]
    That is to say, a change $(p, q)$ is smaller than a change $(r, s)$ whenever
    it \emph{adds fewer elements} ($p \leq r$) and it \emph{removes more
    elements} ($q \geq s$).
\end{proposition}

Under the above order, functions into a Boolean algebra endowed with this change
action in fact have unique maximal and minimal derivatives.
Theorem~\ref{theorem:derivativeCharacterization} then gives us bounds for all
the derivatives on Boolean algebras.

\begin{proposition}
  \label{prop:minimalMaximalDerivatives}
  Let $L$ be a Boolean algebra with the $\cact{L}_{\twist}$ change action, and
  $f: A \rightarrow L$ be a function.
  Then, the following are minus operators:
  \begin{align*}
    a \ominus_\bot b &= (a \wedge \neg b, \neg a)\\
    a \ominus_\top b &= (a, b \wedge \neg a)
  \end{align*}
  Additionally, $\d{f}_{\ominus_{\bot}}$ and $\d{f}_{\ominus_{\top}}$ (defined
  as in Proposition~\ref{prop:minusDerivatives}) are unique least and
  greatest derivatives for $f$ respectively.
\end{proposition}
\begin{proof}
    We show that $\ominus_\bot$ is indeed a minus operator:
    \begin{align*}
        b \oplus_{\twist} (a \ominus_\bot b) 
        &= (b \vee (a \wedge \neg b)) \wedge \neg(\neg a)\\
        &= (b \vee (a \wedge \neg b)) \wedge a\\
        &= (b \wedge a) \vee (a \wedge \neg b)\\
        &= a
    \end{align*}
    Similarly for $\ominus_\top$:
    \begin{align*}
        b \oplus_{\twist} (a \ominus_\top b) 
        &= (b \vee a) \wedge \neg (b \wedge \neg a)\\
        &= (b \vee a) \wedge (\neg b \vee a)\\
        &= a
    \end{align*}
    It remains to prove that the corresponding derivatives are least and
    greatest -- this is a trivial consequence of the following result:
    \begin{lemma}
        For any $a, b \in L$, the change $a \ominus_{\bot} b$ is the least
        element of the set of changes $(p, q) \in L \twist L$ such that 
        $b \oplus_{\twist} (p, q) = a$ (and, conversely, $a \ominus_{\top} b$ is
        the greatest such change).
    \end{lemma}
    \begin{proof}\renewcommand{\qedsymbol}{}
        Suppose $b \oplus_{\twist} (p, q) = a$. Then $(b \vee p) \wedge \neg q =
        a$. On one hand, this implies $a \leq \neg q$ and so we obtain $\neg a \geq q$.
        On the other hand, it also implies $a \leq (b \vee p)$, therefore $a
        \wedge \neg b \leq p$. Hence $(a \wedge \neg b, \neg a) \leq_{\twist}
        (p, q)$ as desired.
    \end{proof}
\end{proof}

As an immediate consequence of this result, we can apply
Theorem~\ref{theorem:derivativeCharacterization} in practice, since it provides
us with concrete bounds for derivatives into Boolean algebras. We will make use
of this property in Section~\ref{sec:datalogDifferentiability} to characterise
all possible derivatives of a Datalog program.

\begin{corollary}
\label{cor:booleanCharacterization}
  Let $L$ be a Boolean algebra with the corresponding change action
  $\cact{L}_{\twist}$, $\cact{A}$ be an arbitrary change action, and $f: A \rightarrow
  L$ be a function. Then the derivatives of $f$ are precisely those functions
  $\d{f}: A \times \changes{A} \rightarrow L \twist L$ such that
  \begin{displaymath}
    \d{f}_{\ominus_{\bot}}
    \changeOrder
    \d{f}
    \changeOrder
    \d{f}_{\ominus_{\top}}
  \end{displaymath}
\end{corollary}

\section{Derivatives for Non-Recursive Datalog}
\label{sec:nonRecursiveDatalog}

We now seek to apply the theory we have developed to the specific case of the
semantics of Datalog. Giving a differentiable semantics for Datalog will lead us
to a strategy for performing incremental evaluation and maintenance of Datalog
programs. To begin with, we will restrict ourselves to the non-recursive
fragment of the language -- the formulae that make up the right hand sides of
Datalog rules. That is to say, we will not incrementalise the computation of an
entire fixed point; rather, we will only show how to incrementalise the logical
semantics of a single predicate.

We should also like to note that the techniques we are using should work for any
language. We have chosen Datalog as a non-trivial case study where the need for
incremental computation is real and pressing, but also because the use of
``derivatives'' for incrementalising Datalog programs is so prevalent in the
literature.

\subsection{Semantics of Datalog Formulae}

Datalog is usually given a logical semantics where formulae are interpreted as
first-order logic predicates and the semantics of a program is the set of models
of its constituent predicates. We will instead give a simple denotational
semantics (as is typical when working with fixed points, see e.g.
\cite{compton1994stratified}) that treats a Datalog formula as directly denoting
a relation\footnote{
    Throughout this section we use the word ``relation'' to denote a set of 
    named tuples, as is usual in database theory, as opposed to the more
    familiar definition of a mathematical relation.
}
with variables ranging over a finite schema.

\begin{definition}
  A \textbf{schema} $\Gamma$ is a finite set of names.
  A \textbf{named tuple}
  over $\Gamma$ is an assignment of a value $v_i$ for each name $x_i$ in
  $\Gamma$. Given disjoint schemata $\Gamma = \{ x_1, \ldots, x_n \}$ and
  $\Sigma = \{ y_1, \ldots, y_m \}$, the \textbf{selection function}
  $\sigma_\Gamma$ is defined as
  \begin{displaymath}
    \sigma_\Gamma(\{x_1 \mapsto v_1, \ldots, x_n \mapsto v_n, y_1 \mapsto w_1, \ldots, y_m \mapsto w_m \})
    \defeq \{ x_1 \mapsto v_1, \ldots, x_n \mapsto v_n \}
  \end{displaymath}
  i.e. $\sigma_\Gamma$ restricts a named tuple over $\Gamma \cup \Sigma$ into a
  tuple over $\Gamma$ with the same values for the names in $\Gamma$. 

  Given signatures $\Gamma_1, \Gamma_2$ with the same cardinality The
  \textbf{renaming function} $\rho_{\Gamma_1 / \Gamma_2}$ transforms a
  $\Gamma_2$-tuple into a $\Gamma_1$-tuple by permuting the names\footnote{
  Strictly speaking, this function $\rho$ is not well-defined, as it assumes
  that both sets are rather ordered sequences. In practice, the free variables
  involved in a Datalog formula are implicitly ordered by the position in which
  they appear in its definition, and so this does not cause any issues.
  }:
  \[
      \rho_{\{x_i\} / \{y_i\}}(\{y_i \mapsto v_i\})
      \defeq \{x_i \mapsto v_i\}
  \]
  
  We also write $\sigma_\Gamma$ and $\rho_{\Gamma_1 / \Gamma_2}$ to denote their
  respective extension to sets of tuples.
\end{definition}

\newcommand{\Formula}{\mathrm{Formula}}
\newcommand{\Rel}{\cat{Rel}}
\newcommand{\universalRel}{\mathcal{U}}
\newcommand{\consq}{\mathcal{I}}

We will adopt the usual closed-world assumption to give a denotation to negation. That
is to say, for every schema $\Gamma$, we will fix a set $\universalRel_\Gamma$ of
named tuples over $\Gamma$, the ``universal relation'', representing the set of
all the $\Gamma$-tuples that are allowed in our programs.

\begin{definition}
  We write $\Rel_\Gamma$ to denote the set of relations over $\Gamma$, that is,
  the set of all subsets of $\universalRel_\Gamma$.

  Negation on relations can then be defined as
  \begin{displaymath}
    \neg R \defeq \universalRel_\Gamma \setminus R
  \end{displaymath}
  This makes $\Rel_\Gamma$, the set of all subsets of
  $\universalRel_\Gamma$, a complete Boolean algebra.
\end{definition}

\newcommand{\semR}[0]{\mathcal{R}}
\newcommand{\diffR}[0]{\change{\semR}}
\begin{definition}
  Given a Datalog formula $T$ whose free term variables are contained in
  $\Gamma$, and which depends on $n$ formula constants $R_i$, its denotation
  $\sem{T}_\Gamma$ is a function 
  \[
      \sem{T}_\Gamma : \Pi_{i=1}^n\Rel_{\Gamma_i} \to \Rel_\Gamma
  \]
  where $\Gamma_i$ is the signature for the constant $R_i$.

  If $\semR = (\semR_1, \ldots, \semR_n)$ is a choice of a relation $\semR_i$ for each of the variables $R_i$,
  $\sem{T}(\semR)$ is inductively defined according to the rules in \ref{fig:datalogSemantics}.
\end{definition}

\begin{figure}
    \begin{minipage}[t]{0.9\textwidth}
      \begin{minipage}[c]{.45\textwidth}
      \begin{align*}
        \sem{\top}_\Gamma(\semR) &\defeq \universalRel_\Gamma\\
        \sem{\bot}_\Gamma(\semR) &\defeq \emptyset
      \end{align*}
    \end{minipage}\hfill\noindent
    \begin{minipage}[c]{.45\textwidth}
      \begin{align*}
        \sem{T \wedge U}_\Gamma(\semR) &\defeq \sem{T}_\Gamma(\semR) \cap \sem{U}_\Gamma(\semR)\\
        \sem{T \vee U}_\Gamma(\semR) &\defeq \sem{T}_\Gamma(\semR) \cup \sem{U}_\Gamma(\semR)\\
        \sem{\neg T}_\Gamma(\semR) &\defeq \neg \sem{T}_\Gamma(\semR)
      \end{align*}
    \end{minipage}
    \begin{align*}
        \sem{R_j(x_1, \ldots, x_n)}_\Gamma(\semR) &\defeq 
        \{ u \in \universalRel_\Gamma \mid \rho_{x_1, \ldots, x_n / \Gamma_j}(\sigma_{x_1, \ldots, x_n}(u)) \in \semR_j\}\\
      \sem{\exists x . T}_\Gamma(\semR) &\defeq \sigma_\Gamma(\sem{T}_{\Gamma \cup \{ x \}}(\semR))
    \end{align*}{}\vspace{-12pt}
    \end{minipage}
  \vspace{8pt}
  \caption{Formula semantics for Datalog}
  \label{fig:datalogSemantics}
  \vspace{-12pt}
\end{figure}

Since $\Rel_\Gamma$ is a complete Boolean algebra, and so is $\Rel_{\Gamma_i}$,
$\sem{T}_\Gamma$ is a function between complete Boolean algebras. For brevity,
we will often omit the schemata whenever they can be inferred from the context.

\subsection{Differentiability of Datalog Formula Semantics}
\label{sec:datalogDifferentiability}

In order to actually perform our incremental computation, we first need to
provide a concrete derivative for the semantics of Datalog formulae. Of course,
since $\sem{T}_\Gamma$ is a function between the complete Boolean algebras
$\Pi\Rel_{\Gamma_i}$ and $\Rel_\Gamma$, and we know that the corresponding
change actions $\cact{\Pi\Rel_{\Gamma_i}}_{\twist}$ and
$\cact{\Rel_\Gamma}_{\twist}$ are complete, this guarantees the existence of a
derivative for $\sem{T}$.

Unfortunately, this does not necessarily provide us with an \emph{efficient}
derivative for $\sem{T}$. The derivatives that we know how to compute (the least
and greatest derivatives defined in Corollary~\ref{cor:booleanCharacterization})
rely on computing $f(a \oplus \change{a})$ itself, which is precisely the
recalculation that incremental computation seeks to avoid. Technically both of
these derivatives \emph{do} correspond to strategies for incremental
computation: in particular, the ``trivial'' strategy where one simply throws
away the partial result $f(a)$ and computes $f(a \oplus \change{a})$ from scratch.

Of course, given a concrete definition of a derivative we can simplify this
expression and hopefully make it easier to compute. But we also know from
Corollary~\ref{cor:booleanCharacterization} that \emph{any} function bounded by
$\d{f}_{\ominus_\bot}$ and $\d{f}_{\ominus_\top}$ is a valid derivative, and we
can therefore optimise anywhere within that range to make a trade-off between
ease of computation and precision.\footnote{The idea of using an approximation
to the precise derivative, and a soundness condition, appears in Bancilhon
\cite{bancilhon1986amateur}.}

There is also the question of how to effectively compute the derivative -- the
notion of derivative is a purely \emph{semantic} one and, while we know that
derivatives do exist, we also need to construct a concrete Datalog program that
computes it. Since the change set for $\cact{\Rel}_{\twist}$ is a subset of
$\Rel \times \Rel$, it is possible and indeed very natural to compute the two
components via a pair of Datalog formulae, which allows us to reuse an existing
Datalog formula evaluator. Indeed, if this process is occurring in an optimising
compiler, the evaluation of the derivative formulae themselves can be optimised.
This is very beneficial in practice, since the initial formulae may be quite
complex.

This does give us additional constraints that the derivative formulae must
satisfy: for starters, we need to be able to evaluate them; and we may wish to
pick formulae that will be easy or cheap for our evaluation engine to compute,
even if they compute a less precise derivative.

The upshot of these considerations is that the optimal choice of derivatives is
likely to be quite dependent on the precise variant of Datalog being evaluated,
and the specifics of the evaluation engine. The version that we present here is
a simplification of the formula actually in use in the Datalog engine developed
by Semmle Ltd..

\subsubsection{A Concrete Datalog Formula Derivative}
\newcommand{\updiff}{\Delta}
\newcommand{\downdiff}{\nabla}
\newcommand{\bothdiff}{{\mathsf X}} 

\begin{figure}[t]
    \begin{minipage}[t]{0.9\textwidth}
    \vspace{-5pt}
    \begin{minipage}[t]{.45\textwidth}
      \begin{align*}
        \updiff(\bot) &\defeq \bot\\
        \updiff(\top) &\defeq \bot\\
        \updiff(R_j) &\defeq \updiff R_j \\
        \updiff(T\vee U) &\defeq \updiff(T) \vee \updiff (U)\\
        \updiff(T\wedge U) &\defeq (\updiff(T)\wedge \bothdiff(U))\\
                            & \vee (\updiff(U) \wedge \bothdiff(T))\\
        \updiff(\neg T) &\defeq \downdiff(T)\\
        \updiff(\exists x.T) &\defeq \exists x.\updiff(T)
      \end{align*}
    \end{minipage}\hfill\noindent
    \begin{minipage}[t]{.45\textwidth}
      \begin{align*}
        \downdiff(\bot) &\defeq \bot\\
        \downdiff(\top) &\defeq \bot\\
        \downdiff(R_j) &\defeq \downdiff R_j \\
        \downdiff(T\vee U) &\defeq (\downdiff(T) \wedge \neg \bothdiff(U))\\
                              & \vee (\downdiff(U) \wedge \neg \bothdiff(T))\\
        \downdiff(T\wedge U) &\defeq (\downdiff(T)\wedge U) \vee (T \wedge \downdiff(U))\\
        \downdiff(\neg T) &\defeq \updiff(T)\\
        \downdiff(\exists x.T) &\defeq \exists x.\downdiff(T) \wedge \neg \exists x.\bothdiff(T)
      \end{align*}
    \end{minipage}
    \vspace{6pt}
    \begin{align*}
      \bothdiff(R) \defeq (R \vee \updiff(R)) \wedge \neg \downdiff(R)
    \end{align*}{}\vspace{-12pt}
    \end{minipage}
  \vspace{8pt}
  \caption{Upwards and downwards formula derivatives for Datalog}
  \label{fig:datalogDerivatives}
  \vspace{-12pt}
\end{figure}

Figure~\ref{fig:datalogDerivatives}, defines a ``symbolic'' derivative operator
as a pair of mutually recursive functions, $\updiff$ and $\downdiff$, which turn
a Datalog formula $T$ into new formulae that compute the upwards and downwards
parts of the derivative, respectively. Our definition uses an auxiliary
function, $\bothdiff$, which computes the ``neXt'' value of a term by applying
the upwards and downwards derivatives. As is typical for a derivative, the new
formulae will have additional free relation variables for the upwards and
downwards derivatives of the free relation variables of $T$, denoted as $\updiff
R_i$ and $\downdiff R_i$ respectively. Evaluating the formula as a derivative means
evaluating it as a normal Datalog formula with the new relation variables set to
the input relation changes.

While the definitions mostly exhibit the dualities we would expect
between corresponding operators, there are a few asymmetries to explain.

The asymmetry between the cases for $\updiff(T \vee U)$ and
$\downdiff(T \wedge U)$ is for operational reasons. The symmetrical version of
$\updiff(T \vee U)$ is $(\updiff(T) \wedge \neg U) \vee (\updiff(U) \wedge \neg
T)$ (which is also precise). The reason we omit the negated conjuncts is simply
that they are costly to compute and not especially helpful to our evaluation engine.
On the other hand, while we could have set $\downdiff(T \wedge U) \defeq \downdiff(T) 
\vee \downdiff(U)$ (and this would lead to a valid derivative), we prefer the present
form as it results in smaller relations at no cost (since the extensions of $T$ and $U$
are already available at this point).

The asymmetry between the cases for $\exists$ is because our
dialect of Datalog does not have a primitive universal quantifier.
If we did have one, the cases for $\exists$ would be dual to the corresponding
cases for $\forall$.

\newcommand{\bothchanges}{\rho}
\begin{theorem}
  \label{thm:datalog-derivative}
  Let $\updiff$, $\downdiff$, $\bothdiff : \Formula \rightarrow \Formula$ be
  mutually recursive functions defined by structural induction as in
  Figure~\ref{fig:datalogDerivatives}.

  Then $\updiff(T)$ and $\downdiff(T)$ are disjoint, and for any schema $\Gamma$
  and any Datalog formula $T$ whose free term variables are contained in $\Gamma$,
  \[
      \d{\sem{T}_\Gamma}(\semR_1\ldots,\delta\semR_1\ldots)
      \defeq (\sem{\updiff(T)}_\Gamma(\semR_i\ldots, \delta\semR_i\ldots),
       \sem{\downdiff(T)}_\Gamma(\semR_i\ldots, \delta\semR_i\ldots))
  \]
  is a derivative for $\sem{T}_\Gamma$.
\end{theorem}
\begin{proof}
\newcommand{\upsem}[1]{\sem{\updiff{#1}}}
\newcommand{\downsem}[1]{\sem{\downdiff{#1}}}
\newcommand{\bothsem}[1]{\sem{\bothdiff{#1}}}
    Let $T$ be a Datalog formula with free relation variables $R_1, \ldots,
    R_n$, a choice of a semantics for the free relation variables $\semR_1,
    \ldots, \semR_n \in \Rel$ and differences $\change{\semR_1}, \ldots,
    \change{\semR_n} \in \Rel\bowtie \Rel$. For brevity, we refer to the tuple
    $(\semR_1, \ldots, \semR_n)$ as $\semR$ and the tuple $(\change{\semR_1},
    \ldots, \change{\semR_n})$ as $\diffR$. We abuse the notation and refer to
    $(\semR_1 \twist \change{\semR_1}, \ldots, \semR_n \twist \change{\semR_n})$
    as $\semR \twist \diffR$. We will also omit the arguments to $\upsem{T}$ and
    $\downsem{T}$ often, as there is never room for ambiguity.

    We want to show:
    \begin{displaymath}
    \sem{T}_\Gamma(\semR \twist \diffR)
    =
    \sem{T}_\Gamma(\semR) \twist 
    (\sem{\updiff{T}}_\Gamma(\diffR), 
    \sem{\downdiff{T}}_\Gamma(\diffR))
    \end{displaymath}
    We proceed by structural induction on $T$. We omit the cases for $\top$, $\bot$
    and relational variables for being trivial.
  
    For the conjunction case, we make use of the following elementary set-theoretic
    identity:
    \begin{proposition}
      \label{prop:capPrecision}
      For any sets $A, B, C$ such that $C \subseteq B$ and $A \subseteq \neg B \cup C$,
      we have $A \cap \neg B = A \cap \neg C$.
    \end{proposition}
    \begin{proof}
      \begin{align*}
      A \cap \neg C &= A \cap \neg C \cap (\neg B \cup C)
      \tag{since $A \subseteq \neg B \cup C$}
      \PROOFSTEP
                    &= A \cap \left((\neg C \cap \neg B) \cup (\neg C \cap C) \right)
                    \tag{distributivity}
      \PROOFSTEP
                    &= A \cap (\neg C \cap \neg B)
                    \tag{cancelling out the empty set}
      \PROOFSTEP
                    &= A \cap \neg (C \cup B)
                    \tag{De Morgan's}
      \PROOFSTEP
                    &= A \cap \neg B
                    \tag{since $C \subseteq B$}
      \end{align*}
      \renewcommand{\qedsymbol}{}
    \end{proof}\vspace{-\baselineskip}

    With the above in mind, the proof of correctness for the conjunctive case follows
    by the following straightforward, if lengthy, derivation.
  \begin{align*}
    &\sem{T \wedge U}_\Gamma(\semR \oplus_{\twist} \diffR)\\
    \eq \sem{T}_\Gamma(\semR \oplus_{\twist} \diffR) 
      \cap \sem{U}_\Gamma(\semR \oplus_{\twist} \diffR) \tag{semantics of $\wedge$}
    \PROOFSTEP
    \eq \sem{T}_\Gamma(\semR) \oplus_{\twist} (\upsem{T}_\Gamma, \downsem{T}_\Gamma) 
      \tag{induction hypothesis}
      \\
      \continued \cap \sem{U}_\Gamma(\semR) 
      \oplus_{\twist} (\upsem{U}_\Gamma, \downsem{U}_\Gamma) 
    \PROOFSTEP
    \eq (\sem{T}_\Gamma(\semR) \cup \upsem{T}_\Gamma) \cap \neg \downsem{T}_\Gamma
      \tag{definition of $\oplus_{\twist}$}
      \\
      \continued \cap (\sem{U}_\Gamma(\semR) \cup \upsem{U}_\Gamma) 
      \cap \neg \downsem{U}_\Gamma 
    \PROOFSTEP
    \eq (\sem{T}_\Gamma(\semR) \cap \neg \downsem{T}_\Gamma \cup \upsem{T}_\Gamma)
      \tag{disjointness of $\upsem{}, \downsem{}$}
      \\
      \continued \cap (\sem{U}_\Gamma(\semR) \cap \neg \downsem{U}_\Gamma \cup \upsem{U}_\Gamma) 
    \PROOFSTEP
    \eq (\sem{T}_\Gamma(\semR) \cap \neg \downsem{T}_\Gamma) 
      \tag{distributing $\upsem{T}$ out}
    \\
    \continued \cap (\sem{U}_\Gamma(\semR) \cap \neg \downsem{U}_\Gamma \cup \upsem{U}_\Gamma) 
    \\
    \continued \cup 
      (\upsem{T}_\Gamma 
      \cap (\sem{U}_\Gamma(\semR) \cup \upsem{U}_\Gamma) 
      \cap \neg \downsem{U}_\Gamma)
    \PROOFSTEP
    \eq \Big(\sem{T}_\Gamma(\semR) \cap \neg \downsem{T}_\Gamma
      \cap \sem{U}_\Gamma(\semR) \cap \neg \downsem{U}_\Gamma\Big)
      \tag{distributing $\upsem{U}$ out}
    \\
    \continued 
      \cup \Big(\upsem{T}_\Gamma \cap (\sem{U}_\Gamma(\semR) \cup \upsem{U}_\Gamma) \cap \neg \downsem{U}_\Gamma\Big)
    \\
    \continued
      \cup \Big(\upsem{U}_\Gamma \cap (\sem{T}_\Gamma(\semR) \cup \upsem{T}_\Gamma) \cap \neg \downsem{T}_\Gamma\Big)
    \PROOFSTEP
    \eq \Big[(\sem{T}_\Gamma(\semR) \cap \sem{U}_\Gamma(\semR))
      \tag{disjointness of $\upsem{}, \downsem{}$}
    \\
    \continued 
      \cup \Big(\upsem{T}_\Gamma \cap (\sem{U}_\Gamma(\semR) \cup \upsem{U}_\Gamma) \cap \neg \downsem{U}_\Gamma\Big)
    \\
    \continued
      \cup \Big(\upsem{U}_\Gamma \cap (\sem{T}_\Gamma(\semR) \cup \upsem{T}_\Gamma) \cap \neg \downsem{T}_\Gamma\Big)\Big]
    \\
    \continued[50pt]
      \cap (\neg \downsem{T}_\Gamma \cap \neg \downsem{U}_\Gamma)
    \PROOFSTEP
    \eq \Big[(\sem{T}_\Gamma(\semR) \cap \sem{U}_\Gamma(\semR))
      \tag{De Morgan's}
    \\
    \continued 
      \cup \Big(\upsem{T}_\Gamma \cap (\sem{U}_\Gamma(\semR) \cup \upsem{U}_\Gamma) \cap \neg \downsem{U}_\Gamma\Big)
    \\
    \continued
      \cup \Big(\upsem{U}_\Gamma \cap (\sem{T}_\Gamma(\semR) \cup \upsem{T}_\Gamma) \cap \neg \downsem{T}_\Gamma\Big)\Big]
    \\
    \continued[50pt]
      \cap \neg (\downsem{T}_\Gamma \cup \downsem{U}_\Gamma)
    \PROOFSTEP
    \eq \Big[(\sem{T}_\Gamma(\semR) \cap \sem{U}_\Gamma(\semR))
      \tag{definition of $\bothdiff{}$}
    \\
    \continued 
      \cup (\upsem{T}_\Gamma \cap \bothsem{U}_\Gamma)
      \cup (\upsem{U}_\Gamma \cap \bothsem{T}_\Gamma)\Big]
    \\
    \continued[50pt]
      \cap \neg (\downsem{T}_\Gamma \cup \downsem{U}_\Gamma)
    \PROOFSTEP
    \eq \Big[(\sem{T}_\Gamma(\semR) \cap \sem{U}_\Gamma(\semR))
      \cup \upsem{(T \wedge U)}_\Gamma \Big]
      \tag{definition of $\upsem{}$}
    \\
    \continued
      \cap \neg \downsem{T}_\Gamma \cup \downsem{U}_\Gamma
    \PROOFSTEP
    \eq \Big[(\sem{T}_\Gamma(\semR) \cap \sem{U}_\Gamma(\semR))
      \cup \upsem{(T \wedge U)}_\Gamma \Big]
      \tag{Proposition~\ref{prop:capPrecision}}
    \\
    \continued
      \cap \neg \Big[(\downsem{T}_\Gamma \cap\sem{U}_\Gamma(\semR)) \cup (\downsem{U}_\Gamma \cap\sem{T}_\Gamma(\semR))\Big]
    \PROOFSTEP
    \eq \Big[(\sem{T}_\Gamma(\semR) \cap \sem{U}_\Gamma(\semR))
      \cup \upsem{(T \wedge U)}_\Gamma \Big] \cap \neg (\downsem{(T \cap U)})
      \tag{definition of $\downsem{}$}
    \\
    \PROOFSTEP
    \eq \sem{T \wedge U}_\Gamma(\semR)
      \oplus_{\twist} (\upsem{(T \wedge U)}_\Gamma, \downsem{(T \wedge U)}_\Gamma) \tag{definition of $\oplus_{\twist}$}
  \end{align*}

  The disjunctive case is identical to conjunction, minus the application of 
  Proposition~\ref{prop:capPrecision}. Negation follows by calculation:
  \begin{align*}
    &\sem{\neg T}_\Gamma(\semR \oplus_{\twist} \diffR)
    \PROOFSTEP
    \eq \neg \sem{T}_\Gamma(\semR \oplus_{\twist} \diffR) 
    \tag{semantics of $\neg$}
    \PROOFSTEP
    \eq \neg ((\sem{T}_\Gamma(\semR) \cup \upsem{T}_\Gamma) \cap \neg \downsem{T}_\Gamma) 
    \tag{induction hypothesis, definition of $\oplus_{\twist}$}
    \PROOFSTEP
    \eq \neg \sem{T}_\Gamma(\semR) \cap \neg \upsem{T}_\Gamma \cup \downsem{T}_\Gamma 
    \tag{De Morgan's}
    \PROOFSTEP
    \eq (\neg \sem{T}_\Gamma(\semR) \cup \downsem{T}_\Gamma) \cap \neg \upsem{T}_\Gamma \tag{disjointness of $\upsem{}, \downsem{}$}
    \PROOFSTEP
    \eq \sem{\neg T}_\Gamma(\semR) \oplus_{\twist} (\downsem{T}_\Gamma, \upsem{T}_\Gamma) \tag{definition of $\oplus_{\twist}$}
    \PROOFSTEP
    \eq \sem{\neg T}_\Gamma(\semR) \oplus_{\twist} (\upsem{\neg T}_\Gamma, \downsem{\neg T}_\Gamma) \tag{definition of $\upsem{}, \downsem{}$}
  \end{align*}

  Before establishing the derivative property for the existential, we state the following 
  straightforward properties of the selection map $\sigma_\Gamma$:
  \begin{lemma}
    \label{prop:sigmaDist}
    For any $A, B \in \Rel_{\Gamma, x}$ we have:
    \begin{enumerate}[\roman{enumi}.,ref={\thelemma.\roman{enumi}}]
      \item $\sigma_\Gamma(A \cup B) = \sigma_\Gamma(A) \cup \sigma_\Gamma(B)$
      \label{lem:sigmaDist-i}
      \item $\sigma_\Gamma (A \cap \neg B) = \sigma_\Gamma(A) \cap (\neg \sigma_\Gamma(B) \cup \sigma_\Gamma(A \cap \neg B))$\\
      \label{lem:sigmaDist-ii}
    \end{enumerate}
  \end{lemma}
  \begin{proof}
    We prove \ref{lem:sigmaDist-ii} explicitly, using a double inclusion argument.
    First:
    \begin{align*}
      \sigma_\Gamma (A \cap \neg B)
      \subseteq \sigma_\Gamma(A) \cap \sigma_\Gamma(A \cap \neg B)
      \subseteq \sigma_\Gamma(A) \cap (\neg \sigma_\Gamma(B) \cup \sigma_\Gamma(A \cap \neg B))
    \end{align*}
    For the second inclusion, we note that the right-hand side of the equation can be
    equivalently written as:
    \[ 
      \Big\lbrack\sigma_\Gamma(A) \cap \neg \sigma_\Gamma(B)\Big\rbrack
      \cup \Big\lbrack\sigma_\Gamma(A) \cap \sigma_\Gamma(A \cap \neg B)\Big\rbrack
    \]
    It suffices to show that both of the bracketed terms are contained in $\sigma_\Gamma (A \cap \neg B)$.
    Clearly this is the case for the term on the right and it
    remains to show that $\sigma_\Gamma(A) \cap \neg \sigma_\Gamma(B) \subseteq \sigma_\Gamma(A \cup \neg B)$.

    For this, consider a tuple $\rho \equiv \{ \Gamma \mapsto w \}$ over $\Gamma$ belonging to 
    $\sigma_\Gamma(A) \cap \neg \sigma_\Gamma(B)$. Since $\rho \in \sigma_\Gamma(A)$, there
    must be some $w_x$ satisfying $\rho_x \equiv \{ x \mapsto w_x, \Gamma \mapsto w \} \in A$. But since
    $\rho \in \neg \sigma_\Gamma(B)$, it follows that $\rho_x$ is in $\neg B$, and hence
    $\rho_x \in A \cup \neg B$. Therefore $\rho \in \sigma_\Gamma(A \cup \neg B)$.
  \end{proof}
  
  Equipped with these, we can proceed with the proof proper:
  \begin{align*}
    &\sem{\exists x . T}_\Gamma(\semR \oplus_{\twist} \diffR)
    \PROOFSTEP
    \eq \sigma_\Gamma(\sem{T}_{\Gamma, x}(\semR \oplus_{\twist} \diffR)) \tag{semantics of $\exists$}
    \PROOFSTEP
    \eq \sigma_\Gamma(\sem{T}_{\Gamma, x}(\semR) \oplus_{\twist} (\upsem{T}_{\Gamma, x}, \downsem{T}_{\Gamma, x})) 
      \tag{induction hypothesis}
    \PROOFSTEP
    \eq \sigma_\Gamma((\sem{T}_{\Gamma, x}(\semR) \cup \upsem{T}_{\Gamma, x}) \cap \neg \downsem{T}_{\Gamma, x}) 
      \tag{definition of $\oplus_{\twist}$}
    \PROOFSTEP
    \eq \sigma_\Gamma(\sem{T}_{\Gamma, x}(\semR) \cup \upsem{T}_{\Gamma, x})
      \tag{Lemma~\ref{lem:sigmaDist-ii}}
    \\
    \continued
      \cap \Big \lbrack \neg \sigma_\Gamma(\downsem{T}_{\Gamma, x})
      \cup \sigma_\Gamma((\sem{T}_{\Gamma, x}(\semR) \cup \upsem{T}_{\Gamma, x}) \cap \neg\downsem{T}_{\Gamma, x})
      \Big\rbrack
    \PROOFSTEP
    \eq \Big\lbrack
      \sigma_\Gamma(\sem{T}_{\Gamma, x}(\semR)) \cup \sigma_\Gamma(\upsem{T}_{\Gamma, x})
      \Big\rbrack
      \tag{Lemma~\ref{lem:sigmaDist-i}}
    \\
    \continued
      \cap \Big \lbrack \neg \sigma_\Gamma(\downsem{T}_{\Gamma, x})
      \cup \sigma_\Gamma((\sem{T}_{\Gamma, x}(\semR) \cup \upsem{T}_{\Gamma, x}) \cap \neg \downsem{T}_{\Gamma, x})
      \Big\rbrack
    \PROOFSTEP
    \eq \Big\lbrack
      \sigma_\Gamma(\sem{T}_{\Gamma, x}(\semR)) \cup \upsem{(\exists x . T)}_\Gamma
      \Big\rbrack
      \tag{definition of $\upsem{}, \bothsem{}$}
    \\
    \continued
      \cap \Big \lbrack \neg \sigma_\Gamma(\downsem{T}_{\Gamma, x})
      \cup \sigma_\Gamma(\bothsem{T}_{\Gamma, x})
      \Big\rbrack
    \PROOFSTEP
    \eq \Big\lbrack
      \sigma_\Gamma(\sem{T}_{\Gamma, x}(\semR)) \cup \upsem{(\exists x . T)}_\Gamma
      \Big\rbrack
      \tag{De Morgan's}
    \\
    \continued
      \cap \neg \Big \lbrack \sigma_\Gamma(\downsem{T}_{\Gamma, x})
      \cap \neg \sigma_\Gamma(\bothsem{T}_{\Gamma, x})
      \Big\rbrack
    \PROOFSTEP
    \eq \Big\lbrack
      \sem{\exists x . T}_{\Gamma}(\semR) \cup \upsem{(\exists x . T)}_\Gamma
      \Big\rbrack
      \tag{semantics of $\exists$}
    \\
    \continued
      \cap \neg \Big \lbrack \sem{\exists x . \downdiff{T}}_\Gamma
      \cap \neg \sem{\exists x . \bothdiff{T}}_\Gamma
      \Big\rbrack
    \PROOFSTEP
    \eq \Big\lbrack
      \sem{\exists x . T}_{\Gamma}(\semR) \cup \upsem{(\exists x . T)}_\Gamma
      \Big\rbrack
      \cap \neg \downsem{(\exists x . T)}_\Gamma
      \tag{definition of $\downsem{}$}
    \PROOFSTEP
    \eq
      \sem{\exists x . T}_{\Gamma}(\semR) \oplus_{\twist}
      (\upsem{(\exists x . T)}_\Gamma, \downsem{(\exists x . T)}_\Gamma)
      \tag{definition of $\oplus_{\twist}$}
  \end{align*}

\end{proof}

We can give a derivative for our $treeP$ predicate by mechanically applying
the recursive functions defined in Figure~\ref{fig:datalogDerivatives}.
\begin{align*}
  &\updiff(treeP(x))\\
  &= p(x) \wedge \exists y. (child(x, y) \wedge \updiff(treeP(y))) \wedge \neg \exists y. (child(x,y) \wedge \neg \bothdiff(treeP(y)))\\
  &\downdiff(treeP(x))\\
  &= p(x) \wedge \exists y. (child(x, y) \wedge \downdiff(treeP(y)))
\end{align*}

The upwards difference in particular is not especially easy to compute. If we
naively compute it, the third conjunct requires us to recompute the whole of the
recursive part. However, the first and second conjuncts act as a guard of sorts:
if either is empty we then the whole formula will be, so we only need to
evaluate the third conjunct if the first and second conjuncts are non-empty, i.e
if there is \emph{some} change in the body of the existential.

This shows that our derivatives aren't a panacea: it is simply \emph{hard} to compute
downwards differences for $\exists$ (and, equivalently, upwards differences for
$\forall$) because we must check that there is no other way of deriving the same
facts.\footnote{The ``support'' data structures introduced by
  \cite{gupta1993maintaining} are an attempt to avoid this issue by
  tracking the number of derivations of each tuple.} However, we can still avoid
the re-evaluation in many cases, and the inefficiency is local to this subformula.

\subsection{Extensions to Datalog}
\label{sec:extensions}

Our formulation of Datalog formula semantics and derivatives is generic and
modular, so it is easy to extend the language with new formula constructs: all
we need to do is add cases for $\updiff$ and $\downdiff$.

In fact, because we are using a transitive change action, we can \emph{always} do
this by using the maximal or minimal derivative. This justifies our claim that
we can support \emph{arbitrary} additional formula constructs: although the
maximal and minimal derivatives are likely to be impractical, they can still be
used as a fall-back option for constructs that do not support incremental evaluation
-- even in those cases, the compositional nature of this approach means that,
while the problematic subformula will use one of these inefficient derivatives,
the rest of the formula can still be incrementalised efficiently.

This power is important in practice: here is a real example from Semmle's
variant of Datalog, which includes a kind of aggregate expressions which have
well-defined recursive semantics. Aggregates have the form
\begin{displaymath}
  r = \mathrm{agg}(p)(vs \mid T \mid U)
\end{displaymath}
where $\mathrm{agg}$ refers to an aggregation function (such as ``sum'' or
``min''), $vs$ is a sequence of variables, $p$ and $r$ are variables, $T$ is a
formula possibly mentioning $vs$, and $U$ is a formula possibly mentioning $vs$
and $p$. The full details can been found in Moor and Baars
\cite{demoor2013aggregates}, but for example this allows us to write
\begin{align*}
  height(n, h) \leftarrow& \neg \exists c. (child(n, c)) \wedge h = 0\\
  &\vee \exists h'. (h' = \mathrm{max}(p)(c \mid child(n, c) \mid height(c, p)) \wedge h = h' + 1)
\end{align*}
which recursively computes the height of a node in a tree.

Here is an upwards derivative for an aggregate formula:
\begin{align*}
  \updiff(r = \mathrm{agg}(p)(vs \mid T \mid U)) \defeq \exists vs. (T \wedge \updiff{U}) \wedge r = \mathrm{agg}(p)(vs \mid T \mid U)
\end{align*}

While this isn't a precise derivative, it is still substantially cheaper than
re-evaluating the whole subformula, as the first conjunct acts as a guard,
allowing us to skip the second conjunct when $U$ has not changed.


\section{Changes on Functions}
\label{sec:changesOnFunctions}
\newcommand{\fix}[0]{\mathbf{fix}}
\newcommand{\lfp}{\mathbf{lfp}}
\newcommand{\adjust}{\mathbf{adjust}}
\newcommand{\iter}{\mathbf{iter}}
\newcommand{\nextiter}{\mathbf{recur}}

So far we have defined change actions for the kinds of objects that typically
make up the relations of a Datalog program, but we would also like to have
change actions on \emph{functions}. This would allow us to define derivatives
for higher-order languages (where functions are first-class); and to give a
derivative for important higher-order maps, like fixed point operators
$\fix : (A \rightarrow A) \rightarrow A$. The latter is of special interest to us as
evaluation of Datalog queries necessitates finding a fixed point, thus the differentiation
of fixed point operators will be the object of Section~\ref{sec:fixpoints}.

Function spaces, however, differ from products and disjoint unions in that there
is no obvious ``best'' change action on $A \rightarrow B$. Therefore instead of
trying to define a single choice of change action, we will instead pick out
subsets of function spaces which have ``well-behaved'' change actions.

\begin{definition}
  \label{def:functionalChanges}
  Given change actions $\cact{A}$ and $\cact{B}$ and a set $U \subseteq A \to
  B$, a change action $\cact{U} = (U, \changes U, \oplus_U, +_U, 0_U)$ is
  \textbf{functional} whenever the evaluation map $\Ev : U \times A \to B$ is
  differentiable, that is to say, whenever there exists a function $\d{\Ev} : (U
  \times A) \times (\changes U \times \changes A) \to \changes{B}$ such that:
  \begin{displaymath}
    (f \oplus_U \change{f})(a \oplus_A \change{a}) = 
    f(a) \oplus_B \d{\Ev}((f, a), (\change{f}, \change{a}))
  \end{displaymath}
  We will write $\cact{U} \subseteq \cact{A} \Rightarrow \cact{B}$ whenever 
  $U \subseteq A \rightarrow B$ and $\cact{U}$ is functional.
\end{definition}

There are two reasons why the definition of a functional change action consists
of a \emph{subset} of $U \subseteq A \rightarrow B$. Firstly, it allows one to
restrict themselves to spaces of monotone or continuous functions. But more
importantly, functional change actions are necessarily made up of differentiable
functions, and thus a functional change action may not exist for the entire
function space $A \rightarrow B$.

\begin{proposition}
  \label{prop:differentiableFunctionalChanges}
  Let $\cact{U} \subseteq \cact{A} \Rightarrow \cact{B}$ be a functional change
  action. Then every $f \in U$ is differentiable, with a derivative $\d{f}$
  given by:
  \begin{displaymath}
    \d{f}(x, \change{x}) = \d{\Ev}((f, x), (0, \change{x}))
  \end{displaymath}
\end{proposition}

Conversely, the derivative of the evaluation map can also be
used to compute the effect of a given function change $\delta f$:
\begin{proposition}
    \label{prop:functionalChangesAreDifferentiable}
  Let $\cact{U} \subseteq \cact{A} \Rightarrow \cact{B}$. Then for every $f \in
  U$ and $\delta f \in \changes U$ we have:
  \begin{displaymath}
    (f \oplus \delta f)(x) = f(x) \oplus \Ev'((f, x), (\delta f, \mathbf{0}))
  \end{displaymath}
\end{proposition}

Even if we were to restrict ourselves to the differentiable functions between
$\cact{A}$ and $\cact{B}$, however, it is hard to find a concrete functional
change action for this set. Fortunately, in many important cases there is a
simple change action on the set of differentiable functions.

\begin{definition}
  \label{def:pointwise-functional}
  Let $\cact{A}$ and
  $\cact{B}$ be change actions. The \textbf{pointwise functional change action}
  $\cact{A} \Ra_{pt} \cact{B}$, when it is defined, is given by 
  \[
\cact{A} \Ra_{pt} \cact{B} \defeq
  (\cact{A}
  \rightarrow \cact{B}, A \rightarrow \changes{B}, \oplus_\rightarrow, +_\ra, 0_\ra)
  \]
  with
  the monoid structure $(A \rightarrow \changes{B}, +_\rightarrow,
  0_\rightarrow)$ and the action $\oplus_\rightarrow$ defined as:
  \begin{align*}
    (f \oplus_\to \delta f)(x) &\defeq f(x) \oplus_B \delta f (x)\\
    (\delta f +_\rightarrow \delta g)(x) &\defeq \delta f(x) +_B \delta g(x)\\
    \mathbf{0}_\rightarrow (x) &\defeq \mathbf{0}_B
  \end{align*}
\end{definition}

That is, a change is given pointwise, mapping each point in the domain to a
change in the codomain.

The above definition is not always well-typed, since given $f : \cact{A}
\rightarrow \cact{B}$ and $\delta f : A \rightarrow \changes{B}$ there is no
guarantee that $f \oplus_\rightarrow \delta f$ is differentiable. We present two
sufficient criteria that guarantee this.

\begin{theorem}
  \label{theorem:functionalCAct}
  Let $\cact{A}$ and $\cact{B}$ be change actions, and suppose that $\cact{B}$
  satisfies one of the following conditions:
  \begin{itemize}
    \item $\cact{B}$ is a transitive change action.
    \item The change action $\cact {\changes{B}} \defeq (\changes{B},
    \changes{B}, +_B, +_B, 0_B)$ is transitive and $\oplus_B : B \times
    \changes{B} \to B$ is differentiable with respect to it.
  \end{itemize}
  Then the pointwise functional change action $(\cact{A} \rightarrow \cact{B}, A
  \rightarrow \changes{B}, \oplus_\rightarrow)$ is well defined\footnote{Either
  of these conditions is enough to guarantee that the pointwise functional
  change action is well defined, but it can be the case that $\cact{B}$
  satisfies neither and yet pointwise change actions into $\cact{B}$ do exist. A
  precise account of when pointwise functional change actions exist is outside
  the scope of this thesis.}.
\end{theorem}
\begin{proof}
    We need to show that, given a differentiable function $f : \cact{A} \to
    \cact{B}$ and a change $\delta f : A \to \Delta B$, the updated function $f
    \oplus_\to \delta f$ is also differentiable. 

    \begin{itemize}
        \item When $\cact{B}$ is transitive, every function into it is
        differentiable and so the above holds trivially.
        \item In the second case, since $\cact{\Delta B}$ is transitive, every
        change $\delta f : A \to \Delta B$ is differentiable. In particular,
        consider a differentiable function $f$ and a change $\delta f$, with
        derivatives $\d{f}, \d{\delta f}$ respectively. Then:
        \begin{align*}
            &(f \oplus_\to \delta f)(x \oplus_A \delta x)\\
            &= (f(x \oplus_A \delta x)) \oplus_B (\delta f(x \oplus_A \delta x))\\
            &= (f(x) \oplus_B \d{f}(x, \delta x)) \oplus_B (\delta f(x) \oplus_B \d{\delta f}(x, \delta x))\\
            &= (f(x) \oplus_B \delta f(x)) \oplus_B \d[\oplus_B](f(x), \delta f(x), \d{f}(x, \delta x), \d{\delta f}(x, \delta x))\\
            &= (f \oplus_\to \delta f)(x) \oplus_B \d[\oplus_B](f(x), \delta f(x), \d{f}(x, \delta x), \d{\delta f}(x, \delta x))
        \end{align*}
        where $\d[\oplus_B]$ denotes the derivative of the map $\oplus_B$. It
        follows that the map
        \[ \d[f \oplus_\to \delta f] \defeq \d[\oplus_B](f(x), \delta f(x),
        \d{f}(x, \delta x), \d{\delta f}(x, \delta x))\]
        is a derivative for $f \oplus_\to \delta f$.\qedhere
    \end{itemize}
\end{proof}

\begin{remark}
  \label{rem:pointwise-change-actions-dont-exist}
  The above result is interesting as it already hints at the difficulties with
  functional change actions, as well as their solutions. As we have shown in
  Proposition~\ref{prop:functionalChangesAreDifferentiable}, every functional
  change action is made of differentiable functions. However, it is in general
  not possible to guarantee that $f \oplus \delta f$ be differentiable, for
  $f$ differentiable.
  
  In order to solve this problem, the previous theorem required that $\delta
  f, \oplus$ were themselves differentiable, but for this approach to work one
  needs to endow the change set itself with the structure of a change action.
\end{remark}

As a direct consequence of Theorem~\ref{theorem:functionalCAct}, it follows that
whenever $L$ is a Boolean algebra (and hence has a complete change action), the
pointwise functional change action $\cact{A} \Ra_{pt} \cact{L}_{\twist}$ is
well-defined.

Pointwise functional change actions are functional in the sense of
Definition~\ref{def:functionalChanges}. Moreover, the derivative of the
evaluation map is quite easy to compute. 
\begin{proposition}
\label{prop:evDerivatives}
  Let $\cact{A}$ and $\cact{B}$ be change actions such that the pointwise functional change action
  $\cact{A} \Ra_{pt} \cact{B}$ is well defined, and let
  $f: \cact{A} \rightarrow \cact{B}$,
  $a \in A$, $\change{a} \in \changes{A}$,
  $\change{f} \in A \rightarrow \changes{B}$.

  Then the following are both derivatives of the evaluation map:
  \begin{align*}
    \d{\Ev}_1((f, a), (\change{f}, \change{a})) 
    &\defeq \d{f}(a, \change{a}) + \change{f}(a \oplus \change{a})\\
    \d{\Ev}_2((f, a), (\change{f}, \change{a})) 
    &\defeq \change{f}(a) + \d[f \oplus \change{f}](a, \change{a})
  \end{align*}
\end{proposition}
\begin{proof}
    Straightforward calculation:
    \begin{align*}
        (f \oplus_\to \delta f)(a \oplus_A \delta a)
        &= f(a \oplus_A \delta a) \oplus_B \delta f(a \oplus_A \delta a)\\
        &= f(a) \oplus_B \big[ \d{f}(a, \delta a) + \delta f(a \oplus_A \delta a)\big]\\
        &= f(a) \oplus_B \d{\Ev}_1((f, a), (\delta f, \delta a))\\
        (f \oplus_\to \delta f)(a \oplus_A \delta a)
        &= (f \oplus_\to \delta f)(a) \oplus_B \d[f \oplus_\to \delta f](a, \delta a)\\
        &= f(a) \oplus_B \big[\delta f(a) + \d[f \oplus_\to \delta f](a, \delta a)\big]
        \qedhere
    \end{align*}
\end{proof}

Functional change actions merely require that a derivative of the evaluation map
exist -- a pointwise change action, on the other hand, actually provides a
definition for it. In practice, this means that we will only be able to use the
results in Section~\ref{sec:fixpoints} (which allow us to differentiate fixed
point h) in settings where we have pointwise change actions, or when we
have some other way of computing a derivative of the evaluation map. 

\section{Directed-Complete Partial Orders and Fixed Points}
\label{sec:dcpos}

As we have seen in Section~\ref{sec:nonRecursiveDatalog}, a Datalog program can
be understood as a function on Boolean algebras mapping the extension of its
free predicate variables to the extension of the entire program. Since
predicates may be recursive, the semantics of the entire program is then
obtained by jointly taking the least fixed point of the denotations of the
predicates that constitute it.

The computation of this fixed point computation is licit, as Datalog predicates
are necessarily monotone\footnote{Even in the case where parity-stratified
negation is allowed, recursion can only happen through an even number of
negations.}, and their denotation lies in the directed-complete partial order of
sets of tuples.

We have already obtained a suitable change action on Boolean algebras in
Proposition~\ref{prop:booleanCAct}, which can be used to interpret the semantics
of a single Datalog predicate; our goal is now to show how to incrementalise the
fixed-point operator in this change action. To do this, we will first define
what it means for a change action to be compatible with a dcpo structure. Then
we will use the functional change actions in
Section~\ref{sec:changesOnFunctions} to show how to differentiate iteration and
fixed point operators. 

\subsection{Continuous Actions on Dcpos}

Before proceeding on to the rest of the section we remind the reader of the
basic definitions of which we will make use throughout. 

\begin{definition}
    Given a partial order $(D, \sqsubseteq)$, a (non-empty) \textbf{directed
    set} is a subset $U \subseteq D$ such that, whenever $u, v \in D$, there is
    some $w \in D$ satisfying $u \sqsubseteq w, v \sqsubseteq w$.
    
    If every directed subset $U \subseteq D$ has a least upper bound, the
    partial order $(D, \sqsubseteq)$ is said to be \textbf{directed-complete}\footnote{
      Some sources refer to directed-complete partial orders as simply ``complete
      partial orders''. We write dcpo instead for clarity and to distinguish them
      from similar objects such as complete lattices and $\omega$-complete partial 
      orders.
    }. We
    will denote such least upper bounds by $\bigsqcup U$ or $\bigsqcup_{i \in I}
    u_i$ when $U = \set{u_i \for i \in I}$.

    A dcpo is \textbf{pointed} whenever it has a least element, which we will
    denote by $\bot$. In what follows, we will assume that every dcpo we work
    with is in fact a pointed dcpo.
\end{definition}

\begin{definition}
    Given dcpos $(D, \sqsubseteq_D)$ and $(E, \sqsubseteq_E)$, a function $f : D
    \to E$ is \textbf{continuous} whenever it is monotone and furthermore it
    preserves least upper bounds, that is, for every directed subset $U
    \subseteq D$:
    \[
        f\big( \bigsqcup U \big) = \bigsqcup f(U)
    \]
\end{definition}

As was the case with partial orders, we can define change actions on dcpos,
rather than sets, as change actions whose base and change sets are endowed with
a dcpo structure, and where the monoid operation and action are
continuous. This is in fact the same thing as a change action internal to the
category $\cat{DCPO}$ of dcpos and continuous maps.

\begin{definition}
  A change action $\cact{A}$ is \emph{continuous} if
  \begin{itemize}
    \item $A$ and $\changes{A}$ are dcpos.
    \item $\oplus$ is a continuous function from $A \times \changes{A} \rightarrow A$.
    \item $+$ is a continuous map from $\changes{A} \times \changes{A} \rightarrow \changes{A}$.
  \end{itemize}
\end{definition}

Unlike posets, the change order $\changeOrder$ does \emph{not}, in general,
induce a dcpo on $\changes{A}$. As a counterexample, consider the change
action $(\NN^*, \NN, +)$, where $\NN^*$ denotes the dcpo
of natural numbers extended with positive infinity. In particular, the corresponding
change order in $\NN$ coincides with the standard order $\leq$, which is not a dcpo.

A key example of a continuous change action is the $\cact{L}_{\twist}$ change
action on Boolean algebras - provided the underlying Boolean algebra $L$ is
itself complete.

\begin{proposition}
  \label{prop:booleanAlgebraContinuous}
  Whenever $L$ is a complete Boolean algebra, the change action
  $\cact{L}_{\twist}$ is continuous.
\end{proposition}
\begin{proof}
  $L$ is a complete lattice, thus in particular it is a dcpo. $L \twist L$ is a
  partial order with the order $\changeOrder$. Now let $U \equiv \{ (p_i, q_i) \mid i
  \in I \} \subseteq L \twist L$ be a $\changeOrder$-directed set. We claim that
  \[
    \bigsqcup U = \left( \bigvee \{p_i \mid i \in I\}
                       , \bigwedge \{q_i \mid i \in I\}\right)
  \]
  is a least upper bound for $U$. To see this, first note that $\bigsqcup U$ is
  indeed the least upper bound for $U$ in the complete lattice $L \times
  L^{op}$, thus all that remains to prove is that $\bigsqcup U$ is in fact an
  element of $L \twist L$, i.e. $\bigvee \{p_i\} \wedge \bigwedge \{ q_i \} =
  \bot$. But this is straightforward: since $\wedge, \vee$ distribute over
  infinite meets and joins, we have:
  \begin{gather*}
    \bigvee \set{p_i \for i \in I} \wedge \bigwedge \set{q_j \for j \in I}
    = \bigvee \left( p_i \wedge \bigwedge \set{q_j \for j \in I }\right)
    \sqsubseteq \bigvee \left(p_i \wedge q_i\right) = \bot
  \end{gather*}

  Continuity of $\oplus_{\twist}$ in its second argument is easy to prove:
  \begin{align*}
    &a \oplus_{\twist} \bigvee_{i \in I} \set{(p_i, q_i)}\\
    &= a \oplus_{\twist} \pa{\bigvee_{i \in I} p_i, \bigwedge_{j \in I} q_j}\\
    &= \pa{a \vee \bigvee_{i \in I} p_i} \wedge \neg \bigwedge_{j \in I} q_j\\
    &= \pa{a \vee \bigvee_{i \in I} p_i} \wedge \bigvee_{j \in I} \neg q_j\\
    &= \bigvee_{i \in I} \bigvee_{j \in I} a \oplus_{\twist} (p_i, q_j)\\
    &= \bigvee_{i \in I} a \oplus_{\twist} (p_i, q_i)
  \end{align*}

  Continuity of $\oplus_{\twist}$ in its first argument and continuity of
  $\twist$ follow easily from their definitions and the continuity of $\vee$ and
  $\wedge$.
\end{proof}

For a general overview of results in domain theory and dcpos, we refer the
reader to an introductory work such as \cite{abramsky1994domain}, but we will
state here some specific results that we find useful, such as the following,
whose proof can be found in \cite[Lemma~3.2.6]{abramsky1994domain}:

\begin{proposition}
  \label{prop:distributivityLimit}
  A function $f : A \times B \rightarrow C$ is continuous iff it is continuous
  in each variable separately.
\end{proposition}

It is a well-known result in standard calculus that the limit of an absolutely
convergent sequence of differentiable functions $\{f_i\}$ is itself
differentiable, and its derivative is equal to the limit of the derivatives of
the $f_i$. A consequence of Proposition~\ref{prop:distributivityLimit} is the
following analogous result:

\begin{theorem}
  \label{thm:diffContinuous}
  Let $\cact{A}$ and $\cact{B}$ be change actions, with $\cact{B}$ continuous
  and let $\{f_i\}$ and $\{\d{f_i}\}$ be $I$-indexed directed sets of functions
  in $A \rightarrow B$ and $A \times \changes{A} \rightarrow \changes{B}$
  respectively.

  Then, if $\d{f_i}$ is a derivative for $f_i$ for every $i \in I$, the least
  upper bound of this set $\bigsqcup_{i \in I} \d{f_i}$ is a derivative for
  $\bigsqcup_{i \in I} f_i $.
\end{theorem}
\begin{proof}
  It suffices to apply $\oplus$ and Proposition~\ref{prop:distributivityLimit}
  to the directed families $\{ f_i(a) \}$ and $\{ \d{f_i}(a, \change{a}) \}$.
\end{proof}

We also state the following additional fixed point lemma. This is a specialisation
of Beki\'c's Theorem \cite[section 10.1]{winskel1993formal}, but it has a
straightforward direct proof.

\begin{proposition}
  \label{prop:factoringFixpoints}
  Let $A$ and $B$ be dcpos, $f : A \rightarrow A$ and $g: A \times B \rightarrow
  B$ be continuous functions, and let
  \begin{displaymath}
    h(a, b) \defeq (f(a), g(a, b))
  \end{displaymath}
  Then
  \begin{displaymath}
    \lfp(h) = (\lfp(f), \lfp(\lambda b . g(\lfp(f), b)))
  \end{displaymath}
\end{proposition}
\begin{proof}
  Let
  \begin{displaymath}
    p(b) = (\lfp(f), g(\lfp(f), b))
  \end{displaymath}
  Then, by induction, $h(h^i(\bot)) \leq p(p^i(\bot))$ and,
  since $h$ is continuous, we have: 
  \[
    \lfp(h) 
    = \bigsqcup_{i \in \NN} h^i(\bot) \leq \bigsqcup_{i \in \NN} p^i(\bot) 
    = \lfp(p)
  \]

  But $h(\lfp(p)) = \lfp(p)$, therefore $\lfp(p)$ is a fixed point 
  for $h$ and so $\lfp(h) \leq \lfp(p)$.

  Hence $\lfp(h) = \lfp(p) = (\lfp(f), \lfp(\lambda b . g(\lfp(f), b)))$.
\end{proof}

\subsection{Fixed Points}
\label{sec:fixpoints}

The usefulness of directed-complete partial orders lies in the ability to apply
Kleene's fixed point theorem to compute least fixed points of monotone maps --
indeed, this is how we will define the full semantics of a recursive Datalog
program.

In order to apply our derivative-based approach to the computation of fixed
points, we now focus on obtaining derivatives for iteration and fixed point
operators. As we shall see, these provide us with a concrete way of applying the
derivative for the (non-recursive) formula semantics of Datalog from
Section~\ref{sec:nonRecursiveDatalog} to the problem of incrementalising the
fixed point computation involved in the evaluation of an entire Datalog program.

For reference, we state here Kleene's fixed point theorem.

\begin{theorem}
    Let $(D, \sqsubseteq)$ a directed-complete partial order (with a least
    element), and $f : D \to D$ a continuous function. Then $f$ has a least
    fixed point $\lfp(f)$ which is the least upper bound of the ascending chain:
    \[
        \bot \sqsubseteq f(\bot) \sqsubseteq f(f(\bot)) \sqsubseteq \ldots
    \]

    Alternatively, this least fixed point can be expressed in terms of the
    finite iteration operator $\iter$ defined by:
    \begin{gather*} 
        \iter : (A \to A) \times \NN \to A\\
        \iter(f, n) \defeq f^n(\bot)
    \end{gather*} 
    Then the least fixed point $\lfp(f)$ can be written as:
    \[
        \lfp(f) = \bigsqcup_{i \in \NN} \iter(f, i)
    \]
\end{theorem}

The iteration function is the basis for all the results in this section. We will
show that it is differentiable; in doing so, we obtain two tools:
differentiating it with respect to the second argument (the number of
iterations) will provide us with a way to incrementally computing $\iter(f, i +
1)$ from $\iter(f, i)$. On the other hand, differentiating it with respect to
the first argument (the function to be iterated) gives us a solution for the
\emph{incremental maintenance} problem, that is to say, it allows one to
incrementally recompute the entire fixed point computation given a change in the
rules of our Datalog program.

The following theorems provide a generalisation of semi-naive evaluation to any
differentiable function over a continuous change action. Throughout this section
we will assume that we have a continuous change action $\cact{A}$, and any
reference to the change action $\cact{\NN}$ will refer to the monoidal change
action on the naturals defined in Example~\ref{ex:monoidal-change-action}.

Since we are trying to incrementalise the iterative step, we start by taking the
partial derivative of $\iter$ with respect to $n$.

\begin{proposition}
  \label{prop:iterDerivatives1}
  Let $\cact{A}$ be a transitive change action and let $f: A \rightarrow A$ be a
  function with derivative $\d{f}$ (note that such a derivative always exists as
  $\cact{A}$ is transitive). Then $\iter$ is differentiable with respect to its
  second argument, and a partial derivative is given by:
  \begin{align*}
    &\partial_2{\iter}: (A \to A) \times \NN \times \changes{\NN} \to \changes{A}\\
    &\partial_2{\iter}(f, n, m) \defeq \begin{cases}
        \iter(f, m) \ominus \iter(f, 0) &\text{ if $n = 0$ }\\
        \d{f}(\iter(f, n - 1), \partial_2{\iter}(f, n - 1, m)) &\text{ otherwise}
    \end{cases}
  \end{align*}
\end{proposition}
\begin{proof}
  The proof proceeds by induction on $n$. We show the inductive step (the base
  step is trivial).
  \begin{align*}
    \iter(f, (n+1) + m)
    &=f(\iter(f, n + m))\\
    &=f(\iter(f, n) \oplus \partial_2{\iter}(f, n, m))\\
    &=\iter(f, n+1) \oplus \d{f}(\iter(f, n), \partial_2{\iter}(f, n, m))\qedhere
  \end{align*}
\end{proof}

Looking back at our leading example from Section~\ref{sec:incremental-1}, we saw
how the semantics of the transitive closure relation can be obtained naively by
repeatedly iterating its semantics until a fixed point is reached, or more
cleverly by incrementalising the iteration procedure by computing a sequence of
increments $\Delta \sem{tc}_i$. Abstractly, this corresponds to computing
$\iter(f, i + 1)$ by updating the previous iteration with its derivative. This
process itself can be elegantly iterated, as the following result shows.

\begin{proposition}
  \label{prop:iterDerivatives2}
  Let $\cact{A}$ be a transitive change action and let $f: A \rightarrow A$ be a
  function with derivative $\d{f}$, and define the recurrence relation
  $\nextiter$ as follows:
    \begin{align*}
        &\nextiter_f : A \times \changes{A} \rightarrow A \times \changes{A}\\
        &\nextiter_f (\bot, \bot) \defeq (\bot, f(\bot) \ominus \bot)\\
        &\nextiter_f (a, \change{a}) \defeq (a \oplus \change{a}, \d{f}(a, \change{a}))
    \end{align*}
    Then $\nextiter$ verifies the equation:
    \[
    \nextiter_f^n (\bot, \bot) = (\iter(f, n), \partial_2{\iter}(f, n, 1))
    \]
\end{proposition}

As we have remarked, this result gives us a way to compute a fixed point
incrementally, by adding successive changes to an accumulator until we reach it.
This is exactly how our semi-naive evaluation example works: by computing the
increment $\Delta\sem{tc}_i$ in lock-step with the accumulator $\sem{tc}_i$, and
adding the delta into the accumulator at each step until converging to the final
output.

\begin{theorem}
\label{theorem:fixpointIter}
  Let $\cact{A}$ be a continuous, transitive change action, with $f: \cact{A}
  \rightarrow \cact{A}$ a continuous function.

  Then $\lfp(f) = \bigsqcup_{n \in \NN}(\pi_1(\nextiter_f^n(\bot,
  \bot)))$.\footnote{Note
    that we have \emph{not} taken the fixed point of $\nextiter_f$, since it may
    not be continuous.}
\end{theorem}

With this we have shown how to use derivatives to compute fixed points more
efficiently, but we also want to consider the derivative of the fixed point operator
itself. A typical use case for this is where we have calculated some fixed point
\begin{displaymath}
  F_{E} \defeq \fix (\lambda X . F(E, X))
\end{displaymath}
then update the parameter $E$ with some change $\change{E}$ and wish to compute the new
value of the fixed point, i.e.
\begin{displaymath}
  F_{E \oplus \change{E}} \defeq \fix (\lambda X . F(E \oplus \change{E}, X))
\end{displaymath}

This process can be understood as applying a change to the \emph{function}
whose fixed point we are taking. We go from computing the least fixed point of $F(E, \_)$ to
computing the least fixed point of $F(E \oplus \change{E}, \_)$. If we have a pointwise
functional change action then we can express this change as a function giving
the change at each point, that is:
\begin{displaymath}
  \lambda X . F(E \oplus \change{E}, X) \ominus F(E, X)
\end{displaymath}

In Datalog this would allow us to update a recursively defined relation given an
update to one of its non-recursive dependencies, or the extensional database.
Continuing with the transitive closure example, we might make changes to the
edge relation $e$ and expect $tc$ to be recalculated efficiently, without
starting from scratch.

However, to compute these examples requires us to provide a derivative for the
fixed point operator $\fix$, as we want to know how the resulting fixed point changes
given a change to its input function.

\begin{definition}
\label{def:fixpointDerivatives}
  Let $\cact{A}$ be a continuous change action, let $\cact{U} \subseteq \cact{A}
  \Rightarrow \cact{A}$ be a functional change action (not necessarily
  pointwise) and suppose $\lfp$ and $\lfp_{\changes{A}}$ are (least) fixed point
  operators on $U$ and $\changes{A}$ respectively.
  
  Then we define
  \begin{align*}
    &\adjust : U \times \changes{U} \rightarrow (\changes{A} \rightarrow \changes{A})\\
    &\adjust(f, \change{f}) \defeq \lambda\change{a} . \d{\Ev}((f, \fix_U(f)), (\change{f}, \change{a}))
  \end{align*}\vspace{-30pt}
  \begin{align*}
    &\d{\lfp} : U \times \changes{U} \rightarrow \changes{A}\\
    &\d{\lfp}(f, \change{f}) \defeq \lfp_{\changes{A}}(\adjust(f, \change{f}))
  \end{align*}
\end{definition}

The suggestively named $\d{\lfp_U}$ will in fact turn out to be a derivative.
The appearance of $\d{\Ev}$, a derivative for the evaluation map, in the
definition of $\adjust$ is also no coincidence: as evaluating a fixed point
consists of many steps of applying the evaluation map, so computing the
derivative of a fixed point consists of many steps of applying the derivative of
the evaluation map.

Since the least fixed point of $f$, $\lfp(f)$ is characterised as the limit of a
chain of functions, Theorem~\ref{thm:diffContinuous} suggests a way to compute
its derivative. It suffices to find a derivative $\d{\iter_n}$ of each iteration
map such that the resulting set $\{ \d{\iter_n} \mid n \in \NN\}$ is directed,
which will entail that $\bigsqcup_{n \in \NN}\d{\iter_n}$ is a derivative of
$\lfp$.

These correspond to the first partial derivative of $\iter$ -- this time with
respect to $f$. As before, we will provide a definition for such derivatives by
induction on $n$.

\begin{proposition}
  \label{prop:iterDerivativesF}
  Let $\cact{A}$ be a change action, let $\cact{U} \subseteq \cact{A}
  \Rightarrow \cact{A}$ be a functional change action and let $\iter|_U$ be the
  restriction of $\iter$ to the set $U$.

  Then $\iter|_U$ is differentiable with respect to its first argument, with a
  partial derivative given by:
  \begin{align*}
    &\partial_1{\iter|_U} : U \times \NN \times \Delta U \to \Delta{A}\\
    &\partial_1{\iter|_U} (f, n, \change{f}) \defeq \begin{cases}
        0 &\text{when n = 0}\\
        \d{\Ev}((f, \iter|_U(f, n - 1)), (\delta f, \d_1{\iter|_U(f, n - 1), \delta f}))
        &\text{otherwise}
    \end{cases}
  \end{align*}
\end{proposition}
\begin{proof}
  The base case is easy to prove. For the inductive step:
  \begin{align*}
    &\iter|_U(f \oplus \change{f}, n + 1)\\
    &=(f \oplus \change{f})(\iter|_U(f \oplus \change{f}, n))\\
    &= (f \oplus \change{f})(
        \iter|_U(f, n)
        \oplus \partial_1{\iter|_U}(f, n, \change{f})
      )\\
    &= f(\iter|_U(f, n)) \oplus \d{\Ev}((f, \iter|_U(f, n)), (\change{f},
      \partial_1{\iter|_U}(f, n, \change{f})))\\
    & =\iter|_U(f, n + 1) \oplus \partial_1{\iter|_U}(f, n, \change{f}) \qedhere
  \end{align*}
\end{proof}

As before, we can now compute $\partial_1{\iter}$ together with $\iter$ by
mutual recursion.\footnote{In
  fact, the recursion here is not \emph{mutual}: the first component does not
  depend on the second. However, writing it in this way makes it
  amenable to computation by fixed point, and we will in fact be able to avoid the
  re-computation of $\iter_n$ when we show that it is equivalent to $\d{\lfp}$.
}
\begin{align*}
  &\nextiter_{f, \change{f}} : A \times \changes{A} \rightarrow A \times \changes{A}\\
  &\nextiter_{f, \change{f}} (a, \change{a}) \defeq (f(a), \d{\Ev}((f, a), (\change{f}, \change{a})))
\end{align*}
Which has the property that
\begin{align*}
  &\nextiter_{f, \change{f}}^n (\bot, \bot) = (\iter(f, n), \partial_1{\iter}(f, \change{f}, n)).
\end{align*}

This provides us with a sequence of derivatives for $\iter|_U$ whose limit we
can take to obtain a derivative for the least fixed point operator, as in
Theorem~\ref{thm:diffContinuous}. If we do so we will discover that it is
exactly $\d{\lfp}$ (defined as in Definition~\ref{def:fixpointDerivatives}),
showing that $\d{\lfp}$ is indeed a derivative for $\lfp$.

\begin{theorem}
  \label{theorem:leastFixpointDerivatives}
  Let
  \begin{itemize}
    \item $\cact{A}$ be a continuous change action
    \item $\cact{U} \subseteq A \Ra A$ be a functional change action where $U$
      is precisely the set of continuous functions $f : A \rightarrow A$.
    \item $f \in U$ be a continuous, differentiable function
    \item $\change{f} \in \changes{U}$ be a function change
    \item $\d{\Ev}$ be a derivative of the evaluation map which is continuous with
      respect to $a$ and $\change{a}$.
  \end{itemize}
  Then $\d{\lfp}$ is a derivative of $\lfp$.
\end{theorem}
\begin{proof}
  $\partial_1{\iter|_U}$ and $\nextiter_{f, \change{f}}$ are continuous since
  $\d{\Ev}$ and $f$ are.
  Hence the set $\{ \partial_1{\iter|_U}(\cdot, \cdot, n) \}$ is directed, and so $\bigsqcup_{i \in \NN}
  \{ \partial_1{\iter|_U}(\cdot, \cdot, n) \}$ is indeed a derivative for $\lfp$.
  We now show that it is equivalent to $\d{\lfp}$:
  \begin{align*}
    &\bigsqcup_{n \in \NN} \partial_1{\iter|_U}\pa{f, \change{f}, n}\\
    &=\bigsqcup_{n \in \NN} \pi_2\pa{\nextiter_{f, \change{f}}^n\pa{\bot}}\\
    &=\pi_2\pa{\bigsqcup_{n \in \NN} \nextiter_{f, \change{f}}^n\pa{\bot}} \tag{$\pi_2$ is continuous}\\
    &= \pi_2 \pa{\lfp\pa{\nextiter_{f, \change{f}}}} \tag{continuity of $\nextiter_{f, \change{f}}$, Kleene's Theorem}\\
    &= \pi_2 \pa{\pa{\lfp\pa{f}, \lfp \pa{\lambda\ \change{a}. \d{\Ev}\pa{\pa{f, \lfp f}, \pa{\change{f}, \change{a}}}}}}
    \tag{by \ref{prop:factoringFixpoints}, and the definition of $\nextiter$}\\
    &= \pi_2 \pa{\lfp\pa{f}, \lfp\pa{\adjust\pa{f, \change{f}}}}\\
    &= \lfp\pa{\adjust\pa{f, \change{f}}}\\
    &= \d{\lfp}\pa{f, \change{f}}\hfill\qedhere
  \end{align*}
\end{proof}

Computing this derivative still requires computing a fixed point (over the poset of
changes) but this may still be significantly less expensive than recomputing the
full new fixed point.

\subsection{Derivatives for Recursive Datalog}
\label{sec:recursiveDatalog}

Given a semantics for the non-recursive fragment of a language, we can extend it
to handle recursive definitions using fixed point operators. Section
\ref{sec:fixpoints} lets us extend our derivative for the non-recursive
semantics to a derivative for the recursive semantics, as well as letting us
compute the fixed points themselves incrementally. 

We have demonstrated the technique with Datalog, although we believe our
approach to be generic enough that it should work in many other settings. First
of all, we define the usual ``immediate consequence operator'' which computes
``one step'' of our program semantics.

\begin{definition}
  Given a Datalog program $\mathbb{P} = (P_1, \dots, P_n)$, where $P_i$ is a
  predicate, with schema $\Gamma_i$, the \textbf{immediate consequence operator}
  $\consq: \Rel^n \rightarrow \Rel^n$ is defined as follows:
  \[
    \consq(\semR_1, \dots, \semR_n) 
    = (\sem{P_1}_{\Gamma_1}(\semR_1, \dots, \semR_n), \dots, \sem{P_n}_{\Gamma_n}(\semR_1, \dots, \semR_n))
    \]
\end{definition}
That is, given a value for the program, we pass in all the relations
to the denotation of each predicate, to get a new tuple of relations.
\begin{proposition}
  \label{prop:consq-diff}
  The immediate consequence operator $\consq$ is differentiable.
\end{proposition}
\begin{proof}
  Straightforward corollary of \ref{thm:datalog-derivative}
\end{proof}

\begin{definition}
  The semantics of a program $\mathbb{P}$ is defined to be
  \begin{displaymath}
    \sem{\mathbb{P}} \defeq \lfp_{\Rel^n}(\consq)
  \end{displaymath}
  and may be calculated by iterative application of $\consq$ to $\bot$ until
  a fixed point is reached.
\end{definition}

Whether or not this program semantics is correct will depend of course on
whether the fixed point exists. Typically this is ensured by constraining the
program such that $\consq$ is monotone (or, in the context of a dcpo,
continuous). In the setting of Datalog, all programs are continuous and, since
the underlying lattices are finite, the fixed point computation terminates in a
finite number of iterations, so we can apply
\ref{theorem:leastFixpointDerivatives}.

We can easily extend a derivative for the formula semantics to a derivative for
the immediate consequence operator $\consq$. Putting this together with the
results from \ref{sec:fixpoints}, we now have created \emph{modular} proofs for
the two main results, which we hope would allow us to preserve them in the face
of changes to the underlying language.

\begin{corollary}
\label{theorem:diffEval}
  Datalog program semantics can be evaluated incrementally.
\end{corollary}
\begin{proof}
  Corollary of Theorem~\ref{theorem:fixpointIter} and
  Proposition~\ref{prop:consq-diff}.
\end{proof}

\begin{corollary}
\label{theorem:diffUpdate}
  Datalog program semantics can be incrementally maintained with changes to
  relations.
\end{corollary}
\begin{proof}
  Corollary of Theorem~\ref{theorem:leastFixpointDerivatives} and
  Proposition~\ref{prop:consq-diff}.
\end{proof}

Neither of these results are novel: indeed the first one is merely the very
well-known method of semi-naive evaluation, while there is already extensive
discussion of incremental maintenance in the literature (see e.g.
\cite{gupta1993maintaining, gupta1995maintenance}). Our contribution lies in the
principled, compositional nature of our approach: our proofs follow trivially
from our derivatives for non-recursive Datalog programs and the derivative for
the least fixed point operator.

%% file: change-action-models.tex
\label{chp:change-action-models}

The category $\ftor{CAct}(\cat{C})$ of change actions offers a natural and
simple model for talking about the notion of ``discrete'' differentiation that
arises in the context of incremental computation. One might object, however, that no
definition of differentiability is complete without a study of smoothness or higher
differentiability of some sort.

Indeed, a differential map $\cact{f} : \cact{A} \ra \cact{B}$ associates a
derivative $\d f$ to a $\cat{C}$-map $f$. But what should the derivative of
$\d f$ be? Its type is $\d f : A \times \Delta A \ra \Delta B$ -- it seems
reasonable, then, that the derivative $\d^2 f$ of such a $\d f$ should have
type
\[
  \d^2 f : (A \times \Delta A) \times \Delta (A \times \Delta A) \to \Delta (\Delta B)
\]
But this type does not immediately make sense, as it is not clear what the
``second-order change spaces'' $\Delta (\Delta A), \Delta (\Delta B)$ should be.

We recall from Remark~\ref{rem:pointwise-change-actions-dont-exist} that a
similar issue arose when we tried to build pointwise functional change actions.
The existence of these was predicated upon changes of type $\delta f : A \to
\Delta B$ being differentiable, but there was, in general, no clear way to pick
a change action on $\Delta B$ with respect to which differentiability could be
decided.

A change action model is precisely a way of endowing each object of a category
$\cat{C}$ with the structure of a change action, and associating every
$\cat{C}$-map to a corresponding differential map in such a way that the chain
rule is satisfied (that is, the derivative associated to the map $g \circ f$ is
precisely the one that would be obtained by applying the chain rule to the
derivatives associated to $f$ and $g$ respectively). 

In a change action model, every object $A$ is assigned a fixed change action
$\cact{A}$ -- and since these change actions are internal to $\cat{C}$, the
change space $\Delta A$ itself carries a change action $\cact{\Delta A}$, thus
the higher change spaces above can be given an interpretation.

This chapter has two main sections: first, in Section~\ref{sec:extrinsic}, we
introduce the formal definition of a change action model and study some basic
properties, such as the structure of products and exponentials and the existence
of a tangent bundle functor. Section~\ref{sec:change-action-models:examples}
then lists a series of concrete examples of change action models of particular
interest.

Many of the results in this chapter have appeare in print in
\cite{alvarez2019change}. Some passages and figures have been reproduced
verbatim with permission.

\section{Coalgebras for \texorpdfstring{$\ftor{CAct}$}{CAct} and Their Properties}
\label{sec:extrinsic}

Recall that a \emph{copointed endofunctor} is a pair $(\ftor{F}, \sigma)$ where
the endofunctor $\ftor{F} : \cat{C} \ra \cat{C}$ is equipped with a natural
transformation $\sigma : \ftor{F} \xrightarrow{.} \ftor{Id}$. A \emph{coalgebra
for a copointed endofunctor $(\ftor{F}, \sigma)$} is an object $A$ of $\cat{C}$
equipped with a section $\alpha$ of $\sigma$, that is, a morphism $\alpha : A
\ra \ftor{F}A$ such that $\sigma_A \circ \alpha = \Id{A}$.

\begin{definition}
  A \emph{change action model} on a Cartesian category $\cat{C}$ is a coalgebra
  $\alpha : \cat{C} \ra \ftor{CAct}(\cat{C})$ for the copointed endofunctor
  $(\ftor{CAct}, \epsilon)$.
  
  Explicitly, a product-preserving functor $\alpha : \cat{C} \ra
  \ftor{CAct}(\cat{C})$ is a change action model whenever $\epsilon \circ \alpha
  = \Id{\cat{C}}$, i.e. whenever $\alpha$ endows every object of $\cat{C}$ with
  the structure of a change action, and associates every map with a choice of
  derivative for it.
\end{definition}

\noindent\emph{Assumption}. Throughout Section~\ref{sec:extrinsic}, we fix a
change action model $\alpha : \cat{C} \to \ftor{CAct}(\cat{C})$.

Given an object $A$ of $\cat{C}$, the coalgebra $\alpha$ specifies a (internal)
change action $\alpha(A) = (A, \changes A, \oplus_A, +_A, 0_A)$ in
$\ftor{CAct}(\cat{C})$. We abuse the notation and write $\changes A$ for the
carrier object of the monoid specified by $\alpha(A)$; similarly for $+_A,
\oplus_A$, and $0_A$ -- we will also write $\d[f]$ for the derivative assigned
to $f$ by $\alpha$\footnote{
  It may seem inconsistent to write $\d f$ for a derivative and $\d[f]$ for
  the derivative that $\alpha$ associates to $f$. We write $\d[f]$ to emphasise
  the role of $\d$ as an \emph{operator}, whereas writing $\d f$ for a derivative
  of $f$ is simply a convention.
}. 

Given a morphism $f : A \ra B$ in $\cat{C}$, there is an
associated differential map $\alpha(f) = (f, \d f) : \alpha(A) \ra \alpha(B)$.
Since 
\[ 
  \d[f] : A \times \changes{A} \ra \changes{B}
\]
is also a $\cat{C}$-morphism, there is a corresponding differential map
$\alpha(\d[f]) = (\d f, \d^2 f)$ in $\ftor{CAct}(\cat C)$, where 
\[
  \d^2[f] : (A \times \changes A) \times (\changes A \times \changes^2A) \ra \changes^2B
\]
is a second derivative for $f$ -- note that $\Delta (A \times \Delta A) = \Delta
A \times \Delta^2 A$, since $\alpha$ is, by definition, product-preserving.
Iterating this process, we obtain an $n$-th derivative $\d^n[f]$ for every
$\cat{C}$-morphism $f$.

An immediate consequence of the definition of a change action model is that the
structure maps $\oplus_A, +_A, 0_A$ associated to any object $A$ are themselves
differentiable, but we can make no further claims about their
derivatives\footnote{This is not entirely true -- we shall show later that the
derivatives of constant maps are always zero, and so $\d[0_A] = 0_{\Delta A}$.}
other that they are indeed derivatives (in the sense that they satisfy
Definition~\ref{def:derivative}). 

\begin{figure}
\begin{center}
  \begin{tikzpicture}
    \node[label=135:{\small $a \oplus_A \delta_1$}] (a) at (-1, 2) {$\bdot$};
    \node[label=45:{\small $(a \oplus_A \delta_1) \oplus_A \delta_2$}] (b) at (3, 3) {$\bdot$};
    \node[label=-45:{\small$a \oplus_A \delta_2$}] (c) at (3, -1) {$\bdot$};
    \node[label=-135:{\small$a$}] (d) at (0, 0) {$\bdot$};
    \draw[->, bend left] (d) to node[below] {\small$\delta_2$} (c);
    \draw[->, bend left] (d) to node[left] {\small$\delta_1$} (a);
    \draw[->, bend left] (a) to node[above] {\small$\delta_2$} (b);
    \draw[->, dashed,bend right, color=blue] (c) to node[right] {\small$\d[\oplus_A]((a, \delta_2),
    (\delta_1, 0_{\Delta A}))$} (b);
  \end{tikzpicture}
\end{center}
\caption{Square-filling property of $\d[\oplus_A]$}
\label{fig:oplus-derivative-1}
\end{figure}
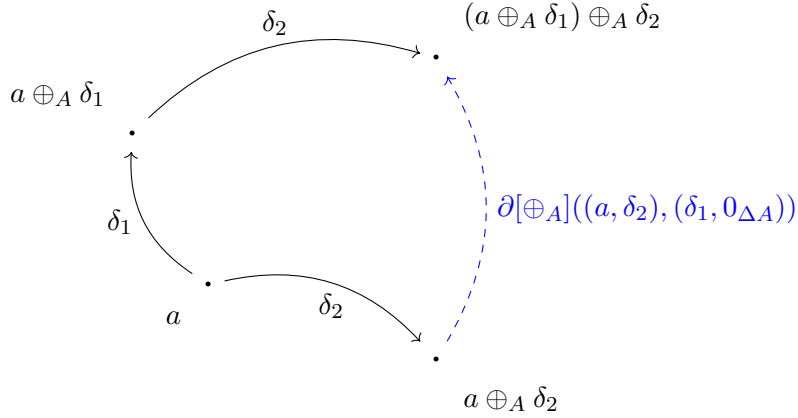

Yet there is some deep geometric intuition to be extracted from the existence of
a derivative for $\oplus$. As mentioned before, one can think of elements of $A$
as points in a space, and a change $\delta$ as a path from $a$ to $a \oplus_A
\delta$. Let us consider, then, a configuration of paths as in Figure~
\ref{fig:oplus-derivative-1}. In a more ``classical'' setting (e.g. if the
$\delta_i$ were simply paths in some geometric space), we could always go from
the point $a \oplus_A \delta_2$ to the point $(a \oplus_A \delta_1) \oplus_A
\delta_2$, by composing $\delta_2^{-1}$ with $\delta_1$. In change actions,
however, changes may not be invertible and so such a $\delta_2^{-1}$ may not
exist. Since $\oplus_A$ is differentiable, however, we can always find a path
that ``fills'' the missing edge in the diagram, as shown in blue.

\begin{figure}[H]
  \begin{center}
    \begin{tikzpicture}
      \node[label=-135:{\small $a$}] (a) at (0, 0) {$\bdot$};
      \node[label=90:{\small $a \oplus_A (\delta \oplus_{\Delta A} \delta^2)$}] (b) at (3, 3) {$\bdot$};
      \node[label=-45:{\small$a \oplus \delta$}] (c) at (4, 0) {$\bdot$};
      \draw[->, bend left=50] (a) to node[above,left] {\footnotesize $\delta \oplus_{\Delta A} \delta^2$} node[below,pos=0.6] (u) {} (b);
      \draw[->, bend left] (a) to node[below] {\footnotesize $\delta$} node[above,pos=0.5] (v) {} (c);
      \draw[->, commutative diagrams/Rightarrow, bend right] (v) to node[right] {\footnotesize$\delta^2$} (u);
      \draw[->, dashed, bend right, color=blue] 
      (c) to node[right] {\small$\d[\oplus_{A}]((a, \delta), (0_A, \delta_2))$} (b);
    \end{tikzpicture}
  \end{center}
  \caption{Triangle-filling property of $\d[\oplus_A]$}
  \label{fig:oplus-derivative-2}
\end{figure}
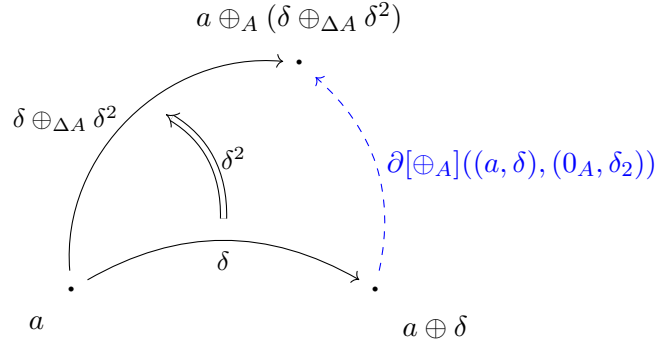

What's more, consider a second-order change $\delta^2$ -- this change maps some
first-order change $\delta$ to the first-order change $\delta \oplus_{\Delta A}
\delta^2$. Stretching the metaphor, such a second-order change can perhaps be
thought of as a (directed) homotopy from path $\delta$ to path $\delta
\oplus_{\Delta A} \delta^2$. Again one can think of the derivative of the
structure map $\oplus_A$ as allowing us to ``fill in'' the missing edge in the
triangle in Figure~\ref{fig:oplus-derivative-2}.

Composition of changes $+_A$ in a change action model is likewise
differentiable, a fact which sheds light on the regularity condition on
derivatives. Indeed, take a point $a$ and changes $\delta_1, \delta_2$ --
derivatives in change actions need not be additive, and so $\d[f] (a, \delta_1
+_A \delta_2) \neq \d[f](a, \delta_1) +_B \d[f](a, \delta_2)$. But
derivatives do satisfy the (weaker) condition of regularity, which states the following:
\[ 
  \d[f](a, \delta_1 +_A \delta_2) = \d[f](a, \delta_1) +_A \d[f](a \oplus_A \delta_1, \delta_2)
\]
Observe, then, that in any change action model $\d[f]$ is itself differentiable,
admitting a derivative $\d^2[f]$. In particular, letting $\omega = \d^2[f]((a,
\delta_2), (\delta_1, 0_{\Delta A}))$, the above equation can be equivalently
written as:
\begin{align*}
  \d[f](a, \delta_1 +_A \delta_2) 
  &= \d[f](a, \delta_1) +_A \d[f](a, \delta_2) \oplus_{\Delta A} \omega\\
  &= \left(\d[f](a, \delta_1) +_A \d[f](a, \delta_2)\right)
  \oplus_{\Delta A} \d[+_A]((\d[f](a, \delta_1), \d[f](a, \delta_2)), (0_{\Delta A}, \omega))
\end{align*}

Independently of the exact value of the second-order term, the above expression
can be read as stating that regular derivatives are additive ``up to some
second-order perturbation''. More generally, whenever we have first-order
changes $\delta_1, \delta_2$ and second-order changes $\delta^2_1, \delta^2_2$,
differentiability of addition defines something akin to a  ``horizontal''
composition $\delta^2_1 \star \delta^2_2 \equiv \d[+_A]((\delta_1, \delta_2),
(\delta^2_1, \delta^2_2))$, represented by the red arrow in the diagram below.

\begin{figure}[H]
  \begin{center}
    \begin{tikzpicture}
      \node (a) at (-3, 0) {};
      \node (a') at (-3, 1) {};
      \node (b) at (0, 0)   {};
      \node (b') at (0, 1) {};
      \node (c) at (3, 0) {};
      \node (c') at (3, 1) {};
      \draw[->, bend right] (a) to node[below] (delta12) {\scriptsize $\delta_1 \oplus_{\Delta A} \delta^2_1$} (b);
      \draw[->, bend right] (b) to node[below] (delta22) {\scriptsize $\delta_2 \oplus_{\Delta A} \delta^2_2$} (c);
      \draw[->, bend left] (a') to node[above] (delta11) {\scriptsize $\delta_1$} (b');
      \draw[->, bend left] (b') to node[above] (delta21) {\scriptsize $\delta_2$} (c');
      \draw[->, bend right=60] (a.south) to node[below] (dst) {\footnotesize$(\delta_1
      \oplus_{\Delta A} \delta^2_1) +_{A}(\delta_2 \oplus_{\Delta A} \delta^2_2)$} (c.south);
      \draw[->, bend left=60] (a'.north) to node[above] (src) {\footnotesize$\delta_1
      +_A \delta_2$} (c'.north);
      \draw[->, shorten >=0.5cm, shorten <=0.5cm, commutative diagrams/Rightarrow] (delta11) to node[left] {\scriptsize $\delta^2_1$}
      (delta12);
      \draw[commutative diagrams/Rightarrow, shorten >=0.5cm, shorten <=0.5cm] (delta21) to node[right] {\scriptsize $\delta^2_2$}
      (delta22);
      \draw[->, commutative diagrams/Rightarrow, color=red, dashed,
      shorten >=0.8cm, shorten <=0.8cm] (src) to node[right]
      {\footnotesize$\delta^2_1 \star \delta^2_2$} (dst);
    \end{tikzpicture}
  \end{center}
  \caption{Horizontal composition of second-order changes}
  \label{fig:plus-derivative-1}
\end{figure}
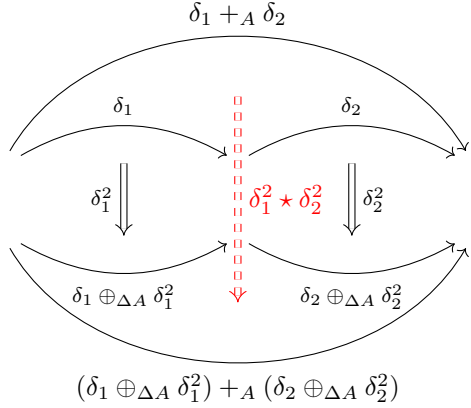

This horizontal composition is strikingly similar to the horizontal composition
of 2-cells in a 2-category -- this is because, as we shall show later, it is
precisely that!

\subsection{Tangent Bundles in Change Action Models}
\label{sec:tangent-bundle-change-action-model}

The tangent bundle functor, which maps every manifold to its tangent bundle and
pairs every function with its differential, is a construction of significant
importance in differential geometry, to the extent that one can formalise much
of it simply in terms of a ``category with tangent structure''
\cite{cockett2014differential}, that is to say, a Cartesian category equipped
with an endofunctor that behaves like the tangent bundle functor in some
suitable sense. 

It is possible to define an analogue of the tangent bundle functor in the
setting of change action models, which exhibits similar properties. In
particular, the derivative condition can be stated very elegantly in terms of
the tangent bundle functor. 

\begin{definition}
  \label{def:tangent-functor}
  The \textbf{tangent bundle functor} $\T : \cat{C} \ra \cat{C}$ is defined
  by:
  \begin{align*}
    \T(A) &\defeq A \times \changes A\\
    \T(f) &\defeq \pair{f \circ \pi_1}{\d[f]}
  \end{align*}
\end{definition}

\begin{notation}
  For clarity and brevity, we will abbreviate $\pi_i \circ \pi_j$ as $\pi_{ij}$.
\end{notation}

\begin{theorem}
  The tangent bundle functor $\T$ is strong monoidal.
  In particular, the map
  \begin{align*}
    \phi_{A, B} &: \T(A \times B) \to \T(A) \times \T(B)\\
    \phi_{A, B} &\defeq \pair{\T(\pi_1)}{\T(\pi_2)}
  \end{align*}
  is an isomorphism.
  Consequently, given maps $f : A \ra B$ and $g : A \ra C$, we have
  \[
    \T(\pair{f}{g}) \circ \phi_{A, B} = \pair{\T(f)}{\T(g)}
  \]
\end{theorem}
\begin{proof}
  Note that $\T(\pi_i) = \pair{\pi_{i1}}{\pi_{i2}}$, hence $\phi_{A, B} =
  \four{\pi_{11}}{\pi_{12}}{\pi_{21}}{\pi_{22}}$. That it is a natural
  isomorphism is immediately clear from its representation as a string diagram:
  \begin{center}
    \begin{tikzpicture}
      \node (Tprod) {$\T(A \times B)$};
      \node[right=1pt of Tprod] (cong1) {$\cong$};
      \node[right=1pt of cong1] (a1) {$A$};
      \node[right=1pt of a1] (tensor1) {$\times$};
      \node[right=1pt of tensor1] (b1) {$B$};
      \node[right=1pt of b1] (tensor2) {$\times$};
      \node[right=1pt of tensor2] (da1) {$\Delta A$};
      \node[right=1pt of da1] (tensor3) {$\times$};
      \node[right=1pt of tensor3] (db1) {$\Delta B$};

      \node[below=3cm of Tprod] (prodT) {$\T(A) \times \T(B)$};
      \node[below=3cm of cong1] (cong2) {$\cong$};
      \node[below=3cm of a1] (a2) {$A$};
      \node[below=3cm of tensor1] (tensor4) {$\times$};
      \node[below=3cm of b1] (b2) {$\Delta A$};
      \node[below=3cm of tensor2] (tensor5) {$\times$};
      \node[below=3cm of da1] (da2) {$B$};
      \node[below=3cm of tensor3] (tensor6) {$\times$};
      \node[below=3cm of db1] (db2) {$\Delta B$};

      \draw (a1) -- (a2);
      \draw (b1) to[out=-90,in=90] (da2);
      \draw (da1) to[out=-90,in=90] (b2);
      \draw (db1) -- (db2);
    \end{tikzpicture}
  \end{center}
  \vspace{-30pt}
  \qedhere
\end{proof}

A consequence of the structure of products in $\ftor{CAct}(\cat{C})$ is that the
map $\oplus_{A \times B}$ is defined pointwise in terms of $\oplus_A, \oplus_B$.
This result can be stated in terms of the previous isomorphism.

\begin{lemma}
  Let $\phi_{A, B} : \T(A \times B) \to \T(A) \times \T(B)$ be the
  canonical isomorphism described above. Then $\oplus_{A \times B} \circ
  \phi_{A, B} = \oplus_A \times \oplus_B$.
\end{lemma}

It will often be convenient to operate directly on the functor $\T$,
rather than on the underlying derivatives. In particular, in
Chapter~\ref{chp:difference-categories} we will make use of the natural
transformations below.

\begin{lemma}
  \label{lemma:tangent-bundle-monad}
  The following families of morphisms are natural transformations:
  \begin{align*}
    \pi_1, \oplus_A &: \T(A) \ra A \\
    \eta_A &: A \ra \T(A)\\
    \eta_A &\defeq \pair{\Id{A}}{0}\\
    \mu_A &: \T^2(A) \to \T(A)\\
    \mu_A &\defeq \pair{\pi_{11}}{\pi_{12} + (\pi_{21} \oplus \pi_{22})}
  \end{align*}

  Furthermore, $\mu_{A} \circ \T(\eta_A) = \mu_{A} \circ \eta_{\T(A)} = \Id{\T(A)}$. 
  However, $\mu_{A} \circ \mu_{\T(A)} \neq \mu_{A} \circ \T(\mu_{A})$ and therefore the 
  triple $(\T, \eta, \mu)$ is \emph{not} a monad (we establish monadicity of $\T$ for the
  special case of difference categories in Theorem~\ref{thm:t-monad}).
\end{lemma}
\begin{proof}
  We shall show naturality of $\eta$ and $\mu$, as we shall make use of it
  later. For $\eta$ we have:
  \begin{salign*}
  \T(f) \circ \eta_A
  = \pair{f \circ \pi_1}{\d[f]} \circ \eta_A
  = \pair{f \circ \pi_1 \circ \pair{\Id{A}}{0}}{
    \d[f] \circ \pair{\Id{A}}{0}}
  = \pair{f}{0}
  = \eta_B \circ f 
  \end{salign*}
  For the naturality of $\mu$, first, it is easy to check that:
  \begin{salign*}
    \T^2(f) = \four{f \circ \pi_{11}}
    {\d[f] \circ \pi_1}
    {\d[f] \circ (\pi_1 \times \pi_1)}
    {\d^2[f]}
  \end{salign*}
  from which we obtain:
  \begin{salign*}
  \mu_B \circ \T^2(f)
  &= \pair{\pi_{11}}{\pi_{12} + (\pi_{21} \oplus \pi_{22})} \circ \T^2(f)\\& =
  \pair{f \circ \pi_{11}}{
    \d[f] \circ (\pi_1 \times \pi_1)
    + \pa{
      (\d[f] \circ \pi_1) \oplus \d^2[f]
    }
  }
  \end{salign*}
  On the other hand:
  \begin{salign*}
    \T(f) \circ \mu_A
    &= \pair{f \circ \pi_1}{
      \d[f]} \circ \pair{\pi_{11}}{\pi_{12} + (\pi_{21} \oplus \pi_{22})
    }
    \\
    &= \pair{
      f \circ \pi_{11}
    }{
      \d[f] \circ \pair{\pi_{11}}{\pi_{12} + (\pi_{21} \oplus \pi_{22})}
    }
    \\
    &= \pair{
      f \circ \pi_{11}
    }{
      \d[f] \circ \pair{\pi_{11}}{\pi_{12}}
      + \d[f] \circ \pair{\pi_{11} \oplus \pi_{12}}{\pi_{21} \oplus \pi_{22}}
    }
    \\
    &= \pair{
      f \circ \pi_{11}
    }{
      \d[f] \circ \pair{\pi_{11}}{\pi_{12}}
      + (\d[f] \circ \pair{\pi_{11}}{\pi_{21}} 
      \oplus (\d^2[f] \circ \four{\pi_{11}}{\pi_{21}}{\pi_{12}}{\pi_{22}}))
    }
    \\
    &= \pair{
      f \circ \pi_{11}
    }{
      \d[f] \circ (\pi_1 \times \pi_1)
      + (\d[f] \circ \pi_1
      \oplus \d^2[f])
    }
  \end{salign*}

  Now we show $\mu_{A} \circ \eta_A = \Id{\T(A)}$.
  \begin{salign*}
    \mu_A \circ \eta_{\T(A)}
    &= \pair{\pi_{11}}{\pi_{12} + (\pi_{21} \oplus \pi_{22})} \circ \pair{\Id{T(A)}}{0}
      \\
    &= \pair{\pi_1}{0 + (\pi_{2} \oplus 0)}
    \\
    &= \pair{\pi_1}{\pi_{2}}
    \\
    &= \Id{\T(A)}
  \end{salign*}
  Finally, we check $\mu_{A} \circ \eta_{\T(A)} = \Id{\T(A)}$.
  \begin{salign*}
    \mu_A \circ T(\eta_{A})
    &= \pair{\pi_{11}}{\pi_{12} + (\pi_{21} \oplus \pi_{22})} 
    \circ \T(\pair{\Id{A}}{0})
      \\
    &= \pair{\pi_{11}}{\pi_{12} + (\pi_{21} \oplus \pi_{22})} 
    \circ \four{\pi_1}{0}{\pi_2}{0}
    \\
    &= \pair{\pi_1}{\pi_2 + (0 \oplus 0)}
    \\
    &= \Id{\T(A)}
  \end{salign*}

  The second monad law, however, fails to hold: intuitively, the term 
  $\mu_{A} \circ \T(\mu_{A})$ involves the derivative $\d[\mu_{A}]$ which in turn makes
  use of the derivatives $\d[\oplus], \d[+]$ which, however, do not appear in the
  expansion of $\mu_{A} \circ \mu_{\T(A)}$. Hence the equation cannot be established
  without making assumptions about the form of $\d[\oplus], \d[+]$.
\end{proof}

\begin{remark}
  Although we have omitted the corresponding proof for being trivial, it is
  an interesting observation that naturality of $\oplus$ is equivalent to the
  derivative condition, in that both propositions amount to the commutativity of
  the diagram below.
  \begin{center}
    \begin{tikzcd}[column sep=75pt, row sep=45pt]
      \T(A)
      \arrow[r, "\pair{f \circ \pi_1}{\d[f]} \equiv \T(f)"]
      \arrow[d, "\oplus_A", swap]
      & \T(B)
      \arrow[d, "\oplus_B"]\\
      A
      \arrow[r, "f"]
      & B
    \end{tikzcd}
  \end{center}
\end{remark}

The tangent bundle functor in a change action model behaves similarly to the
homonymous notion in Cartesian differential categories -- in particular, the
above natural transformations $\pi_1, \epsilon, \mu$ also exist in Cartesian
differential categories (the calculations being essentially identical). An
important property of the tangent bundle in a CDC which does \emph{not} hold for
change action models, however, is that $\epsilon, \mu$ make $\T$ into a
monad. In Chapter~\ref{chp:difference-categories} we will describe a setting of
particularly well-behaved change actions where the tangent bundle functor is
indeed a monad.

\subsection{Exponentials in Change Action Models}
\label{sec:exponentials-change-action-models}

A particularly interesting class of change action models are those that are also
Cartesian closed. Surprisingly, this has as an immediate consequence that the
operation of differentiation is itself internal to the category, in the
following sense:

\begin{lemma}
  \label{lem:internalisation} Whenever $\cat{C}$
  is Cartesian closed, there is a morphism $\arr{d}_{A, B} : (A \Ra B) \ra (A
  \times \changes{A}) \Ra \changes{B}$ such that, for any morphism $f : \terminal \times
  A \ra B$, $\arr{d}_{A, B} \circ \Lambda f = \Lambda (\d[f] \circ
  \pair{\pair{\pi_1}{\pi_{12}}} {\pair{\pi_1}{\pi_{22}}})$.
\end{lemma}
\begin{proof}
  Consider the evaluation map $\arr{ev}_{A, B} : (A \Ra B) \times A \ra B$ in
  $\cat{C}$. Its derivative $\d[\arr{ev}_{A, B}]$ has type
  \begin{gather*}
    \d[\arr{ev}_{A, B}] : ((A \Ra B) \times A) \times (\changes(A \Ra B) \times \changes A) \ra \changes B
  \end{gather*}
  (note that we make no assumptions about the structure of $\changes (A \Ra
  B)$).

  Then consider the following composite:
  \begin{gather*}
    \d[\arr{ev}_{A, B}] \circ \pair{\pair{\pi_1}{\pi_{12}}}{\pair{0_{A \Ra B}}{\pi_{22}}} : (A \Ra B) \times (A \times \changes A) \ra \changes B
  \end{gather*}

  By the universal property of the exponential, we have $\arr{ev}_{A, B} \circ
  (\Lambda f \times \Id{A}) = f$ and therefore $\alpha(\arr{ev}_{A, B} \circ
  (\Lambda f \times \Id{A})) = \alpha(f)$. Thus:
  \begin{salign*}
    \d[f] &= \d[\arr{ev}_{A, B} \circ (\Lambda f \times \Id{A})]\\
         &= \d[\arr{ev}_{A, B} \circ \pair{\Lambda f}{\Id{A}}]\\
         &= \d[\arr{ev}_{A, B}] \circ \pair{\pair{\Lambda f}{\Id{A}} \circ \pi_1}
                                        {\pair{0_{A \Ra B}}{\pi_2}}\\
         &= \d[\arr{ev}_{A, B}]
           \circ \pair{\pair{\pi_1}{\pi_{12}}}{\pair{0_{A \Ra B}}{\pi_{22}}} 
           \circ \pair{\Lambda f}{\pair{\pi_1}{\pi_2}}\\
         &= \d[\arr{ev}_{A, B}]
           \circ \pair{\pair{\pi_1}{\pi_{12}}}{\pair{0_{A \Ra B}}{\pi_{22}}} 
           \circ \pair{\Lambda f}{\Id{A \times \changes A}}
  \end{salign*}
  from which it follows trivially that
  \[
    \arr{d}_{A, B} = \Lambda(\d[\arr{ev}_{A, B}])
    \circ \pair{\pair{\pi_1}{\pi_{12}}}{\pair{0_{A \Ra B}}{\pi_{22}}}
  \]
  is the desired morphism.
\end{proof}

We would like the tangent bundle functor to preserve the exponential structure
in the same way that it preserves the structure of products. In particular, we
would hope the tangent structure of the exponentials to be pointwise, i.e.
$\T(A \Ra B) \cong A \Ra \T(B)$, with $\oplus_{A \Ra B}$ being the
pointwise lifting of $\oplus_B$. Furthermore we aim for a result of the form
$\diffS{(\lambda y . t)}{x}{u} = \lambda y . \pa{\diffS{t}{x}{u}}$, which holds
in Cartesian differential categories and is used to propagate differentiation
through abstractions in the differential $\lambda$-calculus. These results can
in fact be obtained as a consequence of the tangent bundle functor being
representable.

\begin{definition}
  If $\cat{C}$ is Cartesian closed, an \emph{infinitesimal object} \footnote{The
  concept of an ``infinitesimal object'' is borrowed from synthetic differential
  geometry\cite{Kock06}. However, there is nothing intrinsically
  ``infinitesimal'' about these objects in our setting -- we will later see an
  example which is distinctly discrete .} $U$ is an object of $\cat{C}$
  such that the tangent bundle functor $\T$ is represented by the internal
  Hom-functor $U \Ra (\cdot)$, i.e. there is a natural isomorphism $\mathcal{D}
  : (U \Ra (\cdot)) \cong \T$.
\end{definition}

\begin{theorem}
  \label{thm:oplus-lambda-infinitesimal}
  Whenever there is an infinitesimal object in $\cat{C}$, the tangent bundle
  $\T(A \Ra B)$ is naturally isomorphic to $A \Ra \T(B)$. Furthermore,
  this isomorphism respects the structure of $\T$, in the sense
  that the following diagrams commutes:
  \begin{center}
    \begin{tikzcd}
      \T(A \Ra B)
      \arrow[d, "{\oplus_{A \Ra B}}", swap]
      \arrow[r, "{\cong}"]
      & A \Ra \T(B)
      \arrow[ld, "{\Id{A} \Ra \oplus_B}"]
      \\
      A \Ra B
    \end{tikzcd}\quad\quad\quad\quad
    \begin{tikzcd}
      \T(A \Ra B)
      \arrow[d, "{\pi_1}", swap]
      \arrow[r, "{\cong}"]
      & A \Ra \T(B)
      \arrow[ld, "{\Id{A} \Ra \pi_1}"]
      \\
      A \Ra B
    \end{tikzcd}
  \end{center}
  \begin{center}
    \begin{tikzcd}
      \T^2(A \Ra B)
      \arrow[d, "{\mu_{A \Ra B}}", swap]
      \arrow[r, "{\cong}"]
      & A \Ra \T^2(B)
      \arrow[d, "{\Id{A} \Ra \mu_B}"]
      \\
      \T(A \Ra B)
      \arrow[r, "{\cong}"]
      &
      A \Ra \T(B)
    \end{tikzcd}\quad\quad\quad\quad
    \begin{tikzcd}
      A \Ra B
      \arrow[d, "{\epsilon_{A \Ra B}}", swap]
      \arrow[r, "{\cong}"]
      & A \Ra \T(B)
      \arrow[d, "{\Id{A} \Ra \epsilon_B}"]
      \\
      \T(A \Ra B)
      \arrow[r, "{\cong}"]
      &
      A \Ra \T(B)
    \end{tikzcd}
  \end{center}
\end{theorem}
\begin{proof}
  The desired isomorphism can be obtained from the sequence of (natural)
  isomorphisms:
  \[
    \T(A \Ra B) \cong U \Ra (A \Ra B) \cong A \Ra (U \Ra B) \cong A \Ra \T(B)
  \]

  By the Yoneda lemma, the natural transformation
  \begin{gather*}
    \oplus_{A} \circ \mathcal{D} : (U \Ra A) \ra A
  \end{gather*}
  is precisely evaluation at some fixed element $U_\oplus : \terminal \ra U$ (and,
  conversely, evaluating $\mathcal{D}(t)$ at $U_\oplus$ is precisely
  $\oplus_A(t)$).

  In more detail, the map $\oplus_A \circ \mathcal{D}_A : (U \Ra A) \ra A$ is a
  natural transformation from $(U \Ra (\cdot))$ into $\ftor{Id}$ and, therefore,
  corresponds to a natural transformation from $\homset{\cat{C}}{\terminal}{U \Ra
  (\cdot)}$ into $\homset{\cat{C}}{\terminal}{\cdot}$ or, equivalently, a natural
  transformation from $\homset{\cat{C}}{U}{\cdot}$ into
  $\homset{\cat{C}}{\terminal}{\cdot}$. Hence, by the covariant Yoneda lemma, it must
  correspond to some element $U_\oplus \in \homset{\cat{C}}{\terminal}{U}$. By a
  similar argument, $\mu$ corresponds precisely to precomposition with a  map
  $U_\mu : U \to U \times U$, and $\epsilon$ corresponds to sending each $a \in
  A$ to the constant function mapping every $u \in U$ to $a$.

  Commutativity of the above diagrams then can be shown by equational reasoning
  in the internal logic of the CCC $\cat{C}$ (we show the calculations for the
  first diagram, the rest being obtained in a similar manner). Whenever $f
  : \T(A \Ra B)$, we obtain:
  \begin{salign*}
    \lambda a . \oplus_{B} (\mathcal{D}(\lambda u . \mathcal{D}^{-1}(f)(u)(a)))
    &= \lambda a . \mathcal{D}^{-1}(f)(U_\oplus)(a)\\
    &= \mathcal{D}^{-1}(f)(U_\oplus)\\
    &= \oplus_{A \Ra B} (f)
  \end{salign*}
\end{proof}

The above result is quite significant, as it characterises $\T(A \Ra B)$
(almost) entirely in terms of the structure of $\T(B)$, lifted pointwise.
This tells us that, in a change action model with an infinitesimal object, the
change action structure of the exponential objects is essentially the convenient
pointwise structure that we introduced in Chapter~\ref{chp:incremental-computation},
Definition~\ref{def:pointwise-functional}. That is to say, a change on $A \Ra B$
is precisely a function mapping an element of $A$ to a change on $B$, and
applying a functional change is done by lifting $\oplus_B$ pointwise.

Whenever the object $B$ is equipped with a point $b : \terminal \to B$, it is
possible to use the previous result to infer that addition and zero on
the change space $\Delta (A \Ra B)$ are also pointwise as, for example, $+_B$
can be written in terms of $\mu$ as: \[
+_B =  \pi_2 \circ \mu_B \circ \four{a \circ{} !}{\pi_1}{\pi_2}{0_{\Delta B}}
: \Delta B \times \Delta B \to \Delta B
\]

We might expect a property similar to \ref{DlC} to hold in Cartesian closed
change action models or, at least, in those models that admit an infinitesimal
object. While this is not quite true, a weaker result is available: given a map
$f : A \times B \ra C$ we cannot, in general, prove that $\d[\Lambda(f)]$ is
equal to $\Lambda(\d[f] \circ \pair{\pi_1 \times \Id{}}{\pi_2 \times 0})$, but
we can prove that both expressions have the same \emph{effect}, i.e. their
action \emph{qua} changes in $\Delta(B \Ra C)$ is identical.

\begin{theorem}
  \label{thm:oplus-lambda-partial}
  Given objects $A, B$, let $\partial_1$ denote the map 
  \[
    \partial_1 \defeq \pair{\pi_1 \times \Id{B}}{\pi_2 \times 0} : \T(A) \times B
    \ra \T(A \times B)
  \]
  Then for all $f : A \times B \ra C$, we have
  \[
    \oplus \circ \T(\Lambda(f)) = (\Ra \oplus) \circ \Lambda (\T(f) \circ \partial_1)
  \]
\end{theorem}
\begin{proof}
  The map $\partial_1$ can alternatively written as $\phi \circ (\Id{A \times B}
  \times \pair{\Id{\Delta A}}{0})$. 
  We can then apply naturality of $\oplus$ and the fact that $\oplus \circ \phi
  = \oplus \times \oplus$:
  \begin{salign*}
    (\Ra \oplus) \circ \Lambda (\T(f) \circ \partial_1) 
    &= \Lambda (\oplus \circ \T(f) \circ \partial_1)\\
    &= \Lambda (f \circ \oplus \circ \partial_1)\\
    &= \Lambda (f \circ \oplus \circ \phi \circ (\Id{} \times \pair{\Id{}}{0}))\\
    &= \Lambda (f \circ
      ((\oplus \circ \Id{}) \times (\oplus \circ \pair{\Id{}}{0})))\\
    &= \Lambda (f \circ (\oplus \times \Id{}))\\
    &= \Lambda(f) \circ \oplus\\
    &= \oplus \circ \T(\Lambda(f)) \qedhere
  \end{salign*}
\end{proof}

\section{Examples of Change Action Models}

\label{sec:change-action-models:examples}

\subsection{Generalised Cartesian Differential Categories}
\label{sec:change-action-models:examples-differential}

Cartesian differential categories, as described in Section 
\ref{sec:cartesian-differential-categories}, and generalised CDCs, are perhaps
the most well-studied categorical formulation of the ``classical'' notion of derivative
from differential calculus.

\begin{example}[{\cite{blute2009cartesian}}]
  The category $\cat{Smooth}$ of smooth maps between Euclidean spaces $\RR^n$
  is a Cartesian differential category.
\end{example}
\begin{example}[{\cite{cruttwell2017cartesian}}]
  The category $\cat{Smooth}_\subseteq$ of smooth maps between open subsets of $\RR^n$
  is a generalised Cartesian differential category.
\end{example}

We show that change action models generalise GCDC in that GCDCs give
rise to change action models in two different ways.Throughout this subsection
we will take $\cat{C}$ to be a GCDC. The reader may find it helpful to have
Definition~\ref{def:gcdc} at hand for reference.

\paragraph{1.~The Flat Model.}
\label{sec:flat-model}
Define the functor $\alpha : \cat{C} \ra \ftor{CAct}(\cat{C})$ as follows.
Let $f : A \ra B$ be a $\cat{C}$-morphism.
Then $\alpha(A) \defeq (A, L_0(A), \pi_1, +_A, 0_A)$ and $\alpha(f) \defeq (f, \DF{f})$.

\begin{theorem}
  \label{thm:GCDC-change-action-model}
  The functor $\alpha$ is a change action model.
\end{theorem}
\begin{proof}
  We need to check that $\alpha$ is well-defined and a right-inverse to the forgetful
  functor.
  
  First, note that $\alpha(f)$ trivially satisfies the derivative property:
  \begin{gather*}
  f \circ \pi_1 = \pi_1 \circ \pair{f \circ \pi_1}{\DF{f}}
  \end{gather*}
  Furthermore, by the axiom \ref{CDC2} of generalised Cartesian differential categories, we have:
  \begin{salign*}
    \DF{f} \circ \pair{\Id{}}{0_A \circ{} !} &= 0_B\\
    \DF{f} \circ \pair{a}{+ \circ \pair{u}{v}} 
    &= + \circ \pair{\DF{f} \circ \pair{a}{u}}{\DF{f} \circ \pair{a}{v}}
  \end{salign*}
  This entails that the map $\alpha(f) = (f, \DF{f})$ is regular and, therefore,
  a differential map. Functoriality of $\alpha$ is a direct consequence of
  \ref{CDC3} and \ref{CDC5}.
  
  Furthermore, $\alpha$ preserves products (up to isomorphism) since, by definition,
  $L(A \times B) = L(A) \times L(B)$ and by axioms \ref{CDC3} and \ref{CDC4}, 
  and is trivially a right-inverse to the forgetful functor. 
  Therefore $\alpha$ is a change action model.\qedhere
\end{proof}

\paragraph{2.~The Kleisli Model.} GCDCs admit a tangent bundle functor, defined
analogously to the standard notion in differential geometry. Let $f : A \to B$
be a $\cat{C}$-morphism. Define its tangent bundle functor $\T :
\cat{C} \ra \cat{C}$ as: $\T(A) \defeq A \times L_0(A)$, and $\T(f)
\defeq \pair{f \circ \pi_1}{\DF{f}}$.
The functor $\T$ is in fact a monad, with unit $\eta = \pair{\Id{}}{0_A} :
A \ra A \times L_0(A)$ and multiplication $\mu : (A \times L_0(A)) \times
(L_0(A) \times L_0(A)) \ra A \times L_0(A)$ defined by the composite:
\[
(A \times L_0(A)) \times (L_0(A) \times L_0(A))
      \xrightarrow{\pair{\pi_{11}}{\pair{\pi_{21}}{\pi_{12}}}}
      A \times (L_0(A) \times L_0(A))
      \xrightarrow{\Id{} \times +_A}
      A \times L_0(A)
\]
Thus we can work on the Kleisli category of this functor by $\cat{C}_{\T}$
which has geometric significance as a category of generalised vector fields. 

\begin{lemma}
  The tuple $(A, L_0(A), \Id{A} \times \Id{L_0(A)}, \eta \circ +_A, \eta \circ 0_A)$
  is a change action in the category $\cat{C}_{\T}$.
\end{lemma}

\begin{theorem}
We define the functor $\alpha_{\T} : \cat{C}_{\T} \ra
\ftor{CAct}(\cat{C}_{\T})$ as follows. Given an object $A$ in
$\cat{C}_{\T}$, we set:
\[
  \alpha_{\T}(A) \defeq (A, L_0(A), \Id{A} \times \Id{L_0(A)}, \eta \circ +_A, \eta \circ 0_A)
\]
Given a $\cat{C}_{\T}$-map $f : A \ra_{\T} B$, we set:
\[
  \alpha_{\T}(f) \defeq (f, \DF{f})
\]
Thus defined, the functor $\alpha_{\T}$ is a change action model on $\cat{C}$.
\end{theorem}
\begin{proof}
  A proof for a much stronger result will be shown in
  Chapter~\ref{chp:difference-categories}. Although it is written for the
  setting of Cartesian differential categories, it can be easily adapted to
  GCDCs.
\end{proof}

\begin{remark}\label{rem:GCDC-axioms}
  The converse is not true: in general the existence of a change action model on
  $\cat{C}$ does not imply that $\cat{C}$ satisfies the axioms of a Cartesian
  differential category. However, as we shall see in
  Chapter~\ref{chp:difference-categories}, most of the CDC axioms do hold in many
  well-behaved classes of change action models (with the exception that
  derivatives may not be additive).
\end{remark}

\subsection{Finite Differences and Boolean Differential Calculus}
\label{sec:discrete-calculus}
\newcommand{\CGrpS}[0]{\cat{Grp_\star}}

Consider the category $\CGrpS$ whose objects are all the groups (in $\cat{Set}$)
and where morphisms $f : \mathcal{F} \to \mathcal{G}$ are all the
(set-theoretical) functions between the carrier sets $F$ and $G$. This is a
Cartesian closed category which can be endowed with the structure of a
$(\ftor{CAct}, \epsilon)$-coalgebra $\alpha$ in the following way.

\begin{definition}
  Given a group $\mathcal{F} = (F, \cdot, 1)$, define the change action
  $\alpha(\mathcal{G})$ by:
  \[
    \alpha(\mathcal{F}) \defeq  (F, F, \cdot, \cdot, 1).
  \]
  Given an arbitrary function $f : F \ra G$ between the carrier sets of
  $\mathcal{F}$ and $\mathcal{G}$ respectively, define the differential map
  $\alpha(f) \defeq (f, \d[f])$ by:
  \[
    \d[f](x, \change{x}) \defeq f(x)^{-1} \cdot f(x \cdot \change{x}).
  \]
  Notice that the derivative condition is satisfied:
  \[
  f(x) \oplus_\mathcal{G} \d[f](x, \change x) 
  = f(x) \cdot (f(x)^{-1} \cdot f(x \cdot \change{x}))
  = f(x \cdot \change{x})
  = f(x \oplus_\mathcal{F} \change{x});
  \]
  hence $\d[f]$ is a (regular) derivative\footnote{Note that $\d[f]$ need not be
  additive in its second argument, and so derivatives in $\CGrpS$ do not satisfy
  all the axioms of a Cartesian differential category.} for $f$, and therefore
  $\alpha(f)$ is a map in $\ftor{CAct}(\CGrpS)$. The following result is then
  immediate.
\end{definition}

\begin{lemma}
  The mapping $\alpha$ defines a product-preserving functor from $\CGrpS$ into
  $\ftor{CAct}(\CGrpS)$. Furthermore $\epsilon_{\CGrpS} \circ \alpha =
  \ftor{Id}_{\CGrpS}$ and hence $\alpha$ is a coalgebra for the pointed
  endofunctor $\ftor{CAct}$\footnote{
    One can think of the functor $\alpha$ as a ``constructive'' version of the
    functor $\ftor{G}$ from Lemma~\ref{lem:set-to-cact}, making the choice of a group
    structure explicit rather than applying the Axiom of Choice.
  }.
\end{lemma}

\begin{remark}
This category may seem rather strange, as its objects have ``more structure'' than its
arrows do. That is, every object in $\CGrpS$ is isomorphic to every other object with
the same carrier set (in fact, assuming the Axiom of Choice, $\CGrpS$ is 
straightforwardly equivalent, but not isomorphic, to the category of non-empty sets). 
The specifics of the group structure of each object are encoded into the functor $\alpha$,
since addition and subtraction can be recovered from $\alpha$.

For example, given a group $\mathcal{F}$, and an element $x \in F$, the identity of
$\mathcal{F}$ can be obtained by evaluating the derivative $\d (\alpha \kappa_x)(x, x)$,
where $\kappa_x$ is the constant function sending every element of $F$ to $x$. Similarly,
addition can be recovered by differentiating the map $0_x$ which maps $x$ to $0$ and
every other element to itself.
\end{remark}

Despite its seeming novelty, this category is actually a generalisation of the
well-known calculus of finite differences (see \cite{gleich2005finite} for a
quick introduction, or \cite{jordan1965calculus,milne2000calculus} for a more
in-depth treatment of the topic). The calculus of finite differences deals
mainly with the notion of \emph{discrete derivative} (or \emph{discrete
difference operator}) of a function $f : \ZZ \ra \ZZ$, which is defined as
\(
  \delta f(x) \defeq f(x + 1) - f(x)
\).
In fact this discrete derivative $\delta f$ is (an instance of) the derivative of
$f$ \emph{qua} morphism in $\CGrpS$, i.e.~$\delta f(x) = \d[f](x, 1)$.

The calculus of finite differences has found applications in combinatorics and
numerical computation. Our formulation via this change action model on $\CGrpS$ has
several advantages. First it justifies the chain rule, which is an important
novel result. Secondly, it generalises the calculus to arbitrary groups. To
illustrate this, consider the \emph{Boolean differential calculus}
\cite{steinbach2017boolean,thayse1981boolean}, a theory that applies methods
from calculus to the space $\BB^n$ of vectors of elements of some Boolean
algebra $\BB$, with applications to the analysis and synthesis of combinatorial
digital circuits.

\begin{definition}
  Given a Boolean algebra $\BB$ and function $f : \BB^n \ra \BB^m$, the
  \emph{$i$-th Boolean derivative of $f$ at $(u_1, \ldots, u_n) \in \BB^n$} is
  the value
  \[
    \frac{\d[f]}{\d[x_i]}(u_1, \ldots, u_n) \defeq 
    f(u_1, \ldots, u_n) \nlra f(u_1, \ldots, \neg u_i, \ldots, u_n),
    \]
  writing $u \nleftrightarrow v \defeq (u \wedge \neg v) \vee (\neg u \wedge v)$
  for exclusive-or.
\end{definition}

Now $\BB^n$ is clearly a commutative group, with the group operation given by
pointwise exclusive disjunction, and hence it is an object in $\CGrpS$. Set 
\[
  \top_i \defeq (\bot, \overset{i-1}{\ldots}, \bot, \top, \bot, \overset{n - i}{\ldots}, \bot) \in \BB^n
\]
\begin{lemma}
  \label{lem:boolean-derivative}
  The Boolean derivative of $f : \BB^n \ra \BB^m$ coincides with its derivative
  \emph{qua} morphism in $\CGrpS$: $\frac{\d[f]}{\d[x_i]}(u_1, \ldots, u_n) =
  \\d[f]((u_1,
  \ldots, u_n), \top_i)$.
\end{lemma}
\begin{proof}
  Note first that in any Boolean algebra $\BB$, we have $\neg u = u \nlra \top$.
  Moreover
  \begin{gather*}
    (u_1, \ldots, \neg u_i, \ldots, u_n) = (u_1, \ldots, u_n) \oplus (\bot, \ldots, \top, \ldots, \bot)
  \end{gather*}
  Furthermore:
  {\small
  \[
  \begin{array}{ll}
    &\displaystyle f(u_1, \ldots, u_n) \oplus \frac{\d[f]}{\d[x_i]}(u_1, \ldots, u_n)\\
    =& f(u_1, \ldots, u_n) \oplus (f(u_1, \ldots, u_n) \nlra f(u_1, \ldots, \neg u_i, \ldots, u_n))\\
    =& f(u_1, \ldots, u_n) \nlra (f(u_1, \ldots, u_n) \nlra f(u_1, \ldots, \neg u_i, \ldots, u_n))\\
    =& (f(u_1, \ldots, u_n) \nlra f(u_1, \ldots, u_n)) \nlra f(u_1, \ldots, \neg u_i, \ldots, u_n)\\
    =& \bot \nlra f(u_1, \ldots, \neg u_i, \ldots, u_n)\\
    =& f(u_1, \ldots, \neg u_i, \ldots, u_n)\\
    =& f((u_1, \ldots, u_n) \nlra \top_i)\\
    =& f((u_1, \ldots, u_n) \oplus \top_i)
  \end{array}
  \]}
  Thus, since derivatives in $\CGrpS$ are unique, the Boolean derivative 
  \[
  \displaystyle \frac{\d[f]}{\d[x_i]}(u_1, \ldots, u_n)
  \] 
  is precisely the
  derivative $\d[f]((u_1, \ldots, u_n), \top_i)$.
\end{proof}

One surprising fact about $\CGrpS$ is that it is a Cartesian closed category --
and what is more, it has remarkable infinitesimal object: the cyclic group
$\ZZ_2$! Although it does not look very ``infinitesimal'' in any meaningful
geometric sense, it is in fact the smallest (non-trivial) group.

\begin{theorem}
  \label{thm:cgrps-ccc}
  The category $\CGrpS$ is Cartesian closed, with the exponential $\mathcal{F}
  \Ra \mathcal{G}$ being the group of (arbitrary) functions $F \Ra G$, with the
  group operation being the pointwise lifting of $\cdot_\mathcal{G}$.

  For any group $\mathcal{G}$, the group $\ZZ_2 \Ra \mathcal{G}$ is naturally
  isomorphic to the tangent bundle $\T(\mathcal{G}) \equiv
  \mathcal{G}\times\mathcal{G}$.
\end{theorem}
\begin{proof}
  First we show that the group $\mathcal{F} \Ra \mathcal{G}$ is indeed the
  exponential -- this follows immediately from the fact that we can freely take
  the evaluation map $\A{ev} : \mathcal{F} \times \mathcal{F} \Ra \mathcal{G}
  \ra \mathcal{G}$ to be the evaluation map in $\cat{Set}$.

  One might expect the isomorphism $\mathcal{U} : \ZZ_2 \Ra \mathcal{G} \to 
  \mathcal{G} \times \mathcal{G}$ to be defined by $\mathcal{U}(u) = (u(0),
  u(1))$. As natural as this definition is, it would not be a natural
  isomorphism, and so a slightly cleverer mapping is necessary.
  
  We instead define the isomorphism $\mathcal{U}$ by:
  \[
    \mathcal{U}(u) \defeq (u(0), u(0)^{-1} \cdot u(1))
  \]
  To show that this is indeed natural, consider an arbitrary function $f :
  \mathcal{G} \to \mathcal{F}$. Then:
  \begin{salign*}
    (\T(f) \circ \mathcal{U}_{\mathcal{G}})(u)
    &= \T(f) (u(0), u(0)^{-1} \cdot u(1))\\
    &= (f(u(0)), \d[f](u(0), u(0)^{-1} \cdot u(1)))\\
    &= (f(u(0)), f(u(0))^{-1} \cdot f(u(1)))\\
    &= \mathcal{U}_{\mathcal{G}}(f \circ u)\\
    &= (\mathcal{U}_{\mathcal{G}} \circ (\Ra f)) (u)\qedhere
  \end{salign*}
\end{proof}

The existence of a concrete infinitesimal object in $\CGrpS$ lets us illustrate
some of the more abstract results in
Section~\ref{sec:exponentials-change-action-models}. For example, we can verify
that indeed the action $\oplus_{\mathcal{G}}$ in any group $\mathcal{G}$
corresponds precisely to evaluation at a certain point $U_\oplus : \initial
\to \ZZ_2$ (in this case the point $U_\oplus$ being the number $1$). That is
to say:
\[
  a \oplus_{\mathcal{G}} b = (\mathcal{U}(a, b))
\]

\subsection{Polynomials Over Commutative Kleene Algebras}
\label{sec:change-action-models:examples-kleene}

The case of polynomials over a commutative Kleene algebra is yet another setting
where a ``non-standard'' notion of derivative is useful -- in particular Hopkins
and Kozen \cite{DBLP:conf/lics/HopkinsK99} employ it to give a proof of Parikh's
theorem, which can be informally summarised as saying that every context-free
language is ``equivalent'' to a regular language if one is unconcerned about the
order of the characters. Similar sorts of derivatives have been applied to
reconstructing regular expressions from automata
\cite{DBLP:conf/latin/LombardyS04} or accelerating the computation of fixed
points for program analysis \cite{esparza2010newtonian}

We study here the category of polynomials over a (fixed) commutative Kleene
algebra, a remarkable setting where a (non-linear) derivative exists whose
characterising property is that an analogue of the first-order Taylor
approximation holds up to strict equality.

\begin{definition}[Commutative Kleene algebra] 
  Formally a \textbf{Kleene algebra} $\KK$ is a tuple $(K, +, \cdot, {}^\star,
  0, 1)$ such that $(K, +, \cdot, 0, 1)$ is an idempotent semiring under $+$
  satisfying, for all $a, b, c \in K$: 
  \begin{align*}
      1 + a \, a^\star &= a^\star\quad
      1 + a^\star a = a^\star\quad
      b + a \, c \leq c \ra a^\star \, b \leq c
      \quad
      b + c \, a \leq c \ra b \, a^\star \leq c
  \end{align*}
  where ${a \leq b} \defeq a + b = b$. A Kleene algebra is \textbf{commutative}
  whenever $\cdot$ is. Recall that, informally, a Kleene algebra is the algebra
  of regular expressions over some alphabet
  \cite{DBLP:journals/jacm/Brzozowski64,Conway71}.
\end{definition}

In what follows, we fix a commutative Kleene algebra $\KK$. Define its
\textbf{algebra of polynomials} on variables $x_1, \ldots, x_n$, denoted by
$\KK[\overline{x}]$, as the free extension of the algebra $\KK$ with elements
$\overline{x} = x_1, \ldots, x_n$. We write $p(\overline{a})$ for the value of
$p(\overline{x})$ evaluated at $\overline x \mapsto \overline a$. Polynomials,
viewed as functions, are closed under composition: when $p \in
\KK[\overline{x}], q_1, \ldots, q_n \in \KK[\overline{y}]$ are  polynomials, so
is the composite $p(q_1(\overline{y}), \ldots, q_n(\overline{y}))$.

\begin{definition}
  Given a polynomial $p = p(\overline x)$, we define its \textbf{$i$-th derivative} 
  $\der{p}{x_i}(\overline x) \in \KK[\overline x]$ by induction on the structure
  of $p$ according to the following rules:
  \begin{align*}
    \der{a}{x_i} (\overline x) &= 0
    \qquad
    \der{p^\star}{x_i} (\overline x) = p^\star(\overline x) \, \der{p}{x_i} (\overline x)
    \qquad
    \der{x_j}{x_i} (\overline x) = 
    \begin{cases}
    1 \text{ if $i = j$}\\
    0 \text{ otherwise}
    \end{cases}
    \\
    \der{(p+q)}{x_i}(\overline x) &= \der{p}{x_i} (\overline x) + \der{q}{x_i} (\overline x)\qquad
    \der{(p \, q)}{x_i} (\overline x) = p(\overline x) \der{q}{x_i} (\overline x)+ q (\overline x)
                                    \der{p}{x_i}(\overline x)
  \end{align*}
\end{definition}

Most of the previous rules should be familiar from calculus, except for the derivative of
the Kleene star -- informally, this is justified through unfolding its definition
and applying the other rules.
\begin{align*}
  \der{p^\star}{x}(\overline x)
  &= \der{(1 + p \, p^\star)}{x}(\overline x)\\
  &= \der{1}{x}(\overline x) + \der{(p \, p^\star)}{x}(\overline x)\\
  &= p(\overline x) \der{p^\star}{x}(\overline x) + p^\star(\overline x) \der{p}{x}(\overline x)\\
  &= \der{p}{x}(\overline x) + p(\overline x)\pa{\der{p^\star}{x}(\overline x) + p^\star(\overline x) \der {p}{x}(\overline x)}
\end{align*}
Solving the above recurrence relation one then obtains $\der{p^\star}{x}(\overline x)
= p^\star(\overline x) \der{p}{x}(\overline x)$.

\begin{theorem}[Taylor's formula \cite{DBLP:conf/lics/HopkinsK99}]
  \label{thm:taylor-kleene}
  Let $p(x) \in \KK[x]$. 
  For all $a, b \in \KK[x]$, we have
  $p(a + b) = p(a) + b \cdot \der{p}{x}(a + b)$. 
\end{theorem}

\begin{definition}
  Fix a commutative Kleene algebra $\KK$. Its \textbf{category of
  polynomials}, $\cat{\KK}_\times$ has all the natural numbers as objects.
  The morphisms $\KK_\times[m, n]$ are $n$-tuples of polynomials $(p_1, \ldots,
  p_n)$ where $p_1, \ldots, p_n \in \KK[x_1, \ldots, x_m]$. Composition of
  morphisms is the usual composition of polynomials.
\end{definition}

\begin{lemma}
  \label{lem:kleene-change-actions-model}
  The category $\cat{\KK_\times}$ is a Cartesian category, endowed with a change
  action model $\alpha : \cat{\KK_\times} \to \ftor{CAct}(\cat{\KK_\times})$
  whereby $\alpha(\KK) \defeq (\KK, \KK, +, +, 0)$, $\alpha(\KK^i) \defeq
  \alpha(\KK)^i$; for $\overline p = (p_1(\overline{x}), \ldots,
  p_n(\overline{x})) : \KK^m \ra \KK^n$, $\alpha(\overline p) \defeq (\overline
  p, (p_1', \ldots, p_n'))$, where \(  p_i' = p_i' (x_1, \ldots, x_m, y_1,
  \ldots, y_m) \defeq \sum_{j=1}^n y_j \cdot \der{p_i}{x_j}(x_1 + y_1, \ldots,
  x_m + y_m). \)
\end{lemma}
\begin{proof}
  We consider the essential case of $m = n = 1$; the proof of the lemma is then
  a straightforward generalisation.

  We shall make use of the following properties of \emph{commutative} Kleene
  algebras.
  \begin{enumerate}
  \item $(a_1 + \cdots + a_m)^n = \sum \{a_1^{i_1} \cdots a_m^{i_m} \mid i_1 + \cdots + i_m = n; i_1, 
    \cdots, i_m \geq 0\}$.

    Since $(a + b)^\star = a^\star \, b^\star$, we have
    \[
      \big((a_1 + \cdots + a_m)^n\big)^\star = \prod \{(a_1^{i_1} \cdots a_m^{i_m})^\star \mid i_1 +
      \cdots + i_m = n; i_1, \cdots, i_m \geq 0\}.
    \] 
    For example $((a + b + c)^2)^\star = (a \, a)^\star \, (a \, b)^\star \, (a \, c)^\star \, 
    (b \, b)^\star \, (b \, c)^\star \, (c \, c)^\star$.

  \item Pilling's Normal Form Theorem
    \cite{Pilling73,DBLP:conf/lics/HopkinsK99}: every (regular) expression is
    equivalent to a sum $y_1 + \cdots + y_n$ where each $y_i$ is a product of
    atomic symbols and expressions of the form $(a_1 \cdots a_k)^\star$, where
    the $a_i$ are atomic symbols. For example $(((a \, b)^\star c)^\star +
    d)^\star = d^\star + (a \, b)^\star c^\star c \, d^\star$.
  \end{enumerate}

  Take $p(x) \in \KK[x]$, viewed as a function from change action $(\KK, \KK, +,
  +, 0)$ to itself. For $a, b \in \KK$, we have 
  \[
    \d[p](a, b) \defeq \der{p}{x}(a + b) \cdot b.
  \]
  That this satisfies the derivative condition with respect to $p(x)$ is an
  immediate consequence of Theorem~\ref{thm:taylor-kleene}.

  We need to prove that the derivative is regular. Trivially $\d[p] (a, 0) = 0$.
  It remains to prove: for $u, a, b \in \KK$
  \begin{equation}
    \der{p}{x}(u + a + b) \cdot (a + b) =
    \der{p}{x}(u + a) \cdot a + \der{p}{x}(u + a + b) \cdot b
    \label{eq:kleene-regular}
  \end{equation}
  which we argue by structural induction, presenting the cases of $p = q^\star$
  and $p = q \, r$ explicitly.

  Let $p = q^\star$. Thanks to Pilling's Normal Form Theorem, without loss of
  generality we may assume $q = x^{n+1} \, c$. Now $\displaystyle \der{x^{n+1}
  \, c}{x}(x) = x^n \, c$. Then $\displaystyle \der{p}{x}(x) = q^\star (x) \,
  \der{q}{x}(x) = (x^{n+1} \, c)^\star (x^n \, c)$. Clearly
  $\mathrm{RHS}(\ref{eq:kleene-regular}) \leq
  \mathrm{LHS}(\ref{eq:kleene-regular})$. For the opposite containment, it
  suffices to show 
  \[
    \der{p}{x}(u + a + b) \cdot a \ \leq \ \der{p}{x}(u + a) \cdot a + \der{p}{x}(u + a + b) \cdot b
  \]
  I.e.
  \begin{equation}
    (\theta^{n+1} \, c)^\star \, (\theta^n \, c) \, a 
    \ \leq \
    ((u+a)^{n+1} \, c)^\star \, ((u+a)^n \, c) \, a + (\theta^{n+1} \, c)^\star \, (\theta^n \, c) \, b
    \label{eq:kleene-regular-2}
  \end{equation} 
  using the shorthand $\theta = u + a + b$.

  A typical element that matches the left-hand side of
  Equation~\ref{eq:kleene-regular-2} has shape 
  \[
    \Xi \defeq (u^{i'} \, a^{j'} \, b^{k'} \, c)^l \, (u^i \, a^j \, b^k \, c) \, a
  \] 
  satisfying 
  \[
    l \geq 0, \quad i' + j' + k' = n+1, \quad i + j +k = n.
  \]
  It suffices to consider two cases: $l = 0$ and $l = 1$, for if $l > 1$ and
  $(u^{i'} \, a^{j'} \, b^{k'} \, c) \, (u^i \, a^j \, b^k \, c) \, a$ matches
  the right-hand side of Equation~\ref{eq:kleene-regular-2} then so does
  $(u^{i'} \, a^{j'} \, b^{k'} \, c)^l \, (u^i \, a^j \, b^k \, c) \, a$. 
  \begin{itemize}
  \item Now suppose $l = 0$. If $k = 0$ then $\Xi$ matches the first summand of
    the right-hand side of Equation~\ref{eq:kleene-regular-2}; otherwise note
    that $\Xi = (u^i \, a^{j+1} \, b^{k-1} \, c) \, b$ matches the second
    summand of the right-hand side of Equation~\ref{eq:kleene-regular-2}.
  \item Next suppose $l = 1$. If $k = k' = 0$ then $\Xi$ matches the first
    summand of the right-hand side of Equation~\ref{eq:kleene-regular-2};
    otherwise suppose $k' > 0$ then $\Xi = (u^{i'} \, a^{j'+1} \, b^{k'-1} \, c)
    \, (u^i \, a^j \, b^k \, c) \, b$ matches the second summand of the
    right-hand side of Equation~\ref{eq:kleene-regular-2}.
  \end{itemize}

  Let $p (x) = q (x) \, r (x)$. Applying the product rule of 
  derivatives, Equation~\ref{eq:kleene-regular} is equivalent to $\LHS = \RHS$
  where
  {\small
  \[\def\arraystretch{1.9}
    \begin{array}{rll}
      \LHS &\defeq & \displaystyle
                  \left[
                  r(\theta) \, \der{q}{x} (\theta) + 
                  q(\theta) \, \der{r}{x} (\theta)
                  \right] \cdot (a + b)
      \\
      \RHS & \defeq & \displaystyle
                   \left[
                   r(u + a) \, \der{q}{x} (u + a) + 
                   q(u + a) \, \der{r}{x} (u + a)
                   \right] \cdot a \\
        & & \displaystyle + \ \left[
            r(\theta) \, \der{q}{x} (\theta) + 
            q(\theta) \, \der{r}{x} (\theta)
            \right] \cdot b
    \end{array}
  \]}
  using the shorthand $\theta = u + a + b$ as before. Similar to the preceding
  case, clearly $\RHS \leq \LHS$. To show $\LHS \leq \RHS$, it suffices to show:
  \begin{align*}
    r(\theta) \, \der{q}{x} (\theta) \, a &\leq \RHS\\ 
    q(\theta) \, \der{r}{x} (\theta) \, a &\leq \RHS
  \end{align*}
  We consider the first; the same reasoning applies to the second. As before,
  thanks to Pilling's Normal Form Theorem, we may assume that $r(x) = (x^{m+1}
  \, c)^\star$ and $q(x) = (x^{n+1} \, d)^\star$. Then $\displaystyle r(\theta)
  \, \der{q}{x} (\theta) \, a = (\theta^{m+1} \, c)^\star \, (\theta^{n+1} \,
  d)^\star \, (\theta^{n} \, d) \, a$. By considering a typical element $\Xi$
  that matches the preceding expression, and using the same reasoning as the
  preceding case, we can then show that $\Xi$ matches $R$, as desired. 
\end{proof}

In this setting, again, derivatives fail to be additive in their
second argument. Take $p(x) = x^2$, for example: as $\der{p}{x} = x + x = x$, we
have:
\begin{salign*}
  \d[p](a, b + c) &= (b + c) (a + b + c)\\
                 &= b (a + b + c) + c (a + b + c)\\
                 &= b (a + b) + c (a + c) + bc\\
                 &= \d[p](a, b) + \d[p](a, c) + bc\\
                 &\neq \d[p](a, b) + \d[p](a, c)
\end{salign*}
It follows that $\KK[\overline x]$ cannot be modelled by a Cartesian
differential category, generalised or otherwise, as axiom \ref{CDC2} requires that
every derivative be additive.

%% file: difference-categories.tex
\label{chp:difference-categories}

As we saw in Section~\ref{sec:cartesian-differential-categories}, one of the
most notable and general settings for generalised differentiation in the current
literature is that of Cartesian differential categories
\cite{blute2009cartesian}. Introduced by Blute, Cockett and Seely, these provide
an abstract categorical axiomatisation of the directional derivative from
differential calculus. The relevance of Cartesian differential categories lies
in their ability to model both ``classical'' differential calculus (with the
canonical example being the category of Euclidean spaces and smooth functions
between) and the differential $\lambda$-calculus (as every categorical model for
it gives rise to a Cartesian differential category
\cite{manzonetto2012categorical}). 

However, while Cartesian differential categories have proven to be an immensely
successful formalism, they have, by design, some limitations. Firstly, they
cannot account for the ``exotic'' notions of derivative which change actions can
successfully represent, such as the calculus of finite differences (Section
\ref{sec:discrete-calculus}) or the derivatives of programs involved in
incremental computation (Chapter \ref{chp:incremental-computation}).

This is because the axioms of a Cartesian differential category stipulate that
derivatives should be linear in their second argument (in the same way that the
directional derivative is), whereas these aforementioned discrete sorts of
derivative need not be. Additionally, every Cartesian differential category is
equipped with a tangent bundle monad
\cite{cockett2014differential,manzyuk2012tangent} whose Kleisli category can be
intuitively understood as a category of generalised vector fields. This Kleisli
category comes equipped with a natural choice of differentiation operator,
inherited from the underlying category, which comes close to making it a
Cartesian differential category, but again fails the requirement of being linear
in its second argument. 

On the other hand, change action models impose no such restrictions on the
derivatives, but they can be excessively general, in that they do not guarantee
many useful properties that are commonly satisfied -- for example, commutativity
of addition, which holds in important cases such as the calculus of finite
differences.

In Section~\ref{sec:cartesian-difference-categories} we formulate the notion of
\emph{Cartesian difference categories}, a class of categories that bridge the
gap between the generality of change action models and Cartesian differential
categories. Some motivating examples are given in Section \ref{EXsec}. Section
\ref{monadsec} focuses on the tangent bundle functor in Cartesian difference
categories, which has many properties in common with the differential case.
Importantly, we show that the tangent bundle functor is a monad, and that the
induced Kleisli category is itself a Cartesian difference category, something
which is \emph{not} the case of differential categories. Finally, in
Section~\ref{sec:differential-lambda-categories}, we generalize the notion of
differential $\lambda$-categories to our setting, which will later provide the
basis for interpreting a higher-order calculus.

Many of the results in this chapter are the result of a collaboration with
Jean-Simon Lemay and will be appearing in print in \cite{alvarez2020cartesian}.
Some passages and figures have been reproduced verbatim with permission.

\section{Cartesian Difference Categories}
\label{sec:cartesian-difference-categories}

In this section we introduce \emph{Cartesian difference categories}, which are
generalisations of Cartesian differential categories equipped with an operator
that turns a map into an ``infinitesimal'' version of it, in the sense that
every map coincides with its Taylor approximation on infinitesimal elements.

What is meant by this? Consider a smooth function $f : \RR \to \RR$. Taylor's
theorem states that its value can be approximated at a neighbourhood of $a
\in \RR$ by the expression:
\[
  f({a} + b) \approx f({a}) + b \cdot f'({a})
\]

More precisely, given any $b \in \RR$, the value of $f(a + \delta \cdot d)$
tends to $f(a) + \delta \cdot d \cdot f'(a)$ as $\delta$ approaches zero. One
might then say that, when $\delta$ is an infinitesimally small magnitude, the
above is a strict equality. Infinitesimal extensions should then be understood
as an abstraction of the idea of ``multiplying by an infinitesimal $\delta$''.

\begin{definition} 
  A Cartesian left-additive category $\cat{C}$
  (Definition~\ref{def:left-additive-category}) is said to have an
  \textbf{infinitesimal extension} $\varepsilon$ if every hom-set $\cat{C}(A,B)$
  comes equipped with a monoid morphism $\varepsilon: \cat{C}(A,B) \to
  \cat{C}(A,B)$. That is, $\varepsilon$ must satisfy the following conditions:
  
  \begin{align*}
    \epsilon(f + g) &= \epsilon(f) + \epsilon(g)\\
    \epsilon(0) &= 0
  \end{align*}

  Furthermore, we require that $\epsilon$ be compatible with the Cartesian
  structure, in the sense that the following equations hold:

  \begin{align*}
    \epsilon(\pi_1) &= \pi_1 \circ \epsilon(\Id{})\\
    \epsilon(\pi_2) &= \pi_2 \circ \epsilon(\Id{})\\
    \epsilon(\pair{f}{g}) &= \pair{\epsilon(f)}{\epsilon(g)}
  \end{align*}
\end{definition}  

\begin{lemma}
  In any Cartesian left-additive category with infinitesimal extension
  $\epsilon$, the infinitesimal extension of a map $f : A \to B$ 
  is equivalent to post-composition with $\epsilon_B \defeq \epsilon(\Id{B})$. That is to say:
  \[
    \epsilon(f) = \epsilon(\Id{B}) \circ f
  \] 
\end{lemma}

\begin{lemma}
  Whenever $f : A \to B$ is additive (as in
  Definition~\ref{def:left-additive-category}), then so is $\epsilon(f)$.
  As an immediate consequence, $\epsilon_A$ is additive.
\end{lemma}

As a consequence of the above two results, we could have defined an infinitesimal 
extension to be a choice of a monoid homomorphism $\epsilon_A : A \to A$ for every object
$A$ of our category, and used this to formulate the infinitesimal extension of a map 
$f : A \to B$ to be $\epsilon(f) \defeq \epsilon_B \circ f$. Similarly, additive 
categories can be defined by postulating the existence of an (internal) addition map
$+_A : A \times A \to A$ for every object $A$ (in our formulation this map is given
by $\pi_1 + \pi_2$) and defined $f + g$ by $+_B \circ \pair{f}{g}$.
Both the ``internal'' and ``external'' approaches are equivalent and we have chosen
the second one for notational convenience and consistency with Cartesian differential
categories.

\begin{lemma}\label{caelem}
  Let $\cat{C}$ be a Cartesian left-additive category with infinitesimal
  extension $\varepsilon$. For every object $A$, define the maps $\oplus_A: A
  \times A \to A$, $+_A: A \times A \to A$, and $0_A: \terminal \to A$ respectively
  as follows: 
  \begin{align*}
    \oplus_A &= \pi_1 + \varepsilon(\pi_2)\\ 
    +_A &= \pi_1 + \pi_2 
    \\ 0_A &= 0 
  \end{align*}
  Then $(A, A, \oplus_A, +_A, 0_A)$ is a change action internal to $\cat{C}$. 
\end{lemma}
\begin{proof} 
  By \cite[Proposition~1.2.2]{blute2009cartesian}, $(A, +_A, 0_A)$ is a
  commutative monoid was shown On the other hand, that $\oplus_A$ is an action
  follows immediately from the fact that $\varepsilon$ is a monoid homomorphism.
\end{proof} 

Setting $\cact{A} \equiv (A, A, \oplus_A, +_A, 0_A)$, we note that $f
\oplus_{\cact{A}} g = f + \varepsilon(g)$ and $f +_{\cact{A}} g = f + g$, and so
in particular $+_{\cact{A}} = +$. Therefore, from now on we will omit the
subscripts and simply write $\oplus$ and $+$. 

Any Cartesian left-additive category $\cat{C}$ can always be endowed with two
different infinitesimal extensions, that correspond respectively to stating that
the only infinitesimal element is zero, and that every element is infinitesimal.

\begin{lemma} 
  For any Cartesian left additive category $\cat{C}$, 
  \begin{enumerate}
    \item Setting $\varepsilon(f) = 0$ defines an infinitesimal extension on
    $\cat{C}$ and therefore in this case, $\oplus_A = \pi_1$ and $f \oplus g
    = f$.
    \item Setting $\varepsilon(f) = f$ defines an infinitesimal extension on
    $\cat{C}$ and therefore in this case, $\oplus_A = +_A$ and $f \oplus g =
    f + g$.
  \end{enumerate}
\end{lemma}

While these examples of infinitesimal extensions may seem trivial, they are both
very significant as they will give rise to key examples of Cartesian difference
categories. 

\begin{definition}
  A \textbf{Cartesian difference category} is a Cartesian left additive category
  with an infinitesimal extension $\varepsilon$ which is equipped with a
  \textbf{difference combinator} $\d$ of the form:
  \[ \frac{f : A \to B}{\d[f]: A \times A \to B} \] 
  verifying the following coherence conditions (for clarity, we have highlighted the
  terms which differ from the corresponding Cartesian differential axioms):
  \begin{enumerate}[\CdCAx{\arabic*},ref={\CdCAx{\arabic*}},align=left]
  \setcounter{enumi}{-1}
    \item $\mathcolorbox{yellow}{f \circ (x + \varepsilon(u)) = f \circ x + \varepsilon\left( \d[f]
      \circ \langle x, u \rangle \right)}$ \label{CdC0}

    \item $\d[f+g] = \d[f] + \d[g]$, $\d[0] = 0$, and $\d[\varepsilon(f)] =
    \varepsilon(\d[f])$ \label{CdC1}

    \item $\d[f] \circ \langle x, u + v \rangle = \d[f] \circ \langle x, u
      \rangle + \d[f] \circ \langle x \mathcolorbox{yellow}{+ \varepsilon(u)}, v \rangle$ and $\d[f]
    \circ \langle x, 0 \rangle = 0$\label{CdC2}

    \item $\d[\Id{A}] = \pi_2$ and $\d[\pi_1] = \pi_2; \pi_1$ and $\d[\pi_2] =
    \pi_2; \pi_1$\label{CdC3}

    \item $\d[\langle f, g \rangle] = \langle \d[f], \d[g] \rangle$ and $\d[!_A]
    = !_{A \times A}$ \label{CdC4}

    \item $\d[g \circ f] = \d[g] \circ \langle f \circ \pi_1, \d[f] \rangle$
    \label{CdC5}

    \item $\d^2[f] \circ \left \langle \langle  x, u \rangle,
      \langle 0, v \rangle \right \rangle= \d[f] \circ  \langle x \mathcolorbox{yellow}{+
    \varepsilon(u)}, v \rangle$\label{CdC6}

    \item  $ \d^2[f] \circ \left \langle \langle x, u \rangle,
    \langle v, 0 \rangle \right \rangle=  \d^2[f] \circ \left
    \langle \langle x, v \rangle, \langle u, 0 \rangle \right \rangle$
    \label{CdC7}
  \end{enumerate}
\end{definition}

Before giving some intuition on the axioms \ref{CdC0} to \ref{CdC7}, we first
remark that one could have used the language of change actions to express
\ref{CdC0}, \ref{CdC2} and \ref{CdC6} which could then be written as:
\begin{enumerate}[\CdCAx{\arabic*},align=left]
  \setcounter{enumi}{-1}
  \item $f \circ (x \oplus u) = (f \circ x) \oplus \left(\d[f] \circ \langle x,
  u \rangle \right)$ 
  \setcounter{enumi}{1}
  \item  $\d[f] \circ \langle x, u + v \rangle = \d[f] \circ \langle x, u \rangle + \d[f] \circ \langle x \oplus u, v \rangle$ and $\d[f] \circ \langle x, 0 \rangle = 0$
  \setcounter{enumi}{5}
  \item  $\d\left[\d[f] \right] \circ \left \langle \langle  x, u \rangle, \langle 0, v \rangle \right \rangle= \d[f] \circ  \langle x \oplus u, v \rangle$
\end{enumerate}
And also, just like Cartesian differential categories, \ref{CdC6} and \ref{CdC7}
have alternative equivalent expressions. 
\begin{lemma}
  \label{lem:cdc6a}
  In the presence of the other axioms, \ref{CdC6} and \ref{CdC7} are equivalent to:
  \begin{enumerate}[\CdCAx{\arabic*(a)},ref={\CdCAx{\arabic*(a)}},align=left]
  \setcounter{enumi}{5}
    \item $\d\left[\d[f] \right] \circ \left \langle \langle  x, 0 \rangle,
    \langle 0, y \rangle \right \rangle = \d[f] \circ  \langle x, y \rangle$
    \label{CdC6a}
    \item $\d\left[\d[f] \right] \circ \left \langle \langle x, y \rangle, \langle z, w \rangle \right \rangle= \d\left[\d[f] \right] \circ \left \langle \langle x, z \rangle, \langle y, w \rangle \right \rangle$
    \label{CdC7a}
  \end{enumerate}
\end{lemma}
\begin{proof}
  The proof is essentially the same as \cite[Proposition
  4.2]{cockett2014differential}.
\end{proof} 

The keen eyed reader will notice that the axioms of a Cartesian difference
category are very similar to the axioms of a Cartesian differential category.
Indeed, \ref{CdC1}, \ref{CdC3}, \ref{CdC4}, \ref{CdC5}, and \ref{CdC7} are
identical to their counterparts in a Cartesian differential category. Both
definitions, however, differ in axioms \ref{CdC2} and \ref{CdC6}, where the
infinitesimal extension $\varepsilon$ now appears, and of course the inclusion
of the additional axiom \ref{CdC0}. 

On the other hand, interestingly enough, \ref{CdC6a} is the same as \ref{CDC6a}
We also point out that, writing out \ref{CdC0} and \ref{CdC2} using change
action notion, we see that these axioms are precisely an equational version of
the conditions in Definition~\ref{def:derivative}.

In element-like notation, \ref{CdC0} is written as: 
\[
    f(x + \varepsilon(u)) = f(x) + \varepsilon\left(\d[f](x,u) \right)
\]
This condition can be read as a generalisation of the Kock-Lawvere axiom that
characterises the derivative in from synthetic differential geometry
\cite{Kock06}. Broadly speaking, the Kock-Lawvere axiom states that,
for any map $f : \mathcal{R} \to \mathcal{R}$ and any $x \in \mathcal{R}$ and $d
\in \mathcal{D}$, there exists a unique $f'(x) \in \mathcal{R}$ verifying
\[
    f(x + d) = f(x) + d \cdot f'(x)
\]
where $\mathcal{D}$ is the subset of $\mathcal{R}$ consisting of infinitesimal
elements. It is by analogy with the Kock-Lawvere axiom that we refer to
$\varepsilon$ as an ``infinitesimal extension'' as it can be thought of as
embedding the space $A$ into a subspace $\varepsilon(A)$ of infinitesimal
elements. 

\ref{CdC2} is the first of the shared axioms that differs between a Cartesian
difference category and a Cartesian differential category. In a Cartesian
differential category, the differential of a map is assumed to be additive in
its second argument. In a Cartesian difference category, just as derivatives for
change actions, while the differential is still required to preserve zeros in
its second argument, it is only additive ``up to a small perturbation'', that
is:
\[\d[f](x,u+v) = \d[f](x,u) + \d[f](x + \varepsilon(u), v)\]
By taking \ref{CdC0} into consideration, the above can equivalently be written
in a more symmetric form as:
\[
  \d[f](x, u + v) = \d[f](x, u) + \d[f](x, v) + \epsilon(\d[\d[f]](x, v, u, 0))
\]

\ref{CdC6} and \ref{CdC7} tell us how to work with second order
differentials. \ref{CdC6} is expressed as follows: 
\[\d \left[ \d[f] \right]\left( x,y, 0,z \right) =  \d[f](x + \varepsilon(y), z)
\]
and has some remarkable and counter-intuitive consequences.

\begin{lemma}
  \label{lem:d-epsilon}
  Given any map $f : A \to B$ in a Cartesian difference category $\cat{C}$, its
  derivatives satisfy the following equations for any $x, u, v, w : C \to A$:

  \begin{enumerate}[\roman{enumi}.,ref={\thelemma.\roman{enumi}}]
    \item $\d[f] \circ \pair{x}{\epsilon(u)} = \epsilon(\d[f]) \circ
    \pair{x}{u}$
    \label{lem:d-epsilon-i}
    \item
    $\d[f] \circ \pair{x}{u + v}
    = \d[f] \circ \pair{x}{u} + \d[f] \circ \pair{x + \epsilon^2(u)}{v}$
    \label{lem:d-epsilon-ii}
    \item $\epsilon(\d^2[f]) \circ \four{x}{u}{v}{0}
    = \epsilon^2(\d^2[f]) \circ \four{x}{u}{v}{0}$
    \label{lem:d-epsilon-iii}
  \end{enumerate}
\end{lemma}
\begin{proof}\hspace{0pt}
  \begin{enumerate}[i.]
    \Item
    \begin{salign*}
      \d[f] \circ \pair{g}{\epsilon(h)}
      = \d[f]\circ \pair{g}{0}
      + \epsilon(\d^2[f]) \circ \four{g}{0}{0}{h}
      = \epsilon(\d[f]) \circ \pair{g}{h}
    \end{salign*}
  \Item \begin{salign*}
    &\d[f] \circ \pair{x}{u + v}
    \\
    &= \d[f] \circ \pair{x}{u}
    + \d[f] \circ \pair{x + \epsilon(u)}{v}
    \tag{by \ref{CdC2}}
    \\
    &= \d[f] \circ \pair{x}{u}
    + \d^2[f] \circ \four{x}{u}{0}{v}
    \tag{by \ref{CdC6}}
    \\
    &= \d[f] \circ \pair{x}{u}
    + \d^2[f] \circ \four{x}{0}{u}{v}
    \tag{by \ref{CdC7}}
    \\
    &= \d[f] \circ \pair{x}{u}
    + \d^2[f] \circ \four{x}{0}{u}{0}
    + \d^2[f] \circ \four{x}{\epsilon(u)}{0}{v}
    \tag{by \ref{CdC2}}
    \\
    &= \d[f] \circ \pair{x}{u}
    + \d^2[f] \circ \four{x}{u}{0}{0}
    + \d^2[f] \circ \four{x}{\epsilon(u)}{0}{v}
    \tag{by \ref{CdC7}}
    \\
    &= \d[f] \circ \pair{x}{u}
    + \d[f] \circ \pair{x + \epsilon^2(u)}{v}
    \tag{by \ref{CdC2} and \ref{CdC6}}
  \end{salign*}
  \Item
    \begin{salign*}
      &\epsilon(\d^2[f]) \circ \four{x}{u}{v}{0}
      \\
      &= \d^2[f] \circ \four{x}{\epsilon(u)}{v}{0}
      &\text{(by \ref{CdC7} and \ref{lem:d-epsilon-i})}
      \\
      &= \d^3[f] \circ \pair{
        \four{x}{0}{0}{u}
      }{
        \four{0}{0}{v}{0}
      }
      &\text{(by \ref{CdC6})}
      \\
      &= \d^3[f] \circ \pair{
        \four{x}{0}{0}{0}
      }{
        \four{0}{u}{v}{0}
      }
      &\text{(by \ref{CdC7})}
      \\
      &= \d^3[f] \circ \pair{
        \four{x}{0}{0}{0}
      }{
        \four{0}{u}{0}{0}
      }\\
      \continued + \d^3[f] \circ \pair{
        \four{x}{\epsilon^2(u)}{0}{0}
      }{
        \four{0}{0}{v}{0}
      }
      &\text{(by \ref{lem:d-epsilon-ii})}
      \\
      &= \d^3[f] \circ \pair{
        \four{x}{\epsilon^2(u)}{0}{0}
      }{
        \four{0}{0}{v}{0}
      }
      &\text{(by \ref{CdC7} and \ref{CdC2})}
      \\
      &= \d^2[f] \circ \four{x}{\epsilon^2(u)}{v}{0}
      &\text{(by \ref{CdC6})}
      \\
      &= \epsilon^2(\d^2[f]) \circ \four{x}{u}{v}{0}
      &\text{(by \ref{lem:d-epsilon-i})}
    \end{salign*}
  \end{enumerate}
\end{proof}

In particular, the above result lets us work as if the infinitesimal extension was
idempotent -- but only when it acts on a second derivative carrying a zero as its fourth
argument! Indeed, as Section~\ref{moduleex} shows, even though every difference
category satisfies this ``infinitesimal cancellation'' property, the infinitesimal
extension $\epsilon$ may not be idempotent.
These facts will turn out to be crucial once we set out to define a
term calculus (which we shall do in Chapter~\ref{chp:simply-typed-calculus}).
However, one of their consequences is of immediate interest.

\begin{lemma}
  Let $\cat{C}$ be a Cartesian difference category with a nilpotent
  infinitesimal extension. That is to say, there is some $k \in \NN$ such that
  $\epsilon^k(f) = 0$ for any $f$.
  Then every derivative $\d[f]$ is additive in its second argument.
\end{lemma}
\begin{proof}
  Fix a map $f : A \to B$, and consider arbitrary $x, u, v : C \to A$.
  Since $\d[f]$ is regular, it follows that it is additive ``up to a
  second-order term'':
  \begin{salign*}
    &\d[f](x, u + v)\\ 
    &= \d[f](x, u) + \d[f](x + \epsilon(u), v)
    &\text{(by \ref{CdC2})}
    \\
    &= \d[f](x, u) + \d[f](x, v) + \epsilon \d^2[f](x, v, u, 0)
    &\text{(by \ref{CdC0})}
    \\
    &= \d[f](x, u) + \d[f](x, v) + \epsilon^k \d^2[f](x, v, u, 0)
    &\text{(iterating \ref{lem:d-epsilon-iii})}
    \\
    &= \d[f](x, u) + \d[f](x, v)
    &\text{(since $\epsilon$ is nilpotent)}
  \end{salign*}
\end{proof}

\subsection{Differentials as Trivial Differences}
\label{CDCisCdCsec}

From a cursory look at the axioms of a Cartesian difference category, one might
expect that the notion is neither stronger nor weaker than that of a Cartesian
differential category, since neither set of axioms seems to subsume the other.

As we established in Section~\ref{sec:change-action-models:examples}, however,
we know that every Cartesian differential category gives rise to a change action
model. While we have not yet established the connection between change action
models and difference categories, it is clear that the two concepts are closely
related. It should come as no surprise, then, that Cartesian differential
categories are, in fact, a special case of difference categories.

\begin{theorem} 
  \label{thm:CDC-is-CdC}
  Every Cartesian differential category $\cat{C}$ with differential combinator
  $\mathsf{D}$ is a Cartesian difference category with an infinitesimal
  extension defined by $\varepsilon(f) = 0$ and a difference combinator defined
  to be identical to the differential combinator, that is to say, $\d[f] =
  \mathsf{D}[f]$.  
\end{theorem}
\begin{proof} 
Clearly, the first two parts of the \ref{CdC1}, the second part of \ref{CdC2},
\ref{CdC3}, \ref{CdC4}, \ref{CdC5}, and \ref{CdC7} are precisely the same as
their differential counterparts. 

On the other hand, since $\varepsilon(f) =0$, we have that \ref{CdC0} and the
third part of \ref{CdC1} both hold trivially, since they merely state that
$0=0$. The first part of \ref{CdC2} and \ref{CdC6} are exactly equivalent to the
first part of \ref{CDC2} and \ref{CDC6} respectively. Therefore, the
differential combinator satisfies the Cartesian difference axioms.
\end{proof} 

Conversely, one can always build a Cartesian differential category from a
Cartesian difference category by considering the objects for which the
infinitesimal extension is the zero map. 

\begin{proposition}\label{CdtoCD} For a Cartesian difference category $\cat{C}$
with infinitesimal extension $\varepsilon$ and difference combinator $\d$,
define the category $\cat{C}_{0}$ as the full subcategory of objects $A$ such
that $\varepsilon(\Id{A})= 0$. Then $\cat{C}_{0}$ is a Cartesian differential
category with a differential combinator given by $\mathsf{D}[f] = \d[f]$.
\end{proposition}
\begin{proof} 
  First note that if $\varepsilon(\Id{A})=0$ and $\varepsilon(\Id{B})=0$, then
by definition it also follows that $\varepsilon(\Id{A \times B}) = 0$. For the
terminal object we have $\varepsilon(\Id{\terminal}) = !_\terminal = 0$ since maps into
$\terminal$ are necessarily unique. Thus $\cat{C}_{0}$ is closed under finite
products and is therefore a Cartesian left-additive category. 

Furthermore, for any map $f$ in $\cat{C}_0$, we have $\varepsilon(f) =0$. This
implies that, for all such maps, the Cartesian difference axioms are precisely
the Cartesian differential axioms. Therefore, the difference combinator is a
differential combinator for this subcategory, and so $\cat{C}_{0}$ is a
Cartesian differential category.
\end{proof} 

In any Cartesian difference category $\cat{C}$, the terminal object $\terminal$
always satisfies that $\varepsilon(\Id{\terminal}) = 0$, thus $\cat{C}_{0}$ is never
empty. On the other hand, applying Proposition \ref{CdtoCD} to a Cartesian
differential category results in the entire category. 

The above result, however, does not imply that if a difference combinator is a
differential combinator then  infinitesimal extension must be zero. In Section
\ref{moduleex}, we provide such an example of a Cartesian differential category
which comes equipped with a non-zero infinitesimal extension such that the
differential combinator is a difference combinator with respect to this non-zero
infinitesimal extension. 

\subsection{Change Action Models as Difference Categories}\label{CdCisCAsec}

In this section we show how every Cartesian difference category is a
particularly well-behaved change action model, and conversely how every change
action model contains a Cartesian difference category. 

\begin{proposition} Let $\cat{C}$ be a Cartesian difference category with
infinitesimal extension $\varepsilon$ and difference combinator $\d$. Define
the functor $\alpha: \cat{C} \to \ftor{CAct}(\cat{C})$ as $\alpha(A) =
(A, A, \oplus_A, +_A, 0_A)$ (as defined in Lemma \ref{caelem}) and $\alpha(f) =
(f, \d[f])$. Then $(\cat{C}, \alpha: \cat{C} \to
\ftor{CAct}(\cat{C}))$ is a change action model.  
\end{proposition}
\begin{proof} 
By Lemma \ref{caelem}, $(A, A, \oplus_A, +_A, 0_A)$ is a change action and so
$\alpha$ is well-defined on objects. Given a map $f$, $\d[f]$ is a derivative of
$f$ in the change action sense since \ref{CdC0} and \ref{CdC2} are precisely a
restatement of Definition~\ref{def:derivative}, and so $\alpha$ is well-defined
on maps. That $\alpha$ preserves identities and composition follows from
\ref{CdC3} and \ref{CdC5} respectively, and so $\alpha$ is a functor. That
$\alpha$ preserves finite products will follow from \ref{CdC3} and \ref{CdC4}.
Finally, it is evident that $\alpha$ is a section of the forgetful functor, and
therefore the pair $(\cat{C}, \alpha)$ is a change action model. 
\end{proof} 

It is clear that not every change action model is a Cartesian difference
category, since for example, change action models do not require addition to be
commutative. On the other hand, it can be shown that every change action model
contains a Cartesian difference category as a full subcategory, formed of
well-behaved objects.

\begin{definition}
    Let $(\cat{C}, \alpha : \cat{C} \to \ftor{CAct}(\cat{C}))$ be a change
    action model. An object $A$ of $\cat{C}$ is \textbf{flat} whenever the
    following hold:
    \begin{enumerate}[{\bf [F.1]},ref={{\bf [F.{\arabic*}]}},align=left]
    \item $\Delta A = A$
    \label{F1}
    \item $\alpha(\oplus_A) = (\oplus_A, \oplus_A \circ \pi_2)$
    \label{F2}
    \item $0 \oplus_A (0 \oplus_A f) = 0 \oplus_A f$ for any $f : U \to A$.
    \label{F3}
    \item $\oplus_A$ is right-injective, that is, if $\oplus_A \circ \langle f,
    g \rangle = \oplus_A \circ \langle f, h \rangle$ then $g=h$. 
    \label{F4}
    \end{enumerate}

    We define the category $\cat{Flat}_\alpha$ as the full subcategory of
    $\cat{C}$ which contains precisely all the flat objects of $\cat{C}$.
\end{definition}

\begin{example}
  In the category $\CGrpS$ of groups and set-theoretic functions, every object
  satisfies conditions \ref{F1}, \ref{F3} and \ref{F4}. Additionally, $\mathcal{F}$
  satisfies \ref{F2} if and only if it is Abelian. Indeed, in $\CGrpS$:
  \begin{align*}
    \d[\oplus_A](x, u, v, w)
    &= - (x \oplus_A u) + ((x \oplus_A) v \oplus_A (u \oplus_A w))
    \\
    &= - (x + u) + (x + v + u + w)
  \end{align*}
  which is equal to $v + w$ if and only if $+$ is commutative.
\end{example}

Since $\alpha$ preserves finite products, it is straightforward to see that
$\terminal$ is flat and if $A$ and $B$ are flat then so is $A \times B$. The sum of
maps $f: A \to B$ and $g: A \to B$ in $\cat{Flat}_\alpha$ is defined using
addition on the change action $\alpha{B}$, that is, we define $f + g \defeq +_B
\circ \pair{f}{g}$, while the zero map $0: A \to B$ is likewise defined 
as $0 \defeq 0_B \mathrel{\circ}{}!_A$.

\begin{lemma}
  \label{EUCLCLAC} 
  $\cat{Flat}_\alpha$ is a Cartesian left additive category.
\end{lemma}
\begin{proof} 
Most of the Cartesian left-additive structure is immediate. However, since
addition is not required to be commutative for arbitrary change actions, we will
prove that, in any flat object, addition must necessarily be commutative.
Indeed, it suffices to compute. Using set-theoretic notation, we obtain the
following identity, for any $\cat{Flat}_\alpha$-maps $f, g : A \to B$ :
\begin{salign*}
  0 \oplus (f + g)
  &= (0 \oplus f) \oplus g
  &\text{(by the action property of $\oplus$)}
  \\
  &= (0 \oplus g) \oplus \d[\oplus]((0, g), (f, 0))
  &\text{(by the derivative condition)}
  \\
  &= (0 \oplus g) \oplus (f \oplus 0)
  &\text{(by \ref{F2})}
  \\
  &= 0 \oplus (g + f)
  &\text{(by the action property)}
\end{salign*}

By \ref{F4}, we can cancel the $0$ on both sides to obtain $f + g = g + f$.
\end{proof}

We use the action of the change action structure to define the infinitesimal
extension. So for a map $f: A \to B$ in $\cat{Flat}_\alpha$, define
$\varepsilon(f): A \to B$ by:
\[ \varepsilon(f) \defeq \oplus_B \circ \pair{0_B \circ{} !_A}{f} \]
or, using more convenient infix notation, $\varepsilon(f) \defeq 0_B \oplus_B f$.

\begin{lemma} $\varepsilon$ is an infinitesimal extension for
$\cat{Flat}_\alpha$. 
\end{lemma}
\begin{proof} 
  We have to show that $\varepsilon$ preserves the addition. Following the same
  idea as in the proof of Lemma \ref{EUCLCLAC}, we pick arbitrary $f, g : A \to
  B$ and obtain the following:

  \begin{salign*}
      0 \oplus \varepsilon(f + g) &= 0 \oplus (0 \oplus (f + g))
      &\text{(by definition of $\epsilon$)}
      \\
      &= 0 \oplus (0 \oplus f \oplus g)
      &\text{(by the action property)}
      \\
      &= \big(0 \oplus (0 \oplus f)\big)
        \oplus (0 \oplus g)
      &\text{(by the derivative condition, \ref{F2})}
      \\
      &= \big(0 \oplus (0 \oplus f + 0 \oplus g)\big)
      &\text{(by the action property)}
      \\
      &= 0 \oplus (\epsilon(f) + \epsilon(g))
      &\text{(by definition of $\epsilon$)}
  \end{salign*}
  Then, by \ref{F4}, it follows that $\varepsilon(f + g) = \varepsilon(f)
  + \varepsilon(g)$. The remaining infinitesimal extension axioms can be proven
  in a similar fashion.
\end{proof}

Lastly, the difference combinator for $\cat{Flat}_\alpha$ is defined in the
obvious way, that is to say, $\d[f]$ is simply the second component of
$\alpha(f)$.

\begin{theorem}
  \label{thm:flat-is-difference}
  Let $(\cat{C}, \alpha : \cat{C} \to \ftor{CAct}(\cat{C}))$ be a change
  action model. Then $\cat{Flat}_\alpha$ is a Cartesian difference category.
\end{theorem}
\begin{proof}
  \ref{CdC0} and \ref{CdC2} are simply a restatement of the derivative
  condition. \ref{CdC3} and \ref{CdC4} follow immediately from the fact that
  $\alpha$ preserves finite products and from the structure of products in
  $\ftor{CAct}(\cat{C})$ (as per Theorem~\ref{thm:cact-products}), while \ref{CdC5}
  follows from the definition of composition in $\ftor{CAct}(\cat{C})$
  (as per Definition~\ref{def:cact}).
  
  We now prove \ref{CdC1}. First, by definition of $+$ and applying the chain
  rule, we have:
  \begin{salign*}
    \d[f + g] = \d[+ \circ \pair{f}{g}] = \d[+_{B}] \circ \pair{\pair{f}{g}
  \circ \pi_1}{ \pair{\d[f]}{\d[g]}}
  \end{salign*}
  It suffices to show that $\d[+] = + \circ \pi_2$, from which \ref{CdC1}
  will follow. Consider arbitrary maps $u, w : A \to B$. Then (again using
  set-like notation) we have: 
  \begin{salign*}
    0 \oplus \big(u \oplus w)\big)
    = \big(0 \oplus u\big) \oplus (0 \oplus w)
    = 0 \oplus \big( u + (0 \oplus w)\big)
  \end{salign*}
  By \ref{F4} we obtain the identity below:
  \begin{align}
    u \oplus w = u + \epsilon(w)
    \label{oplus-epsilon} 
  \end{align}
  From this, it
  follows that, for any $u, v, w, l : A \to B$, the maps
  $ (u \oplus w) + (v \oplus l) $
  and
  $ u + v + \epsilon (w) + \epsilon(l) $ are identical. But, since $\epsilon$
  is an infinitesimal extension, the second map can also be written as $
  (u + v) + \epsilon (w + l)$, which is precisely $(u + v) \oplus (w + l)$.
  So we have
  \begin{salign*}
     (u + v) \oplus (w + l) = (u \oplus w) + (v \oplus l) = (u + v) \oplus \d[+](u,
  v, w, l))
  \end{salign*}
  Applying \ref{F4} again gives $\d[+] = + \circ \pi_2$ as desired. Axiom
  \ref{CdC7} can be established in a similar way and we omit the calculations,
  proceeding directly to axiom \ref{CdC6a} -- which, per Lemma~\ref{lem:cdc6a}, 
  is equivalent to, and easier to prove than \ref{CdC6}. As before, we pick arbitrary 
  $x, u : A \to B$ and calculate:
  \begin{salign*}
    &f(x) \oplus \d^2[f](x, 0, 0, u)
    \\
    &= f(x) + (0 \oplus \d^2[f](x, 0, 0, u))
    &\text{(by \ref{oplus-epsilon})}
    \\
    &= f(x) + (0 \oplus (0 \oplus \d^2[f](x, 0, 0, u))))
    &\text{(by \ref{F3})}
    \\
    &= f(x) + (0 \oplus (\d[f](x, 0) \oplus \d^2[f](x, 0, 0, u)))
    &\text{(by regularity)}
    \\
    &= f(x) + (0 \oplus \d[f](x, 0 \oplus u))
    &\text{(by the derivative condition)}
    \\
    &= f(x) \oplus \d[f](x, 0 \oplus u)
    &\text{(by \ref{oplus-epsilon})}
    \\
    &= f(x \oplus (0 \oplus u))
    &\text{(by the derivative condition)}
    \\
    &= f(x + 0 \oplus (0 \oplus u))
    &\text{(by \ref{oplus-epsilon})}
    \\
    &= f(x + 0 \oplus u)
    &\text{(by \ref{F3})}
    \\
    &= f(x \oplus u)
    &\text{(by \ref{oplus-epsilon})}
    \\
    &= f(x) \oplus \d[f](x, u)
    &\text{(by the derivative condition)}
  \end{salign*}
  Hence $\d^2[f](x, 0, 0, u) = \d[f](x, u)$ as desired. 
\end{proof}

Note that the proof above goes through without appealing to \ref{F3}, except for
axiom \ref{CdC6}, for which only a weaker version is provable. More precisely,
without \ref{F3}, one can only show
\[
  \epsilon (\d^2[f](x, 0, 0, u)) = \d[f](x, \epsilon(u))
\]
which is strictly weaker than \ref{CdC6} and, in fact, holds trivially whenever
$\epsilon = 0$.

\subsection{Linear and \texorpdfstring{$\varepsilon$}{Epsilon}-Linear Maps}

An important subclass of maps in a Cartesian differential category are the
so-called \emph{linear maps} \cite[Definition 2.2.1]{blute2009cartesian}. These
generalise the usual notion from linear algebra: a map in a Cartesian
differential category is said to be linear whenever it coincides with its
derivative. Notably, linear maps are \emph{not} defined as those maps which
are additive -- additivity of linear maps follows, in fact, as a theorem.

The same definition of linear maps goes through in a Cartesian difference
category $\cat{C}$. Using set-like notation, a map $f$ is linear whenever $\d[f](x,y) =
f(y)$. Linear maps in a Cartesian difference category satisfy many of the same
properties found in \cite[Lemma~2.2.2]{blute2009cartesian} (in particular, they
form a Cartesian subcategory of $\cat{C}$).

\begin{definition} \label{def:linearity}
  In a Cartesian difference category, a map $f$ is said to be \textbf{linear}
  whenever it satisfies the identity $\d[f] = f \circ \pi_2$\footnote{
    Compare and contrast the notion of \emph{self-maintainable derivatives}\cite[Section~4.3]{cai2014changes},
    those which do not depend on the base point but only the change, i.e.
    $\d[f](x, \delta) = g(\delta)$ for some $g$. The notion of a linear map is strictly
    stronger, as we require the map $g$ to coincide with the function itself.
  }.
\end{definition}

\begin{lemma} 
  In any Cartesian difference category, the following properties hold.
\begin{enumerate}[\roman*.,ref={\thelemma.\roman{enumi}}]
  \item If $f: A \to B$ is linear then $\varepsilon(f) = f \circ
  \varepsilon(\Id{A})$.
  \label{epsilon-linear}
  \item If $f: A \to B$ is linear, then $f$ is additive (as per Definition~\ref{def:left-additive-category}).
  \item Identity maps, projection maps, and zero maps are linear.
  \label{id-linear}
  \item The composite, sum, and pairing of linear maps is linear. 
  \item If $f: A \to B$ and $k: C \to D$ are linear, then for any map $g: B \to
  C$, the following equality holds:
  \[ \d[k \circ g \circ f] = k \circ \d[g] \circ (f \times f) \]
  \item If an isomorphism is linear, then its inverse is linear.
  \item For any object $A$, $\oplus_A$ and $+_A$ are linear. 
  \label{struct-linear}
  \item Whenever $\epsilon$ is nilpotent, then every derivative evaluated at
  zero is linear. That is, every map $\d[f] \circ \pair{0}{\Id{}}$ is linear.
  \label{nilpotent-linear}
\end{enumerate}
\end{lemma}
\begin{proof}
  Most of the above results are either immediate or admit a similar proof to the
  ones in \cite[Lemma2.2.2]{blute2009cartesian}. We will prove
  \ref{epsilon-linear} and \ref{nilpotent-linear}, as they differ from the differential
  setting.

  For the first, we note that $f \circ 0 = f \circ \pi_2 \circ \pair{0}{0} = \d[f] \circ \pair{0}{0} = 0$, therefore:
  \begin{salign*}
    \epsilon(f) &= 0 + \epsilon(f) \tag{by the additive axioms}\\
                &= (f \circ 0) + (\epsilon(f \circ \pi_2) \circ \pair{0}{\Id{A}})
                \tag{by the extension axioms}
                \\
                &= (f \circ 0) + (\epsilon(\d[f]) \circ \pair{0}{\Id{A}})
                \tag{by linearity}
                \\
                &= f \circ (0 + \epsilon(\Id{A}))
                \tag{by \ref{CdC0}}
                \\
                &= f \circ \epsilon(\Id{A})
                \tag{by the additive axioms}
  \end{salign*}

  For the second we need to show that $\d[\d[f] \circ \pair{0}{\Id{}}]
  = \d[f] \circ \pair{0}{\Id{}} \circ \pi_2$. The left-hand side expands to:
  \begin{salign*}
    \d[\d[f] \circ \pair{0}{\Id{}}]
    = \d^2[f] \circ \big((\pair{0}{\Id{}}) \times (\pair{0}{\Id{}})\big)
  \end{salign*}
  Fixing arbitrary $u, w$ of the appropriate types and using set-like notation,
  we obtain:
  \begin{salign*}
    \d^2[f](0, u, 0, w)
    &= \d[f](\epsilon(u), w)
    &\text{(by \ref{CdC6})}
    \\
    &= \d[f](0, w) + \epsilon \d^2[f](0, w, u, 0)
    &\text{(by \ref{CdC0})}
    \\
    &= \d[f](0, w) + \epsilon^k \d^2[f](0, w, u, 0)
    &\text{(by \ref{lem:d-epsilon-iii})}
    \\
    &= \d[f](0, w)
    &\text{(since $\epsilon$ is nilpotent)}
  \end{salign*}
\end{proof}

While all linear maps are additive, the converse is not necessarily true (see
\cite[Corollary 2.3.4]{blute2009cartesian}). However, an immediate consequence
of the above lemma is that every Cartesian difference category has a Cartesian
subcategory of linear maps.

Another interesting subclass of maps in a Cartesian difference category are the
$\varepsilon$-linear maps, which are maps whose infinitesimal extension is
linear.

\begin{definition} In a Cartesian difference category, a map $f$ is
\textbf{$\varepsilon$-linear} if $\varepsilon(f)$ is linear.
\end{definition}

\begin{lemma} In a Cartesian difference category, 
\begin{enumerate}[\roman*.,ref={\thelemma.\roman{enumi}}]
  \item If $f: A \to B$ is $\varepsilon$-linear then $f \circ (x + \varepsilon(y)) = f \circ x + \varepsilon(f) \circ y$.
  \item Every linear map is $\varepsilon$-linear.
  \label{linear-is-epsilon-linear}
  \item The composite, sum, and pairing of $\varepsilon$-linear maps is
  $\varepsilon$-linear.
  \item If an isomorphism is $\varepsilon$-linear, then its inverse is again
  $\varepsilon$-linear. 
\end{enumerate}
\end{lemma}

Using element-like notation, the first point of the above lemma says that if $f$
is $\varepsilon$-linear then $f(x + \varepsilon(y)) = f(x) + \varepsilon(f(y))$.
This sheds some light on $\varepsilon$-linear maps: these are those which are
additive on ``infinitesimal'' elements (i.e. elements of the form
$\varepsilon(y)$). 

For a Cartesian differential category considered as a difference category,
linear maps in the Cartesian difference category sense are precisely the same as
Cartesian differential category sense \cite[Definition
2.2.1]{blute2009cartesian}, while every map is trivially $\varepsilon$-linear
since $\varepsilon =0$. 

\section{Examples of Cartesian Difference Categories}\label{EXsec}

\subsection{Euclidean Spaces and Smooth Functions}\label{smoothex}

Every Cartesian differential category is a Cartesian difference category where
the infinitesimal extension is zero. As a particular example, we consider the
category of real smooth functions, which as mentioned above, can be considered
to be the canonical (and motivating) example of a Cartesian differential
category.

Let $\mathbb{R}$ be the set of real numbers and let $\cat{SMOOTH}$ be the
category whose objects are Euclidean spaces $\mathbb{R}^n$ (we take
$\mathbb{R}^0$ to be the single-point space $\lbrace \ast \rbrace$) and whose
maps are all the smooth functions $F: \mathbb{R}^n \to \mathbb{R}^m$.
$\cat{SMOOTH}$ is a Cartesian left additive category where the product
structure is given by the standard Cartesian product of Euclidean spaces and
where the additive structure is defined by point-wise addition, $(F+G)(\vec x) =
F(\vec x) + G(\vec x)$ and $0(\vec x) = (0, \hdots, 0)$, where $\vec x \in
\mathbb{R}^n$. $\cat{SMOOTH}$ is a Cartesian differential category where the
differential combinator is defined by the directional derivative of smooth
functions. Explicitly, for a smooth function $F: \mathbb{R}^n \to \mathbb{R}^m$,
which is in fact a tuple of smooth functions $F= (f_1, \hdots, f_n)$ where $f_i:
\mathbb{R}^n \to \mathbb{R}$, $\mathsf{D}[F]: \mathbb{R}^n \times \mathbb{R}^n
\to \mathbb{R}^m$ is defined as follows: 
\[\mathsf{D}[F]\left(\vec x, \vec y \right) := \left( \sum \limits^n_{i=1}
\frac{\partial f_1}{\partial u_i}(\vec x) y_i, \hdots, \sum \limits^n_{i=1}
\frac{\partial f_n}{\partial u_i}(\vec x) y_i  \right)\] where $\vec x = (x_1,
\hdots, x_n), \vec y = (y_1, \hdots, y_n) \in \mathbb{R}^n$. Alternatively,
$\mathsf{D}[F]$ can also be defined in terms of the Jacobian matrix of $F$.
Therefore $\cat{SMOOTH}$ is a Cartesian difference category with
infinitesimal extension $\varepsilon =0$ and with difference combinator
$\mathsf{D}$. Since $\varepsilon = 0$, the induced action is simply $\vec x
\oplus_{\mathbb{R}^n} \vec y = \vec x$. Also a smooth function is linear in the
Cartesian difference category sense precisely if it is $\mathbb{R}$-linear in
the classical sense, and every smooth function is $\varepsilon$-linear. 

\subsection{Calculus of Finite Differences}\label{discreteex}
\newcommand{\CAbS}[0]{\cat{Ab_\star}}

In Section~\ref{sec:discrete-calculus} we discussed the category $\CGrpS$ of
groups and arbitrary functions, and we showed that it is a change action model.
Clearly $\CGrpS$ fails to be a Cartesian difference category, as the induced
additive structure need not be commutative. The subcategory of
abelian groups, however, is indeed a Cartesian difference category.

Let $\CAbS$ be the full subcategory of $\CGrpS$ whose objects are abelian groups
$(G, +, 0)$. As in $\CGrpS$, a map $f : (F, +, 0) \to (G, +, 0)$ is simply an
arbitrary function $f : \mathcal{F} \to \mathcal{G}$, which does not necessarily
preserve the group structure. Clearly $\CAbS$ is Cartesian left-additive\footnote{
  It would be more appropriate to say that $\CAbS$ can be endowed with the structure of
  a Cartesian left-additive category, since the choice of a group operation for every
  object is somewhat arbitrary.
}, and we
could explicitly prove that it has an infinitesimal extension and a difference
operator. Instead, we will apply Theorem~\ref{thm:flat-is-difference} to show
directly that $\CAbS$ is Cartesian differential.

\begin{proposition}
  The full subcategory $\CAbS$ of $\CGrpS$ consisting of all the abelian groups
  is precisely the category $\cat{Flat}_\CGrpS$ of flat objects in $\CGrpS$.
\end{proposition}
\begin{proof}\belowdisplayskip=-120pt
  Axioms \ref{F1}, \ref{F3}, \ref{F4} hold trivially, as they are true of any
  object in $\CGrpS$. It remains only to prove \ref{F1}. And since functions
  in $\CAbS$ admit exactly one derivative, it suffices to show that 
  $+ \circ \pi_2$ is a derivative for $+$. But this is an immediate consequence
  of commutativity of $+$, since:
  \[\small
    (a \oplus u) \oplus (b \oplus v)
    = (a + u) + (b + v) = (a + b) + (u + v)
    = (a \oplus b) \oplus (u \oplus v)
    \qedhere
  \]
\end{proof}

The induced infinitesimal extension is simply the identity, and the
corresponding difference combinator $\d$ in $\CAbS$ is given, as in $\CGrpS$,
by:
\[
  \d[f](x, u) \defeq f(x + u) - f(x)
\]

On the other hand, $\d$ is not a differential combinator for
$\CAbS$, since it satisfies neither \ref{CDC6} nor \ref{CDC2}.
Thanks to the addition of the infinitesimal extension, $\d$ does satisfy 
\ref{CdC2} and \ref{CdC6}, as well as \ref{CdC0}. 
However, as noted in \cite{FMCS2018}, it is interesting to note that this $\d$
does satisfy \ref{CDC1}, the second part of \ref{CDC2}, \ref{CDC3}, \ref{CDC4},
\ref{CDC5}, \ref{CDC7}, and \ref{CDC6a}. It is worth noting that
in \cite{FMCS2018}, the goal was to drop the addition and develop a
``non-additive'' version of Cartesian differential categories. Cartesian
difference categories can be seen as a stronger form of the ``non-linear
Cartesian differential categories'' presented therein, which simply drop the
requirement that derivatives be additive altogether.

As $\CAbS$ is the paradigmatic example of a Cartesian difference category
(indeed, the axioms of difference categories were engineered precisely to fit
the case of finite differences in abelian groups), it is worth using it to build
some intuition on the less intuitive quirks of difference categories.

For example, in $\CAbS$, axiom~\ref{CdC6} consists of folding a telescoping
expression:
\begin{salign*}
  \d^2[f](x, u, 0, v)
  &= \d[f](x, u + v) - \d[f](x, u)
  \\
  &= \Big(f(x + u + v) - f(x)\Big)
    - \Big(f(x + u) - f(x) \Big)
  \\
  &= f(x + u + v) - f(x + u)
  \\
  &= \d[f](x + u, v)
\end{salign*}
The curious ``duplication of infinitesimals'' from Lemma~\ref{lem:d-epsilon-iii}
holds trivially, since $\epsilon = \Id{}$ is idempotent.

The linear maps in $\CAbS$ (in the sense of Definition~\ref{def:linearity}) are
those which satisfy $\d[f](x, y) = f(y)$ -- that is to say, $f(x + y) - f(x) =
f(y)$. So the linear maps are precisely the group homomorphisms!

\subsection{Module Morphisms}\label{moduleex}

Here we provide a simple example of a Cartesian difference category whose
difference combinator is also a differential combinator, but whose
infinitesimal extension is neither zero nor the identity. 

Let $R$ be a commutative semiring and let $\cat{MOD}_R$ be the category of
$R$-modules and $R$-linear maps between them. $\cat{MOD}_R$ has finite
biproducts and is therefore a Cartesian left additive category where every map
is additive. Every $r \in R$ induces an infinitesimal extension $\varepsilon^r$
defined by scalar multiplication, $\varepsilon^r(f)(m) = r f(m)$. Then
$\cat{MOD}_R$ is a Cartesian difference category with the infinitesimal
extension $\varepsilon^r$ for any $r \in R$ and difference combinator $\d$
defined as:
\[\d[f](m,n)=f(n)\] 

$R$-linearity of $f$ assures that \ref{CdC0} holds, while the remaining
Cartesian difference axioms hold trivially. In fact, $\d$ is also a differential
combinator and therefore $\cat{MOD}_R$ is also a Cartesian differential
category. The induced action is given by $m \oplus_M n = m + rn$. By definition
of $\d$, every map in $\cat{MOD}_R$ is linear and, by
\ref{linear-is-epsilon-linear}, also $\varepsilon$-linear.  

\newcommand{\Ab}[0]{\CAbS}
\subsection{Stream Calculus}\label{streamex}

Here we show how one can extend the calculus of finite differences example to
sets of infinite sequences. This is a particularly interesting model, as the
methods and language of differential calculus have been often used to describe
functions on streams - see \cite{rutten2003behavioural} or
\cite{rutten2005coinductive} for an introduction to this so-called stream
calculus. A particularly important class of stream functions are the so-called
\emph{causal} functions, those where the $n$-th element of the output may only
depend on the first $n$ elements of the input. These are known to arise from
certain classes of ``stream differential equations'' \cite{kupke2011stream} and
have recently been studied in connection to machine learning and backpropagation
\cite{sprunger2019differentiable,sprunger2019differential}.

\begin{definition}
For a set $A$, let $A^\omega$ denote the set of infinite sequences of elements
  of $A$, where we write $\seq{a_i}$ for the infinite sequence $\seq{a_i} =
  (a_1, a_2, a_3, \hdots)$ and $a_{i:j}$ for the (finite) subsequence $(a_i,
  a_{i + 1}, \hdots, a_j)$. A function $f : A^\omega \to B^\omega$ is
  \textbf{causal} whenever the $n$-th element $f\left(\seq{a_i}\right)_n$ of the
  output sequence only depends on the first $n$ elements of $\seq{a_i}$, that
  is, $f$ is causal if and only if whenever $a_{0:n} = b_{0:n}$ then
  $f\left(\seq{a_i}\right)_{0:n} = f\left(\seq{b_i}\right)_{0:n}$.
\end{definition}
  
We now consider streams over abelian groups, so let $\Ab^\omega$ be the category
whose objects are all the abelian groups and where the morphisms between
$\mathcal{G}$ and $\mathcal{G}$ are the causal maps from $G^\omega$ to
$H^\omega$. Every object in $\Ab^\omega$ is itself an abelian group
(internal to $\Ab^\omega$), with addition being defined pointwise, that is to
say, $\seq{a_i} + \seq{b_i} = \seq{a_i + b_i}$.

Hence, $\Ab^\omega$ is a Cartesian left-additive category, where the product is
given by the standard product of abelian groups and where the additive structure
is similarly lifted point-wise from the structure of $\Ab$:
$(f+g)\left(\seq{a_i}\right)_n = f\left(\seq{a_i}\right)_n +
g\left(\seq{a_i}\right)_n$ and $0\left(\seq{a_i}\right)_n = 0$. In order to
define the infinitesimal extension, we first need to define the truncation
operator $\A{z}$. 

\begin{definition}
  Given an abelian group $\mathcal{G}$, we define the \emph{truncation operator}
  $\A{z}_{\mathcal{G}} : G^\omega \to G^\omega$ by
  \begin{align*}
    (\A{z} \seq{a_i})_0 &= 0\\
    (\A{z} \seq{a_i})_{j + 1} &= a_{j + 1}
  \end{align*}
\end{definition}
\begin{lemma}
  $\A{z}_{\mathcal{G}}$ is a causal monoid homomorphism according to the
  pointwise monoid structure on $G^\omega$. Thus all the $\A{z}_\mathcal{G}$
  define an infinitesimal extension\footnote{
  It may seem odd that we should choose this truncation operator as our
  infinitesimal extension, rather than the more natural right shift operator which
  maps an input stream $(a_1, a_2, a_3, \ldots)$ to $(0, a_1, a_2, a_3,
  \ldots)$. But one can show that, with this choice of an infinitesimal
  extension, the derivative of a causal map may itself not be causal, hence the
  category of causal functions would not admit higher derivatives.
  } $\A{z}$ for $\Ab^\omega$.
\end{lemma}

The corresponding action updates every element after the first by
adding it to the corresponding element in the applied change, while leaving the
first element untouched. Formally:
\begin{align*}
  ( \seq{a_i} \oplus \seq{b_i})_0 = a_0 && ( \seq{a_i} \oplus \seq{b_i})_{n+1} = a_{n+1} + b_{n+1}
\end{align*}

Thus defined, the category $\Ab^\omega$ is a Cartesian difference category whose
the infinitesimal extension is given by the truncation operator,
$\varepsilon(f)\left(\seq{a_i}\right) = \A{z} f\left(\seq{a_i}\right)$, and
where the difference combinator $\d$ is defined as follows: 
\begin{align*}
  \d[f]\left(\seq{a_i}, \seq{b_i}\right)_0 
  &= f\left(\seq{a_i} + \seq{b_i}\right)_0 -  f\left(\seq{a_i}\right)_0
  \\
  \d[f]\left(\seq{a_i}, \seq{b_i}\right)_{n + 1} 
  &= f\left(\seq{a_i} + \A{z}\seq{b_i} \right)_{n + 1} 
  - f\left(\seq{a_i}\right)_{n+1}
\end{align*}

\begin{lemma}
  For any arbitrary function $f : G^\omega \to H^\omega$, the map $\d[f]$
  is causal.
\end{lemma}
\begin{proof}
  Follows immediately from the definition of $\d[f]$ -- but we remark that
  $\d[f]$ won't satisfy the derivative condition unless $f$ itself is causal.
\end{proof}

Note the similarities between the difference combinator on $\Ab$ and that on
$\Ab^\omega$. One might suggest that a simpler difference combinator could be
defined by letting $\d[f]\pa{\seq{a_i}, \seq{b_i}}_n = \pa{f(\seq{a_i} +
\A{z}\seq{b_i}) - f(\seq{a_i})_n }$. This, however, does not quite work, as it fails
to satisfy axiom \ref{CdC3}. Indeed, the derivative of the identity computes to 
$\d[\Id{}]\pa{\seq{a_i}, \seq{b_i}} = \A{z}\seq{b_i}$, rather than $\seq{b_i}$.

\begin{lemma}
  The category $\Ab^\omega$ is a Cartesian difference category.
\end{lemma}
\begin{proof}
  \hyperref[CdC1]{\CdCAx{1-7}} are trivial.
  \ref{CdC0} follows
  immediately from the fact that every map in $\Ab^\omega$ is causal 
  (and hence any update to the input which does not change the first element
  gives rise to an update to the output which does not change the first
  element).
\end{proof}

A causal map is linear (in the Cartesian difference category sense) if and only
if it is a group homomorphism. On the other hand, a causal map $f$ is
$\varepsilon$-linear if and only if it is a group homomorphism in every
component except the first.
$\A{z}f(\seq{a_i} + \seq{b_i}) = \A{z}f(\seq{a_i}) + \A{z}f(\seq{b_i})$.

\renewcommand{\T}[0]{\ftor{T}}
\section{Tangent Bundles in Cartesian Difference Categories} 
\label{monadsec}

As we showed in Section \ref{sec:tangent-bundle-change-action-model}, every
change action model is equipped with an endofunctor $\ftor{T}$ that behaves in
some ways like well-known \emph{tangent bundle monad} in Cartesian differential
categories. \cite{cockett2014differential,manzyuk2012tangent}. However, change
action models do not have enough structure to show that this functor is a monad.

In a Cartesian difference category, however, this issue disappears and we
recover the full monad structure. What is more, the Kleisli category of this
monad is again a Cartesian difference category! This is an important result, as
it is not true in the differential setting: the Kleisli category of the tangent
monad in a Cartesian differential category is \emph{not}, itself, Cartesian
differential, but it \emph{will} be a Cartesian difference category.

\subsection{The Tangent Bundle Monad}

Let $\cat{C}$ be a Cartesian difference category with infinitesimal extension
$\varepsilon$ and difference combinator $\d$. Remember that, in
Definition~\ref{def:tangent-functor}, we defined the functor $\T: \cat{C}
\to \cat{C}$ as follows: 
\begin{align*}
  \ftor{T}A &\defeq A \times \changes A\\
  \ftor{T}f &\defeq \pair{f \circ \pi_1}{\d f}
\end{align*}
and define the natural transformations $\eta: \ftor{Id}_\cat{C} \Rightarrow \T$
and $\mu: \T^2 \Rightarrow \T$ by:
\begin{align*}
  \eta_A &\defeq \langle \Id{A}, 0 \rangle \\
  \mu_A &\defeq \left \langle \pi_{11}, \pi_{21} + \pi_{12} + \varepsilon(\pi_{22}) \right \rangle 
\end{align*}

\begin{theorem} 
  \label{thm:t-monad}
  $(\T, \mu, \eta)$ defines a monad. 
\end{theorem} 
\begin{proof}
  Let $f: A \to B$ be an arbitrary map. Naturality of $\eta$ and $\mu$
  as well as the first of the monad
  laws ($\mu \circ \eta_{\T} = \mu \circ \T(\eta)$)
  were
  established in Theorem~\ref{lemma:tangent-bundle-monad}.

  For the last of the monad laws, we first note that, since $\mu$ is linear,
  it follows that $\T(\mu) = \mu \times \mu$. Then it suffices to compute:
  \begin{salign*}
    \mu_A \circ \T(\mu_A)
    &= \mu_A \circ (\mu_A \times \mu_A)
    \\
    &= \pair{\pi_{11}}{\pi_{21} + \pi_{12} + \epsilon(\pi_{22})}
    \circ (\mu_A \times \mu_A)
    \\
    &=
    \pair{\pi_1 \circ \mu_A \circ \pi_1}{\pi_2 \circ \mu_A \circ \pi_1
    + \pi_1 \circ \mu_A \circ \pi_2 + \epsilon(\pi_2 \circ \mu_A \circ \pi_2)}
    \\
    &=
    \pair{\pi_{111}}{
      \pi_{211} + \pi_{121} + \epsilon(\pi_{221})
      + \pi_{112}
      + \epsilon \pa{\pi_{212} + \pi_{122} + \epsilon(\pi_{222})}
    }
    \\
    &=
    \pair{\pi_1 \circ \pi_{11}}{
      \pi_{2} \circ \pi_{11} 
      + \pi_1 \circ \pa{\pi_{21} + \pi_{12} + \epsilon(\pi_{22})}
      + \epsilon\pa{\pi_2 \circ \pa{\pi_{21} + \pi_{12} + \epsilon(\pi_{22})}}
    }
    \\
    &= 
    \pair{\pi_{11}}{\pi_{21} + \pi_{12} + \epsilon(\pi_{22}) }
    \circ \pair{\pi_{11}}{\pi_{21} + \pi_{12} + \epsilon(\pi_{22}) }
    \\
    &= \mu_A \circ \mu_{\T(A)}\qedhere
  \end{salign*}
\end{proof} 

When $\cat{C}$ is a Cartesian differential category with the difference
structure arising from setting $\varepsilon = 0$, this tangent bundle monad
coincides with the standard tangent monad corresponding to its tangent category
structure \cite{cockett2014differential,manzyuk2012tangent}.

\begin{example}
  For a Cartesian differential category, since
  $\varepsilon = 0$, the induced monad is precisely the monad induced by its
  tangent category structure \cite{cockett2014differential,manzyuk2012tangent}.
  For example, in the Cartesian differential category $\mathsf{SMOOTH}$ (as
  defined in Section \ref{smoothex}), one has that $\T(F)(\vec x, \vec y) =
  (F(\vec x), \mathsf{D}[F](\vec x, \vec y))$, $\eta_{\mathbb{R}^n}(\vec x) =
  (\vec x,\vec 0)$, and $\mu_{\mathbb{R}^n}((\vec x,\vec y), (\vec z, \vec w)) =
  (\vec x, \vec y + \vec z)$. 
\end{example}

\begin{example} 
  In the Cartesian difference category
  $\overline{\mathsf{Ab}}$ (as defined in Section \ref{discreteex}), the monad
  is given by $\T(f)(x,y) = (f(x), f(x+y) - f(x))$, $\eta_G(x) = (x,0)$, and
  $\mu_G((x,y),(z,w)) = (x, y + z + w)$. 
\end{example}

\begin{example}
  In the Cartesian difference category
  $\mathsf{MOD}_R$ (as defined in Section \ref{moduleex}) with infinitesimal
  extension $\varepsilon^r$, for $r \in R$, $\T(f)(m,n) = (f(m), f(n))$,
  $\eta_M(m) = (m,0)$, and $\mu_M((m,n),(p,q)) = (m, n + p + rq)$. 
\end{example}

\begin{proposition}
  The family of maps $\oplus_A : \T(A) \to A$, as defined in Lemma \ref{caelem},
  constitute a natural transformation $\oplus : \T \to \ftor{Id}$. Furthermore,
  every pair $(A, \oplus_A)$ is an algebra for the tangent bundle monad.
\end{proposition}

\begin{proposition}
  Every map $f : A \to B$ is a homomorphism of $\T$-algebras from $(A,
  \oplus_A)$ into $(B, \oplus_B)$
\end{proposition}

\subsection{The Kleisli Category of \texorpdfstring{$\T$}{T}}

Recall that the Kleisli category of the monad $(\T, \mu, \eta)$ is defined as
the category $\cat{C}_\T$ whose objects are the objects of $\cat{C}$, and where
a map $A \to B$ in $\cat{C}_\T$ is a map $f: A \to \T(B)$ in $\cat{C}$, which
would be a pair $f = \langle f_0, f_1 \rangle$ where $f_1, f_2: A \to B$. The
identity map in $\cat{C}_\T$ is the monad unit $\eta_A: A \to \T(A)$, while
composition of Kleisli maps $f: A \to \T(B)$ and $g: B \to \T(C)$ is defined as
the composite $\mu_C \circ \T(g) \circ f$. To distinguish between composition in
$\cat{C}$ and $\cat{C}_\T$, we denote Kleisli composition by $g \circ^\T f =
\mu_C \circ \T(g) \circ f$. If $f=\langle f_0, f_1 \rangle$ and $g=\langle g_0,
g_1 \rangle$, then their Kleisli composition can be explicitly computed out to
be: 

\[ g \circ^\T f = \langle g_0, g_1 \rangle \circ^\T \langle f_0, f_1 \rangle =
\left \langle g_0 \circ f_0, \d[g_0] \circ \langle f_0, f_1 \rangle + g_1 \circ
(f_0 + \varepsilon(f_1)) \right \rangle\] 

As noted in \cite{cockett2014differential}, Kleisli maps can be understood as
``generalised'' vector fields. Indeed, $\T(A)$ should be thought of as the
tangent bundle over $A$, and therefore the analogue of a vector field would be a
map of the form $\langle \Id{A}, f \rangle: A \to \T(A)$, 
which is of course a Kleisli map from $A$ to $A$.
For more details on the intuition behind this Kleisli
category see \cite{cockett2014differential}. We now turn to the task of showing
that the Kleisli category of any Cartesian difference category is again a
Cartesian difference category. 

We begin by exhibiting the Cartesian left additive structure of the Kleisli
category.   The product of objects in $\cat{C}_\T$ is defined as $A \times B$
with projections $\pi^{\T}_0: A \times B \to \T(A)$ and $\pi^{\T}_1: A \times B
\to \T(B)$ defined respectively as $\pi^{\T}_0 = \langle \pi_1, 0 \rangle$ and
$\pi^{\T}_1 = \langle \pi_2, 0 \rangle$. The pairing of Kleisli maps $f=\langle
f_0, f_1 \rangle$ and $g=\langle g_0, g_1 \rangle$ is defined as $\langle f, g
\rangle^\T = \langle \langle f_0, g_0 \rangle, \langle f_1, g_1 \rangle
\rangle$. The terminal object is again $\terminal$ and where the unique map to the
terminal object is $!^{\T}_A = \pair{0}{0}$. The sum of Kleisli maps 
$f=\langle f_0, f_1 \rangle$ and $g=\langle, g_0, g_1 \rangle$ is defined as $f
+^\T g = f + g = \langle f_0 + g_0, f_1 + g_1 \rangle$, and the zero Kleisli
maps is simply $0^\T = 0 = \langle 0, 0 \rangle$. Therefore we conclude that the
Kleisli category of the tangent monad is a Cartesian left additive category. 

\begin{lemma} 
  $\cat{C}_\T$ is a Cartesian left additive category. 
\end{lemma} 

For a Kleisli map $f : \pair{f_0}{f_1}$, the infinitesimal extension
$\varepsilon^\T(f)$ is defined as follows:
\[ \varepsilon^\T(f) = \langle 0, f_0 + \varepsilon(f_1) \rangle \]

\begin{lemma} 
  $\varepsilon^\T$ is an infinitesimal extension on $\cat{C}_\T$. 
\end{lemma} 

It is interesting to point out that for an object $A$ the induced action
$\oplus^\T_A$ in the Kleisli category can be computed out to be:
\[ 
  \oplus^\T_A 
  = \pi^\T_1 +^\T \varepsilon^\T(\pi_2) 
  = \pair{\pi_1}{0} + \pair{0}{\pi_2} 
  = \pair{\pi_1}{\pi_2}
  = \Id{\T(A)}
\]
which, perhaps surprisingly, turns out to be the identity map on $\T(A)$ in the base
category $\cat{C}$ (but not the Kleisli identity). In fact, it was the
observation that the identity in $\cat{C}$ had the right type to be an action 
in $\cat{C}_\T$ that initially led to this result.

To define the difference combinator for the Kleisli category, first note that
difference combinators by definition do not change the codomain. That is, if $f
: A \to \T(B)$ is a Kleisli arrow, then the type of its derivative \emph{qua}
Kleisli arrow should be $A \times A \to \T(B) \times \T(B)$, which coincides
with the type of its derivative in $\cat{C}$. Therefore, the difference
combinator $\d^\T$ for the Kleisli category can be defined to be the difference
combinator of the base category, that is, for a Kleisli map $f= \langle f_0, f_1
\rangle$:
\[\d^\T[f] = \d[f] = \langle \d[f_0], \d[f_1] \rangle\] 

\begin{theorem} 
  For a Cartesian difference category $\cat{C}$, the Kleisli
 category $\cat{C}_\T$ is a Cartesian difference category with infinitesimal
 extension $\varepsilon^\T$ and difference combinator $\d^\T$.
\end{theorem}
\begin{proof}
  Let $\phi = \four{\pi_{11}}{\pi_{12}}{\pi_{21}}{\pi_{22}}$ be the
  isomorphism between $\T(A) \times \T(B)$ and $T(A \times B)$.
  We first note that, for any map $f$ in $\cat{C}$, the following equality holds: 
  \[ 
    \T(\d[f]) = 
    \d\brak{\T(f)} 
    \circ \phi
  \]  Since $\phi$ is linear, we also have
  \[
    \d[\phi] = \phi \circ \pi_2
  \]
  This will help simplify many of the calculations to follow, since
  $\T(\d[f])$ appears everywhere due to the definition of Kleisli
  composition.

  We will also make use of the following auxiliary results:
  \begin{lemma}
    Whenever $f : A \to B$ is linear in $\cat{C}$, then so is $\eta_B \circ f$ in
    $\cat{C}_\T$, that is, 
    $\d^\T[\eta_B \circ f] = (\eta_B \circ f) \circ^\T \pi_2^\T$.
  \end{lemma}
  \begin{proof}\let\qed\relax
    \begin{salign*}
      \d^\T[\eta_B \circ f] 
      &= \d[\pair{\Id{B}}{0} \circ f]\\
      &= \pair{\pi_2}{0} \circ \pair{f \circ \pi_1}{f \circ \pi_2}\\
      &= \pair{f \circ \pi_2}{0}\\
      &= \eta \circ f \circ \pi_2\\
      &= \mu \circ \T(\eta_B) \circ \T(f) \circ \eta \circ \pi_2\\
      &= (\eta_B \circ f) \circ^\T \pi_2^\T
    \end{salign*}
  \end{proof}

  \begin{lemma}
    Then the following identities hold:
    \begin{enumerate}[\roman*.]
      \Item $\d^2[f] \circ \phi = \d^2[f]$\\
      \Item $\T^2(f) \circ \phi = \phi \circ \T^2(f)$\\
      \Item $\phi \circ \T(\mu) = \mu \circ \T(\phi) \circ \phi$\\
    \end{enumerate}
  \end{lemma}
  \begin{proof}\let\qed\relax
    The first identity is simply a restatement of \ref{CdC7}. For the second identity:
    \begin{salign*}
      &\hspace{-20pt}\T^2(f) \circ \phi \circ \four{x}{u}{v}{w}\\
      &= \T^2(f) \circ \four{x}{v}{u}{w}\\
      &= \four{f \circ x}{\d[f] \circ \pair{x}{v}}
              {\d[f] \circ \pair{x}{u}}{\d^2[f] \circ \four{x}{v}{u}{w}}\\
      &= \four{f \circ x}{\d[f] \circ \pair{x}{v}}
              {\d[f] \circ \pair{x}{u}}{\d^2[f] \circ \four{x}{u}{v}{w}}\\
      &= \phi \circ \four{f \circ x}{\d[f] \circ \pair{x}{u}}
                         {\d[f] \circ \pair{x}{v}}{\d^2[f] \circ \four{x}{u}{v}{w}}\\
      &= \phi \circ \T^2(f)
    \end{salign*}
    Finally, for the last identity:
    \begin{salign*}
      &\hspace{-20pt}\phi \circ \T(\mu) \circ 
      \pair{\four{x}{u}{v}{w}}{\four{x'}{u'}{v'}{w'}}\\
      &= \phi \circ \pair{\pair{x}{u + v + \varepsilon(w)}}{\pair{x'}{u' + v' + \varepsilon(w')}}\\
      &= \four{x}{x'}{u + v + \varepsilon(w)}{u' + v' + \varepsilon(w')}\\
      &= \pair{\pair{x}{x'}}{\pair{u}{u'} + \pair{v}{v'} + \varepsilon(\pair{w}{w'})}\\
      &= \mu \circ \four{\pair{x}{x'}}{\pair{u}{u'}}{\pair{v}{v'}}{\pair{w}{w'}}\\
      &= \mu \circ \T(\phi) \circ \four{\pair{x}{u}}{\pair{x'}{u'}}{\pair{v}{w}}{\pair{v'}{w'}}\\
      &= \mu \circ \T(\phi) \circ \phi \circ
        \four{\pair{x}{u}}{\pair{v}{w}}{\pair{x'}{u'}}{\pair{v'}{w'}}
    \end{salign*}
  \end{proof}

  We can now prove each of the Cartesian difference category axioms.
  \newcommand{\tcirc}[0]{\mathrel{\circ^\T}}
  \begin{enumerate}[\ref{CdC\arabic*},align=left]
    \setcounter{enumi}{-1}
    \item 
    Pick an arbitrary Kleisli map $f = \pair{f_1}{f_2} : A \to \T(B)$.
    Then:
    {\begin{salign*}
      f \circ^\T \oplus^\T
      &= \mu \circ \T(f) \circ \Id{\T(A)}\\
      &= \pair{f_1 \circ \pi_1}{f_2 \circ \pi_1 + \d[f_1] + \varepsilon(\d[f_2])}
    \end{salign*}}
    On the other hand:
    {\begin{salign*}
      \oplus^\T \circ^\T \pair{f \circ^\T \pi_1^\T}{\d^\T[f]}^\T
      &= \mu \circ \T(\Id{}) \circ \pair{f \circ^\T \pi_1^\T}{\d^\T[f]}^\T\\
      &= \mu \circ \pair{f \circ^\T \pi_1^\T}{\d^\T[f]}^\T\\
      &= \mu \circ \pair{\mu \circ \T(f) \circ \pair{\pi_1}{0}}{\d^\T[f]}^\T\\
      &= \mu \circ \pair{\pair{f_1 \circ \pi_1}{f_2 \circ \pi_1}}{\d^\T[f]}^\T\\
      &= \mu \circ \phi \circ \pair{\pair{f_1 \circ \pi_1}{f_2 \circ \pi_1}}{\d^\T[f]}\\
      &= \mu \circ \phi \circ \four{f_1 \circ \pi_1}{f_2 \circ \pi_1}{\d[f_1]}{\d[f_2]}\\
      &= \pair{f_1 \circ \pi_1}{f_2 \circ \pi_1 + \d[f_1] + \varepsilon(\d[f_2])}
    \end{salign*}}
    \item
      Since $+^\T = \eta \circ +$ and $0^\T = \eta \circ 0$, and both maps are linear
      in $\cat{C}$, it immediately follows that they are linear in $\cat{C}^\T$. It remains
      to show $\d^\T[\varepsilon^\T] = \varepsilon^\T \circ^\T \pi_2^\T$
      \begin{align*}
        \d^\T[\varepsilon^\T]
        = \pair{0}{\varepsilon} \circ \pi_2
        = \mu \circ \eta \circ \pair{0}{\varepsilon} \circ \pi_2
        = \mu \circ \T(\pair{0}{\varepsilon}) \circ \eta \circ \pi_2
        = \varepsilon^\T \circ^\T \pi_2^\T
      \end{align*}
    \item 
      We prove the first condition (the second follows by trivial calculation):
      \begin{salign*}
          &\d^\T[f] \circ^\T \pair{x}{u +^\T v}^\T \\
          \eq \mu \circ \T(\d[f]) \circ \phi \circ \pair{x}{u + v}\\
          \eq \mu \circ \pair{\d[f] \circ \pair{x}{u + v}_0}{\d^2[f] \circ \pair{x}{u + v}}
          \PROOFSTEP
          \eq \mu
          \circ \pair{
            \d[f] \circ \pair{x}{u}_0
            + \d[f] \circ \pair{x + \epsilon(u)}{v}_0
          }{
            \d^2[f] \circ \pair{x}{u} + \d^2[f]\circ \pair{x + \epsilon(u)}{v}
          }
          \PROOFSTEP
          \eq \mu
          \circ \pair{
            \d[f] \circ \pair{x}{u}_0
            + \d[f] \circ \pair{x + u}{v}_0
          }{
            \d^2[f] \circ \pair{x}{u} + \d^2[f]\circ \pair{x + \epsilon(u)}{v}
          }
          \PROOFSTEP
          \eq \mu \circ \Big(
            \pair{\d[f] \circ \pair{x}{u}_0}{\d^2[f] \circ \pair{x}{u}}
          \Big) \\
          \continued + \mu \circ \Big(
            \pair{
              \d[f] \circ \pair{x + \epsilon(u)}{u}_0
            }{
              \d^2[f] \circ \pair{x + \epsilon(u)}{v}
            }
          \Big)
          \PROOFSTEP
          \eq \d^\T[f] \circ \pair{x}{u}
          + \d^\T[f] \circ \pair{x + \epsilon(u)}{v}
      \end{salign*}
    \item
    This axiom follows immediately from the fact that $\Id{A}^\T = \eta_A \circ
    \Id{A}$ and $\pi_i^\T = \eta_{A_i} \circ \pi_i$, with $\Id{A}, \pi_i$ linear in
    $\cat{C}$.

    \item
    Take arbitrary $f = \pair{f_1}{f_2}, g = \pair{g_1}{g_2}$. Then:
    \begin{salign*}
      \d^\T[\pair{f}{g}^\T]
      = \d[\phi \circ \pair{f}{g}]
      = \phi \circ \four{\d[f_1]}{\d[f_2]}{\d[g_1]}{\d[g_2]}
      = \pair{\d^\T[f]}{\d^\T[g]}^\T
    \end{salign*}

    \item
    It follows from the more general fact that the tangent bundle is functorial
    in the Kleisli category: define $\T^\T(f) = \pair{f \circ^\T
    \pi_1^\T}{\d^\T[f]}^\T$. We show that $\T^\T(g \circ^\T f) = \T^\T(g) \circ
    \T^\T(f)$. First, note that:
    \begin{salign*}
      \T^\T(f)
      &= \pair{f \circ^\T \pi_1^t}{\d^\T[f]}^\T\\
      &= \phi \circ \pair{f \circ^\T \pi_1^t}{\d^\T[f]}\\
      &= \phi \circ \pair{\mu \circ \T(f) \circ \eta \circ \pi_1}{\d[f]}\\
      &= \phi \circ \pair{\mu \circ \eta \circ f \circ \pi_1}{\d[f]}\\
      &= \phi \circ \pair{f \circ \pi_1}{\d[f]}\\
      &= \phi \circ \T(f)
    \end{salign*}
    Then:
    \begin{salign*}
      \T^\T(g \circ^\T f)
      &= \phi \circ \T(g \circ^\T f)\\
      &= \phi \circ \T(\mu \circ \T(g) \circ f)\\
      &= \phi \circ \T(\mu) \circ \T(\T(g) \circ f)\\
      &= \mu \circ \T(\phi) \circ \phi \circ \T(\T(g) \circ f)\\
      &= \mu \circ \T(\phi) \circ \T(\T(g)) \circ \phi \circ \T(f)\\
      &= \mu \circ \T(\phi \circ \T(g)) \circ \phi \circ \T(f)\\
      &= \T^\T(g) \circ^\T \T^\T(f)
    \end{salign*}

    \item We prove \ref{CdC6a} instead. For this, the following auxiliary result
    will be of use:
    \[
      \d^3[f] \circ (\phi \times \phi) = \d^3[f]
    \]
    This is a consequence of \ref{CdC7} and the following calculation:
    \begin{salign*}
      \d^3[f] \circ (\phi \times \phi)
      &= \d[\d^2[f]] \circ (\phi \times \phi)
      \\
      &= \d[\d^2[f] \circ \phi] \circ (\phi \times \phi)
      \\
      &= \d[\d^2[f]] \circ (\phi \times \phi) \circ (\phi \times \phi)
      \\
      &= \d^3[f]
    \end{salign*}
    We can then show \ref{CdC6a}:
    \begin{salign*}
      &\d^\T[\d^\T[f]] \circ^\T \four{x}{0}{0}{w}^\T\\
      &= \mu \circ \T(\d^2[f])
      \circ \phi \circ (\phi \times \phi) \circ
      \four{x}{0}{0}{w}
      \PROOFSTEP
      \eq \mu \circ \d\brak{\T(\d[f])} 
      \circ (\phi \times \phi) \circ 
      \four{x}{0}{0}{w}
      \PROOFSTEP
      \eq \mu \circ \pair{\d^2[f] \circ (\pi_1 \times \pi_1)}{\d^3[f]}
      \circ (\phi \times \phi) \circ 
      \four{x}{0}{0}{w}
      \PROOFSTEP
      \eq \mu \circ \pair{\d^2[f] \circ \four{x_0}{0}{0}{w_0}}{
        \d^3[f]
        \circ
        \four{x}{0}{0}{w}
      }
      \PROOFSTEP
      \eq \mu \circ \pair{\d[f] \circ \pair{x_0}{w_0}}{
        \d^2[f] \circ \pair{x}{w}
      }
      \PROOFSTEP
      \eq \d^\T[f] \circ^\T \pair{x}{w}
    \end{salign*}

    \item Follows by similar calculations to \ref{CdC2}. \qedhere 
  \end{enumerate}
\end{proof} 

As a result, the Kleisli category of a Cartesian difference category is again a
Cartesian difference category, whose infinitesimal extension is neither the
identity or the zero map. This allows one to build numerous examples of
interesting and exotic Cartesian difference categories, such as the Kleisli
category of the tangent bundle in any Cartesian differential category (one can,
of course, iterate this process by taking the Kleisli category of the Kleisli
category, \emph{ad infinitum}). We highlight the importance of this construction
for the Cartesian differential case as it does not in general result in a
Cartesian differential category: even if $\varepsilon = 0$, it is always the
case that $\varepsilon^\T \neq 0$. 

We conclude this section by taking a look at the linear maps and the
$\varepsilon^\T$-linear maps in the Kleisli category. A Kleisli map $f=\langle
f_0, f_1 \rangle$ is linear in the Kleisli category if $\d^\T[f] = f \circ^\T
\pi^\T_1$, which amounts to requiring that:
\[ \langle \d[f_0], \d[f_1] \rangle = \langle f_0 \circ \pi_2, f_1 \circ \pi_2 \rangle \]
 Therefore a Kleisli map is linear in the Kleisli category if and only
if it is the pairing of maps which are linear in the base category. On the other
hand, $f$ is $\varepsilon^\T$-linear if $\varepsilon^T(f) = \langle 0, f_0 +
\varepsilon(f_1) \rangle$ is linear in the Kleisli category, which in this case
amounts to requiring that $f_0 + \varepsilon(f_1)$ be linear. Therefore, if
$f_0$ is linear and $f_1$ is $\varepsilon$-linear, then $f$ is
$\varepsilon^\T$-linear. 

\section{Difference \texorpdfstring{$\lambda$}{Lambda}-Categories}
\label{sec:differential-lambda-categories}

In order to give a semantics for the differential $\lambda$-calculus, it does
not suffice to ask for a Cartesian differential category equipped with
exponentials -- the exponential structure has to play well with both the
additive and the differential structure, in the sense of
Definition~\ref{def:differential-lambda-category}.

The same is true of difference categories: if we hope to interpret any
reasonable sort of higher-order calculus, such as the one we will define in the
next chapter, we will require an axiom similar to \ref{DlC}, together with a
condition requiring higher-order functions to respect the infinitesimal
extension. Intuitively, we shall require that $\lambda x . \epsilon(t)$ be equal
to $\epsilon (\lambda x . t)$.

\begin{definition}
  \label{def:difference-lambda-category}
  We remind the reader that a Cartesian left-additive category is
  \textbf{Cartesian closed left-additive} whenever it is Cartesian closed and
  the following equations hold:
  \begin{enumerate}[i.]
    \item $\Lambda(f + g) = \Lambda(f) + \Lambda(g)$
    \item $\Lambda(0) = 0$
  \end{enumerate}

  A Cartesian difference category $\cat{C}$ is a \textbf{difference
  $\lambda$-category} if it Cartesian closed left-additive and satisfies the
  following additional axioms:
  \begin{enumerate}[{\bf[$\d\lambda$C.\arabic*]},ref={\bf[$\d\lambda$C.{\arabic*}]},align=left]
    \item $\d[\Lambda(f)] = \Lambda\pa{\d[f]
      \circ \pair{\pa{\pi_1 \times \Id{}}}{\pa{\pi_2 \times 0}}
    }
    $
    \label{L1}
    \item $\Lambda(\epsilon(f)) = \varepsilon\pa{\Lambda(f)}$
    \label{L2}
  \end{enumerate}

\end{definition}
\begin{remark}
  \label{rem:lambda-d-swap}
  Let $\A{sw}$ denote the composite
  \[
    \pair{\pair{\pi_{11}}{\pi_2}}{\pi_{21}}
    : (A \times B) \times C \to (A \times C) \times B
  \]
  Then the derivative $\d[\Lambda(f)]$ can be
  equivalently written in terms of $\A{sw}$ as:
  \[
    \d[\Lambda(f)] \defeq \Lambda\pa{
      \d[f] 
      \circ \pa{\Id{} \times \pair{\Id{}}{0}}
      \circ \A{sw}
    }
  \]
\end{remark}

Axiom \ref{L1} is identical to its differential analogue \ref{DlC}, and the
previous remark shows that it follows the same broad intuition. Given a map
$f : A \times B \to C$, we usually understand the composite $\d[f] \circ (\Id{A
\times B} \times (\Id{A} \times 0_B)) : (A \times B) \times A \to C$ as a
partial derivative of $f$ with respect to its first argument. Hence, just as it
was with differential $\lambda$-categories, axiom~\ref{L1} states that the
derivative of a curried function is precisely the derivative of the uncurried
function with respect to its first argument (compare with the much weaker
Theorem~\ref{thm:oplus-lambda-partial}, which merely states that both
expressions act on $A \Ra B$ in the same way).

\begin{example}
  Let $\cat{C}$ be a differential $\lambda$-category. Then the corresponding
  Cartesian difference category (setting $\epsilon = 0$) is a difference
  $\lambda$-category.
\end{example}

\begin{example}
  The category $\Ab$ is a difference $\lambda$-category. Given groups $G, H$,
  the exponential $G \Ra H$ is defined pointwise as in Theorem
  \ref{thm:cgrps-ccc}. Evidently the exponential respects the monoidal structure
  and the infinitesimal extension (trivial). We check that it also verifies
  axiom \ref{L1}:
  \begin{salign*}
    \d[\Lambda(f)](x, u)(y) &= \Lambda(f)(x + u)(y) - \Lambda(f)(x)(y)
    \\
    &= f(x + u, y) - f(x, y)
    \\
    &= \Lambda(\d[f] \circ (\Id{} \times \pair{\Id{}}{0}) \circ \A{sw})(x, u)(y)
  \end{salign*}
\end{example}

As is the case in differential $\lambda$-categories, we can define a
``differential substitution'' operator on the semantic side. This operator
is akin to post-composition with a partial derivative, and can be defined
as follows.

\begin{definition}
  \label{def:star-composition}
  Given morphisms $s : A \times B \ra C, u : A \ra B$, we define their
  \textbf{differential composition} $s \star u : A \times B \ra C$ by:
  \[
    s \star u \defeq \d[s] \circ \pair{\Id{A \times B}}{\pair{0_A}{u \circ \pi_1}}
  \]
\end{definition}

We should understand the morphism $s \star u$ as the partial derivative of $s$
in its second argument, pre-composed with the morphism $u$. When we develop our
calculus, this will correspond precisely to the differential substitution of the
top variable in $s$ by the term $u$.

Some technical results follow that will be of use later. These all correspond to
well-known properties of differential $\lambda$-categories (see
e.g.~\cite[Lemma~4.8]{bucciarelli2010categorical}) and their proofs are
identical, but we reproduce them here for reference, and to show that their
statements do not differ in our setting (as opposed to some other technical
lemmas in that work which hinge on derivatives being additive).

\begin{lemma}
  \label{lem:lambda-star-1}
  Let $f : A \to B, g : A \to C, h : (A \times B) \times E \to F$ be arbitrary
  $\cat{C}$-morphisms. Then the following properties hold:
  \begin{enumerate}[i.]
    \item $\d[\A{sw}] = \A{sw} \circ \pi_2$
    \item $(g \circ \pi_1) \star f = 0$
    \item $\Lambda(h) \star f = 
    \Lambda(((h \circ \A{sw}) \star (f \circ \pi_1)) \circ \A{sw})$
    \item $\Lambda^-\pa{\Lambda(h) \star f} = 
    ((h \circ \A{sw}) \star (f \circ \pi_1)) \circ \A{sw}$
  \end{enumerate}
\end{lemma}
\begin{proof}
  \hspace{0pt} 
  \begin{enumerate}[\roman{enumi}.,ref={\thelemma.\roman{enumi}}]
    \Item {\small
    \[
        \pi_2 \star f = \D(\pi_2) \circ \pair{\Id{}}{\pair{0}{f \circ \pi_1}}
        = \pi_{22} \circ \pair{\Id{}}{\pair{0}{f \circ \pi_1}}
        = f \circ \pi_1
        \]}
    \Item \vspace{-9pt}\begin{salign*}
        (g \circ \pi_1) \star f 
        &= \d[g \circ \pi_1] \circ \pair{\Id{}}{\pair{0}{f \circ \pi_1}}\\
        &= \d[g] 
        \circ \pair{\pi_{11}}{\pi_{12}} 
        \circ \pair{\Id{}}{\pair{0}{f \circ \pi_1}}\\
        &= \d[g] \circ \pair{\pi_1}{0}\\
        &= 0
      \end{salign*}
      \vspace{-\parskip}
      \vspace{-\parskip}
      \vspace{-\parskip}
      \vspace{-\parskip}
      \vspace{-\parskip}
      \vspace{-\parskip}
      \vspace{-\parskip}
      \vspace{-\parskip}
    \Item 
      \begin{salign*}
        \Lambda(h) \star f
        &= \d[\Lambda (h)]
        \circ \pair{\Id{}}{\pair{0}{f \circ \pi_1}}
        \PROOFSTEP
        &= \Lambda\Big(
          \d[h] \circ (\Id{} \times \pair{\Id{}}{0}) \circ \A{sw}
        \Big)
        \circ \pair{\Id{}}{\pair{0}{f \circ \pi_1}}
        \PROOFSTEP
        &= \Lambda\Big(
          \d[h] \circ (\Id{} \times \pair{\Id{}}{0}) \circ \A{sw}
          \circ \big(\pair{\Id{}}{\pair{0}{f \circ \pi_1}} \times \Id{}\big)
        \Big)\\
        &= \Lambda\Big(
          \d[h] \circ (\Id{} \times \pair{\Id{}}{0}) 
          \circ \pair{\Id{}}{\pair{0}{f \circ \pi_{11}}}
        \Big)
        \PROOFSTEP
        &= \Lambda\Big(
          \d[h] 
          \circ \pair{\Id{}}{\pair{\pair{0}{f \circ \pi_{11}}}{0}}
        \Big)
        \PROOFSTEP
        &= \Lambda\Big(
          \d[h]
          \circ \pair{\A{sw} \circ \pi_1}{\A{sw} \circ \pi_2}
          \circ \pair{\A{sw}}{\pair{0}{f \circ \pi_{11}}}
        \Big)
        \PROOFSTEP
        &= \Lambda\Big(
          \d[h \circ \A{sw}]
          \circ \pair{\Id{}}{\pair{0}{(f \circ \pi_1) \circ \pi_1}}
          \circ \A{sw}
        \Big)
        \PROOFSTEP
        &= \Lambda\Big(
          ((h \circ \A{sw}) \star (f \circ \pi_1))
          \circ \A{sw}
        \Big)
      \end{salign*}
    \item Immediate consequence of $iii.$ \qedhere
  \end{enumerate}
\end{proof}

A central property of differential $\lambda$-categories is a deep correspondence
between differentiation and the evaluation map. As one would expect, the partial
derivative of the evaluation map gives one a first-class derivative operator
(see, for example, \cite{bucciarelli2010categorical}[Lemma~4.5], which provides
an interpretation for the differential substitution operator in the differential
$\lambda$-calculus). This property
still holds in difference categories, although its formulation is somewhat more
involved.

\begin{lemma}
  \label{lem:lambda-d-ev}
  For any $\cat{C}$-morphisms $\Lambda(f) : A \to (B \Ra C), e : A \to B$, the
  following identities hold:
  \begin{enumerate}[\roman{enumi}.,ref={\thelemma.\roman{enumi}}]
    \item $
    \d[\A{ev} \circ \pair{\Lambda(f)}{e}]
    = \A{ev} \circ \pair{\d[\Lambda(f)]}{e \circ \pi_1}
    +\d[f] \circ \pair{\pair{\pi_1 + \epsilon(\pi_2)}{e \circ
    \pi_1}}{\pair{0}{\d[e]}}$
    \item $
      \d[\A{ev} \circ \pair{\Lambda(f)}{e}]
      = 
      \A{ev} \circ \pair{\d[\Lambda(f)]}{e \circ \pi_1 + \epsilon(\d[e])}
      +\d[f] \circ \pair{\pair{\pi_1}{e \circ \pi_1}}{\pair{0}{\d[e]}}
    $
  \end{enumerate}
\end{lemma}
\begin{proof}
  We prove $i.$ explicitly. The proof for $ii.$ follows the same structure but
  applies regularity of $\d[f]$ in the opposite direction.
  \begin{salign*}
    \d[\A{ev} &\circ \pair{\Lambda(f)}{e}]\\
    &= \d[ f \circ \pair{\Id{}}{e}]
    \PROOFSTEP
    &= \d[f] \circ \pair{\pair{\Id{}}{e} \circ \pi_1}{\pair{\pi_2}{\d[e]}}
    \PROOFSTEP
    &= \d[f] \circ
      \pair{\pair{\pi_1}{e \circ \pi_1}}{\pair{\pi_2}{0} + \pair{0}{\d[e]}}
    \PROOFSTEP
    &= 
    \d[f] \circ \pair{\pair{\pi_1}{e \circ \pi_1}}{\pair{\pi_2}{0}} 
    +
    \d[f] \circ \pair{\pair{\pi_1 + \epsilon(\pi_2)}{e \circ \pi_1}}{\pair{0}{\d[e]}}
    \PROOFSTEP
    &= 
    \d[f] \circ \pair{\pair{\pi_1}{e \circ \pi_1}}{\pair{\pi_2}{0}} 
    +
    \d[f] \circ \pair{\pair{\pi_1 + \epsilon(\pi_2)}{e \circ \pi_1}}{\pair{0}{\d[e]}}
    \PROOFSTEP
    &=
    \d[f] \circ \pair{(\pi_1 \times \Id{})}{(\pi_2 \times 0)} 
    \circ \pair{\Id{}}{e \circ \pi_1}
    +
    \d[f] \circ \pair{\pair{\pi_1 + \epsilon(\pi_2)}{e \circ \pi_1}}{\pair{0}{\d[e]}}
    \PROOFSTEP
    &= 
    \A{ev} 
    \circ \pair{
      \Lambda(\d[f] \circ \pair{(\pi_1 \times \Id{})}{(\pi_2 \times 0)})
    }{e \circ \pi_1}\\
    \continued +
    \d[f] \circ \pair{\pair{\pi_1 + \epsilon(\pi_2)}{e \circ \pi_1}}{\pair{0}{\d[e]}}
    \PROOFSTEP
    &= 
    \A{ev} \circ \pair{\d[\Lambda(f)]}{e \circ \pi_1} +
    \d[f] \circ \pair{\pair{\pi_1 + \epsilon(\pi_2)}{e \circ \pi_1}}{\pair{0}{\d[e]}}
    \qedhere
  \end{salign*}
\end{proof}

The results below also have rough analogues in the theory of Cartesian differential
categories, but the correspondence starts growing a bit more distant as any result
that hinges on derivatives being additive will, in general, only hold up to some
second-order term in the theory of difference categories.

\begin{lemma}
  \label{lem:lambda-star-ev}
  Let $f : A \times B \times C \to D, g : A \to B, g' : A \times B \to B, e : A
  \times B \to C$ be arbitrary $\cat{C}$-morphisms. Then the following
  identities hold:
  \begin{enumerate}[\roman{enumi}.,ref={\thelemma.\roman{enumi}}]
    \item {\small$\pa{\A{ev} \circ \pair{\Lambda(f)}{e}} \star g
      = \A{ev} \circ \pair{
        \Lambda(f \star (e \star g))
      }{e} 
      + \A{ev} 
      \circ \pair{
        \Lambda(f) \star g
      }{
        e \circ \pair{\pi_1}{\pi_2 + \epsilon(g) \circ \pi_1}
      }
    $}
    \item {\small $
      \Lambda (f \star e) \star g =
      \Lambda \Big[
        \Lambda^-(\Lambda(f) \star g) \star (e \circ (\Id{} + \pair{0}{\varepsilon(g)})))
      + \varepsilon (f \star e) \star (e \star g)
      + (f \star (e \star g))
      \Big]
      $}
    \item 
    {\small $\Lambda(f \star e) \circ \pair{\pi_1}{g'} 
      = \Lambda(\Lambda^-(\Lambda(f) \circ \pair{\pi_1}{g'})
      \star (e \circ \pair{\pi_1}{g'}))$}
  \end{enumerate}
\end{lemma}
\begin{proof}
  The proofs follow the same structure as those in
  \cite[Lemma~4.9]{bucciarelli2010categorical}, with the added caveat that
  derivatives in difference categories are not additive.
  \hspace{0pt}
  \begin{enumerate}[i.]
    \item 
    For readability we will abbreviate $\pair{\Id{}}{\pair{0}{g \circ \pi_1}}$ as
    $\psi$, and remark that, for any $s$, $\d[s] \circ \psi = f \star g$. The
    proof then proceeds by straightforward calculation, applying
    Lemma~\ref{lem:lambda-d-ev}[ii.]:
    \begin{salign*}
      \lefteqn{\pa{\A{ev} \circ \pair{\Lambda(f)}{e}} \star g}
      \\
      \eq
      \d[\A{ev} \circ \pair{\Lambda(f)}{e}] \circ \psi
      \PROOFSTEP
      \eq 
      \Big(
        \A{ev} 
        \circ \pair{\d[\Lambda(f)]}{e \circ \pi_1 + \epsilon(\d[e])}
        + 
        \d[f]
        \circ \pair{\pair{\pi_1}{e \circ \pi_1}}{\pair{0}{\d[e]}}
      \Big)
      \circ \psi
      \\
      \eq 
        \Big(
          \A{ev} 
          \circ \pair{
            \d[\Lambda(f)] \circ \psi
          }{
            (e \circ \pi_1 + \epsilon(\d[e]))
            \circ \psi
          }
        \Big)\\
        \continued +
        \Big(
          \d[f]
          \circ \pair{\pair{\Id{}}{e}}{\pair{0}{e \star g}}
        \Big)
      \PROOFSTEP
      \eq 
        \Big(
          \A{ev} 
          \circ \pair{
            \Lambda(f) \star g
          }{
            e \circ (\Id{} + \pair{0}{\epsilon(g)\circ\pi_1})
          }
        \Big)\\
        \continued +
        \Big(
          \d[f]
          \circ \pair{
            \Id{}
          }{
            \pair{0}{(e \star g) \circ \pi_1}
          }
          \circ \pair{\Id{}}{e}
        \Big)
      \PROOFSTEP
      \eq 
        \Big(
          \A{ev} 
          \circ \pair{
            \Lambda(f) \star g
          }{
            e \circ \pair{\pi_1}{\pi_2 + \epsilon(g) \circ \pi_1}
          }
        \Big)\\
        \continued +
        \Big(
          \A{ev}\circ \pair{
            \Lambda(f \star (e \star g))
          }{e}
        \Big)
    \end{salign*}

    \item 
    Before proceeding with the proof we will first eliminate the abstractions on both 
    sides of the goal. To this end, we note we can rewrite the $\LHS$ (the only
    term where the abstraction is not the outermost construct) by applying
    Lemma~\ref{lem:lambda-star-1}:
    \begin{salign*}
      \Lambda (f \star e) \star g 
      &= \Lambda \Big((((f \star e) \circ \A{sw}) \star (g \circ \pi_1)) \circ \A{sw}
      \Big)
      \PROOFSTEP
      \eq \Lambda \Big(
        \d[(f \star e) \circ \A{sw}] \circ \pair{\A{sw}}{\pair{0}{g \circ \pi_{11}}}
      \Big)
    \end{salign*}

    We abbreviate $\pair{\A{sw}}{\pair{0}{g \circ \pi_{11}}}$ as $\varphi$. Then
    we seek to expand the term $\d[(f \star e) \circ \A{sw}] \circ \varphi$ and
    show how it can be decomposed into the following three summands:
    \begin{gather*}
      {\color{green}
        \Lambda^-(\Lambda(f) \star g) \star (e \circ (\Id{} + \pair{0}{\varepsilon(g)}))
      }\\
      {\color{rose}
        \varepsilon (f \star e) \star (e \star g)
      }\\
      {\color{indigo}
        f \star (e \star g)
      }
    \end{gather*}
    Throughout the proof we will colour some terms to indicate into which of the
    three terms above they are intended to expand.
    \begin{salign*}
      \d[(f& \star e) \circ \A{sw}] 
      \circ \varphi
      \PROOFSTEP
      &= \d[ 
        \d[f] \circ \pair{\Id{}}{\pair{0}{e \circ \pi_1}} \A{sw}
      ]
      \circ \varphi
      \PROOFSTEP
      &= \d[ 
        \d[f] \circ \pair{\A{sw}}{\pair{0}{e \circ \pair{\pi_{11}}{\pi_2}}}
      ]
      \circ \varphi
      \PROOFSTEP
      &= \d[ \d [f]]
      \circ \pair{
        \pair{\A{sw}}{\pair{0}{e \circ \pair{\pi_{11}}{\pi_2}}} \circ \pi_1
      }{
        \pair{
          \d[\A{sw}]
        }{
          \pair{0}{\d[e \circ \pair{\pi_{11}}{\pi_2}]}
        }
      }
      \circ \varphi
      \PROOFSTEP
      &= \d[ \d [f]]
      \circ \pair{
        \pair{\A{sw}}{\pair{0}{e \circ \pair{\pi_{11}}{\pi_2}}}
        \circ \A{sw}
      }{
        \pair{
          \d[\A{sw}]
        }{
          \pair{0}{\d[e \circ \pair{\pi_{11}}{\pi_2}]}
        }
        \circ \varphi
      }
      \PROOFSTEP
      &= \d[ \d [f]]
      \circ \pair{
        \pair{\A{sw}}{\pair{0}{e \circ \pair{\pi_{11}}{\pi_2}}}
        \circ \A{sw}
      }{
        \pair{
          \d[\A{sw}] \circ \varphi
        }{
          \pair{0}{
            \d[e \circ \pair{\pi_{11}}{\pi_2}] \circ \varphi
          }
        }
      }
      \PROOFSTEP
      &= \d[ \d [f]]
      \circ \pair{
        \pair{\Id{}}{\pair{0}{e \circ \pair{\pi_{11}}{\pi_{21}}}}
      }{
        \pair{
          \d[\A{sw}] \circ \varphi
        }{
          \pair{0}{
            \d[e] \circ \A{T}(\pair{\pi_{11}}{\pi_2}) \circ \varphi
          }
        }
      }
      \PROOFSTEP
      &= {\color{green}\d[ \d [f]]
      \circ \pair{
        \pair{\Id{}}{\pair{0}{
          e \circ \pi_1
          + \epsilon (\d[e]) \circ \A{T}(\pair{\pi_{11}}{\pi_2}) \circ \varphi
          }
        }
      }{
        \pair{
          \d[\A{sw}] \circ \varphi
        }{
          0
        }
      }}\\
      \continued + {\color{sand}\d[\d[f]]
      \circ \pair{
        \pair{\Id{}}{\pair{0}{e \circ \pi_1}}
      }{
        \pair{
          0
        }{
          \pair{0}{
            \d[e] \circ \pair{\pi_{11}}{\pi_2} \circ \varphi
          }
        }
      }}
    \end{salign*}
    First we simplify the second term (in {\color{sand}sand}), and will
    return to the first term (in {\color{green}green}) later.
    \begin{salign*}
      {\color{sand}\d[\d[f]]}&
      {\color{sand}
      \circ \pair{
        \pair{\Id{}}{\pair{0}{e \circ \pi_1}}
      }{
        \pair{
          0
        }{
          \pair{0}{
            \d[e] \circ \A{T}(\pair{\pi_{11}}{\pi_2}) \circ \varphi
          }
        }
    }}
      \PROOFSTEP
      &=
      \d[f]
      \circ \pair{
        \Id{} + \pair{0}{\epsilon(e) \circ \pi_1}
      }{
        \pair{0}{
          \d[e] \circ \A{T}(\pair{\pi_{11}}{\pi_2}) \circ \varphi
        }
      }
      \PROOFSTEP
      &=
      {\color{indigo}\d[f]
      \circ \pair{
        \Id{} 
      }{
        \pair{0}{
          \d[e] \circ \A{T}(\pair{\pi_{11}}{\pi_2}) \circ \varphi
        }
      }}\\
      \continued + {\color{rose}\epsilon\d[\d[f]] \circ \pair{
        \pair{\Id{}}{
          \pair{0}{
            \d[e] \circ \A{T}(\pair{\pi_{11}}{\pi_2}) \circ \varphi
          }
        }
      }{
        \pair{\pair{0}{e \circ \pi_1}}{0}
      }}
    \end{salign*}
    We now focus on the first summand, in {\color{indigo}indigo} ink.
    \begin{salign*}
      \color{indigo}{\d[f]}&{\color{indigo}
      {}\circ \pair{
        \Id{} 
      }{
        \pair{0}{
          \d[e] \circ \A{T}(\pair{\pi_{11}}{\pi_2}) \circ \varphi
        }
      }}
      \PROOFSTEP
      &=
      \d[f]
      \circ \pair{
        \Id{} 
      }{
        \pair{0}{
          \d[e] 
          \circ \pair{\pair{\pi_{11}}{\pi_2} \circ \pi_1}{\pair{\pi_{11}{\pi_2}} \circ \pi_2}
          \circ \varphi
        }
      }
      \PROOFSTEP
      &=
      \d[f]
      \circ \pair{
        \Id{} 
      }{
        \pair{0}{
          \d[e] 
          \circ \pair{\pair{\pi_{111}}{\pi_{21}}}{\pair{\pi_{112}}{\pi_{22}}}
          \circ \varphi
        }
      }
      \PROOFSTEP
      &=
      \d[f]
      \circ \pair{
        \Id{} 
      }{
        \pair{0}{
          \d[e] 
          \circ \pair{
              \pair{\pi_{11}}{\pi_2} \circ \A{sw}
            }{
              \pair{\pi_{11}}{\pi_2} \circ \pair{0}{g \circ \pi_{11}}
            }
        }
      }
      \PROOFSTEP
      &=
      \d[f]
      \circ \pair{
        \Id{} 
      }{
        \pair{0}{
          \d[e] 
          \circ \pair{
              \pair{\pi_{11}}{\pi_{21}}
            }{
              \pair{0}{g \circ \pi_{11}}
            }
        }
      }
      \PROOFSTEP
      &=
      \d[f]
      \circ \pair{
        \Id{} 
      }{
        \pair{0}{
          \d[e] 
          \circ \pair{
              \pi_1
            }{
              \pair{0}{g \circ \pi_{11}}
            }
        }
      }
      \PROOFSTEP
      &=
      \d[f]
      \circ \pair{
        \Id{} 
      }{
        \pair{0}{
          \d[e] 
          \circ \pair{
              \Id{}
            }{
              \pair{0}{g \circ \pi_{1}}
            }
        }
        \circ \pi_1
      }
      \PROOFSTEP
      &=
      \d[f]
      \circ \pair{
        \Id{} 
      }{
        \pair{0}{
          (e \star g) \circ \pi_1
        }
      }
      \PROOFSTEP
      &=
      f \star (e \star g)
    \end{salign*}
    To simplify the {\color{rose}rose} term, we proceed as follows:
    \renewcommand{\eq}[0]{&\hspace{-20pt}=}
    \begin{salign*}
      \lefteqn{\hspace{-30pt}{\color{rose}\epsilon}{\color{rose}\d[\d[f]] \circ \pair{
        \pair{\Id{}}{
          \pair{0}{
            \d[e] \circ \A{T}(\pair{\pi_{11}}{\pi_2}) \circ \varphi
          }
        }
      }{
        \pair{\pair{0}{e \circ \pi_1}}{0}
      }}}\\
      \eq \epsilon\d[\d[f]] \circ \pair{
        \pair{\Id{}}{
          \pair{0}{(e \star g) \circ \pi_1}
        }
      }{
        \pair{\pair{0}{e \circ \pi_1}}{0}
      }
      \PROOFSTEP
      \eq \epsilon 
      \d[\d[f]]
      \circ
      \pair{
        \pair{\Id{}}{\pair{0}{e \circ \pi_1}}
      }{
        \pair{
          \pair{0}{(e \star g) \circ \pi_1}
        }{
          0
        }
      }
      \PROOFSTEP
      \eq \epsilon 
      \d[\d[f]]
      \circ
      \pair{
        \pair{\Id{}}{\pair{0}{e \circ \pi_1}}
      }{
        \pair{
          \pair{0}{(e \star g) \circ \pi_1}
        }{
          \pair{0}{\d[e] \circ \pair{\pi_{11}}{0}}
        }
      }
      \PROOFSTEP
      \eq \epsilon 
      \d[\d[f]]
      \circ
      \pair{
        \pair{\Id{}}{\pair{0}{e \circ \pi_1}} \circ \pi_1
      }{
        \pair{\pi_2}{\pair{0}{\d[e] \circ (\pi_1 \times \pi_1)}}
      }
      \circ
      \pair{\Id{}}{
        \pair{0}{
          (e \star g) \circ \pi_1
        }
      }
      \PROOFSTEP
      \eq \epsilon 
      \d[
        \d[f] \circ \pair{\Id{}}{\pair{0}{e \circ \pi_1}}
      ]
      \circ
      \pair{\Id{}}{
        \pair{0}{
          (e \star g) \circ \pi_1
        }
      }
      \PROOFSTEP
      \eq \epsilon (f \star e) \star (e \star g)
    \end{salign*}
    It only remains to simplify the {\color{green}green} term. This follows
    mostly the same process as the {\color{rose}rose} term, applying
    axiom~\CdCAx{0} to fold the derivative of $e$. We will write $e'$ for 
    the resulting term $e \circ (\A{id} + \pair{0}{\epsilon(g) \circ \pi_1})$.
    \begin{salign*}
      \lefteqn{\hspace{-40pt}{\color{green}\d[ \d [f]]
      \circ \pair{
        \pair{\Id{}}{\pair{0}{
          e \circ \pi_1
          + \epsilon (\d[e]) \circ \A{T}(\pair{\pi_{11}}{\pi_2}) \circ \varphi
          }
        }
      }{
        \pair{
          \d[\A{sw}] \circ \varphi
        }{
          0
        }
      }}}
      \PROOFSTEP
      \eq
      \d[ \d [f]]
      \circ \pair{
        \pair{\Id{}}{\pair{0}{
          e \circ \pi_1
          + \epsilon (\d[e]) \circ \A{T}(\pair{\pi_{11}}{\pi_2}) \circ \varphi
          }
        }
      }{
        \pair{
          \d[\A{sw}] \circ \varphi
        }{
          0
        }
      }
      \PROOFSTEP
      \eq
      \d[ \d [f]]
      \circ \pair{
        \pair{\Id{}}{\pair{0}{
          e \circ \pi_1
          + \epsilon (\d[e]) 
          \circ \pair{\pi_1}{\pair{0}{g \circ \pi_{11}}}
          }
        }
      }{
        \pair{
          \d[\A{sw}] \circ \varphi
        }{
          0
        }
      }
      \PROOFSTEP
      \eq
      \d[ \d [f]]
      \circ \pair{
        \pair{\Id{}}{\pair{0}{
          (e + \epsilon(\d[e]) \circ \pair{\Id{}}{\pair{0}{g \circ \pi_1}})
          \circ \pi_1
          }
        }
      }{
        \pair{
          \A{sw} \circ \pi_2 \circ \varphi
        }{
          0
        }
      }
      \PROOFSTEP
      \eq
      \d[ \d [f]]
      \circ \pair{
        \pair{\Id{}}{\pair{0}{
          e'
          \circ \pi_1
          }
        }
      }{
        \pair{
          \pair{\pair{0}{g \circ \pi_{11}}}{0}
        }{
          0
        }
      }
      \PROOFSTEP
      \eq
      \d[\d[f]]
      \circ \pair{
        \pair{\Id{}}{\pair{\pair{0}{g \circ \pi_{11}}}{0}}
      }{
        \pair{
          {\pair{0}{e' \circ \pi_1}}
        }{
          0
        }
      }
      \PROOFSTEP
      \eq
      \d[\d[f]]
      \circ \pair{
        \pair{\Id{}}{\pair{\pair{0}{g \circ \pi_{11}}}{0}} \circ \pi_1
      }{
        \pair{\pi_2}{\pair{\pair{0}{\d[g] \circ (\pi_{11} \times \pi_{11})}}{0}}
      }
      \circ \pair{\Id{}}{\pair{0}{e' \circ \pi_1}}
      \PROOFSTEP
      \eq
      \d[
        \d[f] \circ \pair{\Id{}}{\pair{\pair{0}{g \circ \pi_{11}}}{0}}
      ]
      \circ \pair{\Id{}}{\pair{0}{e' \circ \pi_1}}
      \PROOFSTEP
      \eq
      \d[
        \d[f] \circ \pair{\A{sw}}{\pair{\pair{0}{g \circ \pi_{11}}}{0}}
        \circ \A{sw}
      ]
      \circ \pair{\Id{}}{\pair{0}{e' \circ \pi_1}}
      \PROOFSTEP
      \eq
      \d[
        \d[f \circ \A{sw}] \circ \pair{\Id{}}{\pair{0}{g \circ \pi_{11}}}
        \circ \A{sw}
      ]
      \circ \pair{\Id{}}{\pair{0}{e' \circ \pi_1}}
      \PROOFSTEP
      \eq
      (((f \circ \A{sw}) \star (g \circ \pi_1)) \circ \A{sw})
      \star e'
      \PROOFSTEP
      \eq (\Lambda^-(\Lambda(f) \star g)) \star e'
    \end{salign*}
    \item The proof for $iii.$ is simpler and in fact identical to the one in
    \cite{bucciarelli2010categorical}, as additivity of the derivative is never
    assumed. We reproduce it here for completeness and to account for the
    differences in notation.

    First, since $\Lambda(f \star e) \circ \pair{\pi_1}{g'}$ is precisely
    $\Lambda((f \star e) \circ (\pair{\pi_1}{g'} \times \Id{}))$ it suffices to
    show that $(f \star e) \circ (\pair{\pi_1}{g'} \times \Id{}) =
    \Lambda^-(\Lambda(f) \circ \pair{\pi_1}{g'}) \star (e \circ
    \pair{\pi_1}{g'})$, which we do by straightforward calculation.
    \begin{salign*}
      \lefteqn{\hspace{-30pt}(f \star e) \circ (\pair{\pi_1}{g'} \times \Id{})}
      \PROOFSTEP
      \eq \d[f] \circ \pair{\Id{}}{\pair{0}{e \circ \pi_1}} 
      \circ (\pair{\pi_1}{g'} \times \Id{})
      \PROOFSTEP
      \eq
      \d[f]
      \circ \pair{
        (\pair{\pi_1}{g'}\times \Id{})
      }{
        \pair{0}{e \circ \pair{\pi_1}{g'} \circ \pi_1}
      }
      \PROOFSTEP
      \eq
      \d[f]
      \circ \pair{
        \pair{\pair{\pi_{11}}{g' \circ \pi_{1}}}{\pi_{2}}
      }{
        \pair{\pair{0}{\d[g'] \circ (\pi_1 \times 0)}}{e \circ \pair{\pi_1}{g'} \circ \pi_1}
      }
      \PROOFSTEP
      \eq
      \d[f]
      \circ \pair{
        \pair{\pair{\pi_{111}}{g \circ \pi_{11}}}{\pi_{21}}
      }{
        \pair{\pair{\pi_{112}}{\d[g] \circ (\pi_1 \times \pi_1)}}{\pi_{22}}
      }
      \circ \pair{\Id{}}{\pair{0}{e \circ \pair{\pi_1}{g'} \circ \pi_1}}
      \PROOFSTEP
      \eq
      (f \circ \pair{\pair{\pi_{11}}{g \circ \pi_1}}{\pi_2}
      )
      \star (e \circ \pair{\pi_1}{g'})
      \PROOFSTEP
      \eq \Lambda^-(
        \Lambda(f \circ (\pair{\pi_1}{g'} \times \Id{}))
      ) 
      \star (e \circ \pair{\pi_1}{g'})
      \PROOFSTEP
      \eq \Lambda^-(\Lambda(f) \circ \pair{\pi_1}{g'}) 
      \star (e \circ \pair{\pi_1}{g'})
    \end{salign*}
  \end{enumerate}
\end{proof}

%% file: simply-typed-calculus.tex

So far our treatment of change actions and difference categories has been mostly
semantic. Change actions may arise and be applied in the setting of incremental
computation, but they exist merely as a theoretical model for reasoning about it
and deriving algorithms, rather than an executable entity. This is in contrast
with Cartesian differential categories and, in particular, differential
$\lambda$-categories, for which the differential $\lambda$-calculus is known to
be an internal language.

It is reasonable to ask, then, whether there is a language of change actions and
differential maps -- especially since, as we have shown, difference categories
subsume differential categories. The issue is far from trivial, as many of
the properties of the differential $\lambda$-calculus crucially hinge on
derivatives being linear.

Through this chapter we will provide an affirmative answer to this question: in
Section \ref{sec:untyped} we develop $\lambda_\epsilon$, an untyped calculus in
the spirit of the differential $\lambda$-calculus which adds infinitesimal
extensions and relaxes the linearity requirement. In Section
\ref{sec:simple-types} we then introduce a type system for this calculus and
prove a strong normalisation theorem. Finally, in Section \ref{sec:semantics},
we show how $\lambda_\epsilon$-terms can be soundly interpreted in any
difference $\lambda$-category.

\newcommand{\wfParTo}[0]{\mathrel{\ul\parTo}}
\label{chp:simply-typed-calculus}

\newcommand{\permEq}[0]{\sim_+}
\newcommand{\diffEq}[0]{\sim_\epsilon}
\newcommand{\diffEta}[0]{\sim_{\epsilon\eta}}
\newcommand{\canEq}[0]{\sim_\textbf{can}}
\newcommand{\wfReducesTo}[0]{\mathrel{\ul{\reducesTo}}}
\newcommand{\wfTs}[0]{\mathrel{\ul{\ts}}}
\newcommand{\ap}[2]{\textbf{ap}(#1, #2)}
\newcommand{\pri}[1]{\textbf{pri}(#1)}
\renewcommand{\tan}[1]{\textbf{tan}(#1)}
\newcommand{\can}[1]{\textbf{can}\left(#1\right)}
\newcommand{\reg}[1]{\textbf{reg}\left(#1\right)}

\section{An Untyped Calculus of Differences}
\label{sec:untyped}

We proceed in a manner similar to Vaux \cite{vaux2009algebraic} in his treatment
of the algebraic $\lambda$-calculus; that is, we will first define a set of
``unrestricted'' terms $\Lambda_\epsilon$ which we will later consider up to an
equivalence relation arising from the theory of difference categories.

\begin{definition} The set $\Lambda_\epsilon$ of \textbf{unrestricted terms} of
  the $\lambda_\epsilon$-calculus is given by the following inductive
  definition:
  \[
    \begin{array}{rccl}
      \text{Terms: } &s, t, e& \defeq & x
                                        \alter \lambda x . t
                                        \alter (s\ t)
                                        \alter \D(s) \cdot t
                                        \alter \epsilon t
                                        \alter s + t 
                                        \alter 0
    \end{array}
  \]
  assuming a countably infinite set of variables $x, y, z \ldots$ is given.
\end{definition}

Henceforth we will only consider the above terms up to
$\alpha$-equivalence. Since the only binder in the $\lambda_\epsilon$-calculus
is the usual $\lambda$-abstraction, the definition of free and bound variables
is straightforward.

\begin{definition}
  The set of \textbf{free variables} $\text{FV}(t)$ of a term $t \in
  \Lambda_\epsilon$ is defined by induction on the structure of $t$ as follows:
  \[
    \begin{array}{rcl}
      \text{FV}(x) &\defeq& \lbrace x \rbrace\\
      \text{FV}(\lambda x . t) &\defeq& \text{FV}(t) \setminus \lbrace x \rbrace\\
      \text{FV}(s\ t) &\defeq& \text{FV}(s) \cup \text{FV}(t)\\
      \text{FV}(\D(s)\cdot t) &\defeq& \text{FV}(s) \cup \text{FV}(t)\\
      \text{FV}(\epsilon t) &\defeq& \text{FV}(t)\\
      \text{FV}(s + t) &\defeq& \text{FV}(s) \cup \text{FV}(t)\\
      \text{FV}(0) &\defeq& \emptyset\\
    \end{array}
  \]

  As usual, a variable $x$ is \textbf{free} in $t$ whenever $x \in \text{FV}(t)$.
  An occurrence of a variable $x$ in some term $t$ is said to be \textbf{bound}
  whenever it appears in some subterm $t'$ of $t$ with $x \not\in \text{FV}(t)$.

  Two terms are said to be $\alpha$-equivalent if they are identical up to a
  renaming of all their bound variables.
\end{definition}

In what follows, we will speak of terms only up to $\alpha$-equivalence. That
is, we consider the terms $\lambda x . x$ and $\lambda y . y$ to be identical
for all intents and purposes. Since this means we can rename bound variables
freely, we will assume by convention that all bound variables appearing in any
term $t \in \Lambda_\epsilon$ are different from its free variables.

\subsection{Differential Equivalence}

Further to $\alpha$-equivalence, we introduce here the notion of Euclidean
equivalence of terms. The role of this relation is, as in
\cite{vaux2006lambda}, to enforce that the elementary algebraic properties of
sums and actions are preserved. For example, we wish to treat the terms $\lambda
x . (0 + \epsilon (s + t))$ and $(\lambda x . \epsilon t) + (\lambda x .
\epsilon s)$ as if they were equivalent (as it will be the case in the models).
This equivalence relation also has the role of ensuring that the axioms of a
Cartesian difference category are verified, especially regularity of
derivatives.

\begin{definition} A binary relation $\sim{} \subseteq \Lambda_\epsilon \times
  \Lambda_\epsilon$ is \textbf{contextual} whenever it satisfies the conditions
  in Figure~\ref{fig:contextual-relation} below.
  \begin{figure}[H]
    \[
      \begin{array}{ccc}
        t \sim t' & \Rightarrow & \lambda x . t \sim \lambda x . t'\\
        t \sim t' & \Rightarrow & \epsilon t \sim \epsilon t'\\
        s \sim s' \wedge t \sim t' & \Rightarrow & s\ t \sim s'\ t'\\
        s \sim s' \wedge t \sim t' & \Rightarrow & \D(s)\cdot t \sim \D(s')\cdot t'\\
        s \sim s' \wedge t \sim t' & \Rightarrow & s + t \sim s' + t'\\
      \end{array}
    \]
    \caption{Contextuality on unrestricted $\Lambda_\epsilon$-terms}
    \label{fig:contextual-relation}
  \end{figure}
\end{definition}

\begin{lemma}
  \label{lem:contextual-subterm}
  Whenever $\sim$ is contextual, if $t$ is a subterm of $s$ and $t \sim t'$ then
  $s \sim s'$, where $s'$ is the term resulting from substituting the occurrence
  of $t$ in $s$ for $t'$.
\end{lemma}

\begin{definition} 
  \label{defn:differential-equivalence}
  \textbf{Differential equivalence} $\diffEq \subseteq \Lambda_\epsilon \times
  \Lambda_\epsilon$ is the least equivalence
  relation which is contextual and contains the relation $\diffEq^1$ below
  Figure~\ref{fig:differential-equivalence} below.
  \begin{figure}[H]
  \[
    \begin{array}{rcl}
      (s + t) + e           &\diffEq^1& s + (t + e)\\
      s + 0                 &\diffEq^1& s\\
      s + t                 &\diffEq^1& t + s\\
      \epsilon 0            &\diffEq^1& 0\\
      \epsilon (s + t)      &\diffEq^1& \epsilon s + \epsilon t
    \end{array}
    \begin{array}{rcl}
      \lambda x . 0          &\diffEq^1& 0\\
      \lambda x . (s + t)    &\diffEq^1& (\lambda x . s) + (\lambda x . t)\\
      \lambda x . \epsilon t &\diffEq^1& \epsilon (\lambda x . t)\\
      0\ s                   &\diffEq^1& 0\\
      (s + t)\ e             &\diffEq^1& (s\ e) + (t\ e)\\
      (\epsilon s)\ t        &\diffEq^1& \epsilon (s\ t)
    \end{array}
  \]
  \[
    \begin{array}{rcl}
      \D(0) \cdot e          &\diffEq^1& 0\\
      \D(s + t) \cdot e      &\diffEq^1& (\D(s) \cdot e) + (\D(t) \cdot e)\\
      \D(\epsilon t) \cdot e &\diffEq^1& \epsilon (\D(t) \cdot e)
    \end{array}\]\[
    \begin{array}{rcl}
      \D(s) \cdot 0          &\diffEq^1& 0\\
      \D(s) \cdot (t + e)    &\diffEq^1& 
      \D(s) \cdot t
      + \D(s) \cdot e
      + \epsilon (\D(\D(s) \cdot t) \cdot e)\\
      \D(s) \cdot (\epsilon t) &\diffEq^1& \epsilon(\D(s) \cdot t)\\
      \D(\D(s) \cdot t) \cdot e &\diffEq^1& \D(\D(s) \cdot e) \cdot t\\
      \epsilon^2 \D(\D(s) \cdot t) \cdot e &\diffEq^1& \epsilon \D(\D(s) \cdot t) \cdot e\\
      s\ (t + \epsilon e) &\diffEq^1& (s\ t) + \epsilon((\D(s) \cdot e)\ t)
    \end{array}
  \]
  \caption{Differential equivalence on unrestricted $\Lambda_\epsilon$-terms}
  \label{fig:differential-equivalence}
  \end{figure}
\end{definition}

The above conditions can be separated in a number of conceptually distinct
groups corresponding to their purpose. These are as follows:
\begin{itemize}
  \item The first block of equations simply states that $+, 0$ define a
    commutative monoid and that $\epsilon$ defines a monoid homomorphism.
  \item The second block of equations amounts to stating that the monoid and
    infinitesimal extension structure on functions is pointwise.
  \item The third block of equations implies (and is equivalent to) stating that
    addition and infinitesimal extension are ``linear'', in the sense that they
    are equal to their own derivatives (that is, $\d\left[+\right] = + \circ \pi_2$ and
    $\d \left[\epsilon\right] = \epsilon$).
  \item The fourth block of equations states structural properties of the
    derivative, such as the derivative conditions and the commutativity of second
    derivatives. Similar equations are also present in the differential
    $\lambda$-calculus, where they merely state that the derivative is additive
    (as opposed to regular).
\end{itemize}

Most of these equations correspond directly to properties of Cartesian
difference categories, with the only exception being the requirement that $\D(s)
\cdot (\epsilon t) \diffEq \epsilon (\D(s) \cdot t)$ and the ``duplication
of infinitesimals'' in $\epsilon \D (\D(s) \cdot t) \cdot e = \epsilon^2
\D(\D(s) \cdot t) \cdot e$. To understand the logic of the first equation, consider
that the term $\D(\lambda x . s) \cdot \epsilon t$ should roughly correspond to
$\d \sem{s}(x, \epsilon \sem{t})$ which expands to:
\begin{salign*}
  \d \sem{s}(x, \epsilon \sem{t}) &= \d \sem{s}(x, 0 + \epsilon \sem{t})\\
  &= \d \sem{s}(x, 0) + \epsilon \d^2 \sem{s}(x, 0, 0, \sem{t})\\
  &= \epsilon \d^2\sem{s}(x, 0, 0, \sem{t})
\end{salign*}
In the particular setting of Cartesian difference categories, the last term
boils down to $\epsilon \d \sem{s}(x, \sem{t})$ by axiom~\ref{CdC6}, hence the
equality $\D(s) \cdot \epsilon t \diffEq \epsilon (\D(s) \cdot t)$ is justified,
allowing infinitesimal extensions to be ``pulled out'' of differential
application. The duplication of infinitesimals is simply a syntactic version of
Lemma~\ref{lem:d-epsilon-iii}.

\begin{definition}
  The set $\lambda_\epsilon$ of \textbf{well-formed terms}, or simply
  \textbf{terms}, of the $\lambda_\epsilon$-calculus is defined as the quotient
  set $\lambda_\epsilon \defeq \Lambda_\epsilon/\diffEq$. Whenever $t$ is an
  unrestricted term, we write $\ul{t}$ to refer to the well-formed term
  represented by $t$, that is to say, the $\diffEq$-equivalence class of $t$. 
\end{definition}

The notion of differential equivalence allows us to ensure that our calculus
reflects the laws of the underlying models, but has the unintended
consequence that our $\lambda_\epsilon$-terms are equivalence classes, rather
than purely syntactic objects. We will proceed by defining a notion of canonical
form of a term and a canonicalization algorithm which explicitly constructs
the canonical form of any given term, thus proving that $\diffEq$ is decidable.

\newcommand{\canonical}[0]{\text{C}_\epsilon}
\newcommand{\base}[0]{\text{B}_\epsilon}
\newcommand{\additive}[0]{\base^*}
\newcommand{\positive}[0]{\base^+}
\newcommand{\posCanonical}[0]{\canonical^+}
\renewcommand{\b}[0]{\textbf{b}}

\begin{definition}
  We define the sets $\base \subset \positive \subset \additive \subset
  \posCanonical \subset \canonical (\subset \Lambda_\epsilon)$ of \textbf{basic},
  \textbf{positive}, \textbf{additive}, \textbf{positive canonical} and
  \textbf{canonical} terms according to the following grammar:
\[\def\arraystretch{1.3}
  \begin{array}{rccl}
    \text{Basic terms: } &s^\b, t^\b, e^\b \in \base& \defeq & x
                                                  \alter \lambda x . t^\b
                                                  \alter (s^\b\ t^*)
                                                  \alter \D(s^\b) \cdot t^\b\\
    \text{Positive terms: } &s^+, t^+, e^+ \in \positive& \defeq & s^\b \alter s^\b + (t^+)\\
    \text{Additive terms: } &s^*, t^*, e^* \in \additive& \defeq & 0 \alter s^+\\
    \text{Positive canonical terms: } &S^+, T^+ \in 
      \canonical&\defeq& \epsilon^k s^\b \alter \epsilon^k s^\b + (S^+) \\
    \text{Canonical terms: } &S, T \in \canonical&\defeq& 0 \alter S^+
  \end{array}
\] We will sometimes abuse the notation and write $\ul{t^*}$ or $\ul{ t^\b }$ to
  denote well-formed terms whose canonical form is an additive or basic term
  respectively.
\end{definition}

The above definition is somewhat technical, so a more informal description of
canonical forms should be helpful. We observe that all the rules of differential
equivalence can be oriented from left to right to obtain a rewrite system -
intuitively, this rewrite system operates by pulling all the instances of
addition and infinitesimal extension to the outermost layers of a term. Since
every syntactic construct is additive except for application, basic terms may only
contain additive terms as the arguments to a function application. 

As infinitesimal extensions are themselves additive, we also want to disallow
terms such as $\epsilon (s + t)$, instead factoring out the extension
into $\epsilon s + \epsilon t)$. A general canonical term $T \in \canonical$ then has
the form:
\[
  T = \epsilon^{k_1} t^\b_1 + (\epsilon^{k_2}t_2 + (\ldots + \epsilon^{k_n} t^\b_n) \ldots )
\]
That is to say, a canonical term is similar to a polynomial with coefficients in
the set of basic terms and a variable $\epsilon$ (but note that canonical terms
are always written in their ``fully distributed'' form, that is, we write
$\epsilon s + (\epsilon t + \epsilon^2 e)$ rather than $\epsilon ((s + t) + \epsilon
e)$).

We will freely abuse the notation and write $\sum_{i=1}^n \epsilon^{k_i} t^\b_i$ to
denote a general canonical term, as this form is easier to manipulate in many
cases. In particular, the canonical term $0$ is precisely the sum of zero terms.

\begin{definition}
  Given unrestricted terms $s, t$, we define their \textbf{canonical sum} $s
  \cansum t$ by induction as follows:
  \begin{align*}
    0 \cansum t &\defeq t\\
    s \cansum 0 &\defeq s\\
    (s + s') \cansum t &\defeq s + (s' \cansum t)
  \end{align*}
\end{definition}

\begin{lemma}
  The canonical sum $S \cansum T$ of any two canonical terms is a canonical
  term. Furthermore $\cansum$ is associative and has $0$ as an identity
  element. When $S_i$ are canonical terms, we will write $\boxsum_{i=1}^n S_i$
  for the canonical term $S_1 \cansum (S_2 \cansum \ldots S_n)\ldots)$.
\end{lemma}

\begin{definition} Given an unrestricted $\lambda_\epsilon$-term $t \in
  \Lambda_\epsilon$, we define its \textbf{canonical form} $\can{t}$ by structural
  induction on $t$ as follows:
  \begin{itemize}
    \item $\can{0} \defeq 0$
    \item $\can{x} \defeq x$
    \item $\can{s + t} \defeq \can{s} \cansum \can{t}$
    \item $\can{\epsilon t} \defeq \epsilon^* \can{t}$, where 
      \[
        \epsilon^* T \defeq \begin{cases}
          \sum_{i=1}^n \epsilon^*\epsilon^{k_i}t^\b_i
          &\text{ if $T = \sum_{i=1}^n \epsilon^{k_i}t^\b_i$}\\
          T 
          &\text{ if $T = \epsilon \D(\D(e) \cdot u) \cdot v$}\\
          \epsilon T &\text{ otherwise}
        \end{cases}
      \]
    \item If 
      $\can{t} = \sum_{i=1}^{n}\epsilon^{k_i}t^\b_{i}$ 
      then
      \[ \can{\lambda x.t} \defeq \sum_{i=1}^n \epsilon^{k_i} (\lambda^* x.t^\b_i) \]
    \item If
      $\can{s} = \sum_{i=1}^n\epsilon^{k_i}s^\b_i$
      and
      $\can{t} = T$
      then
      \[
        \can{\D(s) \cdot t} \defeq \boxsum_{i=1}^n ((\epsilon^*)^{k_i} \reg{s^\b_i, T})
      \]
      where the \textbf{regularization} $\reg{s, T}$ is
      defined by structural induction on $T$:
      \begin{align*}
        \reg{s, 0} &\defeq 0\\
        \reg{s, \epsilon^{k} t^{\b} + T'} &\defeq
        \brak{\pa{\epsilon^*}^{k} \D\pa{s}\cdot t^\b}
        \cansum \brak{
          \reg{s, T'}
        }\\
        \continued \cansum \brak{
          \pa{\epsilon^*}^{k+1} \D^*\pa{\reg{s, T'}} \cdot t^\b
        }
      \end{align*}
      and $\D^*$ denotes the extension of $\D$ by additivity in its first
      argument, that is to say:
      \begin{align*}
        \D^*\pa{\sum_{i=1}^n \epsilon^{k_i} s^\b_i} \cdot t^\b
        &\defeq \sum_{i=1}^n \epsilon^{k_i} \pa{s^\b_i\ t^\b}
      \end{align*}
      Observe that, whenever $S$ is canonical and $t^\b$ is basic, the term
      $\D^*(S) \cdot t^\b$ is also canonical. Therefore, by induction, the
      regularization $\reg{s^\b, T}$ is indeed a canonical term term, since
      canonicity is preserved by $\epsilon^*, \cansum$.
    \item
      If $\can{s} = \sum_{i=1}^n \epsilon^{k_i} s^\b_i$ and $\can{t} = T$, then
      \begin{align*}
        \can{s\ t} &\defeq \brak{\sum_{i=1}^n \epsilon^{k_i} \pa{s^\b_i \ \textbf{pri}(T)}}
        \cansum
        \brak{
          \epsilon^* \pa{
            \sum_{i=1}^n \textbf{ap}(\reg{s^\b_i, \textbf{tan}(T)}, \textbf{pri}(T))
          }
        }
      \end{align*}
      where the primal $\textbf{pri}$ and tangent $\textbf{tan}$ components of a
      canonical term $T$ correspond respectively to the basic terms with zero and
      non-zero $\epsilon$ coefficients, and $\textbf{ap}$ is the additive
      extension of application.
      \[\def\arraystretch{1.3}
      \begin{array}{rclrcl}
        \textbf{pri}\pa{0} &\defeq& 0 
        &\textbf{tan}\pa{0} &\defeq& 0\\
        \textbf{pri}\pa{\epsilon^{k+1}t^\b + T'} &\defeq& \textbf{pri}\pa{T'}
                                               &\textbf{tan}\pa{\epsilon^{k+1}t^\b + T'} &\defeq& \epsilon^{k+1} t^\b + \textbf{tan}\pa{T'}\\
        \textbf{pri}\pa{\epsilon^{0}t^\b + T'} &\defeq& t^\b + \textbf{pri}\pa{T'}
        &\textbf{tan}\pa{\epsilon^{0}t^\b + T'} &\defeq& \textbf{tan}\pa{T'}\\
      \end{array}
      \]
      \[\def\arraystretch{1.3}
      \begin{array}{rcl}
        \textbf{ap}\pa{\sum_{i=1}^n \epsilon^{k_i} s^\b_i, t^\b} &\defeq&
        \sum_{i=1}^n \epsilon^{k_i} (s^\b_i\ t^\b)
      \end{array}
      \]
  \end{itemize}
\end{definition}

\begin{example}
  The canonicalization algorithm is mostly straightforward, with only the cases
  for differential and standard application being of any interest. Since this is
  the part of the system where we diverge most from the differential
  $\lambda$-calculus, it is worth examining an example of regularization.
  Consider the following unrestricted term on free variables $u, x, y, z$:
  \[
    t \defeq \D (u) \cdot (x + y + \epsilon z)
  \]
  Applying the algorithm above, we compute its canonical form to be:
  \begin{salign*}
    &\hspace{-10pt}\can{\D(u) \cdot (x + y + \epsilon z)}\\
    &= \reg{u, x + y + \epsilon z}\\
    &= 
      (\D(u) \cdot x) 
      \boxplus 
      \reg{u, y + \epsilon z} 
      \boxplus 
      (\epsilon^* \D^*(\reg{u, y + \epsilon z}) \cdot x )\displaybreak[0]\\
    &= 
      (\D(u) \cdot x) \boxplus
      \Big[ (\D(u) \cdot y) \boxplus \reg{u, \epsilon z} \boxplus \epsilon^* (\D^*(\reg{u, \epsilon z}) \cdot y) \Big]
      \\
      \continued \boxplus 
      (\epsilon^* \D^*(\reg{u, y + \epsilon z}) \cdot x )\displaybreak[0]\\
    &= (\D(u) \cdot x) \boxplus 
      \Big[ (\D(u) \cdot y) \boxplus (\epsilon \D(u) \cdot z) \boxplus \epsilon^*(\D^*
      (\epsilon \D(u) \cdot z) \cdot y))\Big]\\
      \continued
      \boxplus (\epsilon^* \D^*(\reg{u, y + \epsilon z}) \cdot x)\displaybreak[0]\\
    &= (\D(u) \cdot x)) \boxplus \Big[ \D(u) \cdot y + \epsilon (\D(u) \cdot z) + \epsilon(\D
      (\D(u) \cdot z) \cdot y) \Big]\\
      \continued
      \boxplus (\epsilon^* \D^*(\reg{u, y + \epsilon z}) \cdot x)
    \PROOFSTEP
    &= \Big[ \D(u) \cdot x + \D(u) \cdot y + \epsilon \pa{ \D(u) \cdot z } + \epsilon\pa{\D
      (\D(u) \cdot z) \cdot y)} \Big]\\
      \continued \boxplus \pa{ 
        \epsilon \pa{
          \D(\D(u) \cdot y) \cdot x
          + \epsilon \pa{
            \D(\D(u) \cdot z) \cdot x
            + \epsilon \pa{
              \D(\D(\D(u) \cdot z) \cdot y) \cdot x
            }
          }
        }
      }
    \PROOFSTEP
    &= \D(u) \cdot y + \D(u) \cdot x\\
    \continued + \epsilon \Big[
      \D(u) \cdot z + \D( \D(u) \cdot y) \cdot x\\
      \continued[60pt] + \epsilon (
        \D(\D(u) \cdot z) \cdot y + {}
        \D(\D(u) \cdot z) \cdot x +
        \epsilon \left( \D(\D(\D(u) \cdot z) \cdot y ) \cdot x\right))\Big]
  \end{salign*}
  This is precisely the result we would expect from fully unfolding the expression 
  $\d(u, x + y + \epsilon(z))$ in a Cartesian difference category and repeatedly
  applying regularity of the derivative!
\end{example}

\begin{theorem}
  \label{thm:canonical-form-differential}
  Every unrestricted $\lambda_\epsilon$-term is differentially equivalent to its
  canonical form. That is to say, for all $t \in \Lambda_\epsilon$, we have $t
  \diffEq \can{t}$.
\end{theorem}
\begin{proof}
  The proof proceeds by straightforward induction on $t$. We explicitly prove
  the case for the canonicalization of differential and standard application, as
  these are the only nontrivial cases. For this we will make use of the
  following results:

  \begin{lemma}
    \label{lem:utils-differential}
    Given canonical terms $S, T$, we have:
    \begin{align*}
      \epsilon^* S &\diffEq \epsilon S\\
      S \cansum T &\diffEq S + T\\
      \D^* (S) \cdot T &\diffEq \D(S) \cdot T\\
      \textbf{ap}\pa{S, T} &\diffEq S\ T
    \end{align*}
  \end{lemma}
  \begin{proof}
    All four results follow by straightforward structural induction on $S$.
  \end{proof}
  \begin{lemma}
    \label{lem:reg-differential}
    Given a basic term $s^\b$ and a canonical term $T$ the differential
    application $\D(s^\b) \cdot T$ is differentially equivalent to its
    regularized version $\reg{s^\b, T}$.
  \end{lemma}
  \begin{proof}
    The proof follows by induction on the structure of $T$.
    \begin{itemize}
      \item When $T = 0$ we have
        $\reg{s^\b, 0} = 0 \diffEq \D(s^*) \cdot 0$
      \item When $T = \epsilon^k t^\b + T'$ we have:
      \begin{salign*}
        \reg{s^\b, \epsilon^k t^\b + T'} &= \brak{(\epsilon^*)^k \D(s^\b) \cdot t^\b} \cansum \brak{\reg{s^\b, T'}}\\
        \continued \cansum \brak{(\epsilon^*)^{k+1} \D^*(\reg{s^\b, T'}) \cdot t^\b}\\
        &\diffEq \epsilon^k \D(s^\b) \cdot t^\b + \D(s^\b) \cdot T' + \epsilon^{k+1} \D(\D(s^\b) \cdot T') \cdot t^\b\qedhere
      \end{salign*}
    \end{itemize}
    \renewcommand{\qedsymbol}[0]{}
  \end{proof}

  Going back to the proof of Theorem~\ref{thm:canonical-form-differential}, the
  case for differential application is obtained as a straightforward corollary
  of Lemma~\ref{lem:reg-differential}. For conventional application, consider
  terms $s, t$, and note that if $\can{t} = T$ then $t \diffEq \textbf{pri}(T) +
  \epsilon \textbf{tan}(T)$. Then, for any basic term $s^\b$, we obtain:
  \begin{salign*}
    \can{s^\b\ t} &= \pa{s^\b\ \textbf{pri}(T)} 
    \cansum \brak{
      \epsilon^* \textbf{ap}\pa{\reg{s^\b, \textbf{tan}(T)}, \textbf{pri}(T)}
    }\\
    &\diffEq \pa{s^\b\ \textbf{pri}(T)} 
    + \epsilon \brak{(\D(s^\b) \cdot \textbf{tan}(T))\ \textbf{pri}(T)}\\
    &\diffEq s^\b\ \brak{
      \textbf{pri}(T) + \epsilon \textbf{tan}(T)
    }\\
    &\diffEq s^\b\ t\qedhere
  \end{salign*}
\end{proof}

Our canonicalization algorithm is a result of orienting the equations in
Figure~\ref{fig:differential-equivalence}. Note, however, that while most of
these equivalences have a ``natural'' orientation to them, two of them are
entirely symmetrical: those being commutativity of the sum and the derivative.
Barring the imposition of some arbitrary total ordering on terms which would
allow us to prefer the term $x + y$ over $y + x$ (or vice versa), we settle for
our canonical forms to be unique ``up to'' these commutativity conditions.

\begin{definition}
  \label{defn:permutative-equivalence}
  \textbf{Permutative equivalence} $\permEq{} \subseteq \Lambda_\epsilon \times
  \Lambda_\epsilon$ is the least equivalence relation which is contextual and
  satisfies the properties in Figure~\ref{fig:permutative-equivalence} below.
  \begin{figure}[H]
    \[
      \begin{array}{rcl}
        s + (t + e) &\permEq{} & (s + t) + e\\
        s + t &\permEq{}& t + s\\
        \D(\D(s) \cdot t) \cdot e &\permEq{}& \D(\D(s) \cdot e) \cdot t
      \end{array}
    \]
    \caption{Permutative equivalence on unrestricted $\Lambda_\epsilon$-terms}
    \label{fig:permutative-equivalence}
  \end{figure}
\end{definition}

We need to include associativity in the definition of permutative
equality, as otherwise the canonical term $x + (y + z)$ would not be
permutatively equivalent to $y + (x + z)$.

\begin{theorem}
  \label{thm:canonicity}
  Given unrestricted terms $s, t \in \Lambda_\epsilon$, they are differentially
  equivalent if and only if their canonical forms are permutatively equivalent.
  More succinctly, $s \diffEq t$ if and only if $\can{s} \permEq \can{t}$
\end{theorem}
\begin{proof}
  As an immediate consequence of Theorem~\ref{thm:canonical-form-differential},
  we know that $s \diffEq t$ if and only if $\can{s} \diffEq \can{t}$. The
  desired result will then follow from the fact that differential equivalence is
  precisely equivalent to permutative equivalence on canonical terms.
  
  Before proving this, we remark that $\canEq$ is reflexive, transitive and
  symmetric, which follows immediately from its definition and reflexivity,
  transitivity and symmetry of $\permEq$. We also prove the following two
  helpful results:

  \begin{lemma}
    \label{lem:can-contextual}
    For any unrestricted terms $s, t, e$, the following equalities hold:
    \begin{align*}
      \can{\can{s} + \can{t}} &= \can{s + t}\\
      \can{\epsilon\can{x}} &= \can{\epsilon(x)}\\
      \can{\can{s}\ \can{t}} &= \can{s\ t}\\
      \can{\D(\can{s}) \cdot \can{t}} &= \can{\D(s) \cdot t}
    \end{align*}
    As a consequence, the relation $\canEq$ is contextual.
  \end{lemma}
  \begin{proof}
    Follows immediately from the definition of $\can{}$ and observing that it
    only depends on the canonicalization of the subterms of the outermost
    syntactic form.
  \end{proof}

  \begin{lemma}
    \label{lem:can-permutative}
    Whenever $S, T$ are canonical terms, $S \diffEq T$ if and only if $S \permEq T$.
  \end{lemma}
  \begin{proof}
    Since $\diffEq$ is the least reflexive, transitive, symmetric and
    contextual relation that verifies the conditions in
    Figure~\ref{fig:differential-equivalence}, it follows that whenever $S
    \diffEq T$ it must be the case that $S = S_1 \diffEq^\star S_2 \diffEq^\star
    \ldots \diffEq^\star S_n = T$, where $\diffEq^\star$ denotes the contextual,
    symmetric (but not transitive) closure of $\diffEq^1$.

    But for each such condition, we have $\can{\text{LHS}} =
    \can{\text{RHS}}$, except for the two commutativity conditions, and in all
    cases we have $\can{\text{LHS}} \permEq \can{\text{RHS}}$.
    Since $\canEq$ is contextual, we have that $S = S_1 \canEq S_2 \canEq \ldots
    \canEq S_n = T$, and thus by transitivity we obtain $S \canEq T$.
  \end{proof}

  Theorem~\ref{thm:canonicity} is then an immediate consequence of
  Lemma~\ref{lem:can-permutative} and
  Theorem~\ref{thm:canonical-form-differential}.
\end{proof}

\begin{corollary}
  Differential equivalence of $\Lambda_\epsilon$-terms is decidable.
\end{corollary}
\begin{proof}
  Permutative equivalence of two terms is decidable, since the set of
  $\permEq$-equivalence classes of any term is finite and can be enumerated
  easily. Canonicalization is also decidable, since it is defined as a clearly
  well-founded recursion.
\end{proof}

\begin{corollary}
  The set $\lambda_\epsilon$ of well-formed terms corresponds precisely to the
  set of canonical terms up to permutative equivalence $\canonical / \permEq$.
\end{corollary}

\subsection{Substitution}

As is usual, our calculus features two different kinds of application: standard
function application, represented as $(s\ t)$; and differential application,
represented as $\D(s) \cdot t$. These two give rise to two different notions of
substitution. The first is, of course, the usual capture-avoiding substitution.
The second, differential substitution, is similar to the equivalent notion in
the differential $\lambda$-calculus, as it arises from the same chain rule that
is verified in both Cartesian differential categories and change action models.

\begin{definition}
  Given terms $s, t \in \Lambda_\epsilon$ and a variable $x$, the
  \textbf{capture-avoiding substitution} of $s$ for $x$ in $t$ (which we write
  as $\subst{t}{x}{s}$) is defined by induction on the structure of $t$ as in
  Figure~\ref{fig:substitution} below.
  \begin{figure}[H]
    \begin{center}
      \[
      \begin{array}{cccl}
        \subst{x}{x}{s} &\defeq& s\\
        \subst{y}{x}{s} &\defeq& y &\text{ if $x \neq y$}\\
        \subst{(\lambda y . t)}{x}{s} &\defeq& \lambda y . (\subst{t}{x}{s}) 
                        &\text{ if $y \not\in \text{FV}(s)$}\\
        \subst{(t\ e)}{x}{s} &\defeq& (\subst{t}{x}{s})(\subst{e}{x}{s})\\
        \subst{(\D(t) \cdot e)}{x}{s} &\defeq& \D(\subst{t}{x}{s}) \cdot (\subst{e}{x}{s}) \\
        \subst{(\epsilon t)}{x}{s} &\defeq& \epsilon (\subst{t}{x}{s})\\
        \subst{(t + e)}{x}{s} &\defeq& (\subst{t}{x}{s}) + (\subst{e}{x}{s})\\
        \subst{0}{x}{s} &\defeq& 0
      \end{array}
      \]
      \caption{Capture-avoiding substitution in $\lambda_\epsilon$}
      \label{fig:substitution}
    \end{center}
  \end{figure}
\end{definition}

\begin{proposition}
  \label{prop:substitution-differential}
  Capture-avoiding substitution respects differential equivalence. That is to
  say, whenever $s \diffEq s'$ and $t \diffEq t'$, it is the case that
  $\subst{t}{x}{s} \diffEq \subst{t'}{x}{s'}$.
\end{proposition}
\begin{proof}
  It suffices to show that $\subst{t}{x}{s} \diffEq^1 \subst{t'}{x}{s}$ (the
  full result will then follow from transitivity and contextuality of
  $\diffEq$), which can be proven by straightforward structural induction on $t$.
\end{proof}

\begin{definition}
  Given terms $s, t \in \Lambda_\epsilon$ and a variable $x$ \emph{which is not
  free in $s$}\footnote{One should emphasise this constraint. Differential
  substitution appears in the reduction of $\D(\lambda x . t) \cdot s$ into
  $\lambda x . \diffS{t}{x}{s}$, and so if $x$ were free in $s$ it would become
  bound by the enclosing $\lambda$-abstraction.}, the \textbf{differential
  substitution} of $s$ for $x$ in $t$, which we write as $\diffS{t}{x}{s}$, is
  defined by induction on the structure of $t$ as in
  Figure~\ref{fig:differential-substitution}.
  \begin{figure}[t]
    \begin{center}
    \[\def\arraystretch{1.7}
      \begin{array}{cccl}
        \diffS{x}{x}{s} &\defeq& s\\
        \diffS{y}{x}{s} &\defeq& 0 &\text{ if $x \neq y$}\\
        \diffS{(\lambda y . t)}{x}{s} &\defeq& \lambda y . \pa{\diffS{t}{x}{s}}
                        &\text{ if $y \not\in \text{FV}(t)$}\\
        \diffS{(t\ e)}{x}{s} &\defeq& \brak{\D(t) \cdot \pa{\diffS{e}{x}{s}}\ e} +
                                      \brak{\diffS{t}{x}{s}\ \pa{\subst{e}{x}{(x
                                      + \epsilon s)}}}\\
        \diffS{(\D(t) \cdot e)}{x}{s} &\defeq& 
          \D(t) \cdot \pa{\diffS{e}{x}{s}} 
          + \D\pa{\diffS{t}{x}{s}} \cdot (\subst{e}{x}{(x + \epsilon s)})\\
        &&
          {}+ \epsilon \D(\D(t) \cdot e ) \cdot \pa{\diffS{e}{x}{s}}\\
        \diffS{(\epsilon t)}{x}{s} &\defeq& \epsilon \pa{\diffS{t}{x}{s}}\\
        \diffS{(t + e)}{x}{s} &\defeq& \pa{\diffS{t}{x}{s}} + \pa{\diffS{e}{x}{s}}\\
        \diffS{0}{x}{s} &\defeq& 0
      \end{array}
    \]
    \caption{Differential substitution in $\lambda_\epsilon$}
    \label{fig:differential-substitution}
    \end{center}
  \end{figure}
  We write 
  \[ 
    \frac{\partial^k t}{\partial (x_1, \ldots, x_k)}(u_1, \ldots, u_k)
   \]
  to denote the sequence of nested differential substitutions
  \[ 
    (\partial (\ldots ((\partial t / \partial x_1)(u_1)) \ldots) / \partial x_k)(u_k)
  \]
\end{definition}

Most of the cases of differential substitution are identical to those in the
differential $\lambda$-calculus (compare the above with
Figure~\ref{fig:differential-substitution-standard}). There are, however, a
number of notable differences which stem from our more general setting. First,
we must point out that this definition in fact coincides exactly with the
original notion of differential substitution in e.g
\cite{ehrhard2003differential}, provided that one assumes the identity $\epsilon
t = 0$ for all terms. This reflects the fact that every Cartesian differential
category is in fact a Cartesian difference category with trivial infinitesimal
extension.

All the differences in this definition stem from the failure of derivatives to
be additive in the setting of Cartesian difference categories. Consider the case
for $\diffS{\D (t) \cdot s}{x}{e}$, and remember that the ``essence'' of a
derivative in our setting lies in the derivative condition, that is to say, if
$t(x)$ is a term with a free variable $x$, we seek our notion of differential
substitution to satisfy a condition akin to Taylor's formula:
\[
  t(x + \epsilon y) \diffEq t(x) + \epsilon \diffS{t}{x}{y}
\]

When the term $t$ is a differential application, and assuming the above
``Taylor's formula'' holds for all of its subterms (which we will show later),
this leads us to the following informal argument:
\begin{salign*}
  \D(t(x + \epsilon y)) \cdot (s(x + \epsilon y))
  &\diffEq \D\pa{t(x) + \epsilon \diffS{t}{x}{y}}  \cdot (s(x + \epsilon y))
  \PROOFSTEP
  &\diffEq \D(t(x)) \cdot (s(x + \epsilon y)) 
    + \epsilon \D\pa{\diffS{t}{x}{y}} \cdot (s (x + \epsilon y))
  \PROOFSTEP
  &\diffEq \D(t(x)) \cdot (s(x))\\
    &\qquad + \epsilon \D(t(x)) \cdot \pa{\diffS{s}{x}{y}}\\
    &\qquad + \epsilon \D\pa{\diffS{t}{x}{y}} \cdot (s (x + \epsilon y))\\
    &\qquad + \epsilon^2\D(\D(t(x)) \cdot (s(x))) \cdot \pa{\diffS{s}{x}{y}}
\end{salign*}

From this calculation, the differential substitution for this case arises
naturally as it results from factoring out the $\epsilon$ and noticing that
the resulting expression has precisely the correct shape to be Taylor's formula
for the case of differential application. The case for standard application can
be derived similarly, although the involved terms are simpler. Differential
substitution verifies some useful properties, which we state below (mechanised
proofs are available, although the details are more cumbersome than
enlightening).

\begin{proposition}
  \label{prop:differential-substitution-differential}
  Differential substitution respects differential equivalence. That is to
  say, whenever $s \diffEq s'$ and $t \diffEq t'$, it is the case that
  $\diffS{t}{x}{s} \diffEq \diffS{t'}{x}{s'}$.
\end{proposition}
\begin{proof}
  See \hyperref[coq:dsubst-diff]{\coqe{Lemma\ dsubst_diff}}.
\end{proof}
\begin{proposition}
  \label{prop:differential-substitution-closed}
  Whenever $x$ is not free in $t$, then $\diffS{t}{x}{u} \diffEq 0$.
\end{proposition}
\begin{proof}
  See \hyperref[coq:dsubst-empty]{\coqe{Lemma dsubst_empty}}.
\end{proof}
\begin{proposition}
  \label{prop:differential-substitution-commute}
  Whenever $x$ is not free in $u, v$, then:
  \[
    \frac{\d^2 t}{\d x^2}(u, v)
    \diffEq
    \frac{\d^2 t}{\d x^2}(v, u)
  \]
\end{proposition}
\begin{proof}
  See \hyperref[coq:dsubst-commute]{\coqe{Lemma dsubst_commute}}
\end{proof}

As we have previously mentioned, the rationale behind our specific definition of
differential substitution is that it should verify some sort of ``Taylor's
formula'' (or rather, Kock-Lawvere formula), in the following sense:
\begin{theorem}
  \label{thm:taylor}
  For any unrestricted terms $s, t, e$ and any variable $x$ which does not
  appear free in $e$, we have
  \[
    \subst{s}{x}{t + \epsilon e} \diffEq \subst{s}{x}{t} 
    + \epsilon\pa{\subst{\pa{\diffS{s}{x}{e}}}{x}{t}}
  \]
  We will often refer to the right-hand side of the above equivalence as the
  \textbf{Taylor expansion} of the corresponding term in the left-hand side.
\end{theorem}
\begin{proof}
  \newcommand{\lhs}[1]{\subst{#1}{x}{t + \epsilon e}}
  \newcommand{\diff}[1]{\subst{\pa{\diffS{#1}{x}{e}}}{x}{t}}
  The proof follows by induction on $s$ and some involved calculations. We also
  provide a mechanised version in \hyperref[coq:taylor]{\coqe{Theorem Taylor}}. 
  \begin{itemize}
    \item When $s = x$ we have $\diffS{x}{x}{e} = e$ and so:
      \begin{align*}
        \lhs{x} = t + \epsilon e = \subst{x}{x}{t} + \epsilon \diff{x}
      \end{align*}
    \item When $s = y \neq x$ we have $\diffS{y}{x}{e} = 0$ and so:
      \begin{align*}
        \lhs{y} = y \diffEq y + 0 = \subst{y}{x}{y} + \epsilon \diff{y}
      \end{align*}
    \item When $s = 0$ we have $\LHS \diffEq 0 \diffEq \RHS$.
    \item When $s = s' + s''$ we have:
    \begin{salign*}
      &\hspace{-20pt}\lhs{(s' + s'')} = \subst{s'}{x}{t + \epsilon e} + \subst{s''}{x}{t + \epsilon e}\\
      &\diffEq \pa{\subst{s'}{x}{t} + \epsilon\pa{\diff{s'}}}
        + \pa{\subst{s''}{x}{t} + \epsilon\pa{\diff{s''}}}\\
      &\diffEq \pa{\subst{s'}{x}{t} + \subst{s''}{x}{t}}
        + \pa{\diff{s'} + \diff{s''}}\\
      &= \subst{(s' + s'')}{x}{t} + \diff{(s' + s'')}
    \end{salign*}
    \item When $s = \epsilon s'$ we have:
    \begin{salign*}
      &\hspace{-30pt}\lhs{(\epsilon s')} = \epsilon \lhs{s'}\\
      \eq[\diffEq] \epsilon\pa{ \subst{s'}{x}{t} + \epsilon \pa{\diff{s'}} } \\
      \eq[\diffEq] \epsilon\pa{\subst{s'}{x}{t}} + \epsilon\pa{\epsilon \pa{\diff{s'}} } \\
      \eq[\diffEq] \subst{(\epsilon s')}{x}{t} + \epsilon \pa{\diff{(\epsilon s')}}
    \end{salign*}
    \item When $s = s'\ s''$ we have:
    \begin{salign*}
      &\hspace{-10pt}\lhs{(s'\ s'')} = (\lhs{s'})\ (\lhs{s''})\\
      &\diffEq \brak{\subst{s'}{x}{t} + \epsilon\pa{\diff{s'}}}\ 
        \brak{\subst{s''}{x}{t} + \epsilon\pa{\diff{s''}}}\displaybreak[0]\\
      &\diffEq \brak{\subst{s'}{x}{t}\ \pa{\subst{s''}{x}{t} + \epsilon\pa{\diff{s''}}}}\\
      \continued + \epsilon\brak{\pa{\diff{s'}}\ \pa{\subst{s''}{x}{t} + \epsilon\pa{\diff{s''}}} }\displaybreak[0]\\
      &\diffEq \pa{\subst{s'}{x}{t}\ (\subst{s''}{x}{t})} \\
      \continued + \epsilon\brak{\pa{\D(\subst{s'}{x}{t}) \cdot \pa{\diff{s''}}}
      \ (\subst{s''}{x}{t})}\\
      \continued + \epsilon\brak{\pa{\diff{s'}}\ \pa{\subst{s''}{x}{t + \epsilon e}}}\displaybreak[0]\\
      &= \subst{(s'\ s'')}{x}{t} \\
      \continued + \subst{\epsilon\brak{\pa{\D(s') \cdot
      \pa{\diffS{s''}{x}{e}}}\ s''}}{x}{t}\\
      \continued + \subst{\epsilon\brak{\pa{\diffS{s'}{x}{e}}\ \pa{\subst{s''}{x}{(x + \epsilon e)}}}}{x}{t}\displaybreak[0]\\
      &\diffEq \subst{(s'\ s'')}{x}{t} + \epsilon \pa{\diff{(s' + s'')}}
    \end{salign*}
    \item When $s = \D(s') \cdot s''$ we have:
      \begin{salign*}
        &\hspace{-10pt}{\lhs{(\D(s') \cdot s'')} = \D(\lhs{s'}) \cdot (\lhs{s''})}\\
        & \diffEq \D\pa{\subst{s'}{x}{t} + \epsilon\pa{\diff{s'}}} \cdot (\lhs{s''})\displaybreak[0]\\
        & \diffEq \D{(\subst{s'}{x}{t})} \cdot (\lhs{s''}) + \epsilon
        \pa{\D\pa{\diff{s'}}\cdot (\lhs{s''})}\displaybreak[0]\\
        & \diffEq \D{(\subst{s'}{x}{t})} \cdot \pa{\subst{s''}{x}{t} + \epsilon\diff{s''}}\\
        \continued + \epsilon\pa{\D\pa{\diff{s'}}\cdot (\lhs{s''})}\displaybreak[0]\\
        & \diffEq \D{(\subst{s'}{x}{t})} \cdot \pa{\subst{s''}{x}{t}}
        + \epsilon\pa{\D{(\subst{s'}{x}{t})} \cdot \pa{\diff{s''}}}\\
        \continued + \epsilon^2 \pa{\D(\D{(\subst{s'}{x}{t})} \cdot \pa{\subst{s''}{x}{t}}) \cdot \pa{\diff{s''}}} \\
        \continued + \epsilon\pa{\D\pa{\diff{s'}}\cdot (\lhs{s''})}\displaybreak[0]\\
        & \diffEq \subst{\pa{\D(s') \cdot s''}}{x}{t}
        + \subst{\epsilon\pa{\D{(s')} \cdot \pa{\diffS{s''}{x}{e}}}}{x}{t}\\
        \continued + \epsilon^2 \subst{\pa{\D(\D{(s')} \cdot s'') \cdot \pa{\diffS{s''}{x}{e}}}}{x}{t} \\
        \continued + \epsilon\subst{\pa{\D\pa{\diffS{s'}{x}{e}}\cdot (\subst{s''}{x}{x + \epsilon e})}}{x}{t}\displaybreak[0]\\
        &\diffEq \subst{\pa{\D(s') \cdot s''}}{x}{t} + \epsilon
        \pa{\diff{(\D{(s')} \cdot s'')}}
        \qedhere
      \end{salign*}
  \end{itemize}
\end{proof}

The following lemmas relate differential substitution and standard
substitution, and will be of much use later.

\begin{lemma}
  \label{lem:standard-differential-substitution}
  Whenever $x, y$ are (distinct) variables then for any unrestricted terms $t,
  u, v$ where $x$ is not free in $v$ we have:
  \[
    \subst{\pa{\diffS{t}{x}{u}}}{y}{v}
    = \diffS{\subst{t}{y}{v}}{x}{\subst{u}{y}{v}}
  \]
\end{lemma}
\begin{proof}
  See \hyperref[coq:replace-dsubst]{\coqe{Lemma replace_dsubst}}.
\end{proof}

\begin{lemma}
  \label{lem:differential-standard-substitution}
  Whenever $x, y$ are (distinct) variables, with $y$ not free in either $u, v$,
  we have:
  \[
    \diffS{\subst{t}{y}{v}}{x}{u} \diffEq 
    \subst{\pa{\diffS{t}{x}{u}}}{y}{(\subst{v}{x}{x + \epsilon u})}
    + \subst{\pa{\diffS{t}{y}{\diffS{v}{x}{u}}}}{y}{v}
  \]
\end{lemma}

One consequence of this ``syntactic Taylor's formula'' is that derivatives in
the difference $\lambda$-calculus can be computed by a sort of quasi-AD
algorithm: given an expression of the form $\lambda x . t$, its derivative at
point $s$ along $u$ can be computed by reducing the differential application
$\pa{\D(\lambda x . t) \cdot (u)}\ s$ which, as we shall see later, reduces (by
definition) to $\subst{\pa{\diffS{t}{x}{u}}}{x}{s}$. Alternatively, we can
simply evaluate $(\lambda x . t)\ (s + \epsilon(u))$ to compute $\subst{t}{x}{s
+ \epsilon(u)}$ which, by Theorem~\ref{thm:taylor} and
Lemma~\ref{lem:standard-differential-substitution}, is equivalent to
$\subst{t}{x}{s} + \epsilon(\subst{\diffS{t}{x}{u}}{x}{s})$. In an appropriate
setting (i.e. one where subtraction of terms is allowed and $\epsilon$ admits an
inverse) the derivative can then be extracted from this result by extracting the
term under the $\epsilon$. This process is remarkably similar to forward-mode
automatic differentiation, where derivatives are computed by adding
``perturbations'' to the program input.

Theorem~\ref{thm:taylor} also allows us to unpack all of the substitutions in
the definition of differential substitution. For example, the term
$\diffS{(t\ e)}{x}{u}$ can be expanded to:
\begin{salign*}
  \diffS{(t\ e)}{x}{u}&= \brak{\D(t)\cdot \pa{\diffS{e}{x}{u}}\ e} 
  + \brak{\diffS{t}{x}{u}\ \pa{\subst{e}{x}{(x + \epsilon u)}}}
  \PROOFSTEP
  &\diffEq \brak{\D(t) \cdot \pa{\diffS{e}{x}{u}}\ e}
  + \brak{
    \diffS{t}{x}{u}\ \pa{
      e + \epsilon \diffS{e}{x}{u}
    }
  }
  \PROOFSTEP
  &\diffEq \brak{\D(t) \cdot \pa{\diffS{e}{x}{u}}\ e}
  + \brak{\diffS{t}{x}{u}\ e}
  + \epsilon \brak{\D\pa{\diffS{t}{x}{u}}\cdot \diffS{e}{x}{u}\ e} 
\end{salign*}
We can generalise this procedure to arbitrary sequences of differential
substitutions, although the terms involved are too complex to give a simple
account.

\begin{lemma}
  \label{lem:differential-substitution-dapp}
  For any basic terms $s^\b, t^\b$, variables $x_1, \ldots, x_n$ and basic terms
  $u^\b_1, \ldots, u^\b_n$ such that none of the $x_i$ appear free in the $u_i$,
  the differential substitution 
  \[ \frac{\partial^k
     (\D(s^\b)\cdot t^\b)}{\partial (x_1, \ldots, x_n)}
     (u^\b_1, \ldots, u^\b_n)
  \]
  is differentially equivalent to a sum of terms of the form
  \[ 
    \epsilon^z \D^l (v) \cdot (w_1, \ldots, w_l)
  \]
  where $v$ is of the form \[
    \frac{\partial^p t^\b}{\partial (x^{(t)}_1, \ldots,
    x^{(t)}_p)}(u^{(t)}_1, \ldots, u^{(t)}_p) 
  \] and every $w_j$ is of the form
    \[\frac{\partial^{q_j} e^\b}{\partial (x^{(w_j)}_1, \ldots,
    x^{(w_j)}_{q_j})}(u^{(w_j)}_1, \ldots, u^{(w_j)}_{q_j}) \] where $1 \leq l
    \leq 2^n$ and each pair of sequences $x^{(t)}_i, u^{(t)}_i$ corresponds to a
    reordering of some subsequence of the $x_i, u^\b_i$.
\end{lemma}
\begin{proof}
  Straightforward induction on $k$ and applying Theorem~\ref{thm:taylor}.
\end{proof}

\begin{lemma}
  \label{lem:differential-substitution-app}
  For any basic term $s^\b$ and additive term $t^*$, variables $x_1, \ldots, x_n$
  and basic terms $u^\b_1, \ldots, u^\b_n$ such that none of the $x_i$ appear
  free in the $u^\b_i$, the differential substitution
  \[ \frac{\partial^k
     (s^\b\ t^*)}{\partial (x_1, \ldots, x_n)}
     (u^\b_1, \ldots, u^\b_n)
  \]
  is differentially equivalent to a sum of terms of the form
  \[ 
    \epsilon^z \D^l (v) \cdot (w_1, \ldots, w_l)
  \]
  where $v$ is of the form \[
    \frac{\partial^p t^\b}{\partial (x^{(t)}_1, \ldots,
    x^{(t)}_p)}(u^{(t)}_1, \ldots, u^{(t)}_p) 
  \] and every $w_j$ is of the form
    \[\frac{\partial^{q_j} e^\b}{\partial (x^{(w_j)}_1, \ldots,
    x^{(w_j)}_{q_j})}(u^{(w_j)}_1, \ldots, u^{(w_j)}_{q_j}) \] where $1 \leq l
    \leq 2^n$ and each pair of sequences $x^{(t)}_i, u^{(t)}_i$ corresponds to a
    reordering of some subsequence of the $x_i, u^\b_i$.
\end{lemma}

The above results may seem overly weak and arcane, but at its core they make a
very simple statement: if one applies any number of differential substitutions
to the term $\D(s) \cdot t$ (or $s\ t$) and ``cranks the lever'', pushing the
substitutions as far down the term as possible, then all the differential
substitutions in the resulting term are applied to either $s$ or $t$, and their
arguments are a reordering of some subsequence of the arguments to the initial
differential substitution.

\newcommand{\dd}[3]{\frac{\partial^2 #1}{\partial {#2}^2}(#3)}
\begin{theorem}
  \label{thm:differential-substitution-regular}
  Differential substitution is regular, that is, for any unrestricted terms $s,
  u, v$ where $x$ does not appear free in either $u$ or $v$, we have:
  \begin{align*}
    \diffS{s}{x}{0} &\diffEq 0\\
    \diffS{s}{x}{u + v} &\diffEq \diffS{s}{x}{u} 
    + \subst{\pa{\diffS{s}{x}{v}}}{x}{x + \epsilon u}
  \end{align*}
\end{theorem}
\begin{proof}
  Both properties follow by induction on $s$. As the proof involves immense
  amounts of tedious calculations, we refer the reader to
  \hyperref[coq:regularity]{\coqe{Theorem Regularity}}
  calculations, 
  We omit the details for the
  first one, and show only the non-trivial inductive cases for the second.
  \begin{itemize}
    \item The case $s = \D(t) \cdot e$ is rather mechanically involved. We first
    expand the right-hand side of the desired result and group the expansion
    into pieces. We will later expand the left-hand side and reconstruct each
    piece from this expansion. We use colourful notation to help the reader
    follow the structure of the proof: terms arising from expanding right-hand
    side of the equivalence will be coloured, whereas terms arising from
    expanding the left-hand side will be highlighted. Whenever a left-hand term
    and a right-hand term match each other exactly, we will use the same colour
    for both and consider them henceforth ``discharged''.
    \begin{salign*}
      &\diffS{s}{x}{u} + \subst{\pa{\diffS{s}{x}{v}}}{x}{x + \epsilon u}
      \PROOFSTEP
      &= \diffS{(\D(t) \cdot e)}{x}{u} 
      + \subst{\pa{\diffS{(\D(t) \cdot e)}{x}{v}}}{x}{x + \epsilon u}
      \PROOFSTEP
      &= \brak{ 
        \D(t) \cdot \pa{\diffS{e}{x}{u}}
        + \D\pa{\diffS{t}{x}{u}} \cdot (\subst{e}{x}{x + \epsilon u})
        + \epsilon \D(\D(t) \cdot e) \cdot \pa{\diffS{e}{x}{u}}
      }\\
      & + \subst{\brak{
        \D(t) \cdot \pa{\diffS{e}{x}{v}}
        + \D\pa{\diffS{t}{x}{v}} \cdot (\subst{e}{x}{x + \epsilon v})
        + \epsilon \D(\D(t) \cdot e) \cdot \pa{\diffS{e}{x}{v}}
      }}{x}{x + \epsilon u}
      \PROOFSTEP
      &\diffEq {\color{rose}\brak{
        \D(t) \cdot \pa{\diffS{e}{x}{u}} 
        + \subst{\pa{\D(t) \cdot \pa{\diffS{e}{x}{v}}}}{x}{x + \epsilon u}
      }}\\
      &+ {\color{indigo}\brak{
        \D\pa{\diffS{t}{x}{u}} \cdot (\subst{e}{x}{x + \epsilon u})
        + \subst{\pa{\D\pa{\diffS{t}{x}{v}} \cdot (\subst{e}{x}{x + \epsilon v})}}{x}{x + \epsilon u}
      }}\\
      &+ {\color{green} \epsilon \brak{
        \D(\D(t) \cdot e) \cdot \pa{\diffS{e}{x}{u}}
        + \subst{\pa{\D(\D(t) \cdot e) \cdot \pa{\diffS{e}{x}{v}}}}{x}{x + \epsilon u}
      }}
    \end{salign*}
    Before we continue, we expand the {\color{rose}rose} summand above by applying
    Theorem~\ref{thm:taylor}, as we will be cancelling these terms later:
    \begin{salign*}
      &{\color{rose}\brak{
        \D(t) \cdot \pa{\diffS{e}{x}{u}} 
        + \subst{\pa{\D(t) \cdot \pa{\diffS{e}{x}{v}}}}{x}{x + \epsilon u}
      }}
      \PROOFSTEP
      &\diffEq \brak{
        \D(t) \cdot \pa{\diffS{e}{x}{u}}
        + \D(t) \cdot \pa{\diffS{e}{x}{v}}
      } + \epsilon \diffS{\pa{\D(t)\cdot \pa{\diffS{e}{x}{v}}}}{x}{u}
      \PROOFSTEP
      &\diffEq \color{rose}\brak{
        \D(t) \cdot \pa{\diffS{e}{x}{u}}
        + \D(t) \cdot \pa{\diffS{e}{x}{v}}
      } \\
      \continued \color{rose}
      + \epsilon \brak{
        {\D(t) \cdot \pa{\dd{e}{x}{v, u}}}
      }\\
      \continued \color{rose}
      + \epsilon \brak{
        \D\pa{\diffS{t}{x}{u}} \cdot \pa{\diffS{(\subst{e}{x}{x + \epsilon u})}{x}{v}}
      }\\
      \continued \color{rose}
      + \epsilon \brak{
        \epsilon \D\pa{\D(t) \cdot \pa{\diffS{e}{x}{v}}}
      }
    \end{salign*}
    We now expand the left-hand side.
    \begin{salign*}
      &\diffS{s}{x}{u + v} = \diffS{\D(t) \cdot e}{x}{u + v}
      \PROOFSTEP
      &= \D(t) \cdot\pa{\diffS{e}{x}{u + v}}
      + \D\pa{\diffS{t}{x}{u + v}} \cdot (\subst{e}{x}{(x + \epsilon (u + v))})\\
      &+ \epsilon \D(\D(t)\cdot e) \cdot \pa{\diffS{e}{x}{u + v}}
      \PROOFSTEP
      &\diffEq \mathcolorbox{teal}{\brak{
        \D(t) \cdot \pa{\diffS{e}{x}{u} + \subst{\pa{\diffS{e}{x}{v}}}{x}{(x + \epsilon u)}}
      }}\\\
      &+ \mathcolorbox{sand}{\brak{
        \D\pa{\diffS{t}{x}{u} + \subst{\pa{\diffS{t}{x}{v}}}{x}{(x + \epsilon u)}}
        \cdot (\subst{e}{x}{(x + \epsilon(u + v))})
      }}\\
      &+ \mathcolorbox{wine}{\white \brak{
        \epsilon \D(\D(t) \cdot e) \cdot \pa{
          \diffS{e}{x}{u} + \subst{\pa{\diffS{e}{x}{v}}}{x}{(x + \epsilon u)}
        }
      }}
    \end{salign*}
    Now the second summand of the above expression, highlighted in \colorbox{sand}{sand}, can be
    expanded even further:
    \begin{salign*}
        &\mathcolorbox{sand}{\D\pa{\diffS{t}{x}{u} + \subst{\pa{\diffS{t}{x}{v}}}{x}{(x + \epsilon u)}}
        \cdot (\subst{e}{x}{(x + \epsilon(u + v))})}
        \PROOFSTEP
        &\diffEq \brak{
          \D\pa{\diffS{t}{x}{u}} \cdot (\subst{\subst{e}{x}{(x + \epsilon u)}}{x}{(x + \epsilon v)})
        }\\
        &+ \subst{\brak{
          \D\pa{\diffS{t}{x}{v}}
          \cdot (\subst{e}{x}{(x + \epsilon v)})
        }}{x}{(x + \epsilon u)}
        \PROOFSTEP
        &\diffEq \brak{
          \D\pa{\diffS{t}{x}{u}} \cdot (\subst{e}{x}{(x + \epsilon u)} 
          + \epsilon \diffS{(\subst{e}{x}{(x + \epsilon u)})}{x}{v})
        }\\
        &+ \subst{\brak{
          \D\pa{\diffS{t}{x}{v}}
          \cdot (\subst{e}{x}{(x + \epsilon v)})
        }}{x}{(x + \epsilon u)}
        \PROOFSTEP
        &\diffEq \mathcolorbox{indigo}{\color{white}\brak{
          \D\pa{\diffS{t}{x}{u}} \cdot (\subst{e}{x}{x + \epsilon u})
        + 
          \subst{\pa{\D\pa{\diffS{t}{x}{v}}
          \cdot (\subst{e}{x}{x + \epsilon v})}}{x}{x + \epsilon u}}}\\
        &+ \mathcolorbox{teal}{\brak{\D\pa{\diffS{t}{x}{u}} \cdot \pa{\epsilon \diffS{(\subst{e}{x}{(x + \epsilon u)})}{x}{v}}}}
        \\
        &+ \mathcolorbox{Orchid}{\brak{\epsilon \D\pa{\D\pa{\diffS{t}{x}{u}} \cdot (\subst{e}{x}{(x + \epsilon u)})} 
              \cdot \pa{\epsilon \diffS{(\subst{e}{x}{(x + \epsilon u)})}{x}{v}}
        }}
    \end{salign*}
    At this point, we note that the summands highlighted in {\colorbox{indigo}{\white indigo}} correspond
    precisely to the {\color{indigo} indigo} summand in the expansion of the right-hand side of our
    original equivalence. Now joining the {\colorbox{teal}{teal}} blocks together we obtain:
    \begin{salign*}
      &\mathcolorbox{teal}{
        \D(t) \cdot \pa{\diffS{e}{x}{u} + \pa{\subst{\diffS{e}{x}{v}}{x}{x + \epsilon u}}
        }}
      + \mathcolorbox{teal}{\D\pa{\diffS{t}{x}{u}} \cdot \pa{\epsilon \diffS{(\subst{e}{x}{x + \epsilon u})}{x}{v}}
      }
      \PROOFSTEP
      &\diffEq \brak{
        \D(t) \cdot \pa{\diffS{e}{x}{u}}
        + \D(t) \cdot \pa{\subst{\pa{\diffS{e}{x}{v}}}{x}{x + \epsilon u}}}\\
      &+ \epsilon \brak{
        \D\pa{ \D(t) \cdot \pa{\diffS{e}{x}{u}} } \cdot \subst{\pa{\diffS{e}{x}{v}}}{x}{x + \epsilon u}
      }
      \\
      &+ \epsilon \brak{\D\pa{\diffS{t}{x}{u}} \cdot \pa{\diffS{(\subst{e}{x}{x + \epsilon u})}{x}{v}}}
      \PROOFSTEP
      &\diffEq \brak{
        \D(t) \cdot \pa{\diffS{e}{x}{u}}
        + \D(t) \cdot \pa{\diffS{e}{x}{u} + \epsilon \pa{\dd{e}{x}{v, u}}}}
      \\
      &+ \epsilon \brak{
        \D\pa{ \D(t) \cdot \pa{\diffS{e}{x}{u}} } \cdot \subst{\pa{\diffS{e}{x}{v}}}{x}{x + \epsilon u}
      }
      \\
      &+ \epsilon \brak{\D\pa{\diffS{t}{x}{u}} \cdot \pa{\diffS{(\subst{e}{x}{x + \epsilon u})}{x}{v}}}
      \PROOFSTEP
      &\diffEq {\mathcolorbox{rose}{\brak{
        \D(t) \cdot \pa{\diffS{e}{x}{u}}
        + \D(t) \cdot \pa{\diffS{e}{x}{u}}}}}
      \PROOFSTEP
      &+ 
        \mathcolorbox{rose}{\brak{\epsilon \D(t) \cdot \pa{\dd{e}{x}{v, u}}
        + \epsilon^2 \D\pa{ \D(t) \cdot \pa{\diffS{e}{x}{u}}} \cdot \pa{\dd{e}{x}{v, u}}
      }}
      \PROOFSTEP
      &+ \mathcolorbox{wine}{\white \epsilon \brak{
        \D\pa{ \D(t) \cdot \pa{\diffS{e}{x}{u}} } \cdot \pa{\subst{\pa{\diffS{e}{x}{v}}}{x}{x + \epsilon u}}
      }}
      \PROOFSTEP
      &+ \mathcolorbox{rose}{\epsilon \brak{\D\pa{\diffS{t}{x}{u}} \cdot \pa{\diffS{(\subst{e}{x}{x + \epsilon u})}{x}{v}}}}
    \end{salign*}
    The terms highlighted in \colorbox{rose}{rose} match the Taylor expansion of the {\color{rose}rose} 
    summand in the
    right-hand side. It remains to assemble the fragments highlighted in \colorbox{wine}{\white wine}
    and show that they are equivalent to the {\color{wine}wine} summand of in the
    right-hand side. To simplify this arduous process, we expand the first \colorbox{wine}{\white wine}
    block and decompose it into three pieces, which will later discharge
    some of our remaining equivalences.
    \begin{salign*}
      &\mathcolorbox{wine}{
        \white
        \epsilon \D(\D(t) \cdot e) \cdot \pa{
          \diffS{e}{x}{u} + \subst{\pa{\diffS{e}{x}{v}}}{x}{x + \epsilon u}
      }}
      \PROOFSTEP
      &\diffEq
        \mathcolorbox{green}{\color{white}\epsilon\D(\D(t) \cdot e) \cdot \pa{\diffS{e}{x}{u}}}
        + \mathcolorbox{cyan}{\color{black}\epsilon \D(\D(t) \cdot e) \cdot \pa{\subst{\pa{\diffS{e}{x}{v}}}{x}{x + \epsilon u}}}
      \\
      &+ \mathcolorbox{olive}{\color{black}\epsilon^2 {
        \D\pa{\D\pa{\D(t) \cdot e} \cdot \pa{\diffS{e}{x}{u}}} \cdot \pa{\subst{\diffS{e}{x}{v}}{x}{x + \epsilon u}}
      }}
    \end{salign*}
    The term highlighted in \colorbox{green}{\white green} already appears in the {\color{green}green} summand of
    the original right-hand expansion and so we will henceforth consider it
    discharged. The remainder of the {\color{green}green} summand Taylor-expands as
    follows:
    \begin{salign*}
      &{\color{green}\epsilon\subst{\pa{\D(\D(t) \cdot e) \cdot \pa{\diffS{e}{x}{v}}}}{x}{x + \epsilon u}}
      \PROOFSTEP
      &\diffEq
      \epsilon\brak{
        \D(\D(t) \cdot e) \cdot \pa{\diffS{e}{x}{v}}
        + \epsilon \diffS{\pa{\D(\D(t) \cdot e) \cdot \pa{\diffS{e}{x}{v}}}}{x}{u}
      }
      \PROOFSTEP
      &\diffEq
        \epsilon \D(\D(t) \cdot e) \cdot \pa{\diffS{e}{x}{v}}
      + \epsilon^2 \Bigg[\\
      &\quad\quad \D(\D(t) \cdot e) \cdot \pa{\dd{e}{x}{v, u}}\\
      &\quad\quad + \D\pa{\diffS{\D(t) \cdot e}{x}{u}} \cdot \pa{\subst{\pa{\diffS{e}{x}{v}}} {x}{x + \epsilon u}}\\
      &\quad\quad + \epsilon \D\pa{\D(\D(t) \cdot e) \cdot \diffS{e}{x}{v}} \cdot \pa{\dd{e}{x}{v, u}}
      \\
      &\quad\Bigg]
      \PROOFSTEP
      &\diffEq
        {\color{cyan} \epsilon \D(\D(t) \cdot e) \cdot \pa{\diffS{e}{x}{v}}}\\
      &\quad\quad \color{cyan}\epsilon^2\D(\D(t) \cdot e) \cdot \pa{\dd{e}{x}{v, u}}\\
      &\quad\quad + \color{wine}\epsilon^2\D\pa{\D(t) \cdot \pa{\diffS{e}{x}{u}}} \cdot \pa{\subst{\pa{\diffS{e}{x}{v}}} {x}{x + \epsilon u}}\\
      &\quad\quad + \color{purple}\epsilon^2{\D\pa{\D\pa{\diffS{t}{x}{u}} \cdot (\subst{e}{x}{x + \epsilon u})} \cdot \pa{\subst{\pa{\diffS{e}{x}{v}}} {x}{x + \epsilon u}}}\\
      &\quad\quad + \color{olive}\epsilon^3 \D\pa{\D\pa{\D (t) \cdot e)} \cdot\pa{\diffS{e}{x}{u}}} \cdot \pa{\subst{\pa{\diffS{e}{x}{v}}} {x}{x + \epsilon u}}\\
      &\quad\quad + \color{cyan} \epsilon^3 \D\pa{\D(\D(t) \cdot e) \cdot \diffS{e}{x}{v}} \cdot \pa{\dd{e}{x}{v, u}}
    \end{salign*}
    The {\color{cyan}cyan} summands are neutralised with the Taylor expansion of the term
    highlighted in \colorbox{cyan}{cyan} (in the expansion of the second \colorbox{wine}{\white wine} block),
    whereas the {\color{wine}wine} summand is exactly the \colorbox{wine}{\white wine-highlighted} term that
    remained from the \colorbox{teal}{teal} blocks. Due to infinitesimal duplication, the
    {\color{olive}olive} summand is equivalent to the term highlighted in \colorbox{olive}{olive}, and the
    term in {\color{purple}purple} ink is equivalent to the remaining \colorbox{purple}{\white purple-highlighted} term.

    \item The case $s = t\ e$ is similar to the previous case, although simpler.
    For brevity's sake, we omit it here, but a fully
    mechanised version of the proof using the \texttt{Coq} proof assistant is
    available. \qedhere
  \end{itemize}
\end{proof}

\subsection{The Operational Semantics of \texorpdfstring{$\lambda_\epsilon$}{Lambda-Epsilon}}

With the substitution operations we have introduced so far, we can now
proceed to give a small-step operational semantics as a reduction system.

\begin{definition}
  The \textbf{one-step reduction relation}  $\reducesTo{} \subseteq \Lambda_\epsilon
  \times \Lambda_\epsilon$ is the least contextual relation satisfying the
  reduction rules in Figure~\ref{fig:reduction} below.
  \begin{figure}[H]
    \[\def\arraystretch{1.3}
    \begin{array}{rcl}
      (\lambda x . t)\ s &\reducesTo_\beta& \subst{t}{x}{s}\\
      \D(\lambda x . t) \cdot s &\reducesTo_\partial& \lambda x . \pa{\diffS{t}{x}{s}}\\
    \end{array}
    \]
    \caption{One-step reduction rules for $\lambda_\epsilon$}
    \label{fig:reduction}
  \end{figure}

  We write $\reducesTo^+$ to denote the transitive closure of $\reducesTo$,
  and $\reducesTo^*$ to denote its transitive, reflexive closure.
\end{definition}

While the one-step reduction rules for $\lambda_\epsilon$ may seem identical to
those in the differential $\lambda$-calculus (see
Figure~\ref{fig:reduction-standard}), they are in fact not equivalent, as our
notions of differential substitution and term equivalence differ substantially.

The above one-step reduction is defined as a relation from unrestricted terms to
unrestricted terms, but it is not compatible with differential equivalence. That
is to say, there may be differentially equivalent terms $t \diffEq t'$ such that
$t'$ can be reduced but $t$ cannot. For example, consider the term $(\lambda x
. x + 0)\ 0$, which contains no $\beta$-redexes that can be reduced. This term
is, however, equivalent to $(\lambda x . x)\ 0$, which clearly reduces to $0$. 

We could lift the one-step reduction relation to well-formed terms by setting
$\ul{t} \reducesTo \ul{t'}$ whenever there exist $s, s'$ such that $t \diffEq
s, t' \diffEq s'$ and $s \reducesTo s'$. This is not very satisfactory,
however, as it would make one-step reduction undecidable. Indeed, in order to
check whether $\ul{s} \reducesTo \ul{s'}$ it would be necessary to check
whether $s \reducesTo s'$ for all their (infinitely many) representatives $s,
s'$!

Another problem with this definition lies in the fact that the term $\ul{0}$
(ostensibly a value which should not reduce) can also be written as $\ul{0\ t}$
for any term $t$. Whenever $t$ reduces to $t'$ in one step, then according to
the previous definition so does $\ul{0\ t}$ reduce to $\ul{0\ t'}$, which is
equivalent to $\ul{0}$. Hence zero reduces to itself, rather than being a normal
form!

Fortunately the canonical form of a term $t$ gives us a representative of
$\ul{t}$ which is ``maximally reducible'', that is to say, whenever any
representative of $\ul{t}$ can be reduced, then so can $\can{t}$, possibly in
zero steps.

\begin{theorem}
  \label{thm:reduction-canonical}
  Reduction is compatible with canonicalization. That is to say, if $s
  \reducesTo s'$, then $\can{s} \reducesTo^* s''$ for some $s'' \diffEq s'$.
\end{theorem}
\begin{proof}
  We prove the theorem by induction on the depth at which reduction happens and
  the number of non-canonical $\beta$-redexes in the term (that is, the number
  of redexes of the form $(\lambda x . s)\ (t + \epsilon u)$). The
  most important cases are the ones where it happens at the outermost level, but
  we explicitly show some of the other cases. Before we do so, however, we state
  the following auxiliary properties:

  \begin{lemma}\label{lem:canonical-lambda}
    Whenever $T \diffEq \lambda x . t$ where $T$ is a canonical term and $t$ is
    an unrestricted term, then $T$ is of the form $\sum_{i=1}^n \epsilon^{k_i}
    (\lambda x. t^\b_i)$, and additionally $t \diffEq \sum_{i=1}^n \epsilon^{k_i}
    t^\b_i$.
  \end{lemma}
  \begin{lemma}\label{lem:reduction-canonical-additive}
    Whenever $s + t \reducesTo^* e$ then $e = s' + t'$ with $s \reducesTo^* s'$
    and $t \reducesTo^* t'$. Whenever $\varepsilon s \reducesTo^* e$ then $e =
    \epsilon s'$ with $s \reducesTo^* s'$. In particular, whenever $\can{t}
    \reducesTo^* t'$ then $t' = \sum_{i=1}^n \epsilon^{k_i} t'_i$, where
    $\can{t} = \sum_{i=1}^n \epsilon^{k_i} t^\b_i$ and $t^\b_i \reducesTo^*
    t'_i$. Note that the $t'_i$ may not be basic terms and thus $t'$ may not be
    canonical.
  \end{lemma}
  \begin{lemma}\label{lem:reduction-cansum}
    Whenever $s \reducesTo^* s'$ and $t \reducesTo^* t'$, then $s \cansum t
    \reducesTo^* s' \cansum t'$.
  \end{lemma}
  \begin{lemma}\label{lem:reduction-reg}
    Whenever $s \reducesTo^* s'$ and $\can{t} \reducesTo^* t'$, then $\reg{s,
    \can{t}} \reducesTo^* \reg{s', t'}$.
  \end{lemma}
  
  We proceed now to prove one of the cases where reduction happens in a subterm
  of $s$, which illustrates the ideas for the other cases:
  \begin{itemize}
    \item Let $s = \D(t) \cdot u$ and $s' = \D(t') \cdot u$, with $t
    \reducesTo t'$. By Lemma~\ref{lem:can-contextual}, we can write $\can{s}$
    as $\can{\D(\can{t}) \cdot \can{u}}$. Let $\can{t} = \sum_{i=1}^{n}
    \epsilon^{k_i} t^\b_i$. By induction and
    Lemma~\ref{lem:reduction-canonical-additive}, we have $t^\b_i \reducesTo^*
    t''_i$ and $t' \diffEq \sum_{i=1}^{n} \epsilon^{k_i} t''_i$. Applying the
    previous auxiliary lemmas, we obtain:
    \begin{salign*}
      \can{s} 
      &= \can{\D(\can{t}) \cdot u}\\
      &= \cansum_{i = 1}^n (\epsilon^*)^{k_i}\reg{t^\b_i, \can{u}}\\
      &\reducesTo^* \cansum_{i = 1}^n (\epsilon^*)^{k_i}\reg{t''_i, \can{u}}\\
      &\diffEq \D\pa{\sum_{i=1}^n \epsilon^{k_i} t''_i} \cdot u\\
      &\diffEq \D(t') \cdot u
    \end{salign*}
    \item Let $s = (t\ e)$ and $s' = (t\ e')$, with $e \reducesTo e'$. The
    result follows from the previous auxiliary lemmas and the fact that the
    primal $\textbf{pri}$ and tangent $\textbf{tan}$ components commute with
    reduction.
    \item Every other case is either immediate or follows from similar arguments.
  \end{itemize}
  
  The more involved cases are those when reduction happens at the outermost level of
  $s$. For brevity we will focus on the non-trivial cases where the underlying
  $\lambda$-abstraction involves only basic terms, as the more general cases
  follow by unfolding the $\lambda$-abstraction into a canonical sum and
  applying the primitive cases below to each summand separately.
  \begin{itemize}
    \item Let $s = \D(\lambda x . s^\b) \cdot t$ and $s' = \lambda x .
    \diffS{s^\b}{x}{t}$, and write $T$ for $\can{t}$. The proof proceeds then by
    induction on the number of summands of $T$.
    \begin{itemize}
      \item When $T = 0$ we have $\can{s} = 0$. On the other hand:
      \[s' = \lambda x . \diffS{s^\b}{x}{t} \diffEq \lambda x .
      \diffS{s^\b}{x}{0} \diffEq 0\]
      \item When $T = \epsilon^k t^\b + T'$, we apply the induction hypothesis
      and Lemma~\ref{lem:canonical-lambda} to obtain 
      \[ 
        \can{\D(\lambda x . s^\b) \cdot T'} 
        \reducesTo^*
        \sum_{i=1}^n\epsilon^{k_i}(\lambda x . w_i)
        \diffEq \lambda x . \diffS{s^\b}{x}{T'}
      \]
      Then the canonical form $\can{s}$ reduces as follows:
      \begin{salign*}
        &\can{s}\\
        &= 
        (\epsilon^*)^k\D(\lambda x . s^\b) \cdot t^\b\\
        &\quad + \brak{\reg{\D(\lambda x . s^\b) \cdot T'}
        \cansum {\epsilon^* \D^*\pa{\reg{\D(\lambda x . s^\b) \cdot T'}} \cdot t^\b}}
        \PROOFSTEP
        &=
        (\epsilon^*)^k\D(\lambda x . s^\b) \cdot t^\b \\
        &\quad + \brak{\can{\D(\lambda x . s^\b) \cdot T'}
        \cansum {\epsilon^* \D^*\pa{\can{\D(\lambda x . s^\b) \cdot T'}} \cdot t^\b}}
        \PROOFSTEP
        &\reducesTo^*
        (\epsilon^*)^k \pa{\lambda x . \diffS{s^\b}{x}{t^\b}}\\
        &\quad + \brak{
        \pa{\sum_{i=1}^n \epsilon^{k_i} (\lambda x . w_i)}
        \cansum \epsilon^* \D^* \pa{
          \sum_{i=1}^n \epsilon^{k_i} (\lambda x . w_i)
        }\cdot t^\b}
        \PROOFSTEP
        &=
        (\epsilon^*)^k \pa{\lambda x . \diffS{s^\b}{x}{t^\b}}\\
        &\quad + \brak{
        \pa{\sum_{i=1}^n \epsilon^{k_i} (\lambda x . w_i)}
        \cansum \epsilon^* \pa{
          \sum_{i=1}^n \epsilon^{k_i} \D(\lambda x . w_i) \cdot t^\b
        }}
        \PROOFSTEP
        &\reducesTo^*
        (\epsilon^*)^k \pa{\lambda x . \diffS{s^\b}{x}{t^\b}}\\
        &\quad + \brak{
        \pa{\sum_{i=1}^n \epsilon^{k_i} (\lambda x . w_i)}
        \cansum \epsilon^* \pa{
          \sum_{i=1}^n \epsilon^{k_i} \lambda x . \diffS{w_i}{x}{t^\b}
        }}
        \PROOFSTEP
        &\diffEq \lambda x . \pa{\diffS{s^\b}{x}{\epsilon^k t^\b}}
        + 
        \lambda x . \pa{\diffS{s^\b}{x}{T'}}
        + \epsilon \pa{
          \lambda x . \diffS{\pa{\sum_{i=1}^n \epsilon^{k_i} w_i}}{x}{t^\b}
        }
        \PROOFSTEP
        &\diffEq \lambda x . \pa{\diffS{s^\b}{x}{\epsilon^k t^\b}}
        + 
        \lambda x . \pa{\diffS{s^\b}{x}{T'}}
        + \epsilon \pa{
          \lambda x . \diffS{\diffS{s^\b}{x}{T'}}{x}{t^\b}
        }
        \PROOFSTEP
        &\diffEq \lambda x . \pa{
          \diffS{s^\b}{x}{\epsilon^k t^\b}
          + \subst{\pa{\diffS{s^\b}{x}{T'}}}{x}{x + \epsilon t^\b}
        }
      \end{salign*}
      By Theorem~\ref{thm:differential-substitution-regular}, this last term is
      precisely the term $\lambda x . \diffS{s^\b}{x}{\epsilon^k t^\b + T'}$,
      which is equivalent to $s'$ and thus the proof is concluded.
    \end{itemize}
    \item Let $s = (\lambda x . s^\b)\ t$ and $s' = \subst{s^\b}{x}{t}$, and
    suppose $\can{t} = t^* + \epsilon^* T$ (that is, $\textbf{pri}(\can{t}) = t^*$
    and $\textbf{tan}(\can{t}) = T$, with $t^*$ additive and $T$ canonical).
    \begin{align*}
      \can{s}
      &= (\lambda x . s^\b)\ t^* \cansum 
      \epsilon^*\textbf{ap}\pa{\reg{\lambda x . s^\b, T}, t^*}
    \end{align*}
    Since the term $(\lambda x . s^\b) \cdot T$ contains one less non-canonical
    $\beta$-redex than the term $s$, we apply our induction hypothesis to obtain
    \[
      \reg{\lambda x . s^\b, T} =
      \can{(\lambda x . s^\b) \cdot T }
      \reducesTo^* \sum_{i=1}^n \epsilon^{k_i} (\lambda x . w_i)
      \diffEq \lambda x . \diffS{s^\b}{x}{T}
    \]
    and therefore $\diffS{s^\b}{x}{T} \diffEq \sum_{i=1}^n \epsilon^{k_i} w_i$. 
    With this in mind we continue to reduce the previous equation:
    \begin{salign*}
      &(\lambda x . s^\b)\ t^* \cansum 
      \epsilon^*\textbf{ap}\pa{\reg{\lambda x . s^\b, T}, t^*}
      \PROOFSTEP
      &\quad \reducesTo^*
      \subst{s^\b}{x}{t^*} 
      \cansum
      \epsilon^*\textbf{ap}\pa{\sum_{i=1}^n \epsilon^{k_i} (\lambda x . w_i), t^*}
      \PROOFSTEP
      &\quad = \subst{s^\b}{x}{t^*}
      \cansum \epsilon^*\pa{\sum_{i=1}^n \epsilon^{k_i} (\lambda x . w_i)\ t^*}
      \PROOFSTEP
      &\quad \reducesTo^* 
      \subst{s^\b}{x}{t^*}
      \cansum \epsilon^*\pa{\sum_{i=1}^n \epsilon^{k_i} (\subst{w_i}{x}{t^*})}
      \PROOFSTEP
      &\quad = 
      \subst{s^\b}{x}{t^*}
      \cansum \epsilon^*\brak{\subst{\pa{\sum_{i=1}^n \epsilon^{k_i} w_i}}{x}{t^*})}
      \PROOFSTEP
      &\quad \diffEq
      \subst{s^\b}{x}{t^*}
      + \epsilon\pa{\subst{\diffS{s^\b}{x}{T}}{x}{t^*}}
      \PROOFSTEP
      &\quad \diffEq
      \subst{s^\b}{x}{t^* + \epsilon T}
      \PROOFSTEP
      &\quad \diffEq \subst{s^\b}{x}{t}\qedhere
    \end{salign*}
  \end{itemize}
\end{proof}

The above result then legitimises our proposed ``existential'' definition of
reduction of well-formed terms, as it shows that, in order to reduce a given
term, it suffices to reduce its canonical form. It also gets rid of the
``reducing zero'' problem, as canonical forms do not contain ``spurious''
representations of zero.

\begin{definition}
  Given well-formed terms $\ul{s}, \ul{s'}$, we say that $\ul{s}$
  \textbf{reduces to} $\ul{s'}$ \textbf{in one step}, and write $\ul{s}
  \mathrel{\ul{\reducesTo}} \ul{t}$, whenever $\can{s}
  \reducesTo s''$ and $s'' \diffEq s'$, for some canonical form $\can{s}$ of
  $\ul{s}$.
\end{definition}

\begin{proposition}
  \label{prop:red-wf-can}
  Whenever $\ul{s} \wfReducesTo \ul{s'}$ then for any term $\ul{t}$ we have
  $\ul{s + t} \wfReducesTo \ul{s' + t}$. 
  
  If $t=t^*$ is an additive term, then additionally $\ul{s\ t^*} \wfReducesTo^+
  \ul{s'\ t^*}$.
  
  Furthermore, when $t=t^\b$ is a basic term (in particular $t^\b$ is not
  differentially equivalent to zero), we also have $\ul{\D(s) \cdot t}
  \wfReducesTo^+ \ul{\D(s') \cdot t}$.

  Conversely, whenever $s$ is not differentially equivalent to zero and $\ul{t}
  \wfReducesTo \ul{t'}$, then $\ul{s\ t} \wfReducesTo^+ \ul{s\ t'}$ and $\ul{\D(s)
  \cdot t} \wfReducesTo^+ \ul{\D(s) \cdot t'}$.
\end{proposition}

The wording of the above definition specifies that a well-formed
term reduces to another whenever \emph{any} of its canonical forms reduces.
As we have shown before, canonical forms are in fact only unique up to
commutativity of addition and derivatives. Addition is not problematic, since
it respects reduction; that is to say, if a sum $s + t$ reduces to $s' + t'$
then its permutation $t + s$ also reduces to $t' + s'$. Symmetry of
derivatives raises a more significant issue: consider the following diagram,
which does \emph{not} commute:
\begin{center}
\begin{tikzpicture}
  \node (a) at (0, 0) {$\D (\D (\lambda x . t) \cdot u) \cdot v$};
  \node (b) at (5, 0) {$\D (\D (\lambda x . t) \cdot v) \cdot u$};
  \node (a') at (0, -2) {$\D \pa{\lambda x . \diffS{t}{x}{u}} \cdot v$};
  \node (b') at (5, -2) {$\D \pa{\lambda x . \diffS{t}{x}{v}} \cdot u$};
  \draw[white] (a) edge node[black]{$\diffEq$} (b);
  \draw[white] (a) edge node[rotate=-90,black]{$\reducesTo$} (a');
  \draw[white] (b) edge node[rotate=-90,black]{$\reducesTo$} (b');
  \draw[white] (a') edge node[black]{$\not\diffEq$} (b');
  \end{tikzpicture}
\end{center}
One-step reduction is still computable, since the set of canonical forms of
any given term is finite, but we will have to keep this behaviour in mind when
showing confluence.

A proof of confluence for $\lambda_\epsilon$ will proceed by the standard
Tait/Martin-L\"of method by introducing a notion of parallel reduction on terms.

\begin{definition}
  The \textbf{parallel reduction} relation between (unrestricted) terms is
  defined according to the deduction rules in Figure~\ref{fig:par-reduction}.
  \begin{figure}[h]
    \begin{center}
    \bpt{
      \AxiomC{}
      \LeftLabel{($\parTo_x$)}
      \UnaryInfC{$x \parTo x$}
    }
    \bpt{
      \AxiomC{}
      \LeftLabel{($\parTo_0$)}
      \UnaryInfC{$0 \parTo 0$}
    }
    \bpt{
      \AxiomC{$t \parTo t'$}
      \LeftLabel{($\parTo_\lambda$)}
      \UnaryInfC{$\lambda x . t \parTo \lambda x . t'$}
    }
    \end{center}
    \begin{center}
    \bpt{
      \AxiomC{$t \parTo t'$}
      \LeftLabel{($\parTo_\varepsilon$)}
      \UnaryInfC{$\epsilon t \parTo \epsilon t'$}
    }
    \bpt{
      \AxiomC{$s \parTo s'$}
      \AxiomC{$t \parTo t'$}
      \LeftLabel{($\parTo_+$)}
      \BinaryInfC{$s + t \parTo s' + t'$}
    }
    \end{center}\begin{center}
    \bpt{
      \AxiomC{$s \parTo s'$}
      \AxiomC{$t \parTo t'$}
      \LeftLabel{($\parTo_{\textbf{ap}}$)}
      \BinaryInfC{$s\ t \parTo s'\ t'$}
    }
    \bpt{
      \AxiomC{$s \parTo s'$}
      \AxiomC{$t \parTo t'$}
      \LeftLabel{($\parTo_\D$)}
      \BinaryInfC{$\D(s)\cdot t \parTo \D(s') \cdot t'$}
    }
    \end{center}
    \begin{center}
    \bpt{
      \AxiomC{$s \parTo \lambda x . s'$}
      \AxiomC{$t \parTo t'$}
      \LeftLabel{($\parTo_\beta$)}
      \BinaryInfC{$s\ t \parTo \subst{s'}{x}{t'}$}
    }
    \bpt{
      \AxiomC{$s \parTo \lambda x . s'$}
      \AxiomC{$t \parTo t'$}
      \LeftLabel{($\parTo_\partial$)}
      \BinaryInfC{$\D(s)\cdot t \parTo \lambda x . \diffS{s'}{x}{t'}$}
    }
    \end{center}
    \caption{Parallel reduction rules for $\lambda_\epsilon$}
    \label{fig:par-reduction}
  \end{figure}

  The parallel reduction relation can be extended to well-formed terms by
  setting $\ul{t} \wfParTo \ul{t'}$ whenever $\can{t} \parTo t''$ with $t''
  \diffEq t'$ for some canonical form of $\ul{t}$.
\end{definition}

\begin{remark}
  Our definition of parallel reduction differs slightly from the usual in the
  rule $(\parTo_\beta)$, which allows reducing a newly-formed
  $\lambda$-abstraction. This is necessary because our calculus contains terms
  of the shape $(\D(\lambda x . s) \cdot u)\ t$, which we need to parallel reduce
  in a single step to $\subst{\pa{\diffS{s}{x}{u}}}{x}{t}$. The original
  presentation of the differential $\lambda$-calculus opted instead for 
  adding an extra parallel reduction rule to allow for the case of reducing an
  abstraction under a differential application. Similarly, our rule
  $(\parTo_\partial)$ allows reducing terms of the form $\D (\D(\lambda x . s) \cdot
  u) \cdot v$ in a single step.
\end{remark}

One convenient property of the parallel reduction relation lies in its relation
to canonical forms. As we saw in Theorem~\ref{thm:reduction-canonical},
canonical forms are ``maximally reducible'', but don't respect the number of
reduction steps. This is no longer the case for parallel reduction: the
process of canonicalization only duplicates regexes ``in parallel'' (that is, by
copying them onto multiple separate summands) or in a ``parallelizable series''
(i.e. a differential application may be regularized into a term of the form 
$\D (\D(\ldots) \cdot u) \cdot v$, which can be entirely reduced in a single
parallel reduction step). 

\begin{theorem}
  \label{thm:parallel-reduction-canonical}
  Whenever $s \parTo s'$, then $\can{s} \parTo s''$ for some $s'' \diffEq s'$.
\end{theorem}
\begin{proof}
  \rff{
    I'm not elaborating this proof any more as it is laborious and brings
    nothing to the table. I never use this theorem but I like it because it
    makes sense of the connection between canonicity and reduction, should I
    leave it in given that I won't write a better proof?
  }
    It suffices to inspect the proof of
    Theorem~\ref{thm:reduction-canonical} and convince oneself that all of the
    reductions introduced by the proof can be lifted into a single instance of
    parallel reduction.
\end{proof}

We also state the following standard properties of parallel reduction, all of which
can be proven by straightforward induction on the term.

\begin{lemma}
  Parallel reduction sits between one-step and many-step reduction. That is to
  say: $\reducesTo {}\subseteq {}\parTo {}\subseteq {}\reducesTo^*$, and
  furthermore $\mathrel{\ul{\reducesTo}}{} \subseteq {} \wfParTo {} \subseteq {} \mathrel{
    \ul{\reducesTo}}^*$.
\end{lemma}

\begin{lemma}
  The parallel reduction relation is contextual. In particular, every term
  parallel-reduces to itself.
\end{lemma}

\begin{lemma}
  Parallel reduction cannot introduce free variables. That is to say: whenever
  $t \parTo t'$, we have $\text{FV}(t') \subseteq \text{FV}(t)$.
\end{lemma}

\begin{lemma}
  Whenever $\lambda x . t \parTo u$, it must be the case that $u = \lambda x .
  t'$ and $t \parTo t'$.
\end{lemma}

\begin{lemma}
  \label{lem:par-to-substitution}
  Whenever $s \parTo s'$ and $t \parTo t'$ then $\subst{s}{x}{t} \parTo
  \subst{s'}{x}{t'}$, and furthermore there is some $w$ with $\diffS{s}{x}{t}
  \parTo w \diffEq \diffS{s'}{x}{t'}$.
\end{lemma}
\begin{proof}
  Both proofs follow by induction on the derivation applied to obtain $s \parTo
  s'$. We explicitly prove some non-trivial cases. First, for standard
  substitution.
  \begin{itemize}
    \item When the last rule applied is $(\parTo_\beta)$, that 
    is: $s = e\ u, s' = \subst{e'}{y}{u'}$, with $e \parTo \lambda y . e', u
    \parTo u'$. By the induction hypothesis, we have $\subst{e}{x}{t} \parTo
    \lambda y . (\subst{e'}{x}{t'})$ and $\subst{u}{x}{t} \parTo \subst{u'}{x}{t'}$, hence:
    \begin{align*}
      \subst{s}{x}{t} &= \subst{e}{x}{t}\ (\subst{u}{x}{t})
      \parTo \subst{(\subst{e'}{x}{t'})}{y}{(\subst{u'}{x}{t'})}
      = \subst{(\subst{e'}{y}{u'})}{x}{t'}
    \end{align*}
    \item When the last rule applied is $(\parTo_\partial)$, that is:
    $s = \D(e) \cdot u$, $s' = \lambda y . \diffS{e'}{x}{u'}$, with
    $e \parTo \lambda y . e', u \parTo u'$. As before, we apply the induction hypothesis and
    obtain:
    \begin{salign*}
      \subst{s}{x}{t} &= D(\subst{e}{x}{t}) \cdot (\subst{u}{x}{t})
      \PROOFSTEP
      &\parTo \lambda y . \pa{\diffS{(\subst{e'}{x}{t'})}{y}{\subst{u'}{x}{t'}}}
      = \subst{\pa{\lambda y . \diffS{e'}{y}{u'}}}{x}{t'}
    \end{salign*}
  \end{itemize}
  The corresponding cases for differential substitution are slightly more technically
  involved.
  \begin{itemize}
    \item When the last rule applied is $(\parTo_\beta)$, that is: $s = (e\ u),
    s' = \subst{e'}{y}{u'}$, with $e \parTo \lambda y . e', u \parTo u'$. By the
    induction hypothesis and applying the definition of differential
    substitution, we have $\diffS{e}{x}{t} \parTo \lambda y . \diffS{e'}{x}{t'}$
    and $\diffS{u}{x}{t} \parTo \diffS{u'}{x}{t'}$. By applying the previous
    proof we also obtain $\subst{u}{x}{x + \epsilon t} \parTo \subst{u'}{x}{x +
    \epsilon t'}$ hence\footnote{
    Observe that the reasoning here would not hold if we had opted to define
    parallel reduction in the ``standard'' way, as differential substitution may
    ``unfold'' an application into a differential application followed by a
    standard one.
    }:
    \begin{salign*}
      \diffS{s}{x}{t} &= 
      \brak{\pa{\D(e) \cdot \pa{\diffS{u}{x}{t}}}\ u}
      + \brak{\diffS{e}{x}{t}\ (\subst{u}{x}{x + \epsilon t})}
      \PROOFSTEP
      &\parTo
      \subst{\brak{\diffS{e'}{y}{\diffS{u'}{x}{t'}}}}{y}{u'}
      + \subst{\pa{\diffS{e'}{x}{t'}}}{y}{(\subst{u'}{x}{x + \epsilon t'})}
    \end{salign*}

    On the other hand, since $y$ is not free in either $u'$ or $t'$, applying
    Lemma~\ref{lem:differential-standard-substitution}, we obtain:
    \begin{salign*}
      \diffS{s'}{x}{t'} &= \diffS{\pa{\subst{e'}{y}{u'}}}{x}{t'}\\
      &\diffEq 
        \subst{\pa{\diffS{e'}{x}{t'}}}{y}{(\subst{u'}{x}{x + \epsilon t'})} 
        + \subst{\pa{\diffS{e'}{y}{\diffS{u'}{x}{t'}}}}{y}{u'}
    \end{salign*}
    \item When the last rule applied is $(\parTo_\partial)$, that is:
    $s = \D(\lambda y . e) \cdot u$, $s' = \lambda y . \diffS{e'}{x}{u'}$, with
    $e \parTo e', u \parTo u'$. As before, we apply the induction hypothesis and
    obtain:
    \begin{salign*}
      \subst{s}{x}{t} &= D(\lambda y . \subst{e}{x}{t}) \cdot (\subst{u}{x}{t})\\
      &\parTo   \lambda y . \pa{\diffS{(\subst{e'}{x}{t'})}{y}{\subst{u'}{x}{t'}}}\\
      &= \subst{\pa{\lambda y . \diffS{e'}{y}{u'}}}{x}{t'}\qedhere
    \end{salign*}
  \end{itemize}
\end{proof}

We first prove that parallel reduction has the diamond property when applied to
canonical terms, taking care that it holds up to differential equivalence (note
that, much like one-step reduction, the result of parallel-reducing a canonical
term need not be canonical). For this, we introduce the usual notion of a full
parallel reduct of a term.

\newcommand{\fpr}[0]{_\da}
\begin{definition}
  Given an unrestricted term $t$, its \textbf{full parallel reduct}
  $t\fpr$ is defined inductively by:
  \[\def\arraystretch{1.5}
  \begin{array}{rcl}
    x\fpr &\defeq& x \\
    (\epsilon t) \fpr &\defeq& \epsilon (t\fpr)\\
    (s + t)\fpr &\defeq& (s\fpr) + (t\fpr)\\
    0\fpr &\defeq& 0\\
    (\lambda x . t) \fpr &\defeq& \lambda x . (t\fpr)\\
    (s\ t)\fpr &\defeq& \begin{cases}
      \subst{e}{x}{t\fpr} &\text{ if $s\fpr = \lambda x . e$}\\
      (s\fpr)\ (t\fpr)&\text{ otherwise}
    \end{cases}\\
    (\D(s)\cdot t)\fpr &\defeq& \begin{cases}
      \lambda x . \diffS{e}{x}{t\fpr} &\text{ if $t\fpr = \lambda x . e$}\\
      D(s\fpr)\cdot (t\fpr) &\text{ otherwise}
    \end{cases}
  \end{array}
  \]
\end{definition}

\begin{lemma}
  \label{lem:full-reduction-lambda}
  Whenever $s \parTo \lambda x . v$, then $s\fpr$ is of the form $\lambda x .
  w$, for some term $w$.
  \rff{I don't like this proof. Is it formal enough?}
\end{lemma}
\begin{proof}
  The proof follows by inspection of the parallel reduction rules. Consider a
  derivation of $s \parTo s'$, which will be of the form
  $\lambda x_1.\lambda x_2.\ldots \lambda x_n . t$ (with $n$ possibly equal to
  $0$). In general the amount of abstractions at the outermost level of the term
  depends on our choice of a derivation for $s \parTo s'$. Suppose then that we
  pick a derivation for which $n$ is maximal. If this derivation does not
  use the rules $\parTo_\textbf{ap}$ or $\parTo_\D$, then it is already
  the case that $s' = s\fpr$. On the other hand, if it contains either rule, it
  is straightforward to see that replacing the last application of
  $\parTo_\textbf{ap}$ or $\parTo_\D$ by $\parTo_\beta$ or $\parTo_\partial$
  respectively the resulting term has at least as many $\lambda$-abstractions at
  the outermost level as the previous one. Iterating this process, we obtain
  that the number of outermost abstractions is maximised precisely whenever $s'
  = s\fpr$.
\end{proof}

\rff{
  Here I make use of the fact that $\lambda x . u \diffEq \lambda x . v$ 
  implies $u \diffEq v$. Is this intuitive enough, or would a full proof be
  useful?
}
\begin{theorem}
  \label{thm:full-reduction}
  For any unrestricted terms $s, s'$ such that $s \parTo s'$, there is an
  unrestricted term $w$ such that $s' \parTo w$ and $w \diffEq s\fpr$.
\end{theorem}
\begin{proof}
  The proof follows by induction on the derivation of $s \parTo s'$. Most cases
  are straightforward, and the rest follow as a corollary of
  Lemma~\ref{lem:par-to-substitution}, as we now show.
  \begin{itemize}
    \item The last applied rule is $\parTo_\beta$, that is, $s = (t\ e)$, 
      $s' = \subst{t'}{x}{e'}$, with $t \parTo \lambda x . t', e \parTo e'$.
    
      By the induction hypothesis we have $e' \parTo w_e \diffEq e\fpr$ and
      $(\lambda x . t') \parTo (\lambda x . w_t) \diffEq t\fpr$. 
      By Lemma~\ref{lem:full-reduction-lambda}, it follows that $t\fpr$ has the
      form $\lambda x . v_t$, and since $(\lambda x . w_t) \diffEq (\lambda x .
      v_t)$ we also have $w_t \diffEq v_t$.
      Then: \[
        s' = \subst{t'}{x}{e'} \parTo \subst{w_t}{x}{w_e}
        \diffEq \subst{v_t}{x}{e\fpr} = (t\ e)\fpr
      \]

      \item The last applied rule is $\parTo_\partial$, that is,
      $s = (\D(t)\cdot u), s' = \lambda x . \diffS{t'}{x}{u'}$, with
      $t \parTo \lambda x . t', u \parTo u'$.

      By the induction hypothesis we have $u' \parTo w_u \diffEq u\fpr$ and $\lambda x . t'
      \parTo \lambda x . w_t \diffEq t\fpr$. Again we must have $t\fpr = \lambda
      x . v_t$ with $w_t \diffEq v_t$. Then: \[
        s' = \diffS{t'}{x}{u'} \parTo \diffS{w_t}{x}{w_u}
        \diffEq \diffS{v_t}{x}{u\fpr} = (\D(t) \cdot u)\fpr\qedhere
      \]
  \end{itemize}
\end{proof}

\begin{corollary}
  Parallel reduction has the diamond property up to differential equivalence.
  That is to say, for any unrestricted term $t$ and terms $t_1, t_2$ such that $t
  \parTo t_1$ and $t \parTo t_2$, there are terms $u, v$ making the following
  diagram commute:

  \begin{center}
  \begin{tikzpicture}
    \node (a) at (0, 0) {$T$};
    \node (t1) at (-1, -1) {$t_1$};
    \node (t2) at (1, -1) {$t_2$};
    \node (u) at (-1, -2.5) {$u$};
    \node (v) at (1, -2.5) {$v$};
    \draw[white] (a) edge node[rotate=225,black]{$\parTo$} (t1);
    \draw[white] (a) edge node[rotate=-45,black]{$\parTo$} (t2);
    \draw[white] (t1) edge node[rotate=-90,black]{$\parTo$} (u);
    \draw[white] (t2) edge node[rotate=-90,black]{$\parTo$} (v);
    \draw[white] (u) edge node[black]{$\diffEq$} (v);
    \end{tikzpicture}
  \end{center}
\end{corollary}

\begin{lemma}
  \label{lem:perm-eq-fpr}
  Given unrestricted terms $s \permEq s'$ which are permutatively equivalent,
  that is, which differ only up to a reordering of their additions and
  differential applications, their full parallel reducts are differentially
  equivalent.
\end{lemma}
\begin{proof}
  The proof follows by straightforward induction on the proof of permutative
  equivalence. The only involved case is $s = \D (\D(\lambda x . t) \cdot u)
  \cdot v \permEq \D(\D(\lambda x . t) \cdot v)\cdot u = s'$. But in this case
  the result follows as a corollary of
  Proposition~\ref{prop:differential-substitution-commute}
  \begin{salign*}
    s\fpr &= (\D (\D(\lambda x . t)\cdot u)\cdot v)\fpr
    \PROOFSTEP
    &=       \lambda x . \dd{t}{x}{u, v}
    \PROOFSTEP
    &\diffEq \lambda x . \dd{t}{x}{v}{u}\\
    &= s'\fpr\qedhere
  \end{salign*}
\end{proof}

\begin{theorem}
  The reduction relation $\wfParTo$ has the diamond property. That is, whenever $\ul
  s \wfParTo \ul u$ and $\ul s \wfParTo \ul v$ there is a term
  $\ul c$ such that $\ul u \wfParTo \ul c$ and $\ul v \wfParTo \ul c$.
\end{theorem}
\begin{proof}
  Consider a well-formed term $\ul{s}$, and suppose that $\ul{s} \wfParTo
  \ul{u}$ and $\ul{s} \wfParTo \ul{v}$. In particular, this means there are two
  canonical forms $\can{s}_1, \can{s}_2$ of $s$ such that $\can{s}_1 \parTo u$
  and $\can{s}_2 \parTo v$. These canonical forms $\can{s}_1, \can{s}_2$
  are equivalent up to permutative equivalence, and so their full parallel
  reducts are differentially equivalent as per Lemma~\ref{lem:perm-eq-fpr}.
  Denote their $\diffEq$-equivalence class by $\ul{c}$.
  Therefore since $\can{s}_1 \parTo \can{s}_{1\da} \diffEq c$ and $\can{s}_2 \parTo 
  \can{s}_{2\da} \diffEq c$ it follows that $\ul{u} \wfParTo \ul{c}$ and $\ul{v}
  \wfParTo \ul{c}$.
\end{proof}

\begin{corollary}
  The reduction relation $\ul{\reducesTo}$ is confluent.
\end{corollary}

\subsection{Encoding the Differential \texorpdfstring{$\lambda$}{Lambda}-Calculus}

It is immediately clear, from simply inspecting the operational semantics for
$\lambda_\epsilon$, that it is closely related to the differential
$\lambda$-calculus -- indeed, every Cartesian differential category is a
Cartesian difference category, and this connection should also be reflected in
the syntax.

As it turns out, there is a clean translation that embeds $\lambda_\epsilon$
into the differential $\lambda$-calculus, which proceeds by deleting every term
that contains an $\epsilon$. The intuition behind this scheme should be
apparent: every single differential substitution rule in $\lambda_\epsilon$ is
identical to the corresponding case for the differential $\lambda$-calculus,
once all the $\epsilon$ terms are cancelled out.

\begin{definition}
  Given an unrestricted $\lambda_\epsilon$ term $t$, its
  \textbf{$\epsilon$-erasure} is the differential $\lambda$-term $\ceil{t}$ defined
  as follows:

  \begin{figure}[H]
  \[
  \begin{array}{rcl}
    \ceil{x}&\defeq&x\\
    \ceil{0}&\defeq&0\\
    \ceil{s+t}&\defeq&\ceil{s} + \ceil{t}\\
    \ceil{\epsilon t}&\defeq& 0\\
    \ceil{s\ t}&\defeq&\ceil{s}\ \ceil{t}\\
    \ceil{\D(s)\cdot t}&\defeq&\D(\ceil{s})\cdot \ceil{t}
  \end{array}{rcl}
  \]
  \label{def:erasure}
  \caption{$\epsilon$-erasure of a term $t$}
\end{figure}
\end{definition}

\rff{this is confusing because it's hard to tell what is an $\epsilon$-term and
what is a differential $\lambda$-term. Any ideas on how to distinguish them?
Color would be nice}

\begin{proposition}
  The erasure $\ceil{t}$ is invariant under differential equivalence. That is to
  say, whenever $t\diffEq t'$, it is the case that $\ceil{t} = \ceil{t'}$.
\end{proposition}
\begin{proof}
  Follows immediately from inspecting the differential equivalence rules
  in Figure~\ref{fig:differential-equivalence} and noticing that the erasure of
  both sides coincides.
  
  Note that the standard presentation of the differential
  $\lambda$-calculus does not distinguish between equivalent terms, so terms
  like $s + t$ and $t + s$ are not merely equivalent but in fact identical.
\end{proof}

\begin{proposition}
  Erasure is compatible with standard and differential substitution. That is to
  say, for any terms $s, t$ and a variable $x$, we have:
  \[
    \ceil{\subst{s}{x}{t}} = \subst{\ceil{s}}{x}{\ceil{t}}\\
    \ceil{\diffS{s}{x}{t}} = \diffS{\ceil{s}}{x}{\ceil{t}}\\
  \]
\end{proposition}

\begin{corollary}
  Whenever $s \reducesTo s'$, then $\ceil{s} \reducesTo^* \ceil{s'}$.
\end{corollary}

These results form the syntactic obverse to the purely semantic
Theorem~\ref{thm:CDC-is-CdC}: the former exhibits the differential
$\lambda$-calculus as an instance of $\lambda_\epsilon$ where the $\epsilon$
operator is ``degenerate'', whereas the later shows that every Cartesian
differential category can be understood as a ``degenerate'' Cartesian difference
category, in the same sense that the corresponding infinitesimal extension is
just the zero map.

\section{Simple Types for \texorpdfstring{$\lambda_\epsilon$}{Lambda-Epsilon}}
\label{sec:simple-types}

Much like the differential $\lambda$-calculus, $\lambda_\epsilon$ can be endowed
with a system of simple types, built from a set of basic types using the usual
function type constructor. 

\begin{definition} The set of \textbf{types} and \textbf{contexts} of the
  $\lambda_\epsilon$-calculus is given by the following inductive definition:
  \[
    \begin{array}{rccl}
      \text{Types: } &\sigma, \tau& \defeq & \mathbf{t} \alter \sigma \Ra \tau\\
      \text{Contexts: } &\Gamma&\defeq& \emptyset \alter \Gamma, x : \tau
    \end{array}
  \]
  assuming a countably infinite set of basic types $\mathbf{t}, \mathbf{s}
  \ldots$ is given. 
\end{definition}

The typing rules for the $\lambda_\epsilon$-calculus are given in
Figure~\ref{fig:typing-rules} below, and should not be in the least surprising,
as they are identical to the typing rules for the differential
$\lambda$-calculus, with the addition of a typing rule for the infinitesimal
extension of a term. As one would expect, our type system  enjoys all the
``usual'' structural properties and their proofs follow by straightforward
induction on the typing derivation. Note, however, that all of these typing
rules operate on unrestricted terms, rather than on well-formed terms, for
reasons that we will clarify later.

\begin{figure}[h]
  \begin{center}
  \bpt{
    \AxiomC{}
    \UnaryInfC{$\Gamma, x : \tau \ts x : \tau$}
  }
  \bpt{
    \AxiomC{$\Gamma \ts s : \tau \Ra \sigma$}
    \AxiomC{$\Gamma \ts t : \tau$}
    \BinaryInfC{$\Gamma \ts (s\ t) : \sigma$}
  }
  \bpt{
    \AxiomC{$\Gamma, x : \tau \ts t : \sigma$}
    \UnaryInfC{$\Gamma \ts \lambda x . t : \sigma \Ra \tau$}
  }
  \end{center}
  \begin{center}
  \bpt{
    \AxiomC{}
    \UnaryInfC{$\Gamma \ts 0 : \tau$}
  }
  \bpt{
    \AxiomC{$\Gamma \ts s : \tau$}
    \AxiomC{$\Gamma \ts t : \tau$}
    \BinaryInfC{$\Gamma \ts s + t : \tau$}
  }
  \bpt{
    \AxiomC{$\Gamma \ts t : \tau$}
    \UnaryInfC{$\Gamma \ts \epsilon t : \tau$}
  }
  \end{center}
  \begin{center}
  \bpt{
    \AxiomC{$\Gamma \ts s : \tau \Ra \sigma$}
    \AxiomC{$\Gamma \ts t : \tau$}
    \BinaryInfC{$\Gamma \ts \D (s) \cdot t : \tau \Ra \sigma$}
  }
  \end{center}
  \caption{Simple types for $\lambda_\epsilon$}
  \label{fig:typing-rules}
\end{figure}

According to the above rules, typing derivations are
\emph{invertible}, that is to say, whenever $\Gamma \ts t : \tau$ and $t$ is of
the form $s + e$, then it must be the case that $\Gamma \ts s : \tau$ and
$\Gamma \ts e : \tau$, and so on. One property that fails to hold is uniqueness
of typings: indeed the term $0$ admits any type, as do terms such as $0 + 0$ or
$(\lambda x . 0)\ y$.

The following ``standard'' properties also hold, and can be proven by
straightforward induction on the relevant typing derivation.

\begin{proposition}[Weakening]
  Whenever $\Gamma \ts t : \tau$, then for any context $\Sigma$ which is
  disjoint with $\Gamma$ it is also the case that $\Gamma, \Sigma \ts t : \tau$.
\end{proposition}

\begin{proposition}[Substitution]
  Whenever $\Gamma, x : \tau \ts s : \sigma$ and $\Gamma \ts t : \tau$, we have:
  \begin{enumerate}[(i)]
    \item $\Gamma \ts \subst{s}{x}{t} : \sigma$
    \item $\Gamma, x : \tau \ts \diffS{s}{x}{t} : \sigma$
  \end{enumerate}
\end{proposition}

\begin{theorem}[Subject reduction]
  \label{thm:subject-reduction}
  Whenever $\Gamma \ts t : \tau$ and $t \reducesTo t'$ then $\Gamma \ts t' : \tau$.
\end{theorem}

Since we have defined well-formed terms as equivalence classes of unrestricted
terms, we might ask if typing is compatible with this equivalence relation. The
answer is unfortunately no, that is to say, there are ill-typed terms that are
differentially equivalent to well-typed terms. In particular, the term $(0\ t)$
is differentially equivalent to the term $0$, but while the later is trivially
well-typed, the former will not be typable for many choices of $t$ (for example,
whenever $t = (x\ x)$). A weaker version of this property does hold, however,
that makes use of canonicity.

\begin{proposition}
  \label{prop:typing-canonical}
  Whenever $\Gamma \ts t : \tau$, then $\Gamma \ts \can{t} : \tau$, and
  furthermore whenever $\Gamma \ts \can{t} : \tau$ then every canonical form of
  $t$ admits the same type. 
\end{proposition}
\begin{proof}
  The proof proceeds by induction on the typing derivation, by noting that every
  operation involved in canonicalization respects the typing rules. \rff{Should I go
  deeper into this? It is a laborious proof to do in detail, and does not add much.}
\end{proof}

\begin{remark}
  The above issue could have been entirely avoided by circumventing the untyped
  calculus altogether and instead defining and operating on well-typed (unrestricted) 
  terms directly. We have preferred to work out the untyped case first for two reasons:
  first, to mimic the development of the differential $\lambda$-calculus. Second, since
  differentiation of control and fixpoint operators is suspect (in that there is not
  an ``obvious'' choice of a derivative for them), we hope that working in an untyped
  calculus featuring Church encodings and a Y combinator can illustrate what the
  ``natural'' choice for their derivatives should be.
\end{remark}

Before stating a progress theorem for $\lambda_\epsilon$, we must point out one
small subtlety, as the definition of reduction of unrestricted terms depends on
the particular representation chosen for the term. For example, the terms 
$((\lambda x . x) + 0)\ 0$ and $(\lambda x . x)\ 0$ are equivalent, but the first
one contains no $\beta$-redexes, whereas the second one reduces to $0$ in one
step. We can prove that progress holds for canonical terms, however, as those
are ``maximally reducible''.

\begin{definition}
  A canonical term $T$ is a \textbf{canonical value} whenever it is of the form
  \[ T = \sum_{i=1}^i \varepsilon^{k_i}(\lambda x_i . t_i)
  \]
\end{definition}
\rff{
  I've only proven progress for closed terms. It is easy enough to extend this to
  terms with free variables by adding a notion of neutral term etc. but it takes
  time and space and adds nothing - should I write it anyways?
}
\begin{theorem}[Progress]
  Whenever a canonical term $T$ admits a typing derivation $\ts T :
  \tau$, then either $T$ is a canonical value or there is some term $t'$ with $T
  \reducesTo t'$.
\end{theorem}
\begin{proof}
  The proof proceeds by induction on the structure of $T$.
  \begin{itemize}
    \item When $T = 0$ then $T$ is trivially a canonical value.
    \item When $T = \epsilon^k s^\b$, then $s^\b$ has the form $\lambda x .
    e^\b$ or $\D(e^\b) \cdot u^\b$. In the first case $T$ is already a 
    canonical value. In the second case, note that the term $e^\b$ is itself a
    canonical term and a strict subterm of $T$. By inversion of the typing
    rules, we have that $\ts e^\b : \tau$ (note that the type of a differential
    application is the same as the type of its body, unlike the case of standard
    application). Hence either $e^\b reduces$ (and therefore so does $T$),
    or it is a canonical value, i.e. $e^\b$ is of the form $\lambda x . w^\b$.
    But in the last case then $T = \epsilon^k \D(\lambda x . w^\b) \cdot s^\b$
    and therefore $T \reducesTo \epsilon^k \lambda x . \diffS{w^\b}{x}{s^\b}$.
    \item When $T = T_1 + T_2$, then either both $T_1, T_2$ are canonical values,
    and then so is $T$, or one of $T_1, T_2$ reduces, in which case so does $T$.
    \qedhere
  \end{itemize}
\end{proof}

\begin{definition}
  We extend typing judgements to well-formed terms by setting $\Gamma \mathrel{\ul{\ts}} \ul{t} 
  : \tau$ whenever $\Gamma \ts \can{t} : \tau$.
\end{definition}

\begin{corollary}[Subject reduction for well-formed terms]
  \label{cor:well-formed-subject-reduction}
  Whenever $\Gamma \wfTs \ul{t} : \tau$ and $\ul{t} \wfReducesTo
  \ul{t'}$, then $\Gamma \wfTs \ul{t'} : \tau$.
\end{corollary}
\begin{proof}
  Since $\ul{t} \wfReducesTo \ul{t'}$, there is some canonical form $T =
  \can{\ul{t}}$ such that $T \reducesTo t'' \diffEq t'$, and furthermore $\Gamma
  \ts T : \tau$. By definition of 
  Theorem~\ref{thm:subject-reduction}, we have that $\Gamma \ts t'' : \tau$ and
  therefore by Proposition~\ref{prop:typing-canonical} $\Gamma \ts \can{t''} :
  \tau$, from which it follows that $\Gamma \wfTs \ul{t'} : \tau$.
\end{proof}

\begin{corollary}[Progress for well-formed terms]
  \label{cor:well-formed-progress}
  Whenever $\Gamma \wfTs \ul{t} : \tau$ then either $\ul{t} \wfReducesTo
  \ul{t'}$ or every canonical form $\can{\ul{t}}$ is a canonical value. 
\end{corollary}

\subsection{Strong Normalisation}

With our typing rules in place, we set out to show that $\lambda_\epsilon$ is
strongly normalising. Our proof follows the structure of Ehrhard and Regnier's
\cite{ehrhard2003differential} and Vaux's\cite{vaux2006lambda}, which use an
adaptation of the well-known argument by reducibility candidates. Our proof will
be somewhat simpler, however, due to two main reasons: first, we are not
concerning ourselves with terms with coefficients on some general rig; and
second, we have defined unrestricted and canonical terms as inductive types, and
so we can freely use induction on the syntax of our terms. We will need some
auxiliary results, which we prove now.

\begin{lemma}
  Given an unrestricted term $t$, there are only finitely many terms $t'$ such that
  $t \reducesTo t'$.
\end{lemma}
\begin{proof}
  Since our reduction relation is defined by simple induction on the syntax of
  $t$, it suffices to observe that any term $t$ will only contain a finite number
  of applications where reduction may take place.
\end{proof}

\begin{lemma}
  Given a well-formed term $\ul{t}$, there are only finitely many canonical
  terms $T$ such that $T \diffEq \ul{t}$.
\end{lemma}
\begin{proof}
  By Theorem~\ref{thm:canonicity}, we know that any two canonical forms for
  $\ul{t}$ must be permutatively equivalent. But any term has a finite number of
  permutative equivalence classes, hence a term $\ul{t}$ only has finitely many
  canonical forms.
\end{proof}
\DeclarePairedDelimiter\abs{\lvert}{\rvert}%

As a corollary of the two previous results, whenever a well-formed term $\ul{t}$
is strongly normalising, by K\"onig's lemma there is a longest sequence of
well-formed terms $\ul{t} = \ul{t_1}, \ul{t_2}, \ldots, \ul{t_n}$ such that
$\can{t_i} \reducesTo t'_i \diffEq t_{i+1}$. We write $\abs{\ul{t}}$ to indicate
the length of this sequence. The following result is then immediate.

\begin{lemma}
  Whenever $\ul{t}$ is strongly normalising and $\ul{t} \reducesTo \ul{t'}$, we
  have $\abs{\ul{t}} > \abs{\ul{t'}}$.
\end{lemma}

\begin{lemma}
  \label{lem:sum-strongly-normalising}
  A term $\ul{s + t}$ is strongly normalising if and only if
  $\ul{s}, \ul{t}$ are strongly normalising.
\end{lemma}
\begin{proof}
  The proof in the first direction proceeds by induction on $\abs{s + t}$.

  Suppose $\ul{s + t}$ is strongly normalising.  and suppose $S, T$
  are canonical forms for $\ul{s}, \ul{t}$ respectively. Then $S + T$ is
  a canonical form for $\ul{s + t}$ (up to associativity of addition), and
  any sequence of reductions 

  If $\abs{\ul{s + t}} = 0$ it follows that its canonical form $S +
  T$ does not reduce, and hence neither do the separate $S + T$, thus
  $\ul{s}, \ul{t}$ are normal forms.

  On the other hand, suppose $S \reducesTo s', T \reducesTo t'$ then
  $\ul{s + t} \wfReducesTo \ul{s' + t'}$ which is therefore
  strongly normalising, with $\abs{\ul{s' + t'}} < \abs{t}$. Hence, by
  induction, $\ul{s'}$ and $\ul{t'}$ are strongly normalising for any
  choices of canonical forms $S, T$ and reducts $s', t'$.

  In the opposite direction, the proof follows similarly by induction on
  $\abs{s} + \abs{t}$. The base case is equally trivial. For the inductive step,
  since any canonical form for $\ul{s + t}$ is (up to commutativity of addition)
  of the form $S + T$, with $S, T$ being canonical forms for $\ul{s}, \ul{t}$
  respectively, it follows that if $\ul{s + t} \wfReducesTo \ul{e}$. Without
  loss of generality, we assume that $S \wfReducesTo S'$ with $\can{\ul{e}} = S'
  + T$. Then $\ul{s} \wfReducesTo \ul{S'}$ and $\ul{e} = \ul{S' + t}$. Now
  $\abs{\ul{S'}} < \abs{\ul{s}}$, and so we apply the induction hypothesis and
  obtain that $\ul{e}$ must be strongly normalising, and therefore so is $\ul{s
  + t}$.
\end{proof}

\begin{lemma}
  \label{lem:epsilon-strongly-normalising}
  A term $\ul{\epsilon s}$ is strongly normalising if and only if $\ul{s}$ is.
\end{lemma}

\newcommand{\rc}[1]{\mathcal{R}_{#1}}

\begin{definition}
  For every
  type $\tau$ we introduce a set $\rc{\tau}$ of well-formed terms of type
  $\tau$. We do so by induction on $\tau$.
  \begin{itemize}
    \item Whenever $\tau = \mathbf{t}$ is a primitive type, $\ul{s} \in
    \rc{\mathbf{t}}$ if and only if $\ul{s}$ is strongly normalising.
    \item Whenever $\tau = \sigma_1 \Ra \sigma_2$, $\ul{s} \in \rc{\sigma_1 \Ra
    \sigma_2}$ if and only if for any additive term $\ul{t^*}
    \in \rc{\sigma_1}$ and for any sequence
    $\ul{v^\b_1}, \ldots, \ul{v^\b_n}$ of 
    basic terms $\ul{v^\b_i} \in \rc{\sigma_1}$ 
    of length $n \geq 0$
    we have $\ul{\pa{\D^n(s) \cdot (v^\b_1, \ldots, v^\b_i)}\ t^*} \in \rc{\sigma_2}$ 
  \end{itemize}

  If $\ul{t} \in \rc{\tau}$ we will often just say that $\ul{t}$
  is \textbf{reducible} if the choice of $\tau$ is clear from the context.
\end{definition}

\begin{lemma}
  \label{lem:rc-renaming}
  Whenever $\ul{t} \in \rc{\tau}$, then for any two distinct variables $x, y$
  the renaming $\ul{\subst{t}{x}{y}}$ is also in $\rc{\tau}$.
\end{lemma}
\begin{proof}
  Straightforward induction on $\tau$.
\end{proof}

\begin{lemma}
  \label{lem:rc-strongly-normalising}
  Whenever $\ul{t} \in \rc{\tau}$, then $\ul{t}$ is strongly normalising.
\end{lemma}
\begin{proof}
  By induction on $\tau$. When $\tau$ is a primitive type, the result follows
  trivially. Let $\tau = \sigma_1 \Ra \sigma_2$ and $\ul{t} \in
  \rc{\tau}$. By the induction hypothesis we know that for all $\ul{u} \in
  \rc{\sigma_2}$ the application $\ul{t\ u}$ is strongly normalising.

  Now suppose some canonical form of $\ul{t}$ reduces, that is, $T \reducesTo
  t'$. Since $\can{t\ u^*} = \ap{\can{t}}{\pri{u^*}} = \ap{T}{{\can{u^*}}}$, it
  follows that $\can{t\ u} \reducesTo^+ \ap{t'}{\pri{u}}$. Hence if there were
  any infinite sequence of reductions starting from $\ul{t}$, so would there be
  an infinite sequence of reductions starting from $\ul{t\ u}$. Since $\ul{t\ u}$
  is strongly normalising, this must be impossible and so $\ul{t}$ must be
  strongly normalising as well.
\end{proof}

\begin{lemma}
  \label{lem:rc-sum}
  Whenever $\ul{s}, \ul{t} \in \rc{\tau}$, then both $\ul{s + t},
  \ul{\varepsilon{s}}$ are in $\rc{\tau}$. Conversely, whenever $\ul{s + t}$
  is in $\rc{\tau}$ then so are $\ul{s}, \ul{t}$.
\end{lemma}
\begin{proof}
  When $\tau$ is a primitive type the proof is a straightforward corollary of 
  Lemmas~\ref{lem:sum-strongly-normalising}
  and~\ref{lem:epsilon-strongly-normalising}.

  When $\tau = \sigma_1 \Ra \sigma_2$, consider an additive term $\ul{u^*} \in
  \rc{\sigma_1}$. We ask whether the application $\ul{(s + t)\ u^*}$ is in
  $\rc{\sigma_2}$. But note that $\can{(s + t)\ u^*}$ is equal to $\can{s\ u^*} +
  \can{t\ u^*}$ (modulo commutativity of the sum) and therefore $\ul{(s + t)\ u^*}
  = \ul{s\ u^*} + \ul{t\ u^*}$. Since $\ul{s\ u^*}$ and $\ul{t\ u^*}$ are both in
  $\rc{\sigma_2}$, it follows by the induction hypothesis that so is $\ul{(s +
  t)\ u^*}$. The same reasoning shows that $\ul{(\D(s + t) \cdot v^\b)\ u^*}$ is
  in $\rc{\sigma_2}$.
  
  The proof for $\varepsilon$ follows by a simpler but otherwise identical
  procedure.

  On the opposite direction, consider a sum $\ul{s + t} \in \rc{\tau}$. When
  $\tau$ is a primitive type then $\ul{s + t}$ is strongly normalising and
  therefore so are $\ul{s}, \ul{t}$, hence they are regular. On the other hand,
  if $\tau = \sigma_1 \Ra \sigma_2$, we have that for any $e^* \in
  \rc{\sigma_1}$ the reducible $\ul{(s + t)\ e^*}$ is equal to $\ul{(s\ e^*) +
  (t\ e^*)}$. By the induction hypothesis, both $(s\ e^*)$ and $(t\ e^*)$ are
  regular. A similar argument proves that differential applications of $s$ and
  $t$ are also regular, thus $s, t \in \rc{\sigma_1 \Ra \sigma_2}$.
\end{proof}

\begin{lemma}
  \label{lem:rc-d}
  Whenever $\ul{s} \in \rc{\sigma \Ra \tau}$ and $\ul{t} \in \rc{\sigma}$ then
  $\ul{\D (s) \cdot t} \in \rc{\sigma \Ra \tau}$.
\end{lemma}
\begin{proof}
  Pick a canonical form $T$ of $t$. The proof proceeds by induction on the
  number of summands of $T$. If $T = 0$ or $T = \epsilon^k t^\b$ the result
  follows directly by definition of $\rc{\sigma \Ra \tau}$.

  Now suppose $T = \epsilon^k t^\b + T'$. Then
  \begin{salign*}
    \D(s) \cdot t &\diffEq \D(s) \cdot (t^\b + T')\\
    &\diffEq \D(s) \cdot t^\b + \D(s) \cdot T'
    + \varepsilon \D(\D(s) \cdot T') \cdot t^\b
  \end{salign*}
  Now $\D(s) \cdot t^\b$ is evidently reducible (as $s$ is reducible by
  hypothesis and, by Lemma~\ref{lem:rc-sum}, $t^\b$ is also reducible), and by
  the induction hypothesis so is $\D(s) \cdot T'$, from which also follows that
  $\D(\D(s) \cdot T') \cdot t^\b$ is reducible as well. By
  Lemma~\ref{lem:rc-sum} it follows that $\ul{t}$ is reducible.
\end{proof}

\begin{corollary}
  A well-formed term $\ul{t}$ is in $\rc{\tau}$ if and only if some canonical
  form $T = \can{\ul{t}}$ is of the form $\sum_{i=1}^n\varepsilon^{k_i}t^\b_i$
  with $\ul{t^\b_i} \in \rc{\tau}$ for each $1 \le i \le n$.
\end{corollary}

\begin{lemma}
  Whenever $\ul{t} \in \rc{\tau}, \ul{t} \wfReducesTo^+ \ul{t'}$, then $\ul{t'}
  \in \rc{\tau}$.
\end{lemma}
\begin{proof}
  We proceed by induction on $\tau$.
  When $\tau$ is a primitive type, we have that $\ul{t}$ is strongly normalising
  and therefore so is $\ul{t'}$, hence $\ul{t'} \in \rc{\tau}$.

  When $\tau = \sigma_1 \Ra \sigma_2$, we pick some additive reducible term
  $\ul{e^*}$ and a sequence of reducible basic terms $\ul{u^\b_1}, \ldots,
  \ul{u^\b_k}$, and establish that $\pa{\D^k(t') \cdot(u^\b_1, \ldots,
  u^\b_k)}\ e^*$ is reducible. But this is immediate: since $\ul{t}$ reduces to
  $\ul{t'}$, then so does $\ul{\D^k(t) \cdot (u^\b_1, \ldots, u^\b_k)\ e^*}$
  reduce to $\ul{\D^k(t') \cdot (u^\b_1, \ldots, u^\b_k)\ e^*}$ and, by induction,
  $\ul{\D^k(t') \cdot(u^\b_1, \ldots, u^\b_k)\ e^*}$ is reducible.
\end{proof}

\begin{definition}
  A basic term $t^\b$ is \textbf{neutral} whenever it is not a
  $\lambda$-abstraction. In other words, a basic term is neutral whenever it is
  of the form $x, (s\ t)$ or $\D(s)\cdot u$.

  A canonical term $T$ is neutral whenever it is of the form
  $\sum_{i=1}^n\varepsilon^{k_i} s_i^\b$, where each of the $s_i^\b$ are
  neutral. In particular, $0$ is a neutral term.

  A well-formed term $\ul{t}$ is neutral whenever some canonical form (and
  therefore all of its canonical forms) is neutral.
\end{definition}

\begin{lemma}
  \label{lem:rc-neutral}
  Whenever $\ul{t}$ is neutral and every $\ul{t'}$ such that $\ul{t}
  \wfReducesTo^+ \ul{t'}$ is in $\rc{\tau}$, then so is $\ul{t}$.
\end{lemma}
\begin{proof}
  When $\tau$ is a primitive type the proof is immediate, as $\ul{t'} \in
  \rc{\tau}$ implies $\ul{t'}$ is strongly normalising and therefore so is
  $\ul{t}$.

  When $\tau = \sigma_1 \Ra \sigma_2$, we show the reasoning for standard
  application first. We select arbitrary $\ul{e^*}, \ul{u^\b_1}, \ldots,
  \ul{u^\b_k} \in \rc{\sigma_1}$ and show that whenever $\pa{\D^k(t) \cdot
  (u^\b_1, \ldots, u^\b_k)}\ e^*$ reduces then it reduces to a reducible term
  (hence our desired result will follow by induction on $\tau$). 

  We prove this property by induction on $Q \defeq \abs{\ul{e^*}} +
  \abs{\ul{u^\b_1}} + \ldots + \abs{\ul{u^\b_k}}$ which is well-defined since,
  by hypothesis, all of the involved terms are strongly normalising. 
  
  When $Q = 0$ then all of our chosen terms are normal, and so, since $t$ is
  neutral, if $\ul{\pa{\D^k(t)\cdot(u^\b_1, \ldots, u^\b_k)}\ e^*}$ reduces it
  must be that $\ul{t} \wfReducesTo^+ \ul{t'}$. By hypothesis, $\ul{t'} \in
  \rc{\sigma_1 \Ra \sigma_2}$ and therefore $\ul{\pa{\D^k(t')\cdot(u^\b_1,
  \ldots, u^\b_k)}\ e^*}$ is reducible.

  When $Q > 0$, then a reduction may occur in $t$, in which case we apply the
  previous reasoning, or in one of the applied terms, in which case we apply the
  induction hypothesis on $Q$.
  induction hypothesis.
\end{proof}

\begin{lemma}
  \label{lem:rc-lambda}
  If, for all $\ul{t^*} \in \rc{\sigma_1}$ where $x$ does not appear free, the
  term $\ul{\subst{s}{x}{t^*}}$ is in $\rc{\sigma_2}$ and, for all $\ul{u^\b}$
  where $x$ does not appear free, the term
  $\ul{\subst{\pa{\diffS{s}{x}{u^\b}}}{x}{t^*}}$ is in $\rc{\sigma_2}$, then the
  term $\ul{\lambda x . s}$ is in $\rc{\sigma_1 \Ra \sigma_2}$.
\end{lemma}
\begin{proof}
  As a corollary of Lemma~\ref{lem:rc-sum}, it suffices to check the case
  when $s$ is some basic term $s = s^\b$. 
  
  Pick any variable $y \neq$ z. Since the variable $y$ itself is an additive
  term in $\rc{\sigma_1}$, by hypothesis we have $\ul{\subst{s}{x}{y}} \in
  \rc{\sigma_2}$. But since reducible terms are closed under renaming as per
  Lemma~\ref{lem:rc-renaming}, this means that so is $\ul{s}$.

  Now consider an arbitrary $\ul{e^*} \in \rc{\sigma_1}$. We show that the
  application $\ul{(\lambda x . s^\b)\ e^*}$ is in $\rc{\sigma_2}$. As it is a
  neutral term, by Lemma~\ref{lem:rc-neutral} it suffices to prove that every
  one-step reduct of $(\lambda x . s^\b)\ e^*$ is in $\rc{\sigma_2}$. We do so by
  induction on $\abs{s^\b} + \abs{e^*}$. The term $(\lambda x . s^\b)\ e^*$
  reduces to one of:
  \begin{itemize}
    \item $\subst{s^\b}{x}{e^*}$, which by hypothesis is a representative of a
    term in $\rc{\sigma_2}$.

    \item $(\lambda x . s')\ e^*$ with $s^\b \reducesTo s'$. Then $s' \in
    \rc{\sigma_2}$ and since $\abs{s'} < \abs{s^\b}$ we apply our induction on
    $\abs{s^\b}$ to obtain that $(\lambda x . s')\ e^* \in \rc{\sigma_2}$.

    \item $(\lambda x . s^\b)\ e'$ with $e^* \reducesTo e'$. Then $e' \in
    \rc{\sigma_1}$ and since $\abs{e'} < \abs{e^*}$ we apply our induction on
    $\abs{e^*}$ to obtain that $(\lambda x . s^\b)\ e' \in \rc{\sigma_2}$.
  \end{itemize}

  By a similar argument we can show that $\ul{
    \pa{\D^k(\lambda x . s^\b)\cdot(u^\b_1, \ldots, u^\b_k)}\ e^*
  }$ is reducible, applying Lemma~\ref{lem:rc-neutral} and using induction
  in $\abs{e^*} + \abs{u^\b_1} + \ldots + \abs{u^\b_k}$. 
\end{proof}

\begin{theorem}
  Consider a well-formed term $\ul{t}$ which admits a typing of the form $x_1 :
  \sigma_1, \ldots, x_n : \sigma_n \wfTs \ul{t} : \tau$ and assume given the
  following data:
  \begin{itemize}
    \item A sequence of basic terms 
      $\ul{d^\b_1} \in \rc{\sigma_1}, \ldots, \ul{d^\b_n} \in \rc{\sigma_n}$.
    \item An arbitrary sequence of indices 
      $i_1, \ldots, i_k \in \{1, \ldots, n\}$ (possibly with repetitions).
    \item A sequence of additive terms 
      $\ul{s^*_1} \in \rc{\sigma_{i_1}}, \ldots, \ul{s^*_k} \in \rc{\sigma_{i_k}}$.
  \end{itemize}
  such that none of the variables $x_1, \ldots, x_i$ appear free in the $d^\b_i,
  s^*_i$. Then the term 
  \[
    \ul{t'} = \ul{
      \pa{
        \frac{\partial^k t}{\partial(x_{i_1}, \ldots, x_{i_k})}
        (d^\b_1, \ldots, d^\b_k)}
      \left[s^*_1, \ldots, s^*_n / x_1, \ldots, x_n\right]}
  \]
  is in $\rc{\tau}$.
\end{theorem}
\begin{proof}
  \newcommand{\thediff}[1]{
      \pa{\partial^k t / \partial x_{i_1} \ldots \partial x_{i_k}}
      \pa{d^\b_1, \ldots, d^\b_k}
  }
  \newcommand{\thesubs}[1]{
    {#1}\brak{s^*_1, \ldots, s^*_n / x_1, \ldots, x_n}
  }
  \newcommand{\thesubss}[1]{
    \subst{#1}{\overline{x_i}}{\overline{s^*_i}}
  }
  \newcommand{\thesd  }[1]{
    \thesubs{\pa{\thediff{#1}}}
  }
  Throughout the proof we will write $\overline{x_i}, \overline{s^*_i},
  \overline{d^\b_i}$ as a shorthand for the corresponding sequences $x_1,
  \ldots, x_n$, etc.

  By definition of $\wfTs$, we know that there is some canonical form $T$ (in
  fact any canonical form) of $\ul{t}$ such that $x_1 : \sigma_1, \ldots, x_n :
  \sigma_n \ts T : \tau$. We prove our property holds by induction on this
  typing derivation. Furthermore, by Lemma~\ref{lem:rc-sum}, it suffices to
  consider the case when $T$ is in fact some basic term $t^\b$. 
  We proceed now by case analysis on the last rule of the typing derivation:

  \begin{itemize}
    \item $t^\b = x_i$ (and therefore $\tau = \sigma_i$)
    
      If the sequence of indices $i_1, \ldots, i_k$ is empty then the
      substitution $\ul{t'}$ is exactly equal to $\ul{s^*_i}$ and therefore
      $\ul{t'} \in \rc{\sigma_i}$.

      If the sequence of indices $i_1, \ldots, i_k$ is exactly the sequence
      containing only $i$ then since $x_i$ does not appear free in the
      substituted term $d^\b$ then $t'$ is differentially equivalent to
      $\pa{\diffS{x_i}{x_i}{d^\b_i}}$ and therefore $\ul{t'} = \ul{d^\b_i} \in
      \rc{\sigma_i}$.

      If the list of indices contains two or more indices, or does not contain
      $i$, then $t' \diffEq 0$ and therefore $\ul{t'}$ is trivially in $\in
      \rc{\sigma_i}$ (either the derivative)

    \item $t^\b = \D(s^\b) \cdot e^\b$
    
      Applying Lemma~\ref{lem:differential-substitution-dapp}, we know that
      the term \[ \thesd{t^\b} \] is equivalent to a sum of terms of the
      form
      \[
        \pa{\epsilon^z \D^l \pa{\thesubss{v}}
        \cdot \pa{\thesubss{w_1}, \ldots, \thesubss{w_l}}}
      \]
      Again by Lemmas~\ref{lem:rc-sum} and~\ref{lem:rc-d}, it suffices to show
      that each of the $\thesubss{v}$, $\thesubss{w_j}$ are reducible. But
      by Lemma~\ref{lem:differential-substitution-app}, we know that $v$ has
      the form
      \[
        \frac{\d^p s^\b}{\d (x_{j_1}, \ldots, x_{j_m})}\pa{
          d^\b_{j_1}, \ldots,
          d^\b_{j_m}
        }
      \]

      Since $s^\b$ is a subterm of $t^\b$ its typing derivation is therefore a
      sub-derivation of the one for $t^\b$. We apply the induction
      hypothesis, obtaining that $\ul{\thesubss{v}}$ is reducible
      (as each of the $d^\b_i$ are reducible). By a similar argument,
      each of the $\ul{\thesubss{w_j}}$ are reducible as well, and therefore
      so is $\ul{t^\b}$.

    \item $t^\b = (s^\b\ e^*)$
    
      Applying Lemma~\ref{lem:differential-substitution-dapp}, we know that
      the term \[ \thesd{t^\b} \] is equivalent to a sum of terms of the
      form
      \[
        \pa{\epsilon^z \D^l \pa{\thesubss{v}}
        \cdot \pa{\thesubss{w_1}, \ldots, \thesubss{w_l}}}\ \pa{\thesubss{e^*}}
      \]
      Again by Lemma~\ref{lem:rc-sum} it suffices to show that every such term
      is reducible. First, by Lemma~\ref{lem:differential-substitution-app}, we
      know that $v$ has the form
      \[
        \frac{\d^p s^\b}{\d (x_{j_1}, \ldots, x_{j_m})}\pa{
          d^\b_{j_1}, \ldots,
          d^\b_{j_m}
        }
      \]
      Since $s^\b$ is a subterm of $t^\b$ its typing derivation is therefore a
      sub-derivation of the one for $t^\b$. We apply the induction hypothesis,
      obtaining that $\ul{\thesubss{v}}$ is reducible (as each of the $d^\b_i$
      are reducible). By a similar argument, each of the $\ul{\thesubss{w_j}}$
      are reducible as well, as is $\ul{\thesubss{e^*}}$. Thus, by
      Lemma~\ref{lem:rc-d}, the entire differential application is reducible.
      
      But since $t^\b$ is an application of the reducible term
      \[ \epsilon^z \D^l \pa{\thesubss{v}} \cdot \pa{\thesubss{w_1}, \ldots,
      \thesubs{w_l}} \] to the reducible term $\thesubs{e^*}$, it follows then
      that $\ul{t^\b}$ is itself reducible.

    \item $t^\b = \lambda y . s^\b$
    
      Pick fresh variables $z_1, \ldots, z_n$. By the induction hypothesis, we
      know that both of the following substitutions are reducible:
      \begin{gather*}
        \subst{\pa{t^\b}}{x_1, \ldots, x_n, y}{z_1, \ldots, z_n, e^*}\\
        \subst{\pa{\diffS{t^\b}{y}{d^\b}}}{x_1, \ldots, x_n, y}{z_1, \ldots, z_n, e^*}
      \end{gather*}
      But since reducible terms are closed under renaming, then the terms
      $\subst{t^\b}{y}{e^*}$ and $\subst{\pa{\diffS{t^\b}{y}{d^\b}}}{y}{e^*}$
      are also reducible. Hence, by Lemma~\ref{lem:rc-lambda}, the term 
      $\lambda y . s^\b$ is reducible.\qedhere
  \end{itemize}
\end{proof}

\begin{corollary}[Strong normalisation]
  Whenever a closed well-formed term is
  typable with type $\wfTs \ul{t} : \tau$, it is strongly normalising.
\end{corollary}
\begin{proof}
  By the previous result, we know that any such term satisfies $\ul{t} \in
  \rc{\tau}$, and therefore $\ul{t}$ is strongly normalising.
\end{proof}

\section{Semantics}
\label{sec:semantics}

It is a well-known result that the differential $\lambda$-calculus can be
soundly interpreted in any differential $\lambda$-category, that is to say, any
Cartesian differential category where differentiation ``commutes with''
abstraction (in the sense of \cite[Definition~4.4]{bucciarelli2010categorical}).

The exact same result holds for the difference $\lambda$-calculus and difference
$\lambda$-categories. In what follows we will consider a fixed difference
$\lambda$-category $\cat{C}$, and proceed to define interpretations for the
types, contexts and terms of the simply-typed $\lambda_\epsilon$-calculus.

\begin{definition}
  Given a $\mathbf{t}$-indexed family of objects $O_\mathbf{t}$, we define the
  interpretation $\sem{\tau}$ of a type $\tau$ by induction on its structure by
  setting $\sem{\mathbf{t}} \defeq O_\mathbf{t}$, $\sem{\sigma \Ra \tau} \defeq
  \sem{\sigma} \Ra \sem{\tau}$. We lift the interpretation of types to contexts
  in the usual way. Or, more formally, we have: $\sem{\cdot} \defeq \terminal$,
  $\sem{\Gamma, x : \tau} \defeq \sem{\Gamma} \times \sem{\tau}$.
\end{definition}

\begin{definition}
  Given a well-typed unrestricted $\lambda_\epsilon$-term
  $\Gamma \ts t : \tau$, we define its interpretation $\sem{t} : \sem{\Gamma}
  \ra \sem{\tau}$ inductively as in Figure~\ref{fig:semantics} below. When
  $\Gamma$ and $\tau$ are irrelevant or can be inferred from the context, we
  will simply write $\sem{t}$.
  \begin{figure}[H]
  \[\def\arraystretch{1.3}
    \begin{array}{rccl}
      \sem{(x_i : \tau_i)_{i = 1}^n \ts x_k : \tau_k} 
      &\defeq& 
      \pi_2 \circ \pi_1^{n - k} 
      &: \prod_{i=1}^n\sem{\tau_i} \to \sem{\tau_k}
      \\
      \sem{\Gamma \ts 0 : \tau} &\defeq&
      0 
      &: \sem{\Gamma} \to \sem{\tau}
      \\
      \sem{\Gamma \ts s + t : \tau} &\defeq&
      \sem{s} + \sem{t} 
      &: \sem{\Gamma} \to \sem{\tau}
      \\
      \sem{\Gamma \ts \epsilon t : \tau} &\defeq&
      \epsilon \sem{t} 
      &: \sem{\Gamma} \to \sem{\tau}
      \\
      \sem{\Gamma \ts \lambda x . t : \sigma \Ra \tau} &\defeq&
      \Lambda \sem{t} 
      &:\sem{\Gamma} \to \sem{\sigma} \Ra \sem{\tau}
      \\
      \sem{\Gamma \ts (s\ t) : \tau} &\defeq&
      \A{ev} \circ \pair{\sem{s}}{\sem{t}} 
      &:\sem{\Gamma} \to \sem{\tau}
      \\
      \sem{\Gamma \ts \D(s) \cdot t : \sigma \Ra \tau} &\defeq&
      \Lambda(\d[\Lambda^-(\sem{s})] \circ \pair{\Id{}}
      {\pair{0}{\sem{t} \circ \pi_1}})
      &: \sem{\Gamma} \to \sem{\tau}
    \end{array}
  \]
  \caption{Interpreting $\lambda_\epsilon$ in $\cat{C}$}
  \label{fig:semantics}
  \end{figure}
\end{definition}

\begin{lemma}
  \[
    \d[\Lambda^-f] \circ \pair{\Id{}}{\pair{0}{g \circ \pi_1}}
    = \d[\A{ev}] \circ \pair{}{} \circ
  \]
\end{lemma}

\newcommand{\semEq}[0]{\sim_{\llbracket\rrbracket}}
\begin{lemma}
  Define the relation $\semEq \subseteq \Lambda_\epsilon$ by letting $s \semEq
  t$ whenever there exist $\Gamma, \tau$ such that
  $\sem{\Gamma \ts s : \tau} = \sem{\Gamma \ts t : \tau}$. Then the relation
  $\semEq{}$ is contextual.
\end{lemma}

\begin{theorem}
  Whenever $s \diffEq t$ are equivalent unrestricted terms that admit typing
  derivations $\Gamma \ts s : \tau, \Gamma \ts t : \tau$, then their
  interpretations are identical, that is to say:
  \[
    \sem{\Gamma \ts s : \tau} = \sem{\Gamma \ts t : \tau}
  \]
\end{theorem}
\begin{proof}
  The result can be equivalently stated as ``$\semEq$ contains $\diffEq$''.
  Since $\diffEq$ is defined as the least contextual equivalence relation that
  contains $\diffEq^1$, and $\semEq$ is both contextual and an equivalence
  relation, it suffices to prove that it contains $\diffEq^1$.

  We examine the rules in Figure~\ref{fig:differential-equivalence} and show
  that they all hold, by checking that each rule verifies $\sem{\LHS} =
  \sem{\RHS}$. The first and second blocks are trivial, as the rules correspond
  precisely to stating that $\cat{C}$ is a Cartesian closed left-additive
  category with an infinitesimal extension which is compatible with the
  Cartesian structure. The third block is a trivial consequence of \CdCAx{1} and
  so we will omit it as well, but we give explicit proofs for some
  of the properties of the fourth block.

  First we show a general equivalence between syntactic and semantic second
  derivatives which will simplify the task considerably.
  \begin{salign*}
    &\hspace{-10pt}\Lambda^-(\sem{\D(\D(s) \cdot u) \cdot v})
    \\
    &= \d[\Lambda^-(\sem{\D(s)\cdot u})] \circ \pair{\Id{}}{\pair{0}{\sem{v} \circ \pi_1}}
    \\
    &= 
    \d[\d[\Lambda^-(\sem{s})] \circ \pair{\Id{}}{\pair{0}{\sem{u} \circ \pi_1}}]
    \pair{\Id{}}{\pair{0}{\sem{v} \circ \pi_1}}
    \\
    &=
    \d[\d[\Lambda^-(\sem{s})]] 
    \circ 
    \pair{\pair{\Id{}}{\pair{0}{\sem{u} \circ \pi_1}} \circ \pi_1}
    {\d[\pair{\Id{}}{\pair{0}{\sem{u}\circ \pi_1}}]}
    \circ
    \pair{\Id{}}{\pair{0}{\sem{v}\circ \pi_1}}
    \\
    &=
    \d[\d[\Lambda^-(\sem{s})]]
    \\
    &\quad
    \circ 
    \pair
    {\pair{\Id{}}{\pair{0}{\sem{u} \circ \pi_1}}}
    {\pair{\d[\Id{}] \circ \pair{\Id{}}{\pair{0}{\sem{v} \circ \pi_1}}}
          {\d[\sem{v} \circ \pi_1] \circ \pair{\Id{}}{\pair{0}{\sem{v} \circ \pi_1}}}}
    \\
    &= 
    \d[\d[\Lambda^-(\sem{s})]]
    \circ 
    \pair
    {\pair{\Id{}}{\pair{0}{\sem{u} \circ \pi_1}}}
    {\pair{\pair{0}{\sem{v} \circ \pi_1}}
          {\d[\sem{v}] \circ \pair{\pi_1}{0}}}
    \\
    &= 
    \d[\d[\Lambda^-(\sem{s})]]
    \circ 
    \pair
    {\pair{\Id{}}{\pair{0}{\sem{u} \circ \pi_1}}}
    {\pair{\pair{0}{\sem{v} \circ \pi_1}}{0}}
  \end{salign*}

  With the above, most of the conditions become trivial. For example, we can
  prove that regularity of the syntactic derivative follows from semantic
  regularity (that is to say, \CdCAx{2}) with the following calculation:
  \begin{itemize}
    \item $\D(s) \cdot (u + v) \diffEq^1 \D(s) \cdot u + \D(s) \cdot v
    + \epsilon(\D(\D(s) \cdot u)\cdot v)$
    \begin{salign*}
      \sem{\D(s) \cdot (u + v)} &=
      \Lambda(\d[f] \circ \pair{\Id{}}{\pair{0}{\sem{u} \circ \pi_1 + \sem{v} \circ \pi_1}})
      \\
      &=
      \Lambda(\d[f] \circ \pair{\Id{}}{\pair{0}{\sem{u} \circ \pi_1} + \pair{0}{\sem{v} \circ \pi_1}})\\
      &=
      \Lambda\big[
                (\d[f] \circ \pair{\Id{}}{\pair{0}{\sem{u} \circ \pi_1}})\\
      &\quad
        + (\d[f] \circ \pair{\Id{} + \epsilon(\pair{0}{\sem{u} \circ \pi_1})}{\pair{0}{\sem{v} \circ \pi_1}})
      \big]
      \\
      &=
      \Lambda\brak{
                \d[f] \circ \pair{\Id{}}{\pair{0}{\sem{u} \circ \pi_1}} }
      \\&\quad
        + \Lambda \brak{(\d[f] \circ \pair{\Id{}}{\pair{0}{\sem{v} \circ \pi_1}})}
      \\&\quad
        + \epsilon\Lambda\brak{\d[\d[f]] \circ 
        \pair{\pair{\Id{}}{\pair{0}{\sem{v} \circ \pi_1}}}{\pair{\pair{0}{\sem{u} \circ \pi_1}}{0}}
        }
      \big]
      \\
      &=
      \Lambda\brak{
                \d[f] \circ \pair{\Id{}}{\pair{0}{\sem{u} \circ \pi_1}} }
      \\&\quad
        + \Lambda \brak{(\d[f] \circ \pair{\Id{}}{\pair{0}{\sem{v} \circ \pi_1}})}
      \\&\quad
        + \epsilon\Lambda\brak{\d[\d[f]] \circ 
        \pair{\pair{\Id{}}{\pair{0}{\sem{u} \circ \pi_1}}}{\pair{\pair{0}{\sem{v} \circ \pi_1}}{0}}
        }
      \big]
      \\
      &= \sem{\D(s) \cdot u)} + \sem{\D(s) \cdot v} + \sem{\varepsilon(\D (\D(s) \cdot u) \cdot v)}
    \end{salign*}
  \end{itemize}
  Most of the other conditions follow from similar arguments. For example,
  commutativity of the syntactic second derivative follows from commutativity of
  the semantic second derivative. The third and fifth conditions, dealing with
  infinitesimal extensions, may seem harder to prove, but they are both
  corollaries of axiom~\CdCAx{6}, as we showed in Lemma~\ref{lem:d-epsilon}.
  It remains to show
  that the syntactic derivative condition holds; this is not hard, but we do
  it explicitly as the derivative condition is such a central notion.

  \begin{itemize}
    \item $s\ (t + \varepsilon e) \diffEq^1 (s\ t) + \varepsilon((\D(s) \cdot
    e)\ t)$
    \begin{salign*}
      &\hspace{-10pt}\sem{s\ (t + \varepsilon e)}\\
      &= \A{ev} \circ \pair{\sem{s}}{\sem{t} + \varepsilon\sem{e}}\\
      &= \Lambda^-\sem{s} \circ \pair{\Id{}}{\sem{t} + \varepsilon\sem{e}}\\
      &= \pa{\Lambda^-\sem{s} \circ \pair{\Id{}}{\sem{t}}}
        + \epsilon\pa{\d[\Lambda^-\sem{s}] \circ 
        \pair{\pair{\Id{}}{\sem{t}}}{\pair{0}{\sem{e}}}}
      \\
      &= \sem{s\ t}
        + \epsilon\brak{
        \pa{\d[\Lambda^-\sem{s}]
        \circ \pair{\Id{}}{\pair{0}{\sem{e} \circ \pi_1}}}
        \circ \pair{\Id{}}{\sem{t}}
      }
      \\
      &= \sem{s\ t}
        + \epsilon\brak{
          \A{ev} \circ \pair{
        \Lambda\pa{\d[\Lambda^-\sem{s}]
        \circ \pair{\Id{}}{\pair{0}{\sem{e} \circ \pi_1}}}}
        {\sem{t}}
      }
      \\
      &= \sem{s\ t}
        + \epsilon\brak{
          \A{ev} \circ \pair{
            \sem{\D(s) \cdot e}
          }
        {\sem{t}}
      }
      \\
      &= \sem{s\ t} + \epsilon\sem{(\D(s) \cdot e)\ t}\qedhere
    \end{salign*}
  \end{itemize}
\end{proof}

\begin{lemma}
  Let $t$ be some unrestricted $\lambda_\epsilon$-term. The following properties
  hold:
  \begin{enumerate}[i.]
    \item If $\Gamma \ts t : \tau$ and $x$ does not appear in $\Gamma$ then 
    $\sem{\Gamma, x : \sigma \ts t : \tau} = \sem{\Gamma \ts t : \tau} \circ
    \pi_1$
    \item If $\Gamma, x : \sigma_1, y : \sigma_2 \ts t : \tau$ then
    $\sem{\Gamma, y:\sigma_2, x:\sigma_1 \ts t : \tau}
    = \sem{\Gamma, x : \sigma_1, y : \sigma_2 \ts t : \tau} \circ \A{sw}$
  \end{enumerate}
  The morphism $\A{sw}$ above is the obvious isomorphism between $(A \times B)
  \times C$ and $(A \times C) \times B$, which we can define explicitly by:
  \[
    \A{sw} \defeq \pair{\pair{\pi_{11}}{\pi_2}}{\pi_{21}}
    : (A \times B) \times C \ra (A \times C) \times B
  \]
\end{lemma}

\begin{lemma}
  Let $\Gamma, x : \tau \ts s : \sigma$, with $s$ some unrestricted
  $\lambda_\epsilon$-term. Then:
  
  \begin{enumerate}[i.]
  \item Whenever $\Gamma, x : \tau \ts t : \tau$, then
  $\sem{\subst{s}{x}{t}}_\Gamma = \sem{s}_{\Gamma, x : \tau} \circ
  \pair{\pi_1}{\sem{t}_{\Gamma, x : \tau}}$
  
  \item Whenever $\Gamma \ts t : \tau$, then $\sem{\diffS{s}{x}{t}}_{\Gamma,
  x : \tau} =
  \d[\sem{s}_{\Gamma, x : \tau}] 
  \circ \pair{\Id{}}{\pair{0}{\sem{t}_\Gamma \circ \pi_1}}$. Or, using the
  notation in Definition~\ref{def:star-composition}, $\sem{\diffS{s}{x}{t}} =
  \sem{s} \star \sem{t}$.
  \end{enumerate}
\end{lemma}
\begin{proof}
  The proof follows roughly the structure of
  \cite[Theorem~4.11]{bucciarelli2010categorical}, taking into account the
  differences in our notion of differential substitution. Note also that we
  prove substitution in the case that the variable $x$ is not free in $t$. This
  is because we require this (stronger) form of substitution to write
  $\subst{e}{x}{x + \varepsilon(t)}$ in some cases of differential substitution.
  
  Both properties will follow by induction on the typing derivation of $s$. The
  only non-trivial case for the first one is differential application. For this,
  we must show that $\sem{\D(\subst{s}{x}{t}) \cdot (\subst{u}{x}{t})}$ is equal
  to $\sem{\D(s) \cdot u} \circ \pair{\Id{}}{\sem{t}}$. Expanding the term we obtain:
  \begin{salign*}
    \sem{\D(\subst{s}{x}{t}) \cdot (\subst{u}{x}{t})}
    &= \Lambda \Big(
      \Lambda^-(\sem{\subst{s}{x}{t}}) \star \sem{\subst{u}{x}{t}}
    \Big)
    \PROOFSTEP
    &= \Lambda \Big(\Lambda^-(\sem{s} \circ \pair{\pi_1}{\pair{0}{\sem{u}}})
     \star
      (\sem{u} \circ \pair{\pi_1}{\pair{0}{\sem{u}}})
    \Big)
  \end{salign*}
  By Lemma~\ref{lem:lambda-star-ev}(iii.), the above expression can be written
  as:
  \[ \Lambda \Big( (\Lambda^-\sem{s}) \star \sem{u} \Big) \circ \pair{\pi_1}{\sem{t}} \]
  which concludes the proof.

  We show now the cases for differential substitution.
  \begin{itemize}
    \item $s = x$
      Then $\sem{s} = \pi_2$ and 
      \begin{salign*}
        \hspace{-30pt}
        \d[\sem{s}] \circ \pair{\Id{}}{\pair{0}{\sem{t} \circ \pi_1}}
        = \pi_{22}
        \circ \pair{\Id{}}{\pair{0}{\sem{t} \circ \pi_1}}
        = \sem{t} \circ \pi_1
        = \sem{\Gamma, x : \tau \ts t : \tau}
      \end{salign*}

    \item $s = y \neq x$
    
      Then $\sem{s} = \pi_2 \circ \pi_1^n \circ \pi_1$ and
      \begin{salign*}
        \hspace{-30pt}
        \d[\sem{s}] \circ \pair{\Id{}}{\pair{0}{\sem{t} \circ \pi_1}}
        = \pi_{2} \circ \pi_1^n \circ \pi_1 \circ \pi_2
        \circ \pair{\Id{}}{\pair{0}{\sem{t} \circ \pi_1}}
        = 0
        = \sem{\Gamma, x : \tau \ts 0 : \tau}
      \end{salign*}
    
    \item $s = \epsilon s_1$

      Then $\sem{s} = \epsilon\sem{s_1}$ and
      \begin{salign*}
        \hspace{-30pt}
        \d[\sem{s}] \circ \pair{\Id{}}{\pair{0}{\sem{t} \circ \pi_1}}
        = \epsilon (\d[\sem{s_1}] \circ \pair{\Id{}}{\pair{0}{\sem{t} \circ \pi_1}})
        = \epsilon \sem{\diffS{s_1}{x}{t}}
        = \sem{\diffS{(\epsilon s_1)}{x}{t }}
      \end{salign*}

    \item The case $s = \sum_{i=1}^n s_i$ follows by a similar argument as the
    previous one.
    
    \item $s = \lambda y . s_1 : \sigma_1 \Ra \sigma_2$

      Then $\Gamma, x : \tau, y : \sigma_1 \ts s_1 : \sigma_2$ and therefore,
      by the induction hypothesis, we know that:
      \begin{salign*}
        \hspace{-30pt}
        \sem{\diffS{s_1}{x}{t}}_{\Gamma, x : \tau, y : \sigma_1}
        &=
        \sem{\diffS{s_1}{x}{t}}_{\Gamma, y : \sigma_1, x : \tau}
        \circ \A{sw}\\
        &=
        \d[\sem{s_1}_{\Gamma, y : \sigma_1, x : \tau}] 
        \circ \pair{\Id{}}
                   {\pair{0}{\sem{t}_{\Gamma, y:\sigma_1, x:\tau}}}
        \circ \A{sw}\\
        &=
        \d[\sem{s_1}_{\Gamma, x : \sigma_1, y : \tau}] 
        \circ (\A{sw} \times \A{sw})
        \circ \pair{\Id{}}
                   {\pair{0}{\sem{t}_{\Gamma} \circ \pi_{11}}}
        \circ \A{sw}
        \PROOFSTEP
        &=
        \d[\sem{s_1}_{\Gamma, x : \sigma_1, y : \tau}] 
        \circ \pair{\A{sw}}
                   {\pair{\pair{0}{\sem{t}_\Gamma \circ \pi_{11}}}{0}}
        \circ \A{sw}
        \PROOFSTEP
        &=
        \d[\sem{s_1}_{\Gamma, x : \sigma_1, y : \tau}] 
        \circ \pair{\Id{}}
                   {\pair{\pair{0}{\sem{t}_\Gamma \circ \pi_{11}}}{0}}
      \end{salign*}
      Obtaining the final result is just a matter of applying this identity
      and Remark~\ref{rem:lambda-d-swap}.
      \begin{salign*}
        \d[\sem{s}] \mathrel{\circ}{}& \pair{\Id{}}{\pair{0}{\sem{t} \circ \pi_1}}
        =\d[\Lambda(\sem{s_1})] \circ \pair{\Id{}}{\pair{0}{\sem{t} \circ \pi_1}}
        \PROOFSTEP
        &= \Lambda \Big[
          \d[\sem{s_1}] 
          \circ \pa{\Id{} \times \pair{\Id{}}{0}}
          \circ \A{sw}
        \Big]
        \circ \pair{\Id{}}{\pair{0}{\sem{t} \circ \pi_1}}
        \PROOFSTEP
        &= \Lambda \Big[
          \d[\sem{s_1}] 
          \circ \pa{\Id{} \times \pair{\Id{}}{0}}
          \circ \A{sw}
          \circ \pair{\pair{\Id{}}{\pair{0}{\sem{t} \circ \pi_1}} \circ \pi_1}{\pi_2}
        \Big]
        \PROOFSTEP
        &= \Lambda \Big[
          \d[\sem{s_1}] 
          \circ \pa{\Id{} \times \pair{\Id{}}{0}}
          \circ \A{sw}
          \circ \pair{\pair{\pi_1}{\pair{0}{\sem{t} \circ \pi_{11}}}}{\pi_2}
        \Big]
        \PROOFSTEP
        &= \Lambda \Big[
          \d[\sem{s_1}] 
          \circ \pa{\Id{} \times \pair{\Id{}}{0}}
          \circ \pair{\Id{}}{\pair{0}{\sem{t} \circ \pi_{11}}}
        \Big]
        \PROOFSTEP
        &= \Lambda \Big[
          \d[\sem{s_1}] 
          \circ \pair{\Id{}}{\pair{\pair{0}{\sem{t} \circ \pi_{11}}}{0}}
        \Big]
        \PROOFSTEP
        &= \Lambda \Big[
          \d[\sem{s_1}] 
          \circ (\A{sw} \times \A{sw})
          \circ \pair{\Id{}}{\pair{0}{\sem{t} \circ \pi_{11}}}
        \Big]
        \PROOFSTEP
        &= \Lambda \Big[
          \d[\sem{s_1}] 
          \circ \pair{\A{sw}}{\pair{\pair{0}{\sem{t} \circ \pi_{11}}}{0}}
        \Big]
        \PROOFSTEP
        &= \Lambda \Big[
          \d[\sem{s_1}] 
          \circ \pair{\Id{}}{\pair{\pair{0}{\sem{t} \circ \pi_{11}}}{0}}
          \circ \A{sw}
        \Big]
        \PROOFSTEP
        &= \Lambda \sem{\Gamma, x : \tau, y : \sigma_1 \ts 
          \diffS{s_1}{x}{t} : \sigma_2
        }
        \PROOFSTEP
        &= \sem{\Gamma, x : \tau \ts 
          \lambda y . \diffS{s_1}{x}{t} : \sigma_1 \Ra \sigma_2
        }
      \end{salign*}

      \newcommand{\theterm}[0]{\mathfrak{s}}
    \item $s = s_1\ e$
      To simplify the calculations, we write $\theterm$ for
      $\Lambda^-(\sem{s_1})$. 
      The result follows as
      a consequence of Lemma~\ref{lem:lambda-star-ev}[iii.].
      \begin{salign*}
        \lefteqn{\sem{\diffS{(s_1\ e)}{x}{t}}}\\
        \eq \sem{
            \D(s_1) \cdot \pa{\diffS{e}{x}{t}}\ e
        } + \sem{
          \pa{\diffS{s_1}{x}{t}}\ (\subst{e}{x}{x + \varepsilon t})
        }
        \PROOFSTEP
        \eq \A{ev} \circ \pair{\Lambda\pa{
          \theterm \star (\sem{e} \star \sem{t})
        }}{\sem{e}}\\
        \continued + 
          \A{ev} \circ \pair{
            \sem{s_1} \star \sem{t}
          }{
            \sem{e} \circ \pair{\pi_1}{\pi_2 + \varepsilon(\sem{t})\circ \pi_1}
          }
        \PROOFSTEP
        \eq (\A{ev} \circ \pair{\sem{s_1}}{\sem{e}}) \star \sem{t}
        \PROOFSTEP
        \eq \sem{s_1\ e} \star \sem{t}
      \end{salign*}

    \item $s = \D(s_1)\cdot u$

      We will again abbreviate $\Lambda^-(\sem{s_1})$ as $\theterm$ to make the
      subsequent calculations more readable.
      The result then follows
      by applying of Lemma~\ref{lem:lambda-star-ev}[ii.].
      \begin{salign*}
        \lefteqn{\sem{\diffS{\D(s_1) \cdot u}{x}{t}}}\\
        \eq \sem{
          \D(s_1) \cdot \diffS{u}{x}{t}
        }
        + \sem{
          \D\pa{\diffS{s_1}{x}{t}} \cdot (\subst{u}{x}{x + \epsilon(t)})
        }\\
        \continued + \epsilon\sem{
          \pa{\D(\D(s_1) \cdot u) \cdot \pa{\diffS{u}{x}{t}}}
        }
        \PROOFSTEP
        \eq \Lambda\pa{
          \theterm \star \sem{\diffS{u}{x}{t}}
        }
        + \Lambda\pa{
          \pa{\Lambda^-\sem{\diffS{s_1}{x}{t}}}
          \star {\sem{\subst{u}{x}{x + \varepsilon(t)}}}
        }\\
        \continued + \varepsilon\Lambda\pa{
          \pa{\Lambda^-\sem{\D(s_1) \cdot u}} \star \sem{\diffS{u}{x}{t}}
        }
        \PROOFSTEP
        \eq \Lambda\pa{
          \theterm \star (\sem{u} \star \sem{t})
        }
        + \Lambda\pa{
          \pa{\Lambda^-(\Lambda(\sem{s_1}) \star \sem{t})}
          \star {\sem{\subst{u}{x}{x + \varepsilon(t)}}}
        }\\
        \continued + \varepsilon\Lambda\pa{
          \pa{\Lambda^-(\theterm \star \sem{u})} \star (\sem{u} \star \sem{t})
        }
        \PROOFSTEP
        \eq \Lambda\pa{
          \theterm \star (\sem{u} \star \sem{t})
        }
        + \Lambda\pa{
          \pa{\Lambda^-(\Lambda(\sem{s_1}) \star \sem{t})}
          \star \pa{{\sem{u} \circ \pair{\pi_1}{\pi_2 + \varepsilon(\sem{t}) \circ \pi_1}}}
        }\\
        \continued + \varepsilon\Lambda\pa{
          \pa{\Lambda^-(\theterm \star \sem{u})} \star (\sem{u} \star \sem{t})
        }
        \PROOFSTEP
        \eq \pa{\Lambda\pa{\theterm \star \sem{u}}} \star \sem{t}
        \PROOFSTEP
        \eq \pa{\Lambda\pa{\Lambda^-(\sem{s_1}) \star \sem{u}}} \star \sem{t}
        \PROOFSTEP
        \eq (\sem{\D(s_1) \cdot u}) \star \sem{t}
      \end{salign*}
  \end{itemize}
\end{proof}

\begin{definition}
  Given well-formed terms $\ul{s}, \ul{s'}$, we define the equivalence relation
  $\sim_{\beta\d}$ as the least contextual equivalence relation that contains
  the one-step reduction relation $\wfReducesTo$.
\end{definition}

\begin{corollary}
  The interpretation $\sem{\cdot}$ is \emph{sound}, that is to say, whenever
  $\ul{s} \sim_{\beta\d} \ul{s'}$ then $\sem{s} = \sem{s'}$, independently of
  the choice of representatives $s, s'$.
\end{corollary}

%% file: conclusions.tex
\section{Summary of Results}

The main contribution of this thesis is the notion of a differential map between
change actions. Initially motivated by their applications to incremental
computation, we believe differential maps are an important generalisation of
many natural definitions of derivative. We have shown that they arise from, and
connect, many seemingly disparate areas, providing a formal basis for the idea
that incrementalisation is indeed a sort of differentiation.

We believe change actions compare favorably to Cai and Giarrusso's change
structures, having three main advantages: first, the formulation of a change
action is not dependently-typed, and so it can be applied in many more settings,
e.g. categories where one might not be able to interpret dependent type theory.
Second, the presence of the difference operator $\ominus$ means that the theory
of change structures is, in a sense, ``trivial'': every function into a change
structure always admits a derivative. This also rules out interesting models,
such as the ones arising from Cartesian differential categories. Finally, the
requirement that the change space $\Delta A$ be a monoid, together with the
regularity property of derivatives, allow for a rich theory (for example, the
proof that a multivariate function is differentiable if and only if partial
derivatives exist crucially hinges on the monoid structure of $\Delta A$).

On the other hand, change actions, through difference categories, are a strict
generalisation of Cartesian difference categories which can account for important
notions of derivative that do not fit in the Cartesian differential mold. As we
have shown, many theorems about Cartesian differential categories also hold in
the difference setting, and one can likewise construct a calculus out of
difference categories.

\section{Future Work}

The writing of this thesis has been an exercise in restraint, with many research
leads having been postponed due to time constraints. We compile here some of the
most promising ones, in the hopes that they will serve as a suggestion for the
interested reader.

\subsection{The Higher-Categorical Story}

As we showed in Section~\ref{sec:change-actions-as-categories}, a change action
is nothing more and nothing less than a particular kind of internal category,
with the category $\ftor{CAct}(\cat{C})$ of $\cat{C}$-change actions embedding
fully and faithfully into the category $\ftor{Cat}(\cat{C})$ of
$\cat{C}$-categories. A natural question arising from this result is:
exactly what sort of categories correspond to change actions? Or, in other
words, is there a meaningful property characterising those categories that are
in the image of the previous embedding?

The above statement also has interesting implications for the higher-order
setting: taking it into consideration, a change action model $\alpha : \cat{C}
\to \ftor{CAct}(\cat{C})$ is simply a particular kind of functor $\alpha : \cat{C}
\to \ftor{Cat}(\cat{C})$, and thus every object in a change action model on
$\cat{C}$ is a category internal to $\cat{C}$ -- but, by iterating $\alpha$, a
much stronger statement is also true: every object in $\cat{C}$ is a
particularly well-behaved internal $\infty$-fold category (see
\cite{nlab:n-fold-category} for a discussion of $n$-fold categories, and
\cite{nlab:double-category} for the $2$-fold case). In
particular, every object is an $\infty$-fold category where the horizontal and
vertical morphisms coincide, which leads us to conjecture that change action
models are in fact strict $\omega$-categories. The significance and implications
of these observations are yet to be understood.

\subsection{Iteration and Integration}

Through this work we have focused on differentiation, while ommitting any
mentions of integrals or differential equations. The categorical theory of
integration is, in general, much less developed than that of differentiation,
only recently having received more attention
\cite{cockett2019integral,lemay2019exponential}.

Similarly, we have developed a simply-typed calculus which does not include any
mechanisms for iteration. Iteration is itself problematic, as it is not
immediately clear that iteration combinators are differentiable (although some
of the results in this thesis already indicate that fixed-point combinators can
be differentiated under reasonable conditions, see Theorem
\ref{theorem:leastFixpointDerivatives}).

These two missing pieces are more closely related than one might imagine.
Indeed, consider a hypothetical extension of the difference $\lambda$-calculus
equipped with a type of natural numbers (with the identity as its corresponding
infinitesimal extension, that is to say, $\epsilon_\NN = \Id{\NN}$). How should
an iteration operator $\textbf{iter}$ be defined? The straightforward option
would be to give it the usual behavior, that is to say:
\[
\begin{array}{rcl}
  \textbf{iter}\ \A{Z}\ z\ s &\reducesTo & z\\
  \textbf{iter}\ (\A{S}\ n)\ z\ s &\reducesTo & s\ (\textbf{iter}\ n\ z\ s)
\end{array}
\]

While this is not unreasonable, an unexpected consequence of this reduction rule
is that it implies that every change action involved must be complete, that is
to say, for every $s, t : A$, there must be a change $u : \Delta A$ with 
$x \mathrel{+}{} \epsilon(u) = y$ -- such a change can be obtained by the term 
$((\D (\lambda n . \textbf{iter}\ n\ s\ (\lambda x . t)) \cdot (\A{S}\ \A{Z}))\
\A{Z})$, which has the property that:
\begin{salign*}
  t &\approx \textbf{iter}\ (\A{S}\ \A{Z})\ s\ (\lambda x . t)\\
  &\approx (\lambda n . \textbf{iter}\ n\ s\ (\lambda x . t))\ (\A{S}\ \A{Z})\\
  &\approx 
  \brak{ (\lambda n . \textbf{iter}\ n\ s\ (\lambda x . t))\ \A{Z})}
  + \epsilon \brak{
    ((\D (\lambda n . \textbf{iter}\ n\ s\ (\lambda x . t)) \cdot (\A{S}\ \A{Z}))\
\A{Z})
  }\\
  &\approx 
  s + \epsilon \brak{
    ((\D (\lambda n . \textbf{iter}\ n\ s\ (\lambda x . t)) \cdot (\A{S}\ \A{Z}))\
\A{Z})
  }
\end{salign*}

This would rule out a number of interesting models -- especially ones where the
set $\Delta A$ is to be interpreted as a set of infinitesimals -- and so it
seems rather unsatisfactory. An alternative is to define the iteration
operator by:
\[
\begin{array}{rcl}
  \textbf{iter}\ \A{Z}\ z\ s &\reducesTo & z\\
  \textbf{iter}\ (\A{S}\ n)\ z\ s &\reducesTo & (\textbf{iter}\ n\ z\ s) 
  + \epsilon(s\ (\textbf{iter}\ n\ z\ s))
\end{array}
\]
or, in terms of actions rather than the associated infinitesimal extensions:
\[
\begin{array}{rcl}
  \textbf{iter}\ \A{Z}\ z\ s &\reducesTo & z\\
  \textbf{iter}\ (\A{S}\ n)\ z\ s &\reducesTo & (\textbf{iter}\ n\ z\ s) 
  \oplus (s\ (\textbf{iter}\ n\ z\ s))
\end{array}
\]
where $s : A \to \Delta A$. If one extends the Cartesian difference category
$\Ab$ to cover all commutative monoids where addition is injective (which
includes monoids such as $(\NN, +, 0)$) and all infinitely differentiable maps,
it turns out that this $\textbf{iter}$ operator is itself an arrow in this
difference category, that is, the above definition of iteration is always
smooth, without requiring that every change action be complete. Fixed $z, s$,
define the map $\mu(n) \defeq \textbf{iter}\ n\ z\ s$. What is the derivative
$\D[\mu](n, \A{S}\ \A{Z})$? Nothing more and nothing less than $s(\mu(n))$. Then
the function $\mu : \NN \to A$ defines a ``curve''\footnote{
  It seems adequate
  that curves in difference categories should be maps from $\NN$, rather than
  $\RR$, as befits a discrete setting.
} which starts at $z$ and whose derivative at a given point $n$ is $s(\mu(n))$
-- if one squints, this is nothing more and nothing less than the requirement
that the curve $\mu$ be an integral curve for the smooth vector field $s$
satisfying the initial condition $\mu(\A{Z}) = z$! Hence one can think of
iteration as a discrete counterpart of the Picard-Lindel\"of theorem, which
states that such integral curves always exist (locally).

It would be of great interest, then, to extend $\lambda_\epsilon$ with an
interation operator and give its semantics in terms of differential (or
difference) equations. Studying recurrence equations using the language of
differential equations is a very useful tool in discrete analysis; for example,
one can treat the recursive definition of the Fibonacci sequence as a discrete
ODE and use differential equation methods to find a closed-form solution. We
believe that in a language which frames iteration in such terms may be
amenable to optimisation by similar analytic methods.

\subsection{Gradients and Coderivatives}

One of the main applications of automatic differentiation nowadays is computing
gradients for the purpose of optimising functions by gradient descent. This
immediately raises the question of whether a similar notion of gradient can be
defined ``synthetically'' for change actions, and whether it leads to
optimisation algorithms for discrete spaces.

An initial attempt at this task led us to proposing the notion of \emph{coderivatives}.
\begin{definition}
  Given a map $f : A \ra B$ and change actions $\cact{A}, \cact{B}$, a map
  $\codiff{f} : A \times \changes B \ra \changes A$ is a \emph{coderivative} for
  $f$ whenever it satisfies the following condition:
  \begin{gather*}
    f \circ \oplus_A \circ \pair{\pi_1}{\codiff{f}} = \oplus_B \circ (f \times \Id{\Delta B}) 
  \end{gather*}
  We also require that coderivatives preserve the zero change\footnote{One may
    further require that coderivatives preserve sums in a similar sense to regularity
    but the usefulness of this is debatable.}, i.e.
  \begin{gather*}
    \codiff{f} \circ \pair{\Id{A}}{0_B} = 0_A 
  \end{gather*}
\end{definition}

Coderivatives do not quite correspond to gradients -- in fact, it can be shown
that, whenever a map $f$ admits an invertible derivative, its inverse is a
coderivative for $f$. They are, however, not without importance: a
generalisation of Newton's method to change actions can be framed in terms of
coderivatives, and it can be shown that a specific kind of codifferential maps
corresponds to lenses \cite{bohannon2006relational,hofmann2011symmetric}, a
well-known concept in the functional programming community where they provide a
simple and elegant solution to the problem of updating deeply nested
data-structures.

Furthermore, the following theorem shows that coderivatives satisfy a reverse
version of the chain rule, where changes are propagated backwards and values are
propagated forwards.
\begin{theorem}
  Let $\cact{A}, \cact{B}, \cact{C}$ be change actions and consider maps $f : A \ra B, g : B \ra C$
  with coderivatives $\codiff{f}, \codiff{g}$ respectively. Then the map 
  \[
    \codiff{(g \circ f)} \defeq \codiff{f} \circ \pair{\pi_1}{\codiff{g} \circ (f \times \Id{\Delta C})}
  \]
  or, more intuitively,
  \[
    \codiff{(g \circ f)}(a, \delta c) \defeq \codiff f(a, \codiff g(f(a), \delta c))
  \]
  is a coderivative for $g \circ f$.
\end{theorem}
The above chain rule looks strikingly similar to the operational behavior of
reverse-mode automatic differentiation, where computation is first performed in
the usual fashion and perturbations are then propagated backwards through a
stored trace of the computation, and one hopes that a theory of coderivatives
could likewise lead to a theory and calculus of reverse-mode automatic
differentiation.

%% file: mechanized-proofs.tex
\label{chp:mechanised-proofs}


\section{Module \texttt{Definitions.v}}
\begin{lstlisting}[language=Coq]
Require Import Unicode.Utf8.
Require Import Coq.Program.Equality.
Require Import Coq.Relations.Relation_Operators.
Require Import Coq.Relations.Relation_Definitions.
Require Import Coq.Arith.EqNat.
Require Import Coq.Arith.Compare_dec.

(* The type of (unrestricted) terms of the difference lambda-calculus *)
Inductive Term :=
| Var  : nat -> Term
| LAM    : Term -> Term
| App  : Term -> Term -> Term
| Zero : Term
| Add  : Term -> Term -> Term
| EPSILON    : Term -> Term
| Dif  : Term -> Term -> Term
.

Notation "0" := Zero.
Notation "s + t" := (Add s t) (at level 50, left associativity).
Notation "'D' s @ t" := (Dif s t) (at level 40).
Notation "{ s | t }" := (App s t).

Section Contextual_Closure.

  Variable ρ : relation Term.

  Inductive Ctx : relation Term :=
  | CVar  n    : Ctx (Var n) (Var n)
  | CZero      : Ctx 0 0
  | CAbs  t t' : Ctx t t' -> Ctx (LAM t) (LAM t')
  | CE    t t' : Ctx t t' -> Ctx (EPSILON t) (EPSILON t')

  | CApp  s s' t t' : Ctx s s' -> Ctx t t' -> Ctx {s|t} { s' | t' }
  | CAdd  s s' t t' : Ctx s s' -> Ctx t t' -> Ctx (s + t) (s' + t')
  | CD    s s' t t' : Ctx s s' -> Ctx t t' -> Ctx (D s @ t) (D s' @ t')

  | CRho  t t' : ρ t t' -> Ctx t t'
  .
  Hint Constructors Ctx : diff.

  Theorem Ctx_refl : forall t, Ctx t t.
  Proof. intros t; induction t; eauto with diff.
  Qed.
End Contextual_Closure.
Hint Constructors Ctx : diff.
Hint Resolve Ctx_refl : diff.

(* Contextual relations on terms *)
Definition Contextual ρ := forall s t, Ctx ρ s t -> ρ s t.

(* One-step differential equivalence on terms *)
Reserved Notation "s TILDE1 t" (at level 90).
Inductive Equiv1 : Term -> Term -> Prop :=
(* Commutative monoid conditions *)
| Eq_add_0 t         : t + 0 TILDE1 t
| Eq_add_comm  s t   : s + t TILDE1 t + s
| Eq_add_assoc s t e : (s + t) + e TILDE1 s + (t + e)
| Eq_add_E s t       : EPSILON (s + t) TILDE1 EPSILON s + EPSILON t
| Eq_E_0             : EPSILON 0 TILDE1 0

(* Linearity conditions *)
| Eq_app_0   t     : {0 | t} TILDE1 0
| Eq_app_add s t e : {s + t | e} TILDE1 {s|e} + {t|e}
| Eq_app_E s t     : {EPSILON s | t} TILDE1 EPSILON {s | t}
| Eq_lam_0         : LAM 0 TILDE1 0
| Eq_lam_add s t   : LAM (s + t) TILDE1 LAM s + LAM t
| Eq_lam_E s       : LAM (EPSILON s) TILDE1 EPSILON (LAM s)
| Eq_D_0 u         : D 0 @ u TILDE1 0
| Eq_D_add s t e   : D (s + t) @ e TILDE1 D s @ e + D t @ e
| Eq_D_E s t       : D (EPSILON s) @ t TILDE1 EPSILON (D s @ t)

| Eq_D_0_r t   : D t @ 0 TILDE1 0
| Eq_D_E_r t u : D t @ (EPSILON u) TILDE1 EPSILON (D t @ u)

(* Regularity *)
| Eq_regularity s u v : D s @ (u + v) TILDE1 D s @ u + D s @ v + EPSILON (D (D s @ v) @ u)

(* Derivative condition *)
| Eq_derivative s t u : {s | t + EPSILON u} TILDE1 {s | t} + EPSILON {D s @ u | t}

(* Commutativity of derivatives *)
| Eq_D_comm s u v : D (D s @ u) @ v TILDE1 D (D s @ v) @ u

(* Infinitesimal cancellation *)
| Eq_D_inf s u v : EPSILON (EPSILON (D (D s @ u) @ v)) TILDE1 EPSILON (D (D s @ u) @ v)
where "s TILDE1 t" := (Equiv1 s t).
Hint Constructors Equiv1 : diff. 

(* Differential equivalence of terms *)
Definition Equiv (s : Term) (t : Term) := 
  clos_refl_sym_trans Term (Ctx Equiv1) s t.
Notation "s ≈ t" := (Equiv s t) (at level 90).

(* Avoid using rst_* as that replaces ≈ by clos_refl_sym_trans... and messes
with the pattern matching in tactics *)
Definition eq_step : forall s t, Ctx Equiv1 s t -> s ≈ t.
Proof. constructor; trivial. Qed.
Definition eq_trans : forall x y z, x ≈ y -> y ≈ z -> x ≈ z.
Proof. apply rst_trans. Qed.
Definition eq_sym : forall x y, x ≈ y -> y ≈ x.
Proof. apply rst_sym. Qed.
Definition eq_refl : forall x, x ≈ x.
Proof. apply rst_refl. Qed.
Opaque Equiv.

Hint Resolve eq_step eq_trans eq_sym eq_refl : diff.

(* Differential equivalence is contextual *)
Theorem eq_contextual :
  Contextual Equiv.
Proof with (auto with diff).
  intros s s' cx; induction cx; auto.
  constructor...
  constructor...
  - induction IHcx...
    + apply eq_trans with (y := LAM y)...
  - induction IHcx...
    + apply eq_trans with (y := EPSILON y)...
  - apply eq_trans with (y := { s' | t }).
    + induction IHcx1...
      * eapply eq_trans... 
    + induction IHcx2...
      * eapply eq_trans... 
  - apply eq_trans with (y := s' + t).
    + induction IHcx1...
      * eapply eq_trans... 
    + induction IHcx2...
      * eapply eq_trans... 
  - apply eq_trans with (y := D (s') @ t).
    + induction IHcx1...
      * eapply eq_trans... 
    + induction IHcx2...
      * eapply eq_trans...
Qed.
Hint Resolve eq_contextual : diff.

(* 
  The tactic Cx tries to prove a goal of the form Rho t1 t2, 
  where t1, t2 are terms and Rho is a contextual relation.
  It does so by applying contextuality and adding the resulting
  subcomponents as goals.
*)
Ltac Cx := 
  let cx := fresh "cx" in
  lazymatch goal with
  | |- ?ρ (EPSILON ?u) (EPSILON ?v) => 
      assert (Contextual ρ) as cx; 
      [auto with diff | apply cx; apply (CE ρ u v)]

  | |- ?ρ (LAM ?u) (LAM ?v) => 
      assert (Contextual ρ) as cx; 
      [auto with diff | apply cx; apply (CAbs ρ u v)]

  | |- ?ρ 0 0 =>
      assert (Contextual ρ) as cx; 
      [auto with diff | apply cx; apply (CZero ρ)]

  | |- ?ρ (Var ?n) (Var ?m) => 
      assert (Contextual ρ) as cx; 
      [auto with diff | apply cx; apply (CVar ρ n)]

  | |- ?ρ (?s + ?t) (?s' + ?t') => 
      assert (Contextual ρ) as cx; 
      [auto with diff | apply cx; apply (CAdd ρ s s' t t')]

  | |- ?ρ { ?s | ?t } { ?s' | ?t' } =>
      assert (Contextual ρ) as cx; 
      [auto with diff | apply cx; apply (CApp ρ s s' t t')]

  | |- ?ρ (D ?s @ ?t) (D ?s' @ ?t') => 
      assert (Contextual ρ) as cx; 
      [auto with diff | apply cx; apply (CD ρ s s' t t')]

  | |- ?g => fail "Not a contextual goal " g
  end; clear cx; try apply CRho.

Ltac Cxx := repeat Cx.

Ltac eq_step ctor := (
  eapply eq_trans; [| apply eq_step; apply CRho; apply ctor]
).

Ltac eq_rstep ctor := (
  apply eq_sym; eq_step ctor; apply eq_sym
).

(* 
  Term rewriting modulo differential equivalence.
  This should probably be moved to setoid rewriting.

  This tactic should never be called. Instead prefer the "local rewrite"
  and "local rewrite <-" macros defined later.
*)
Ltac local_rewrite u v eqn := match goal with
| [|- ?t1 ≈ ?t2] => 
  lazymatch t1 with
  | context tm [u] =>
    let qu := context tm[u] in
    let qv := context tm[v] in
    let lhs := fresh "lhs" in
    let rhs := fresh "rhs" in
    let eq := fresh "eq" in
    pose (lhs := u);
    pose (rhs := v);
    assert (u ≈ v -> qu ≈ qv) as eq by (
      let hyp := fresh "hyp" in
      fold rhs lhs;
      intro hyp; Cxx; try apply eq_refl; 
      try exact hyp; 
      match goal with
      | |- ?H => idtac "I can't prove contextuality in local rewrite: " H;
        idtac "Falling back on evars, I hope someone can prove this later";
        let hack := fresh "hack" in
        evar (hack : H); exact hack
      end); 
    apply eq_trans with (x := qu) (y := qv); 
    fold lhs rhs; 
    [apply eq; apply eqn|];
    unfold lhs, rhs;
    (*try (solve [ auto with diff | try apply eq_refl]; fail);*)
    clear lhs rhs eq
  | _ =>
    lazymatch t2 with
    | context tm [u] => apply eq_sym; local_rewrite u v eqn; apply eq_sym
    | _ => fail "I can't find lhs" u "in goal" t1 "≈" t2
    end
  end
end.

Ltac local_rewrite_eq eq :=
  lazymatch type of eq with 
  | ?u ≈ ?v => (local_rewrite u v eq)
  | clos_refl_sym_trans Term (Ctx Equiv1) ?u ?v 
      => local_rewrite u v (eq_step u v eq)
  | Equiv1 ?u ?v 
      => (local_rewrite u v (eq_step u v (CRho Equiv1 u v eq)))
  | ?u TILDE1 ?v 
      => (local_rewrite u v (eq_step u v (CRho Equiv1 u v eq)))
  | forall y, @?w y => (
      let eqname := fresh "eq" in
      epose (eq _) as eqname; simpl in eqname;
      let the_expr := (eval hnf in eqname) in
      local_rewrite_eq the_expr;
      clear eqname
    )
  | _ => fail "I do not understand " eq " as an equation"
  end.

Ltac local_rewrite_rev eq :=
  lazymatch type of eq with 
  | ?u ≈ ?v => (local_rewrite v u eq)
  | clos_refl_sym_trans Term (Ctx Equiv1) ?u ?v 
      => local_rewrite v u (eq_sym u v (eq_step u v eq))
  | Equiv1 ?u ?v 
      => local_rewrite v u (eq_sym u v (eq_step u v (CRho Equiv1 u v eq)))
  | forall y, @?w y => (
      let eqname := fresh "eq" in
      epose (eq _) as eqname; simpl in eqname;
      let the_expr := (eval hnf in eqname) in
      local_rewrite_rev the_expr;
      clear eqname
    )
  | _ => fail "I do not understand " eq " as an equation"
  end.

Tactic Notation "local" "rewrite" constr(eq) :=
  local_rewrite_eq eq.

Tactic Notation "local" "rewrite" "<-" constr(eq) :=
  local_rewrite_rev eq.

Lemma Eq_0_add : forall t, 0 + t ≈ t.
Proof. intro.
  local rewrite Eq_add_comm.
  local rewrite Eq_add_0.
  apply eq_refl.
Qed.

(* Simplifying a term by applying every equivalence rule once *)
Ltac simplify :=
  try local rewrite Eq_add_0;
  try local rewrite Eq_0_add;
  try local rewrite Eq_add_assoc;
  try local rewrite Eq_add_E;
  try local rewrite Eq_E_0;
  try local rewrite Eq_app_0;
  try local rewrite Eq_app_add;
  try local rewrite Eq_app_E;
  try local rewrite Eq_lam_0;
  try local rewrite Eq_lam_add;
  try local rewrite Eq_lam_E;
  try local rewrite Eq_D_0;
  try local rewrite Eq_D_add;
  try local rewrite Eq_D_E;
  try local rewrite Eq_D_0_r;
  try local rewrite Eq_D_E_r;
  try local rewrite Eq_regularity;
  try local rewrite Eq_derivative;
  try local rewrite Eq_D_inf.

Ltac simplify2 := try simplify; apply eq_sym; try simplify; apply eq_sym.

(* 
  Reduce a term to a (near) canonical form by iterating 
  one-step simplification. This is VERY SLOW. Also note
  that sometimes it won't get to a canonical form, 
  depending on how reduction happens.
*)
Ltac can := repeat simplify; apply eq_sym; repeat simplify; apply eq_sym.

(*
  Right-associate all the additions in a term.
*)
Ltac associate := repeat local rewrite Eq_add_assoc.

(*
  Turn a term of the form u + (v + w) into v + (w + u).
  Usually Cx gets called immediately after this.
*)
Ltac rotate :=
match goal with
| [|- ?u + ?v ≈ _] => local rewrite (Eq_add_comm u v); repeat associate
end.

Tactic Notation "rot" integer(i) := do i rotate.

Tactic Notation "r" tactic(t) := apply eq_sym; t; apply eq_sym.

(*
  Rotate the term in the left-hand side until the leftmost 
  summand of both terms is the same. Note: very expensive.
  Prefer the listify combinators in the next file.
*)
Ltac align n :=
match eval compute in n with
| O => idtac
| S ?n' => match goal with
          | [|- ?u + _ ≈ ?u + _ ] => idtac
          | [|- _ ] => rotate; align n'
          end
end.

Example test_cx : forall t, (t + t + EPSILON ({t | t})) ≈ t + t + EPSILON {t | t + 0}.
intro.
  local rewrite (Eq_add_comm t 0).
  local rewrite (Eq_0_add).
  Cxx; apply eq_refl.
Qed.
\end{lstlisting}

\newpage
\section{Module \texttt{Substitution.v}}

\begin{lstlisting}[language=Coq]
Require Import Unicode.Utf8.
Require Import Coq.Relations.Relation_Operators.
Require Import Coq.Relations.Relation_Definitions.
Require Import Coq.Arith.EqNat.
Require Import Coq.Arith.Compare_dec.
Require Import Coq.Arith.PeanoNat.
Require Import Omega.

Add LoadPath "./".
Load Definitions.

Lemma lt_leb : forall m n, m < n -> m <=? n = true.
Proof. intros; assert (m ≤ n) by omega; now apply leb_correct. Qed.

Lemma lt_sym : forall m n, m > n -> n < m.
Proof. intros; omega. Qed.

(*
  A tactic for turning inequalities in Prop to 
  equalities in bool
*)
Ltac bool_from_nat eq :=
  lazymatch type of eq with
  | ?m < ?n =>
    lazymatch goal with
    | |- ?m <=? ?n = true => apply lt_leb
    | |- ?m <? ?n = true => apply Nat.ltb_lt
    | |- ?m ?= ?n = Lt => apply nat_compare_lt
    | |- ?n <=? ?m = false => apply leb_iff_conv
    | |- ?n ?= ?m = Gt => apply nat_compare_gt; apply lt_sym
    end; exact eq
  | ?m = ?n =>
    lazymatch goal with
    | |- ?m <=? ?n = true => (rewrite eq; apply Nat.leb_refl)
    | |- ?n <=? ?m = true => (rewrite eq; apply Nat.leb_refl)
    | |- ?m ?= ?n = Eq => apply Nat.compare_eq_iff; exact eq
    end
  end.

Tactic Notation "bool" constr(target) "from" constr(eq) "as" ident(eqname) :=
  assert target as eqname by bool_from_nat eq.
Tactic Notation "bool" "from" constr(eq) := bool_from_nat eq.
Tactic Notation "bool" constr(target) "from" constr(eq) :=
  let eqname := fresh "beq" in bool target from eq as eqname.

Tactic Notation "bool" "rewrite" constr(target) "from" constr(eq) 
                "in" hyp_list(HS) "|-" "*" :=
  let beq := fresh "beq" in bool target from eq as beq;
  repeat try rewrite beq in HS|-; repeat try rewrite beq; clear beq.
Tactic Notation "bool" "rewrite" constr(target) "from" constr(eq) 
                "in" "*" :=
  let beq := fresh "beq" in bool target from eq as beq;
  repeat try rewrite beq in *|-*; repeat try rewrite beq; clear beq.
Tactic Notation "bool" "rewrite" constr(target) "from" constr(eq) :=
  bool rewrite target from eq in |-*.

(* 
  Prove a boolean goal whenever the propositional version
  can be proven by Omega.
  Slow and overkill, prefer bool beq from eq 
*)
Ltac decide_nat :=
  match goal with
  | [|- ?m <=? ?n = true ] => 
    assert (m <= n) by omega; now apply leb_correct
  | [|- ?m <=? ?n = false ] =>
    assert (m > n) by omega; now apply leb_correct_conv
  | [|- ?m <? ?n = true ] =>
    assert (m < n) by omega; now apply Nat.ltb_lt
  | [|- ?m <? ?n = false ] =>
    assert (n <= m) by omega; now apply Nat.ltb_ge
  | [|- ?m =? ?n = true ] =>
    assert (m = n) by omega; now apply Nat.eqb_eq
  | [|- ?m =? ?n = false ] =>
    assert (m ≠ n) by omega; now apply Nat.eqb_neq
  | [|- ?m ?= ?n = Lt ] =>
    assert (m < n) by omega; now apply nat_compare_lt
  | [|- ?m ?= ?n = Eq ] =>
    assert (m = n) by omega; now apply Nat.compare_eq_iff
  | [|- ?m ?= ?n = Gt ] =>
    assert (m > n) by omega; now apply nat_compare_gt
  end.

(* Try to prove any goal from an absurd arithmetic premise *)
Ltac prune eq :=
  lazymatch type of eq with
  | ?m < O => solve [exfalso; apply (Nat.nlt_0_r m eq)]
  | ?m < ?m => solve [exfalso; apply (Nat.lt_irrefl m eq)]
  | ?m < ?n =>
    lazymatch goal with
    | [h:(n < m) |- _] => 
      solve [exfalso; apply (Nat.lt_asymm n m h eq)]
    | [h:(m > n) |- _] =>
      solve [exfalso; apply (Nat.lt_asymm n m (lt_sym m n h) eq)]
    | [h:(n = m) |- _] => 
      solve [exfalso; apply (Nat.lt_neq m n eq); rewrite h; auto]
    | [h:(m = n) |- _] => 
      solve [exfalso; apply (Nat.lt_neq m n eq); rewrite h; auto]
    | _ => idtac
    end
  | ?n = ?m =>
    lazymatch goal with
    | [h:(n < m) |- _] => 
      solve [exfalso; apply (Nat.lt_neq n m h); rewrite eq; auto]
    | [h:(n > m) |- _] => 
      solve [exfalso; apply (Nat.lt_neq m n (lt_sym _ _ h)); rewrite eq; auto]
    | [h:(m > n) |- _] => 
      solve [exfalso; apply (Nat.lt_neq n m (lt_sym _ _ h)); rewrite eq; auto]
    | [h:(m < n) |- _] => 
      solve [exfalso; apply (Nat.lt_neq m n h); rewrite eq; auto]
    | _ => idtac
    end
  end.

(* 
  Compare two naturals and generate one goal for each reasonable case.
  Uses already-existing information to prune absurd branches.
*)
Ltac compare m n :=
  let name := fresh "name" in
  let cmp := fresh "cmp" in
  let eq := fresh "eq" in
  destruct (lt_eq_lt_dec m n) as [cmp|eq]; 
  [destruct cmp as [eq|eq];
    [ bool rewrite (m <=? n = true) from eq in *;
      bool rewrite (m ?= n = Lt) from eq in *;
      prune eq
    | try bool rewrite (m <=? n = true) from eq in *;
      try bool rewrite (m ?= n = Eq) from eq in *;
      prune eq
  ] | bool rewrite (m <=? n = false) from eq in *;
      bool rewrite (m ?= n = Gt) from eq in *;
      prune eq
  ]; simpl.

(*
  Unfold every arithmetic comparison in the goal into its 
  corresponding branches
*)
Ltac trivial_arithmetic :=
  match goal with
  | [|- ?p ] => 
    match p with
    | context foo [?u ?= ?v] => compare u v
    | context foo [?u <=? ?v] => compare u v
    end
  | [ H:?h |- _] =>
    match h with
    | context foo [?u ?= ?v] => compare u v
    | context foo [?u <=? ?v] => compare u v
    end
  end; try (omega).

Arguments leb n m : simpl never.
Arguments Nat.compare n m : simpl never.

(*
  Increase every DeBruijn index above n by one.
*)
Fixpoint open (n : nat) (u : Term) : Term :=
match u with
| Var m   => Var (if n <=? m then S m else m)
| 0       => 0
| {s|t}   => {open n s|open n t}
| s + t   => open n s + open n t
| D s CDOT t => D (open n s) CDOT (open n t)
| EPSILON t     => EPSILON (open n t)
| LAM t     => LAM (open (S n) t)
end.

Lemma open_open_le
  :  forall t x y
  ,  x ≤ y
  -> open x (open y t) = open (S y) (open x t).
Proof with auto with arith.
  intro t; induction t; intros x y le; simpl;
  try rewrite IHt; try rewrite IHt1; try rewrite IHt2...
  - repeat trivial_arithmetic...
Qed.
Hint Resolve open_open_le : diff.

(*
  Decrease every DeBruijn index above n by one.
*)
Fixpoint close (n : nat) (u : Term) : Term :=
match u with
| Var m   => Var (if n <=? m then pred m else m)
| 0       => 0
| {s|t}   => {close n s|close n t}
| s + t   => close n s + close n t
| D s CDOT t => D (close n s) CDOT (close n t)
| EPSILON t     => EPSILON (close n t)
| LAM t     => LAM (close (S n) t)
end.

Lemma close_close_le
  :  forall t x y
  ,  x ≤ y
  -> close x (close (S y) t) = close y (close x t).
Proof with auto with arith.
  intro t; induction t; intros x y le; simpl;
  try rewrite IHt; try rewrite IHt1; try rewrite IHt2...
  - repeat trivial_arithmetic... f_equal. omega.
Qed.

Lemma close_open_eq
  :  forall t x
  ,  close x (open x t) = t.
Proof with auto with arith.
  intro t; induction t; intro x; simpl;
  try rewrite IHt; try rewrite IHt1; try rewrite IHt2...
  - repeat trivial_arithmetic...
Qed.

Lemma close_open_eq_S
  :  forall t x
  ,  close (S x) (open x t) = t.
Proof with auto with arith.
  intro t; induction t; intro x; simpl;
  try rewrite IHt; try rewrite IHt1; try rewrite IHt2...
  - repeat trivial_arithmetic...
Qed.

Lemma close_open_lt
  :  forall t x y
  ,  x < y
  -> close x (open (S y) t) = open y (close x t).
Proof with auto with arith.
  intro t; induction t; intros x y lt; simpl;
  try rewrite IHt; try rewrite IHt1; try rewrite IHt2...
  - repeat trivial_arithmetic... 
    f_equal; omega.
    f_equal; omega.
Qed.

Lemma close_open_gt
  :  forall t x y
  ,  x > y
  -> close (S x) (open y t) = open y (close x t).
Proof with auto with arith.
  intro t; induction t; intros x y ge; simpl;
  try rewrite IHt; try rewrite IHt1; try rewrite IHt2...
  - repeat trivial_arithmetic... all: f_equal; try omega.
Qed.

(* 
  Shifting DeBruijn indices is compatible with differential equivalence
*)
Lemma open_diff
  :  forall t t'
  ,  t ≈ t'
  -> forall x
  ,  open x t ≈ open x t'.
Proof with (auto with diff).
  intros t t' eq; induction eq...
  - induction H; intros; simpl...
    induction H; simpl...
  - intro; eapply eq_trans...
Qed.
Hint Resolve open_diff : diff.

Lemma close_diff
  :  forall t t'
  ,  t ≈ t'
  -> forall x
  ,  close x t ≈ close x t'.
Proof with (auto with diff).
  intros t t' eq; induction eq...
  - induction H; intros; simpl...
    induction H; simpl...
  - intro; eapply eq_trans...
Qed.
Hint Resolve close_diff : diff.
  
(* Like substitution, except it doesn't actually bind the variable *)
Fixpoint replace (t : Term) (x : nat) (u : Term) : Term :=
match t with
| Var m   =>
  (match Nat.compare x m with
  | Lt => Var m
  | Eq => u
  | Gt => Var m
  end)
| 0       => 0
| {s | t} => { replace s x u | replace t x u }
| D s CDOT t => D (replace s x u) CDOT (replace t x u)
| s + t   => (replace s x u) + (replace t x u)
| EPSILON t     => EPSILON (replace t x u)
| LAM t     => LAM (replace t (S x) (open 0 u))
end.

Lemma replace_open_eq
  :  forall t
  ,  forall x u
  ,  replace (open x t) x u = (open x t).
Proof with auto.
  intro t; induction t; intros; simpl; auto;
  try rewrite IHt; try rewrite IHt1; try rewrite IHt2...
  - repeat trivial_arithmetic...
Qed.

Lemma replace_open_le
  :  forall t
  ,  forall x y u
  ,  x ≤ y
  -> replace (open x t) (S y) (open x u) = open x (replace t y u).
Proof with auto.
  intro t; induction t; intros; simpl; auto;
  try rewrite IHt; try rewrite IHt1; try rewrite IHt2...
  - repeat trivial_arithmetic...
  - f_equal; rewrite open_open_le; auto with arith; rewrite IHt.
Qed.

Lemma replace_replace
  :  forall t x y u v
  ,  x ≠ y
  -> replace (replace t x u) y (open x v)
     = replace (replace t y (open x v)) x (replace u y (open x v)).
Proof with auto with diff arith.
  intro t; induction t; intros x y u v ne; simpl in *|-*;
  try rewrite IHt1; try rewrite IHt2; try rewrite IHt...
  - repeat trivial_arithmetic...
    all: rewrite replace_open_eq...
  - repeat rewrite <- replace_open_le...
    repeat rewrite (open_open_le _ O)...
    rewrite IHt...
Qed.

Lemma replace_diff_new
  :  forall t
  ,  forall x s s'
  ,  s ≈ s' 
  -> replace t x s ≈ replace t x s'.
Proof with (auto with diff).
  intro t; induction t; intros x s s' eq; simpl...
  compare x n... Cx...
Qed.

Lemma replace_diff_base
  :  forall t t'
  ,  t ≈ t'
  -> forall x s
  ,  replace t x s ≈ replace t' x s.
Proof with (auto with diff).
  intros t t' eq; induction eq...
  - induction H; intros; simpl; Cxx...
    induction H; simpl...
  - intros; eauto using eq_trans with diff.
Qed.

(* Replacing a variable by a term respects differential equivalence *)
Lemma replace_diff 
  :  forall t t'
  ,  t ≈ t'
  -> forall s s'
  ,  s ≈ s'
  -> forall x
  ,  replace t x s ≈ replace t' x s'.
Proof with (auto with diff).
  intros.
  local rewrite replace_diff_base. 
  local rewrite replace_diff_new.
  apply eq_refl.
Unshelve. all: auto.
Qed.
Hint Resolve replace_diff : diff.

(* 
  Standard, capture-avoiding substitution -- we make sure that the
  variable that we are substituting does not appear free in the
  new sub-term by opening it beforehand.
*)
Definition subst t x u := close (S x) (replace t x (open x u)).
Notation "t [ x := u ]" := (subst t x u) (at level 85).
Reserved Notation "t [ u / x ]" (at level 90).
Hint Unfold subst : diff.

Theorem subst_diff
  :  forall s s'
  ,  s ≈ s'
  -> forall t t'
  ,  t ≈ t'
  -> forall x
  ,  (s[x := t]) ≈ (s'[x := t']).
Proof with (auto with diff).
  intros; simpl; unfold subst...
Qed.

(*
  Differential substitution
*)
Reserved Notation "( ∂ t // ∂ x ) [ u ]" (at level 85).
Fixpoint dsubst (t : Term) (x : nat) (u : Term) : Term :=
match t with
| Var m   =>
  (match Nat.compare x m with
  | Lt => 0
  | Eq => u
  | Gt => 0
  end)
| 0       => 0
| {t | e} => 
    { D t CDOT ((∂e // ∂x)[u]) | e } 
    + { (∂t // ∂x)[u] | (replace e x (Var x + EPSILON u)) }
| D t CDOT e => 
    D (t) CDOT ((∂e // ∂x)[u]) 
    + D((∂t // ∂x)[u]) CDOT (replace e x (Var x + EPSILON u))
    + EPSILON(D (D t CDOT e) CDOT ((∂e // ∂x)[u]))
| s + t   => ((∂ s // ∂ x)[u]) + ((∂ t // ∂ x)[u])
| EPSILON t     => EPSILON ((∂t // ∂x)[u])
| LAM t     => LAM ((∂ t // ∂ (S x))[open 0 u])
end
where "( ∂ t // ∂ x ) [ u ]" := (dsubst t x u).

Lemma open_injective
  :  forall s t
  ,  forall m
  ,  open m s = open m t 
  -> s = t.
Proof with auto.
  intro s; induction s; intro t; destruct t; intros m eq; 
  simpl in eq; try inversion eq; try f_equal; eauto.
  - compare m n; compare m n0; auto; try omega.
Qed.

Lemma open_subst_ge
  :  forall t x y u
  ,  x ≥ y
  -> (open (S x) t) [y := open x u] = open x (t[y := u]).
Proof with (auto with arith).
  intro t; induction t; intros x y u le; simpl; [idtac|f_equal..];
  unfold subst in *|-*; simpl in *|-*;
  try (rewrite IHt1); try (rewrite IHt2); try rewrite (IHt)...
  - destruct n.
    + repeat trivial_arithmetic; auto;
      do 2 rewrite close_open_eq_S...
    + destruct n; repeat trivial_arithmetic; auto;
      do 2 rewrite close_open_eq_S...
  - f_equal; repeat rewrite (open_open_le _ 0 )...
Qed.

Tactic Notation "and" tactic(t) := [>  t|..].
Tactic Notation "and" "finally" tactic(t) := [> t.. ] .

Lemma subst_subst_le
  :  forall t x u y v
  ,  x ≤ y 
  -> (t [x := u])[y := v] 
     = ((t[S y := open x v])[x := u[y:=v]]).
Proof with (auto with arith).
  intro t; induction t; intros; 
  unfold subst in *|-*; simpl in *|-*; auto with diff;
  try rewrite IHt1; try rewrite IHt2; try rewrite IHt...
  - repeat trivial_arithmetic...
    all: do 2 rewrite close_open_eq_S; auto;
      rewrite replace_open_eq;
      rewrite close_open_eq_S...
  - f_equal. repeat rewrite (open_open_le _ O); auto with arith.
    rewrite (IHt (S x));(
    and do 3 f_equal;
    and (rewrite <- (open_open_le _ 0)); 
    and rewrite replace_open_le;
    and rewrite close_open_gt)...
Qed.

Lemma subst_subst_gt
  :  forall t x u y v
  ,  x > y
  -> (t [S x := u])[y := v] = ((t[y := open x v])[x := u[y := v]]).
Proof with auto with arith.
  intro t; induction t; intros; simpl; unfold subst in *|-*; simpl in *|-*;
  try rewrite IHt; try rewrite IHt1; try rewrite IHt2...
  - repeat trivial_arithmetic...
    all: do 2 rewrite close_open_eq_S...
      rewrite replace_open_eq; rewrite close_open_eq_S...
  - f_equal; repeat rewrite (open_open_le _ O); auto with arith; (
      rewrite (IHt (S x));
      and do 3 f_equal;
      and rewrite <- close_open_gt;
      and rewrite <- open_open_le;
      and rewrite replace_open_le)...
Qed.

Lemma replace_E : forall t x u, EPSILON (replace t x u) = replace (EPSILON t) x u. 
Proof. now simpl. Qed.
Lemma replace_0 : forall x u, 0 = replace 0 x u.
Proof. now simpl. Qed.
Lemma replace_add : forall s t x u, 
  replace s x u + replace t x u = replace (s + t) x u.
Proof. now simpl. Qed.
Lemma replace_app : forall s t x u,
  { replace s x u | replace t x u } = replace { s | t } x u.
Proof. now simpl. Qed.
Lemma replace_D : forall s t x u,
  D (replace s x u) CDOT (replace t x u) = replace (D s CDOT t) x u.
Proof. now simpl. Qed.

(* 
  A tactic for automatically folding calls to replace
  when simpl or unfold go too far.
*)
Ltac fold_replace :=
  repeat (progress (
      try rewrite replace_E;
      try rewrite replace_0;
      try rewrite replace_add;
      try rewrite replace_app;
      try rewrite replace_D
    )
  ).

Lemma open_E : forall x t, EPSILON (open x t) = open x (EPSILON t).
Proof. now simpl. Qed.
Lemma open_0 : forall x, 0 = open x 0.
Proof. now simpl. Qed.
Lemma open_add : forall x s t, (open x s) + (open x t) = open x (s + t).
Proof. now simpl. Qed.
Lemma open_app : forall x s t, ({open x s | open x t}) =  (open x {s | t}).
Proof. now simpl. Qed.
Lemma open_D : forall x s t, D (open x s) CDOT (open x t) = open x (D s CDOT t).
Proof. now simpl. Qed.

(* 
  A tactic for automatically folding calls to open
  when simpl or unfold go too far.
*)
Ltac fold_open :=
  progress (
    try rewrite open_E;
    try rewrite open_0;
    try rewrite open_add;
    try rewrite open_app;
    try rewrite open_D
  ).

(* Replacing a variable by itself is the identity *)
Lemma replace_trivial
  :  forall t x
  ,  t = replace t x (Var x).
Proof with (auto with diff).
  intro t; induction t; intros; simpl; f_equal...
  trivial_arithmetic...
Qed.

Lemma replace_update
  :  forall t x u v
  ,  replace (replace t x u) x v = replace t x (replace u x v).
Proof with auto with diff. intro t; induction t; intros; simpl; f_equal...
  - repeat trivial_arithmetic; auto; rewrite replace_open_eq...
  - rewrite IHt; rewrite replace_open_le; auto with arith.
Qed.
\end{lstlisting}

\begin{lstlisting}[language=Coq,label=coq:dsubst-empty]

(* 
  Differential substitution of a variable that does not appear free
  in a term is zero.
*)
Lemma dsubst_empty
  :  forall t x u
  ,  (∂ (open x t) // ∂ x)[u] ≈ 0.
Proof with (auto with diff). 
  intro t; induction t; intros; simpl;
  try local rewrite IHt; try local rewrite IHt1;
  repeat local rewrite IHt2; can...
  - repeat trivial_arithmetic...
Qed.

(*
  A handy tactic for introducing shorthands in proofs with
  lots of partial derivatives.
*)
Tactic Notation "differential" ident(name) ":=" 
  "d" constr(t) "/" "d" ident(x) "[" constr(u) "]"
  := pose (name := (∂ t // ∂ x)[open x u]); fold name.

(*
  A tactic for commuting the arguments to a differential
  application.
*)
Ltac drot i expr :=
  lazymatch expr with
  | D (D ?t CDOT ?u) CDOT ?v =>
    match eval compute in i with
    | O => local rewrite (Eq_D_comm t u v)
    | S ?i' => drot i' (D t CDOT u)
    end
  | _ => fail "Not a differential goal"
  end.

Tactic Notation "d" "rot" constr(pos) :=
  match goal with
  | |- ?w ≈ _ => drot pos w
  end.

(*
  Handling raw terms is slow because of how Coq handles pattern
  matching. The combinators below allow for splitting a term into
  a list of its summands and operating on that instead. Avoids a lot
  of canonicalisation headaches.
*)
Require Import List.

Definition sum (l : list Term) := fold_right (fun s t => s + t) 0 l.

Definition eqs (l1 l2 : list Term) := sum l1 ≈ sum l2.

Notation "l1 [≈] l2" := (eqs l1 l2) (at level 80).

Lemma eqs_trans : forall l1 l2 l3, l1 [≈] l2 -> l2 [≈] l3 -> l1 [≈] l3.
intros; eapply eq_trans; eauto.
Qed.
Lemma eqs_sym : forall l1 l2, l1 [≈] l2 -> l2 [≈] l1.
intros; apply eq_sym; auto.
Qed.

Lemma eqs_start : forall t t',  (t :: nil) [≈] (t' :: nil) -> t ≈ t'.
unfold eqs; unfold sum; intros; simpl in H; local rewrite <- Eq_add_0.
r local rewrite <- Eq_add_0. auto.
Qed.

Ltac local_rewrite_eqs eq prf :=
  lazymatch type of eq with
  | ?ll1 [≈] ?ll2 => 
    eapply eqs_trans with (l2 := ll2); [eapply prf|]
  | forall y, @?w y => (
      let eqname := fresh "eq" in
      epose (eq _) as eqname; simpl in eqname;
      let the_expr := (eval hnf in eqname) in
      local_rewrite_eqs the_expr prf;
      clear eqname
    )
  end.

Tactic Notation "[rewrite]" constr(eq) :=
  local_rewrite_eqs eq eq; simpl "++".

Tactic Notation "[l]" tactic(tac) :=
  tac.

Tactic Notation "[r]" tactic(tac) :=
  apply eqs_sym; tac; apply eqs_sym.

Lemma sum_diff : forall l a a', a ≈ a' ->
  fold_right (fun s t => s + t) a l ≈ fold_right (fun s t => s + t) a' l.
Proof with auto with diff. intro l; induction l; intros; unfold sum; simpl...
Qed.

Lemma sum_extra : forall l s t, 
  fold_right (fun s t => s + t) (s + t) l
  ≈ fold_right (fun s t => s + t) s l + t.
Proof with auto with diff. intro l; induction l; intros; unfold sum; simpl...
  - can; Cx...
Qed.

Lemma fold_append : forall (l1 : list Term) l2 f (x : Term),
  fold_right f x (l1 ++ l2) = fold_right f (fold_right f x l2) l1.
Proof with auto. intro l1; induction l1; intros; simpl...
  - rewrite IHl1...
Qed.

Lemma split_eqs : forall l1 l2, sum (l1 ++ l2) ≈ (sum l1) + (sum l2).
Proof with auto with diff.
  intro l1; induction l1; intros; unfold sum; simpl.
  - fold (sum l2); can; apply eq_refl.
  - fold (sum (l1 ++ l2)) (sum l1) (sum l2); can; Cx...
Qed.

Theorem rots : forall l t, t :: l [≈] l ++ (t :: nil).
Proof with auto with diff. 
  intros; unfold eqs; local rewrite (split_eqs l (t :: nil)); 
  unfold sum; simpl... can; local rewrite Eq_add_comm; Cx...
Qed.

Tactic Notation "[rot]" integer(i) := do i ([rewrite] rots).

Theorem splits : forall l t1 t2, (t1 + t2) :: l [≈] t1 :: t2 :: l.
Proof with auto with diff. intros; unfold eqs; unfold sum; simpl...
Qed.

Theorem eqs_diff : forall t t', t ≈ t' -> forall l, t :: l [≈] t' :: l.
intros; unfold eqs; unfold sum; simpl; local rewrite H; apply eq_refl.
Qed.

Tactic Notation "[*]" tactic(tac) := [rewrite] eqs_diff; 
  and tac; and eapply eq_refl.

Ltac splits :=
  match goal with
  | |- (?s + ?t) :: ?l [≈] _ => [rewrite] (splits l s t)
  end.

Tactic Notation "[can]" := [*] can.

Ltac stomp := [can]; 
  ([*] repeat local rewrite <- Eq_add_assoc);
  repeat splits.
Lemma eqs_refl : forall l, l [≈] l.
Proof with auto with diff. 
  induction l; unfold eqs; unfold sum; simpl... 
Qed.
Tactic Notation "[refl]" := apply eqs_refl.

Ltac listify := apply eqs_start.

Tactic Notation "[aligns]" :=
  timeout 1 repeat match goal with
  | |- ?t :: _ [≈] ?t :: _ => idtac "found" t
  | |- ?u :: _ [≈] ?v :: _ => [rot] 1
  end.

Lemma pop : forall t t' l l', t ≈ t' -> l [≈] l' -> (t :: l) [≈] (t' :: l').
intros; unfold eqs; unfold sum; simpl; Cx; auto using eq_refl.
Qed.
Ltac pop := apply pop.

Lemma replace_open_lt
  :  forall t x y u
  ,  x < y
  -> replace (open y t) x (open y u) = open y (replace t x u).
Proof with auto with diff.
  intro t; induction t; intros; simpl; [|repeat f_equal..]...
  - repeat trivial_arithmetic...
  - rewrite open_open_le; and rewrite IHt... all: auto with arith.
Qed.

Lemma replace_open_ge
  :  forall t x y u
  ,  x ≥ y
  -> replace (open y t) (S x) (open y u) = open y (replace t x u).
Proof with auto with diff.
  intro t; induction t; intros; simpl; [|repeat f_equal..]...
  - repeat trivial_arithmetic...
  - rewrite open_open_le; and rewrite IHt... all: auto with arith.
Qed.

Lemma dsubst_open_lt
  :  forall t x y u
  ,  x < y
  ->  dsubst (open y t) x (open y u) = open y (dsubst t x u).
Proof with auto with diff arith.
  intro t; induction t; intros; simpl...
  - repeat trivial_arithmetic...
  - f_equal; rewrite open_open_le; and rewrite IHt... 
  - f_equal.
    + rewrite IHt2...
    + rewrite IHt1... rewrite <- replace_open_lt; simpl...
      trivial_arithmetic...
  - f_equal; auto with diff; rewrite <- replace_open_lt; simpl;
    trivial_arithmetic...
  - f_equal; rewrite IHt...
  - f_equal; and f_equal.
    + rewrite IHt2...
    + rewrite IHt1... rewrite <- replace_open_lt; simpl...
      trivial_arithmetic...
    + rewrite IHt2... 
Qed.

Lemma dsubst_open_ge
  :  forall t x y u
  ,  x ≥ y
  -> dsubst (open y t) (S x) (open y u) = open y (dsubst t x u).
Proof with auto with diff arith.
  intro t; induction t; intros; simpl;
  try f_equal; try rewrite IHt; try rewrite IHt1; try rewrite IHt2...
  - repeat trivial_arithmetic...
  - rewrite open_open_le; and rewrite IHt...
  - rewrite <- replace_open_ge; simpl... trivial_arithmetic...
  - rewrite <- replace_open_ge; simpl... trivial_arithmetic...
Qed.
\end{lstlisting}

\newpage
\section{Module \texttt{Theorems.v}}

\begin{lstlisting}[language=Coq]
(* 
  Mechanically checked proofs of Taylor's theorem, regularity of
  differential substitution, and other "hard" results.
*)
Require Import Unicode.Utf8.
Require Import Coq.Relations.Relation_Operators.
Require Import Coq.Relations.Relation_Definitions.
Require Import Coq.Arith.EqNat.
Require Import Coq.Arith.Compare_dec.
Require Import Coq.Arith.PeanoNat.
Require Import Omega.

Add LoadPath "./".

Require Import Substitution.

\end{lstlisting}

\begin{lstlisting}[language=Coq, label=coq:taylor]
(* Syntactic Taylor's Theorem *)
Theorem Taylor
  :  forall s x t u
  ,  replace s x (t + EPSILON (open x u)) 
     ≈ replace (s + EPSILON ((∂ s // ∂ x)[ open x u ])) x t.
Proof with (auto with diff).
  intro s; induction s; intros; simpl...
  - trivial_arithmetic; simplify2...
    rewrite replace_open_eq...
  - rewrite open_open_le; and (local rewrite IHs; simpl; can)...
    auto with arith.
  - local rewrite IHs1; local rewrite IHs2; simpl; can... Cx...
    fold_replace; apply replace_diff_base; local rewrite IHs2.
    simpl; can; repeat rewrite <- replace_trivial; Cxx...
  - can...
  - local rewrite IHs1; local rewrite IHs2; simpl; can.
    Cx... rot 1; Cx...
  - local rewrite IHs; simpl...
  - local rewrite IHs1; local rewrite IHs2; simpl; can; Cx...
    Cx... r rot 1; Cx...
    local rewrite IHs2; simpl; can; repeat rewrite <- replace_trivial.
    Cxx...
Unshelve. apply replace_diff...
Qed.

Lemma dsubst_congruent_base
  :  forall t t'
  ,  t ≈ t'
  -> forall x u
  ,  (∂ t // ∂ x)[open x u] ≈ (∂ t' // ∂ x)[open x u].
Proof with auto with diff.
  intros t t' eq; induction eq.
  - induction H; try solve [intros; simpl; Cxx; auto with diff].
    + intros; simpl; Cxx; rewrite open_open_le; and apply IHCtx. 
      auto with arith.
    + induction H; intros; try solve [simpl; auto with diff];
      simpl; can; try solve [Cx ; [ auto with diff ..]].
      * repeat (align 5; Cx; try apply eq_refl).
      * repeat (align 5; Cx; try apply eq_refl). 
      * differential du := d u / d x [u0].
        differential dv := d v / d x [u0].
        differential ds := d s / d x [u0].
        listify; stomp; [r] stomp.
        repeat ([aligns]; pop; and apply eq_refl).
        ([r] [rot] 1); [aligns]. pop...
        ([r] [rot] 1); pop...
        apply eqs_sym. 
        [*] local rewrite Taylor; rewrite <- replace_trivial.
        [rot] 1; [*] local rewrite Taylor; rewrite <- replace_trivial. 
        stomp. fold du dv ds. [rot] 3; stomp.
        pop... [rot] 2; pop... [r] [rot] 1. pop...
        pop... Cx; d rot O... pop... pop; and Cx; d rot O; Cx...
        [refl].
      * differential dt := d t / d x [u0].
        differential ds := d s / d x [u0].
        differential du := d u / d x [u0].
        repeat (align 10; Cx; [solve [apply eq_refl]|]).
        r rot 6; Cx...
        rot 2; r rot 1; Cx; and do 2 Cx...
        repeat local rewrite Taylor; simpl; repeat rewrite <- replace_trivial;
        can. fold dt ds du.
        rot 2; Cx... Cx... Cx.
        do 3 Cx; and d rot O...
        do 4 Cx; and d rot O...
      * differential du := d u / d x [u0];
        differential dv := d v / d x [u0];
        differential ds := d s / d x [u0].
        rot 2; r rot 2; Cx...
        repeat local rewrite Taylor; simpl; repeat rewrite <- replace_trivial.
        fold du dv ds; can.
        r rot 3; Cx. Cx; d rot O...
        r rot 3; Cx; and Cx; and d rot O; and d rot 1...
        r rot 2; Cx; and Cx; and d rot 2; and d rot 1; and d rot O; and Cx;
        and d rot 1; and d rot O; and Cx...
        r rot 3; Cx; and (Cx; and d rot O; and Cx)...
        r rot 2; Cx...
        rot 1; Cx... Cx... Cx; d rot 1; d rot O...
  - intros...
  - intros...
  - intros... eapply eq_trans; [eapply IHeq1 | eapply IHeq2].
Unshelve. all: try apply eq_refl.
Qed.

Theorem dsubst_congruent_new
  :  forall s
  ,  forall t t'
  ,  t ≈ t'
  -> forall x
  ,  (∂s // ∂x)[t] ≈ (∂s // ∂x)[t'].
Proof with (auto with diff).
  intro s; induction s; intros; simpl; try Cxx...
  - destruct (Nat.compare x n)...
  - apply replace_diff...
  - apply replace_diff...
Qed.

\end{lstlisting}

\begin{lstlisting}[language=Coq,label=coq:dsubst-diff]
(* Differential substitution respects differential equivalence *)
Lemma dsubst_diff
  :  forall t t'
  ,  t ≈ t'
  -> forall x u u'
  ,  u ≈ u'
  -> (∂ t // ∂ x)[open x u] ≈ (∂ t' // ∂ x)[open x u'].
Proof with auto with diff.
  intros; eapply eq_trans.
  apply dsubst_congruent_base; and exact H.
  apply dsubst_congruent_new; and apply open_diff; and exact H0.
Qed.
\end{lstlisting}

\begin{lstlisting}[language=Coq,label=coq:dsubst-commute]
Lemma dsubst_commute
  :  forall t x u v
  ,  dsubst (dsubst t x (open x u)) x (open x v)
     ≈ dsubst (dsubst t x (open x v)) x (open x u).
Proof with auto with diff arith.
  intro t; induction t; intros; simpl...
  - repeat trivial_arithmetic... repeat local rewrite dsubst_empty...
  - Cx. repeat rewrite (open_open_le _ 0)...
  - can. repeat local rewrite Taylor. simpl. repeat rewrite <- replace_trivial.

    specialize (IHt1 x u v); specialize (IHt2 x u v).
    differential d2u := d t2 / d x [u].
    differential d2v := d t2 / d x [v].
    differential d1u := d t1 / d x [u].
    differential d1v := d t1 / d x [v].
    differential d1uv := d d1u / d x [v].
    differential d1vu := d d1v / d x [u].
    differential d2uv := d d2u / d x [v].
    differential d2vu := d d2v / d x [u].
    fold d1u d1v in IHt1; fold d1uv d1vu in IHt1.
    fold d2u d2v in IHt2; fold d2uv d2vu in IHt2.
    repeat local rewrite IHt1. repeat local rewrite IHt2. Cx...
    listify.
    stomp; [r] stomp.
    [aligns]; pop...
    [r] [rot] 1. [aligns]; pop...
    repeat ([aligns]; pop; [auto with diff|])...
    pop; and repeat local rewrite <- Eq_app_E; can...
    repeat ([aligns]; pop; [auto with diff|])...
    [r] [rot] 8. pop... Cx... Cx... repeat local rewrite Eq_add_assoc; Cx...
    [r] [rot] 1. pop... [r] [rot] 1.
    repeat (pop; [auto with diff|])...
    + repeat local rewrite <- Eq_app_E; repeat local rewrite Eq_D_inf; can...
    + do 3 Cx... d rot 1; d rot O...
    + [refl].
    Optimize Proof.
  - can. repeat local rewrite Taylor. simpl. repeat rewrite <- replace_trivial.
    specialize (IHt1 x u v); specialize (IHt2 x u v).
    differential d2u := d t2 / d x [u];
    differential d2v := d t2 / d x [v];
    differential d1u := d t1 / d x [u];
    differential d1v := d t1 / d x [v];
    differential d2uv := d d2u / d x [v];
    differential d2vu := d d2v / d x [u];
    differential d1uv := d d1u / d x [v];
    differential d1vu := d d1v / d x [u].
    fold d1u d1v in IHt1; fold d1uv d1vu in IHt1.
    fold d2u d2v in IHt2; fold d2uv d2vu in IHt2. 
    repeat local rewrite IHt1. repeat local rewrite IHt2.
    listify. Optimize Proof.
    [*] repeat local rewrite <- Eq_add_assoc. repeat splits.
    [r] ([*] repeat local rewrite <- Eq_add_assoc); repeat splits.
    pop... [rot] 2; pop... ([r] [rot] 2); pop...
    Cx... repeat local rewrite Eq_add_assoc. Cx... 
    repeat local rewrite Eq_add_E. repeat local rewrite <- Eq_add_assoc; Cx...
    repeat ([aligns]; pop; [auto with diff|]).
    stomp; [r] stomp. pop... [rot] 1; [r] [rot] 1. pop... 
    Cx; and d rot 1; and d rot O; Cx...
    [rot] 3; [r] [rot] 5. pop...
    [rot] 4; [r] [rot] 4.
    pop... pop... [r] [rot] 2. pop...
    stomp; [r] stomp.
    pop; and Cx; and d rot O...
    [rot] 2; pop... [rot] 1; pop... Cx; and d rot 1; and d rot O...
    pop... [refl].
Unshelve.
  all: try apply eq_refl.
  all: try apply dsubst_diff...
  all:
    apply replace_diff; auto with diff;
    Cx; [apply replace_diff|]; auto with diff;
    Cx; apply dsubst_diff...
Qed.
\end{lstlisting}

\begin{lstlisting}[language=Coq,label=coq:regularity]
(* Regularity of differential substitution *)
Theorem Regularity
  :  forall s x u v
  ,  (∂ s // ∂ x)[ open x (u + v) ] ≈
      ((∂ s // ∂ x)[ open x u ] 
      + ((∂ s // ∂ x)[ open x v ] 
      + EPSILON ((∂ ((∂ s // ∂ x)[ open x v ]) // ∂ x)[ open x u ]))).
Proof with (auto with diff).
  intro s; induction s; intros; simpl.
  - compare x n; can; Cx...
    local rewrite dsubst_empty; can...
  - repeat rewrite (open_open_le _ 0); try auto with arith.
    fold_open;
    local rewrite IHs.
    local rewrite <- Eq_lam_E.
    repeat local rewrite <- Eq_lam_add...
  - fold_open; repeat local rewrite IHs1; repeat local rewrite IHs2;
    repeat local rewrite Taylor; simpl; repeat rewrite <- replace_trivial;
    repeat local rewrite Taylor; simpl; repeat rewrite <- replace_trivial.
    fold_open; local rewrite IHs2.
    differential d2u := d s2 / d x [u].
    differential d2v := d s2 / d x [v].
    differential d1u := d s1 / d x [u].
    differential d1v := d s1 / d x [v].
    differential d2vu := d d2v / d x [u].
    differential d1vu := d d1v / d x [u].
    listify; stomp;
    [r] stomp.
    pop... [r] [aligns]. pop...
    [r] [aligns]. pop...
    [aligns]. pop... [r] [rot] 7. pop; and do 4 Cx...
    [rot] 3. [r] [rot] 25. pop; and do 3 Cx...
    [r] [rot] 18. pop...
    [*] repeat local rewrite Eq_derivative. stomp.
    repeat (([r] [aligns]); pop; [auto with diff|]).
    [*] repeat local rewrite Eq_derivative. stomp.
    repeat (([r] [aligns]); pop; [auto with diff|]).
    pop; and do 3 Cx...
    pop; and do 3 Cx...
    [rot] 1; pop...
    ([r] [rot] 1); pop. 
    repeat local rewrite <- Eq_app_E;
    repeat local rewrite Eq_D_inf; Cx...
    [rot] 1. [r] [rot] 8. pop.
    repeat local rewrite <- Eq_app_E;
    repeat local rewrite Eq_D_inf; Cx...
    [*] repeat local rewrite Eq_derivative. stomp.
    repeat ([aligns]; pop; [auto with diff|]).
    [refl].
  - can... 
  - fold_open; repeat local rewrite IHs1; repeat local rewrite IHs2;
    repeat local rewrite Taylor; simpl; repeat rewrite <- replace_trivial.
    differential d1u := d s1 / d x [u].
    differential d1v := d s1 / d x [v].
    differential d2u := d s2 / d x [u].
    differential d2v := d s2 / d x [v].
    can; repeat (align 10; Cx; [auto with diff|])...
  - fold_open; repeat local rewrite IHs; can; Cxx...
  - fold_open. repeat local rewrite IHs1; repeat local rewrite IHs2;
    repeat local rewrite Taylor.
    differential d1u := d s1 / d x [u].
    differential d1v := d s1 / d x [v].
    differential d2u := d s2 / d x [u].
    differential d2v := d s2 / d x [v].
    differential d1uv := d d1u / d x [v].
    differential d1vu := d d1v / d x [u].
    differential d2uv := d d2u / d x [v].
    differential d2vu := d d2v / d x [u].
    repeat rewrite <- replace_trivial.
    simpl; repeat local rewrite Taylor.
    fold_open; local rewrite IHs2.
    fold d1u d1v d2u d2v. fold d1uv d1vu d2uv d2vu.
    repeat rewrite <- replace_trivial.
    listify. repeat splits.
    [r] ([*] repeat local rewrite <- Eq_add_assoc); repeat splits.
    [*] can; repeat local rewrite <- Eq_add_assoc.
    repeat splits.
    repeat ([aligns];pop;[auto with diff|]).
    repeat ([r][aligns]; pop; [auto with diff|]).
    [rot] 6. stomp.
    [r] [rot] 2. [r] ([*] do 10 simplify; repeat local rewrite <- Eq_add_assoc).
    [r] repeat splits.
    repeat ([aligns];pop;[auto with diff|]).
    [r] [rot] 3. pop...
    [rot] 4; [r] [aligns]. pop...
    [r] [aligns]. pop...
    [rot] 8. [*] do 10 simplify; repeat local rewrite <- Eq_add_assoc.
    repeat splits.
    repeat ([aligns]; pop;[auto with diff|]).
    pop. repeat local rewrite Eq_D_inf...
    [rot] 8; pop...
    Optimize Proof.
    [rot] 1; pop... can...
    [rot] 1; stomp. [rot] 1; pop... 
    [r] [rot] 2; [*] repeat local rewrite Eq_add_E. [r] repeat splits.
    [r] [rot] 6. [r] repeat splits.
    [rot] 4. [r] [rot] 4. pop; and can...
    [r] [rot] 10. [rot] 22. pop...
    [rot] 3. [r] stomp. pop... pop... pop... Cx; d rot O...
    [r] [rot] 2; stomp. pop... pop...
    [rot] 7. [r] [rot] 1; stomp. pop... pop...
    [r] [rot] 4. [rot] 1; pop; and can...
    [r] [rot] 1; stomp. [rot] 1; stomp.
    [rot] 2; pop...
    [r] [rot] 1. pop; and can; and Cx; and d rot 1; and d rot O...
    [rot] 10. [r] [rot] 3. pop; and can...
    [rot] 6; pop...
    [rot] 15; ([r] [rot] 10); pop; and can...
    pop; and can...
    [rot] 12; pop; and can...
    [r] [rot] 2; stomp.
    [rot] 4; pop...
    pop...
    pop; and Cx; and d rot O...
    [r] [rot] 3; stomp.
    [*] can. [r] [rot] 1. pop...
    [*] can. [r] [rot] 2. pop...
    [r] [rot] 2. pop; and can; and Cx; and d rot O...
    [rot] 2; stomp. [rot] 2; pop; and Cx; d rot 1; d rot O; Cx; and r d rot O...
    [r] [rot] 2; stomp.
    [rot] 1; pop; and Cx; and d rot O...
    ([r] [rot] 1); pop; and Cx; and d rot O; and Cx; and r d rot O...
    [rot] 1; stomp. pop... [rot] 2; pop; and Cx; and r d rot O...
    [r] [rot] 1. pop; and Cx; and r d rot O...
    [r] [rot] 3. pop...
    [rot] 1; pop...
    [*] can. pop; and Cx; and d rot O...
    [*] do 10 simplify; repeat local rewrite <- Eq_add_assoc. repeat splits.
    [r] ([*] do 10 simplify; repeat local rewrite <- Eq_add_assoc); repeat
    splits.
    pop; and can...
    repeat (pop; [solve [auto with diff]|]).
    pop; and can...
    Optimize Proof.
    stomp. [r] stomp. repeat (pop; [solve [auto with diff]|]).
    stomp. repeat (pop; [solve [auto with diff]|]).
    [refl].
Unshelve. 
  (* There is no reason this should have to be done manually*)
  all: try apply eq_refl.
  all: try apply dsubst_diff...
Qed.
\end{lstlisting}

\begin{lstlisting}[language=Coq,label=coq:replace-dsubst]
Lemma replace_dsubst
  :  forall t x e y v
  ,  x ≠ y
  -> replace (dsubst t x e) y (open x v)
     ≈ dsubst (replace t y (open x v)) x (replace e y (open x v)).
Proof with auto with arith diff.
  intro t; induction t; intros x e y v le; simpl in *|-*;
  try rewrite IHt1; try rewrite IHt2; try rewrite IHt...
  - repeat trivial_arithmetic...
    all: local rewrite dsubst_empty...
  - Cx; repeat rewrite (open_open_le _ O)...
    local rewrite IHt;
    rewrite <- replace_open_le...
    repeat rewrite (open_open_le _ O)...
  - Cxx...
    + rewrite replace_replace; simpl...
      trivial_arithmetic...
  - Cxx...
    + rewrite replace_replace; simpl...
      trivial_arithmetic...
Unshelve.
  auto with arith.
Qed.
\end{lstlisting}